\documentclass[PhD]{ucfthes}

\newcommand\includeFirst{y}
\newcommand\includeSecond{y}
\newcommand\includegTEA{y}
\newcommand\includeBART{y}

\usepackage{amsmath,amssymb,amsfonts} 	
\usepackage[dvips]{graphics}			
\usepackage{epsfig}
\usepackage{astjnlabbrev-jh}
\usepackage{deluxetable}
\usepackage[colon,sectionbib]{natbib}
\usepackage{chapterbib}
\usepackage{bibentry}
\nobibliography{Blecic}
\usepackage{longtable}
\usepackage{pdflscape}
\usepackage{setspace}
\usepackage{indentfirst}
\usepackage{subfig}
\usepackage{fltpage}

\newcommand\bf{\bfseries}

\newcommand\degr{$^{\circ}$}
\newcommand\degree{\degr}
\newcommand\degrees\degree

\DeclareSymbolFont{UPM}{U}{eur}{m}{n}
\DeclareMathSymbol{\umu}{0}{UPM}{"16}
\let\oldumu=\umu
\renewcommand\umu{\ifmmode\oldumu\else\math{\oldumu}\fi}
\newcommand\micro{\umu}
\newcommand\micron{\micro m}
\newcommand\microns \micron

\let\oldsim=\sim
\renewcommand\sim{\ifmmode\oldsim\else\math{\oldsim}\fi}
\let\oldpm=\pm
\renewcommand\pm{\ifmmode\oldpm\else\math{\oldpm}\fi}
\newcommand\by{\ifmmode\times\else\math{\times}\fi}
\newcommand\ttt[1]{10\sp{#1}}
\newcommand\tttt[1]{\by\ttt{#1}}

\newbox{\wdbox}
\renewcommand\c{\setbox\wdbox=\hbox{,}\hspace{\wd\wdbox}}
\renewcommand\i{\setbox\wdbox=\hbox{i}\hspace{\wd\wdbox}}

\newcount\timect
\newcount\hourct
\newcount\minct
\newcommand\now{\timect=\time \divide\timect by 60
         \hourct=\timect \multiply\hourct by 60
         \minct=\time \advance\minct by -\hourct
         \number\timect:\ifnum \minct < 10 0\fi\number\minct}

\newcommand\mctc{\multicolumn{2}{c}}

\catcode`@=11

\newcommand\comment[1]{}
\newcommand\commenton{\catcode`\%=14}
\newcommand\commentoff{\catcode`\%=12}

\renewcommand\math[1]{$#1$}
\newcommand\mathshifton{\catcode`\$=3}
\newcommand\mathshiftoff{\catcode`\$=12}

\comment{the backslash is necessary}

\comment{alignment tab}

\let\atab=&
\newcommand\atabon{\catcode`\&=4}
\newcommand\ataboff{\catcode`\&=12}

\let\oldmsp=\sp
\let\oldmsb=\sb
\def\sp#1{\ifmmode
           \oldmsp{#1}%
         \else\strut\raise.85ex\hbox{\scriptsize #1}\fi}
\def\sb#1{\ifmmode
           \oldmsb{#1}%
         \else\strut\raise-.54ex\hbox{\scriptsize #1}\fi}
\newbox\@sp
\newbox\@sb
\def\sbp#1#2{\ifmmode%
           \oldmsb{#1}\oldmsp{#2}%
         \else
           \setbox\@sb=\hbox{\sb{#1}}%
           \setbox\@sp=\hbox{\sp{#2}}%
           \rlap{\copy\@sb}\copy\@sp
           \ifdim \wd\@sb >\wd\@sp
             \hskip -\wd\@sp \hskip \wd\@sb
           \fi
        \fi}
\def\msp#1{\ifmmode
           \oldmsp{#1}
         \else \math{\oldmsp{#1}}\fi}
\def\msb#1{\ifmmode
           \oldmsb{#1}
         \else \math{\oldmsb{#1}}\fi}
\def\supon{\catcode`\^=7}
\def\supoff{\catcode`\^=12}
\def\subon{\catcode`\_=8}
\def\suboff{\catcode`\_=12}
\def\supsubon{\supon \subon}
\def\supsuboff{\supoff \suboff}

\DeclareOldFontCommand{\rm}{\normalfont\rmfamily}{\mathrm}
\DeclareOldFontCommand{\tt}{\normalfont\ttfamily}{\mathtt}
\DeclareOldFontCommand{\bf}{\normalfont\bfseries}{\mathbf}
\DeclareOldFontCommand{\it}{\normalfont\itshape}{\mathit}

\newcommand\actcharon{\catcode`\~=13}
\newcommand\actcharoff{\catcode`\~=12}

\newcommand\paramon{\catcode`\#=6}
\newcommand\paramoff{\catcode`\#=12}

\comment{And now to turn us totally on and off...}

\newcommand\reservedcharson{\commenton \mathshifton \atabon \supsubon \actcharon
	\paramon}

\newcommand\reservedcharsoff{\commentoff \mathshiftoff \ataboff
	\supsuboff \actcharoff \paramoff}

\catcode`@=12
\reservedcharsoff

\reservedcharson

\comment{ Must have ONLY ONE of these... trust these macros, they work

}

\newcommand{\squishlist}{
 \begin{list}{$\bullet$}
  { \setlength{\itemsep}{1pt}
     \setlength{\parsep}{0pt}
     \setlength{\topsep}{3pt}
     \setlength{\partopsep}{0pt}
     \setlength{\leftmargin}{2.0em}
     \setlength{\labelwidth}{1.5em}
     \setlength{\labelsep}{0.5em} } }

\newcommand{\squishend}{
  \end{list}  }

\reservedcharsoff

\actcharon

\reservedcharson
\actcharon

\renewcommand{\bibname}{\section{LIST OF REFERENCES}}

\usepackage[dvipdfm, breaklinks=true]{hyperref}

\hypersetup{
 pdfauthor = {JASMINA BLECIC},
 pdftitle = {OBSERVATIONS, THERMOCHEMICAL CALCULATIONS, AND MODELING OF EXOPLANETARY ATMOSPHERES},
 pdfsubject = {Phd Dissertation, Fall 2015, Univ. of Central Florida},
 pdfkeywords = {exoplanets, photometry, thermochemical equilibrium, radiative transfer, WASP-14b, WASP-43b, TEA, BART},
 pdfcreator = {LaTeX with hyperref package},
 pdfdisplaydoctitle = true,
 bookmarksnumbered = true,
 pdfborder = {0 0 0},
 }

\title {OBSERVATIONS, THERMOCHEMICAL CALCULATIONS, \\ 
         AND MODELING OF EXOPLANETARY ATMOSPHERES}

\author         {JASMINA BLECIC}

\prevdegreei 	{B.Sc. University of Belgrade, Serbia, 2007}
\prevdegreeii 	{M.Sc. University of Belgrade, Serbia, 2007}

\department	{Physics, Planetary Sciences Track}
\college	{Sciences}
\term		{Fall}

\degreeyear     {2015}

\professor	{Joseph Harrington}

\chair          {Joseph Harrington}
\member         {Daniel Britt}
\member         {Robert Peale}
\member         {Jonathan J. Fortney}

\dedication {\textsl{To my dearest family and friends, and those who have believed in me.}}

\acknowledgments 
{
This was quite a journey, exciting and satisfying on many levels, but not without obstacles. I knew the minute I entered grad school that life would change, but still I dared to have a child. Faced with these challenges, I have learned a great deal about myself and discovered potential I never thought I had. Still, this was not a one-(wo)man success. First, I have to thank my advisor, Dr.\ Joseph Harrington, who has acted as a mentor and a friend. He taught me how to think like a scientist and how to write concisely and to the point. He supported me from the very beginning, helping me shape my ideas into beautiful projects, which opened up bright prospects for my future. I also want to thank Dr.\ Yan Fernandez for his kindness and thoughts that often soothed me. I would like to thank Dr.\ Daniel Britt, Dr.\ Ahlam Al-Rawi, Dr.\ Joseph Harrington, and Dr.\ Yan Fernandez for the exquisite recommendation letters. As for my fellow students, first I have to thank my closest colleague and friend Patricio Cubillos, without whom our BART project would never have succeed. We spent many hours discussing science in-person and over email, from which I have learned immensely. I would also like to thank Oliver Bowman, a great programmer and my teammate on the TEA project; Kevin Stevenson who unknowingly drove me to make the right decisions in my research; Carthik Sharma for his support and kindness in the first two years of my Ph.D.; A. J. Foster for his help in the last two hectic months; the exoplanet research group for proofreading; and Erika Rasso for an exquisite editing job. I also want to thank my colleagues from Belgrade, Darko and Damlja, who helped me decide to go to grad school. On a personal level, I must thank my dearest friend Vladimir, without whom I would not have had the courage to pursue this field. Throughout my doctoral studies, he was beside me to share his vast knowledge, give advice, show patience, and offer a warm and calm word. His goodness, selflessness, and wisdom inspired me so many times. I want to express my deepest appreciation to my mom and dad, and my two brothers, who taught me to be responsible, hard-working, and forgiving, and embedded the strongest human principles in my character. Above all, I want to thank my husband, Djordje, who took it upon himself to work two jobs, take care of our boy and me, help me proofread, design my presentations, draw figures, and often forgive my tiredness and absence. His love and tenderness fill my heart. Foremost, I want to thank my love and my life, my baby boy Filip, who has shown an unbelievable reason and immense love in the very beginning of his life, allowing his mom to fight her battles with more confidence and support.
}
\notes {NOTES ABOUT CONTRIBUTIONS FROM COAUTHORS}
{
The work presented in this dissertation consists of published, submitted, and  journal articles in preparation, each resulting from a collaborative effort of the coauthors listed in each paper. The paragraphs below describe specific contributions made in each paper using the initials of each author.

In Chapter \ref{chap:WASP14b}, J.B wrote the paper with contributions from J.H., N.M., K.B.S., R.A.H. and P.F.L.M; N.M. produced the atmospheric models; P.F.L.M analyzed the star activity; J.B. reduced the data; J.B., K.B.S, and J.H. analyzed the results; R.A.H. produced the orbital parameter results; D.R.A. provided the system parameters; M.H. produced the IRSA tables and light curves; T.J.L. provided statistical advice; J.H., K.B.S., P.E.C, and C.C. wrote the analysis pipeline; and S.N., R.A.H., W.C.B. contributed code to the POET pipeline.

In Chapter \ref{chap:WASP43b}, J.B. wrote the paper with contributions from J.H., N.M., and R.A.H.; N.M. produced the atmospheric models; J.B. reduced the data; R.A.H. produced the orbital parameter results; M.H. produced the IRSA tables and light curves; K.B.S., P.C., J.H., M.O.B., S.N., and M.H. contributed code to the POET pipeline; and D.R.A., C.H., A.M.S.S., and A.C.C. discovered the planet.

\newpage
In Chapter \ref{chap:TEA}, J.B. wrote the paper with contributions from J.H. and M.O.B; J.B. performed the analysis; J.B. and M.O.B. wrote the TEA code.

In Chapter \ref{chap:BART}, J.B. wrote the paper with contributions from P.E.C.; J.H. designed BART architecture; J.B. performed the analysis; J.B. and M.O.B. wrote the TEA code; J.B wrote the best-fit and contribution function modules; J.B. and P.E.C. wrote the initialization routines; P.R. wrote the original {\em Transit} code; J.B. wrote the eclipse module in {\em Transit}; P.C. implemented the opacity-grid calculation, opacity-grid interpolation, Voigt-function pre-calculation, spectrum integrator;  P.E.C. wrote {\em MCcubed} module with contributions from N.B.L and M.S; J.B., P.E.C., and M.S. wrote the atmospheric generator; M.S. and P.E.C. wrote MPI modules with contributions from A.S.D.F;  P.C., J.B., and R.C.C wrote the Transit module documentation with contribution from A.J.F.; P.C and S.D.B. implemented the VO partition-function code; P.C and D.B. implemented the HITRAN reader; T.J.L. provided advise about statistical factors.
}

\abstract{

This dissertation as a whole aims to provide the means to better understand hot-Jupiter planets through observing, performing thermochemical calculations, and modeling their atmospheres. We used {\em Spitzer} multi-wavelength secondary-eclipse observations to characterize planetary atmospheres. We chose targets with high signal-to-noise ratios, as their deep eclipses allow us to detect signatures of spectral features and assess planetary atmospheric structure and composition with greater certainty. 

Chapter \ref{chap:intro} gives a short introduction. Chapter \ref{chap:WASP14b} presents the {\em Spitzer} secondary-eclipse analysis and atmospheric characterization of WASP-14b. The decrease in flux when a planet passes behind its host star reveals the planet dayside thermal emission, which, in turn, tells us about the atmospheric temperature and pressure profiles and molecular abundances. WASP-14b is a highly irradiated, transiting hot Jupiter. By applying a Bayesian approach in the atmospheric analysis, we found an absence of thermal inversion contrary to theoretical predictions. 

Chapter \ref{chap:WASP43b} describes the infrared observations of WASP-43b's {\em Spitzer} secondary eclipses, data analysis, and atmospheric characterization. WASP-43b is one of the closest-orbiting hot Jupiters, orbiting one of the coolest stars with a hot Jupiter. This configuration provided one of the strongest signal-to-noise ratios. The eclipse timings improved the estimate of the orbital period compared to previous analyses. The atmospheric analysis ruled out a strong thermal inversion in the dayside atmosphere of WASP-43b and put a nominal upper limit on the day-night energy redistribution.

Chapter \ref{chap:TEA} presents an open-source Thermochemical Equilibrium Abundances (TEA) code and its application to several hot-Jupiter temperature and pressure models. TEA calculates the abundances of gaseous molecular species using the Gibbs free-energy minimization method within an iterative Lagrangian optimization scheme. The thermochemical equilibrium abundances obtained with TEA can be used to initialize atmospheric models of any planetary atmosphere. The code is written in Python, in a modular fashion, and it is available to the community via {\em http://github.com/dzesmin/TEA}. 

Chapter \ref{chap:BART} presents my contributions to an open-source Bayesian Atmospheric Radiative Transfer (BART) code, and its application to WASP-43b. BART characterizes planetary atmospheres based on the observed spectroscopic information. It initializes a planetary atmospheric model, performs radiative-transfer calculations to produce models of planetary spectra, and using a statistical module compares models with observations. We describe the implementation of the initialization routines, the atmospheric profile generator, the eclipse module, the best-fit routines, and the contribution function module. We also present a comprehensive atmospheric analysis of all WASP-43b secondary-eclipse data obtained from the space- and ground-based observations using BART.

}

\begin {document}

\makeintropages

\chapter{INTRODUCTION}
\label{chap:intro}

The exoplanetary field began its endeavor in 1992 with the first discovery of planets outside of our Solar system \citep{wolszczan1992planetary}. Soon after, in 1995, the detection of an extrasolar planet around a Sun-like star \citep{MayorQueloz1995Nature-fisrtExoplanet-mainSequence} led to the discovery that planets similar to the ones in our solar neighborhood are common \citep{FressinEtal2013ApJ-frequency}. In 2002, the first discovery of an exoplanetary atmosphere \citep{CharbonneauEtal2002apjnadet}, gave us hope that we have come closer to answering the ultimate question: ``Is there life out there?" 

Since then, the exoplanetary field has become one of the most rapidly developing fields in astronomy. Today, there are around 1500 confirmed planets and more than 3500 candidates (August 2015, \href{www.exoplanets.org}{www.exoplanets.org}). The response of the scientific community to such a large number of exoplanets is the number of different techniques employed to analyze their data \citep[e.g., ][]{MoralesEtal2006-intraPixel,  DemingEtal2006-ramp, Knutson2007,  PontEtal2008-HST, PontEtal2009-HST, SwainEtal2009-HD209458b, SwainEtal2009-HD189733b, Desert2009-intraPixel,KnutsonEtal2009ApJ-Tres4Inversion, CarterWinn2010, StevensonEtal2012-BLISS, deWitEtal2012aapFacemap, DemingEtal2015-PixelDecorelation}. In addition, serious efforts have been made to characterize their atmospheres \citep[e.g., ][]{CharbonneauEtal2005apjTrES1, DemingEtal2005natHD209458b, HarringtonEtal2007natHD149026b, Knutson08, Fortney2008, Tinetti2007Nature, TinettiEtal2010-XO-1b, BurrowsEtal2006, Anderson2011-Ch24-WASP17b, ShowmanEtal2009ApJ-Circulation, MadhusudhanSeager2009ApJ-AbundanceMethod, LeeEtal2012-CF, BennekeSeager2012-Retrieval, LineEtal2014-Retrieval-I}.  The analyses revealed that the diversity of extrasolar planets is astonishing, and surprisingly, some of them are very different from the ones in our Solar system.

In the past decade, exoplanetary research was mostly centered on hot Jupiters, gaseous giant planets. These planets, located very close to their host stars, are usually found on circular orbits, expected to be facing their stars with the same side. Although hot Jupiters are in fact less frequent than hot Neptunes and super Earths \citep{HowardEtal2012-panetOccurance}, current technological limitations and instrumental systematics have directed us to use them to teach ourselves how to analyze exoplanets until more advanced instruments become available \citep{SeagerDeming2010, hansen2014features, Burrows2014-review}. Due to their large radii and small semimajor axes, these objects have provided the strongest signal-to-noise ratios and become the best targets for ground- and space-based observations, allowing us to get the first insights into exoplanetary atmospheres \citep{FortneyEtal2005apjlhjmodels, HarringtonEtal2007natHD149026b, Knutson08, SwainEtal2008NatureHD189733bspec, MadhusudhanSeager2009ApJ-AbundanceMethod, StevensonEtal2010Natur, AndersonEtal2010A&A-WASP19bHband, LineEtal2013-Retrieval-II, benneke2015strict}. 

To learn about an exoplanetary atmosphere, we need to study the planetary spectrum, a  thermal emission caused by the planet's intrinsic temperature, shaped as a Planck function (a black-body curve), which carries spectral features of molecular species present in its atmosphere. Each molecule has its unique signature (a set of absorption and emission lines) connected to specific wavelength locations. Thus, by distinguishing the spectral features in the planetary spectrum, we can say which molecules are present in the planetary atmosphere. In addition, the depth and the width of each feature tells us the abundance of a certain molecule and the local temperature in the atmosphere. 

The most fruitful technique for observing and studying planetary atmospheres, thus far, is the transit method. As seen from Earth, transiting planets pass in front of and/or behind their stars, revealing the most fundamental parameters of the planet-star system and providing a view into the planetary atmosphere. Being very far away from us (the closest confirmed exoplanet is \sim4 light years away, \citealp{Dumusque2012Nature-closestExoplanet}), the majority of planet-star systems appear as point sources in most instruments. The transit method allows us to separate the planet from its star by observing the planet-star system in and out of a transit event. During transit, the planet passes in front of its star and obstructs some of its light. Comparing the flux before and during the transit, we see a small dip when the planet is in transit. During transit, the stellar light passes through the terminator, the transition zone between the day and the night side. Superimposed on the stellar spectrum, we see the spectral features of the molecules present in the upper part of the planetary atmosphere. During secondary eclipse, the planet is passing behind its star, looking at the star with its dayside. The dip is observed when the planet is hidden by the star. Comparing the flux in and out of eclipse, we obtain information about the planetary dayside thermal emission.

Planetary spectra can be observed using either spectroscopy or photometry. In spectroscopy, we measure many spectral channels at ones. In photometry, we measure the flux only in certain wavelengths. However, we choose wavelength regions where we know that certain molecular species have spectral features. Thus, combining photometry measurements in multiple broad wavelength regions (bandpasses), we get useful information about the species present in the planetary atmosphere. The majority of the most abundant molecular species have spectral features in the infrared. 
 
Two techniques have been proposed to date to characterize exoplanetary atmospheres. One employs the direct method, where the theoretical spectra are generated based on several physical parameters and comparisons with the observations are done by eye \citep[e.g., ][]{FortneyEtal2005apjlhjmodels}. The other is called the inverse approach, where the properties of the planetary atmosphere are determined based on the available observations. This approach uses a statistical exploration algorithm that seeks the best combination of physical parameters that matches the data \citep{MadhusudhanSeager2010, BennekeSeager2012-Retrieval, LineEtal2014-Retrieval-I, Waldmann2015-TAU}. Both approaches provide information about the atmospheric temperature and pressure profile, chemical composition, and energy redistribution. However, the inverse approach allows for better exploration of plausible ranges for physical parameters and a more accurate match to the data.

Interested in characterizing planetary atmospheres, for my dissertation I have focused on transiting hot Jupiters.  I chose to observe hot Jupiters during secondary eclipse, as this geometric configuration provides the best way to study the temperature and pressure structures and chemical compositions of planetary dayside atmospheres.

I have used the {\em Spitzer Space Telescope} to observe and analyze their atmospheres, employing broadband photometry in multiple bandpasses. Although not designed for exoplanetary observations, {\em Spitzer} enabled some revolutionary achievements in the exoplanetary field. It was the first telescope that detected the emission from a planetary atmosphere \citep{CharbonneauEtal2005apjTrES1, DemingEtal2005natHD209458b} and the first that found a carbon-rich planet \citep{MadhusudhanEtal2011natWASP12batm}. A large number of characterized exoplanetary atmospheres followed \citep[e.g., ][]{TodorovEtal2010ApJ-HAT-P-1b, ChristiansenEtal2010ApJ-HAT-P-7b, O'DonovanEtal2010ApJ-Tres-2, MachalekEtal2010-XO-3b, DemingEtal2011ApJ-CoRoT-1-2, KnutsonEtal2011ApJ-GJ436b, CrossfieldEtal2012ApJ-HD209458b, StevensonEtal2010Natur, CubillosEtal2013apjWASP8b, BlecicEtal2013ApJ-WASP-14b, BlecicEtal2014ApJ-WASP-43b}. Trained to target and observe exoplanets with {\em Spitzer}, among dozens of planets, I have selected WASP-14b and WASP-43b as the main objects of my study. Both planets, having unusually high signal-to-noise ratios, provided excellent opportunities to study their gaseous envelopes in more detail. 

In the beginning of my Ph.D. studies, our research group did not have an atmospheric modeling tool. Thus, we had to call on an external theorist to help us characterize planetary atmospheres. In 2011, I proposed to my committee to develop modeling methods that would allow us to do the atmospheric characterization within our research group. At the time, only one theorist employed the aforementioned inverse technique to infer exoplanetary thermal structures and chemical compositions \citep{MadhusudhanSeager2009ApJ-AbundanceMethod, MadhusudhanSeager2010}. Wishing to provide others with the tools to do the atmospheric analysis themselves, our research group decided to make an open-source code that incorporates thermochemical, radiative, and statistical algorithms within the inverse approach. This tool allows us to calculate thermochemical states in planetary atmospheres, and put constraints on the atmospheric chemical composition, temperature profile, energy redistribution, equilibrium and non-equilibrium processes, and the presence or absence of thermal inversion using either transit or eclipse geometry. We call this project BART, Bayesian Atmospheric Radiative Transfer. To develop the thermochemical algorithm, (TEA, Thermochemical Equilibrium Abundances) and parts of BART, I received a NASA Earth and Space Science Fellowship in 2012 and continued to be funded through the end of my Ph.D. studies.  

In this dissertation, through observations, thermochemical calculations, and atmospheric modeling, we have gained new insights into the equilibrium chemistry, thermal structure, and composition of several hot Jupiters. During this process, chemical and modeling tools have been developed that our research group is sharing with the community. 

Chapter \ref{chap:WASP14b} presents the {\em Spitzer} observations, secondary eclipse analysis, and atmospheric characterization of WASP-14b. Chapter \ref{chap:WASP43b} describes the secondary eclipse observations and atmospheric analysis of WASP-43b. Chapter \ref{chap:TEA} presents the open-source Thermochemical Equilibrium Abundances (TEA) code that calculates equilibrium abundances of gaseous molecular species and its application to several hot Jupiters, and Chapter \ref{chap:BART} describes my contributions to the open-source Bayesian Atmospheric Radiative Transfer (BART) code and its application to WASP-43b.

The legacy of this dissertation are the two {\em Spitzer} secondary eclipse analyses of WASP-14b and WASP-43b, a comprehensive atmospheric analysis of WASP-43b, and two open-source codes, TEA and BART, that we hope will provide useful contributions to the field and grow as the science progresses.

\newpage
\bibliographystyle{apj}
\bibliography{chap-intro}

\chapter{THERMAL EMISSION OF WASP-14b REVEALED WITH THREE SPITZER ECLIPSES}
\label{chap:WASP14b}

{\singlespacing
\noindent{\bf Jasmina Blecic\sp{1}, Joseph Harrington\sp{1, 2}, Nikku Madhusudhan\sp{3}, Kevin B.\ Stevenson\sp{1}, Ryan A.\ Hardy\sp{1}, Patricio E. Cubillos\sp{1}, Matthew Hardin\sp{1}, Christopher J.\ Campo\sp{1},  William C.\ Bowman\sp{1}, Sarah Nymeyer\sp{1}, Tomas J.\ Loredo\sp{4}, David R.\ Anderson\sp{5},  and Pierre F.\ L.\ Maxted\sp{5}
}

\vspace{1cm}

\noindent{\em
\sp{1} Planetary Sciences Group, Department of Physics, University of Central Florida, Orlando, FL 32816-2385, USA\\
\sp{2} Max-Planck-Institut f\"{u}r Astronomie, D-69117 Heidelberg, Germany\\
\sp{3} Department of Physics and Department of Astronomy, Yale University, New Haven, CT 06511, USA\\
\sp{4} Center for Radiophysics and Space Research, Space Sciences Building, Cornell University Ithaca, NY 14853-6801, USA\\
\sp{5} Astrophysics Group, Keele University, Keele, Staffordshire ST5 5BG, UK
}

\vspace{1cm}

\centerline{Received 6 November 2011.}
\centerline{Accepted 25 September 2013.}
\centerline{Published in {\em The Astrophysical Journal} 15 November 2013.}

\vspace{1cm}

\centering{Blecic, J., Harrington, J., Madhusudhan, N., Stevenson, K. B., Hardy, R. A., Cubillos,\\
P. E., Hardin, M., Campo, C. J., Bowman, W. C., Nymeyer, S., Loredo, T. J., Anderson,\\
D. R., \& Maxted, P. F. L. 2013, ApJ, 779, 5}
\vspace{0.2cm} 

\centering{http://adsabs.harvard.edu/abs/2013ApJ...779....5B \\
doi:10.1088/0004-637X/779/1/5}

\vspace{0.5cm}
\centerline{\copyright AAS. Reproduced with permission.}

}\clearpage

\if \includeFirst y
    
\section{ABSTRACT}

Exoplanet WASP-14b is a highly irradiated, transiting hot Jupiter. Joshi et al. calculate an equilibrium temperature (\math{T\sb{\rm eq}}) of 1866 K for zero albedo and reemission from the entire planet, a mass of 7.3 {\pm} 0.5 Jupiter masses (\math{M\sb{\rm J}}) and a radius of 1.28 {\pm} 0.08 Jupiter radii (\math{R\sb{\rm J}}). Its mean density of 4.6 g\,cm\sp{-3} is one of the highest known for planets with periods less than 3 days. We obtained three secondary eclipse light curves with the {\em Spitzer Space Telescope}. The eclipse depths from the best jointly fit model are 0.224\% {\pm} 0.018\% at 4.5 {\micron} and 0.181\% {\pm} 0.022\% at 8.0 {\micron}. The corresponding brightness temperatures are 2212 {\pm} 94 K and 1590 {\pm} 116 K. A slight ambiguity between systematic models suggests a conservative 3.6 {\micron} eclipse depth of 0.19\% {\pm} 0.01\% and brightness temperature of 2242 {\pm} 55 K. Although extremely irradiated, WASP-14b does not show any distinct evidence of a thermal inversion. In addition, the present data nominally favor models with day night energy redistribution less than \sim30\%. The current data are generally consistent with oxygen-rich as well as carbon-rich compositions, although an oxygen-rich composition provides a marginally better fit. We confirm a significant eccentricity of \math{e} = 0.087 {\pm} 0.002 and refine other orbital parameters.

\section{INTRODUCTION}
\label{intro}

The {\em Spitzer Space Telescope} \citep{Werner2004} is the most widely used facility for measuring thermal properties of extrasolar planets. {\em Spitzer} systematics are well studied and modeled, providing an invaluable resource for exoplanet characterization \citep{SeagerDeming2010}. This has enabled the measurement of tens of atmospheres, using the detection of primary and secondary eclipses as the most prolific method of investigation to date. 

The planet-to-star flux ratio is enhanced in the infrared due to the rising planetary thermal emission and the dropping stellar emission, enabling detection of planetary emission through high-precision photometric measurements. Combining several secondary-eclipse observations measured in broad {\em Spitzer} bandpasses with the Infrared Array Camera \citep[IRAC; ][]{Fazio2004IRAC}, a low-resolution dayside spectrum from the planet can be reconstructed, revealing key atmospheric and physical parameters. These measurements can further be used to constrain atmospheric composition, thermal structure, and ultimately the formation and evolution of the observed planet.

WASP-14b represents an intriguing object for such an analysis, having characteristics not so common for close-in, highly irradiated giant planets. \citet{Joshi2009-WASP14b} discovered it as a part of the SuperWASP survey \citep[Wide-Angle Search for Planets; ][]{Pollocco06, Cameron2006-SuperWASP, Cameron2007-SuperWASP}. Photometric and radial-velocity observations revealed a planetary mass of 7.3 {\pm} 0.5 \math{M\sb{\rm J}} and a radius of 1.28 {\pm} 0.08 \math{R\sb{\rm J}}. Its density (\math{\rho} = 4.6 g\,cm\sp{-3}) is significantly higher than typical hot-Jupiter densities of 0.34--1.34 g\,cm\sp{-3} \citep{Loeillet2008}. The planet is also very close to its star (semi-major axis 0.036 {\pm} 0.001 AU), and has a significant orbital eccentricity, refined slightly to \math{e} = 0.087 {\pm} 0.002 in this work.

Detailed spectroscopic analyses of the stellar atmosphere determined that the star belongs to the F5 main-sequence spectral type with a temperature of 6475 {\pm} 100 K and high lithium abundance of log \math{N}(Li) = 2.84 {\pm} 0.05. F-type stars with this temperature should have depleted Li, being close to the Li gap or ``Boesgaard gap'' \citep{BoesgaardTripicco1986, Balachandran1995}. However, the high amount of Li and a relatively high rotational speed of \math{v\sin(i)} = 4.9 {\pm} 1.0 km\,s\sp{-1} indicate that WASP-14 is a young star. Comparing these results with models by \citet{FortneyEtal2007apjPlanetRadii} for the range of planetary masses and radii led \citet{Joshi2009-WASP14b} to constrain the age of the system to 0.5--1.0 Gyr. 

\citet{Joshi2009-WASP14b} also discuss the high eccentricity of the planet. Because WASP-14b has a very small orbital distance, probable scenarios for such a significant eccentricity (their \math{e} = 0.091 {\pm} 0.003) would be either that the system age is comparable to the tidal circularization time scale or there is a perturbing body.

\citet{Husnoo2011-WASP14b} performed long-term radial-velocity measurements to discover or reject the presence of a third body. They refined the orbital eccentricity to \math{e} = 0.088 {\pm} 0.003.  They argue that this planet has undergone some degree of orbital evolution, but that it is still subject to strong tidal forces. They state that since there is no observable unambiguous trend in residuals with time, there is no firm evidence for a planetary companion. This would establish a new lower limit for the semimajor axis at which orbital eccentricity can survive tidal evolution for the age of the system.
We obtained three secondary eclipse light curves at 3.6 {\micron}, 4.5 {\micron}, and 8.0 {\micron} using {\em Spitzer}. We present analytic light-curve models that incorporate corrections for systematic effects that include the new \citet{StevensonEtal2012apjHD149026b} pixel sensitivity mapping technique, a Keplerian orbital model, estimates of infrared brightness temperatures, and constraints on atmospheric composition and thermal structure.

In Section \ref{sec:obs} we describe our observations. Section \ref{sec:data_red} discusses data reduction procedures. Section \ref{sec:Phot} presents our photometry and Section \ref{sec:SecEclResults} discusses the modeling techniques and results from each dataset. Section \ref{sec:orbit} presents constraints on the orbit of WASP-14b, and Section \ref{sec:atm} reveals the atmospheric structure and composition. In Section \ref{sec:discus} we discuss our results and in Section \ref{sec:concl} we present our conclusions. Data files containing the light curves, best-fit models, centering data, photometry, etc., are included as electronic supplements to this article.

\section{OBSERVATIONS}
\label{sec:obs}

The {\em Spitzer} IRAC instrument observed two events; one at 3.6 {\microns} in 2010 March (Knutson's program 60021, Warm {\em Spitzer}) and one observation simultaneously in two wavelength bands (4.5 and 8.0 {\microns}) in 2009 March (Harrington's program 50517, {\em Spitzer} cryogenic mission). The observation at 3.6 {\microns} (channel 1) was made in subarray mode with 2 s exposures, while the observations at 4.5 and 8.0 {\microns} (channels 2 and 4) were made in stellar mode (2\math{\times}2,12) with pairs of 2 s frames taken in the 4.5 {\micron} band for each 12 s frame in the 8.0 {\microns} band. This mode was used to avoid saturation in channel 2.


\begin{table}[ht]
\vspace{15pt}
\caption{\label{table:ObsDates1} 
Observation Information}
\atabon\strut\hfill\begin{tabular}{lccc@{ }c@{ }c@{ }l}
    \hline
    \hline 
    Channel                     & Observation          & Start Time   & Duration      & Exposure           & Number of                   \\
                                & Date                 & (JD)        & (s)            & Time (s)           & Frames                      \\
    \hline
    \multicolumn{6}{c}{Main science observation}                                                                                         \\
    \hline 
    Ch1                         & 2010 Mar 18             & 2455274.4707   & 28055.4         & 2                     & 13760              \\
    Ch2                         & 2009 Mar 18             & 2454908.8139   & 19998.7         & 2 \math{\times} 2       & 2982               \\
    Ch4                         & 2009 Mar 18             & 2454908.8139   & 19998.7         & 12                    & 1481               \\ 
    \hline
    \multicolumn{6}{c}{Pre-observation}                                                                                                  \\ 
    \hline 
    Ch2+4                            & 2009 Mar 18              & 2454908.7877   & 2019            & 2                     & 213          \\ 
    \hline
    \multicolumn{6}{c}{Post-observation}                                                                                                 \\ 
    \hline
    Ch2+4                             & 2009 Mar 18             & 2454909.0455    & 367             & 2 \math{\times} 2,12  & 10            \\
   \hline
    
\end{tabular}\hfill\strut\ataboff
\end{table}

We have pre- and post-observation calibration frames for the 4.5 and 8.0 {\microns} observation. Prior to the main observation, we exposed the array to a relatively bright source (see Section\ \ref{sec:ch2}). That quickly saturated charge traps in the detector material, reducing the systematic sensitivity increase during the main observation. Post-eclipse frames of blank sky permit a check for warm pixels in the aperture. The {\em Spitzer} pipeline version used for the 3.6 {\micron} observation is S.18.14.0 and for the 4.8 and 8.0 {\microns} observation is S18.7.0. The start date of each observation, duration, exposure time and total number of frames are given in Table\ \ref{table:ObsDates1}. 

\section{DATA REDUCTION}
\label{sec:data_red}

\subsection{Background}
\label{sec:backgr}

Our analysis pipeline is called Photometry for Orbits, Eclipses and Transits (POET). It produces light curves from {\em Spitzer} Basic Calibrated Data (BCD) frames, fits models to the light curves, and assesses uncertainties. The derived parameters constrain separate orbital and atmospheric models. In this section we give a general overview of POET. Subsequent sections will provide details as needed.

Each analysis starts by identifying and flagging bad pixels in addition to the ones determined by the {\em Spitzer} bad pixel mask (see Section\ \ref{sec:Phot}). Then we perform centering. Due to the \sim 0.1\% relative flux level of secondary-eclipse observations and {\em Spitzer's} relative photometric accuracy of 2\% \citep{Fazio2004IRAC}, we apply a variety of centering routines, looking for the most consistent. We test three methods to determine the point-spread function (PSF) center precisely: center of light, two-dimensional Gaussian fitting, and least asymmetry (see Supplementary Information of \citealp{StevensonEtal2010Natur} and \citealp{LustEtal2013apjCentering}). The routines used for each data set are given below. We then apply 5\math{\times}-interpolated aperture photometry \citep{HarringtonEtal2007natHD149026b}, where each image is re-sampled using bilinear interpolation. This allows the inclusion of partial pixels, thus reducing pixelation noise \citep{StevensonEtal2012apjHD149026b}. We subtract the mean background within an annulus centered on the star and discard frames with bad pixels in the photometry aperture.

{\em Spitzer}\/ IRAC has two main systematics, which depend on time and the sub-pixel position of the center of the star.  To find the best time-dependent model (the ``ramp''), we fit a variety of systematic models from the literature, and some of our own, using a Levenberg--Marquardt \math{\chi\sp{2}} minimizer \citep{Levenberg1944, Marquardt1963}. We use our newly developed \citep{StevensonEtal2012apjHD149026b} BiLinearly Interpolated Subpixel Sensitivity (BLISS) mapping technique to model intrapixel sensitivity variation (see Section\ \ref{sec:sys}).  The BLISS method can resolve structures inaccessible to the widely used two-dimensional polynomial fit \citep{Knutson08, MachalekEtal2009ApJ-XO2b, FressinEtal2010ApJ-Tres3}.  It is faster and more accurate than the mapping technique developed by \citet{BallardEtalr2010PASP-NewIntrapixel}, which uses a Gaussian-weighted interpolation scheme and is not feasibly iterated in each step of Markov-Chain Monte Carlo (MCMC; see next section for details on modeling systematics).

To determine the best aperture size, we seek the smallest standard deviation of normalized residuals (SDNR) among different aperture sizes for the same systematic model components. The best ramp model at that aperture size is then determined by applying the Bayesian (BIC) and Akaike (AIC) information criteria \citep{Liddle2008}, which compare models with different numbers of free parameters (see Section \ref{sec:LightCurves}). The BIC and AIC cannot be used to compare BLISS maps with differing grid resolutions, or BLISS versus polynomial maps (see Section \ref{sec:sys}), but BLISS has its own method for optimizing its grid \citep{StevensonEtal2012apjHD149026b}.

To explore the parameter space and to estimate uncertainties, we use an MCMC routine (see Section \ref{sec:LightCurves}).  We model the systematics and the eclipse event simultaneously, running four independent chains until the \citet{GelmanRubin1992} convergence test for all free parameters drops below 1\%. Our MCMC routine can model events separately or simultaneously, sharing parameters such as the eclipse midpoint, ingress/egress times or duration.

Finally, we report mid-times in both BJD\sb{UTC} (Barycentric Julian Date, BJD, in Coordinated Universal Time) and BJD\sb{TT} (BJD\sb{TDB},  Barycentric Dynamical Time), calculated using the Jet Propulsion Laboratory (JPL) Horizons system, to facilitate handling discontinuities due to leap seconds and to allow easy comparison of eclipse mid-times (see \citealp{EastmanEtal2010apjLeapSec} for discussion of timing issues).

\subsection{Modeling Systematics}
\label{sec:sys}

Modeling systematics is critical to recovering the extremely weak signal of an exoplanetary atmosphere against the stellar and/or background noise, particularly when using instrumentation not specifically built for the job.  Several re-analyses of early {\em Spitzer}\/ eclipse data sets underscore this.  For example, our group's initial analysis of an HD 149026b lightcurve \citet{HarringtonEtal2007natHD149026b} found two \math{\chi\sp{2}} minima, with the deeper eclipse having the deeper minimum.  This analysis used the bootstrap Monte Carlo technique as described without statistical justification and too simplistically by \citet{PressEtalNumRec}.  The re-analysis by \citet{KnutsonEtal2009apjHD149026bphase}, using MCMC, preferred the lower value, which additional observations confirmed.  Our own re-analysis, by \citealp{StevensonEtal2012apjHD149026b}, agreed with \citeauthor{KnutsonEtal2009apjHD149026bphase}.  Another example is the \citet{Desert2009} re-analysis of the putative detection of H\sb{2}O on HD 189733b by \citet{Tinetti2007Nature}.  \citeauthor{Desert2009} found a shallower transit that did not support the detection.  Although the number of such discrepancies in the {\em Spitzer}\/ eclipse and transit literature is not large compared to the many dozens of such measurements, they serve as cautionary tales.  It is critical to use only the most robust statistical treatments (e.g., MCMC rather than bootstrap), to compare dozens of systematic models using objective criteria (like BIC), and to worry about minutiae like the differences between various centering and photometry methods.  Re-analyses of photometric work done with such care have uniformly been in agreement.  Most of these appear as notes in original papers stating that another team confirmed the analysis (e.g., \citealp{StevensonEtal2012apjGJ436c}).

{\em Spitzer's}\/ IRAC channels can exhibit both time-dependent and position-dependent sensitivity variations.  These variations can be up to \sim 3\%, much more than typical (0.01\%--0.5\%) eclipse depths.  The 3.6 and 4.5 {\micron} bands use InSb detectors, and the 5.8 and 8.0 {\micron} bands use Si:As detectors.  Although each type of systematic is strongest in a different set of channels, many authors reported both systematics in both sets of channels \citep{StevensonEtal2010Natur, Reach2005-IRACCalibration, CharbonneauEtal2005apjTrES1, CampoEtal2011apjWASP12b}, so we test for them all in each observation.

The time-varying sensitivity (``ramp'') is most pronounced at 8.0 {\micron} \citep{CharbonneauEtal2005apjTrES1, HarringtonEtal2007natHD149026b} and is very weak, often nonexistent, in the InSb channels. It manifests as an apparent increase in flux with time, and at 8.0 {\microns} it is attributed to charge trapping. Observing a bright ($>$250 MJy\,sr\sp{-1} in channel 4), diffuse source (``preflashing'') saturates the charge traps and produces a flatter ramp \citep{KnutsonEtal2009apjHD149026bphase}.  An eclipse is easily separated from the ramp by fitting, but not without adding uncertainty to the eclipse depth. Model choice is particularly important for weak eclipses, where a poor choice can produce an incorrect eclipse depth. To model the ramp effect, we test over 15 different forms of exponential, logarithmic, and polynomial models (see \citealp{StevensonEtal2012apjHD149026b}, Equations\ (2)--(11)).

InSb detectors can have intrapixel quantum efficiency variations, which strongly affects {\em Spitzer's}\/ underresolved PSF and requires accurate (\sim 0.01-pixel) determination of the stellar center location. This intrapixel sensitivity is greatest at pixel center and declines toward the edges by up to 3.5\% \citep{Morales-Calderon2006}. It is also not symmetric about the center and the amplitude of the effect varies from pixel to pixel. Over the total duration of the observation, the position varies by several tenths of a pixel. Since the stellar center oscillates over this range frequently, this systematic is adequately sampled during a single eclipse observation. Observing with fixed pointing minimizes the effect \citep{Reach2005-IRACCalibration, CharbonneauEtal2005apjTrES1, HarringtonEtal2007natHD149026b, StevensonEtal2010Natur}.

Our BLISS method \citep{StevensonEtal2012apjHD149026b} maps a pixel's sensitivity on a fine grid of typically over 1000 ``knots'' within the range of stellar centers.  It then uses bilinear interpolation to calculate the sensitivity adjustment for each observation from the nearest knot values (\math{M(x,y)} in Equation (\ref{eqn:full})).  To compute the map, we divide the observed fluxes by the eclipse and ramp models, and assume that any residual fluxes are related to the stellar center's position in the pixel (hence the need for accurate stellar centering; see above).  We average the residuals near each knot to calculate its value.  Each data point contributes to one knot, and each knot comes from a small, discontiguous subset of the data.  The map is recalculated after each MCMC iteration and is used to calculate \math{\chi\sp{2}} in the next iteration.  The MCMC does not directly vary the knot values, but the values change slightly at each iteration.  This method quickly converges.

The crucial setup item in BLISS is determining the knot spacing (i.e., bin size or resolution). The bin size must be small enough to catch any small-scale variation, but also large enough to ensure no correlation with the eclipse fit (see Section \ref{sec:ch1}).  Either bilinear (BLI) or nearest-neighbor (NNI) interpolation can generate the sensitivities from the knots.  Assuming accurate centering, BLI should always outperform NNI. The bin size where NNI outperforms BLI thus indicates the centering precision and determines the bin size for that particular data set. If NNI always outperforms BLI, that indicates very weak intrapixel variability, and intrapixel modeling is unnecessary.

Compared to polynomial intrapixel models, the SDNR improves with BLISS mapping, but this would be expected of any model with more degrees of freedom.  Previously, we have used BIC and AIC to evaluate whether a better fit justifies more free parameters. Both BIC and AIC are approximations to the Bayes factor, which is often impractical to calculate.  Both criteria apply a penalty to \math{\chi\sp{2}} for each additional free parameter (\math{k}, in Equations\ (\ref{eqn:BIC}) and (\ref{eqn:AIC})), allowing comparison of model goodness-of-fit to the same dataset for different models.  However, both criteria assume that every data point contributes to each free parameter.  That is, they assume that changing any data point potentially changes {\em all} of the free parameters, as do all other information criteria we have researched. However, each BLISS knot value comes from only a specific, tiny fraction of the data.  Changing any individual data point changes exactly one BLISS knot.  Thus, the knots each count for much less than one free parameter in the sense of the assumptions of BIC and AIC, but not zero (i.e., they each increment \math{k} by much less than 1).  Because BLISS violates their assumptions, BIC and AIC are inappropriate for comparing models using BLISS to models that do not use it.  It is still possible to compare two models using BLISS maps with the same knot grid because the increment in the penalty terms would be the same for both grids and would thus not affect the comparison.  See Appendix A of \citet{StevensonEtal2012apjHD149026b} for a more statistically rigorous discussion.

At this point in BLISS's development, we are still working on an appropriate comparison metric. What we do know is that BLISS resolves fine detail in pixel sensitivity that, in many cases, is not compatible with any low-order polynomial form.  For example, \citet{StevensonEtal2012apjHD149026b} show (and compensate for) the effects of pixelation in digital aperture photometry, and demonstrate how our interpolated aperture photometry reduces pixelation bias.  For this paper, the eclipse-depth values are similar between BLISS and non-BLISS analyses, and the residuals are smaller with BLISS, since it is taking out some of these effects in a way that low-order polynomial models cannot (see Figure \ref{fig:SensMap} and examples in \citealp{StevensonEtal2010Natur}).  We have a large excess of degrees of freedom, so we adopt the BLISS results. We continue to use BIC for ramp-model selection.

\subsection{Modeling Light Curves and the Best Fit Criteria}
\label{sec:LightCurves}

To find the best model, for each aperture size we systematically explore every combination of ramp model and intrapixel sensitivity model.  The final light curve model is:
\begin{eqnarray}
\label{eqn:full}
F(x, y, t) = F\sb{s}R(t)M(x,y)E(t),
\end{eqnarray}
\noindent where \math{F(x,y,t)} is the aperture photometry flux, \math{F\sb{\rm s}} is the constant system flux outside of the eclipse, \math{R(t)} is the time-dependent ramp model, \math{M(x, y)} is the position-dependent intrapixel model and \math{E(t)} is the eclipse model \citep{MandelAgol2002ApJtransits}.  We fit each model with a Levenberg--Marquardt \math{\chi\sp{2}} minimizer and calculate SDNR, BIC, and AIC (note that parameter uncertainties, and hence MCMC, are not needed for these calculations).

To estimate uncertainties, we use our MCMC routine with the Metropolis--Hastings random walk algorithm, running at least 10\sp{6} iterations to ensure accuracy of the result. This routine simultaneously fits eclipse parameters and {\em Spitzer} systematics.  It explores the parameter phase space, from which we determine uncertainties fully accounting for correlations between the parameters. The depth, duration, midpoint, system flux, and ramp parameters are free. Additionally, the routine can model multiple events at once, sharing the eclipse duration, midpoint and ingress/egress times. These joint fits are particularly appropriate for channels observed together (see \citealp{CampoEtal2011apjWASP12b} for more details about our MCMC routine).

To avoid fixing any model parameter during MCMC, we use Bayesian priors \citep[e.g.,][]{Gelman2002-prior}. This is particularly relevant for noisy or low signal-to-noise (S/N) datasets where some parameters like ingress and egress times are not well constrained by the observations. For them we use informative priors taken from the literature (see Section \ref{sec:SecEclResults} for the values used in this analysis).

Photometric uncertainties used in our analyses are derived by fitting an initial model with a Levenberg--Marquardt \math{\chi\sp{2}} minimizer and re-scaling it so reduced \math{\chi\sp{2}} = 1.  This is needed because {\em Spitzer} pipeline uncertainties have often been overestimated \citep{HarringtonEtal2007natHD149026b}, sometimes by a factor of two or three. Along with the BCD frames, the Spitzer Science Center provides images giving the uncertainties of the BCD pixels.  The calculations behind these images include uncertainty in the absolute flux calibration, which effect we divide out.  The {\em Spitzer}-provided errors are thus too large for exoplanet eclipses and transits, but they do contain information about the relative noisiness of different pixels.

Most workers ignore the uncertainty frames and calculate a single per-frame uncertainty from their root mean square (rms) model residuals, sometimes taken over just a short time span.  This has the effect of fixing the reduced chi-squared to 1, and possibly ignoring red noise, depending on the time span of residuals considered.  We do use the {\em Spitzer}-provided uncertainties, resulting in slightly differing uncertainties per frame.  However, this approach can produce reduced chi-squared values of 0.3, and sometimes 0.1, as the {\em Spitzer}\/ uncertainties are computed with absolute calibration in mind.  So, we also re-scale the per-frame uncertainties to give a reduced chi-squared of 1.  As a practical matter, the variation in our uncertainties is a few percent and the typical uncertainty is the same as with the rms method applied to the entire dataset, which accounts for a global average of red noise.

Rescaling the uncertainties is changing the dataset, and BIC can only compare different models applied to a single dataset.  So, we use just one rescaling per aperture size, and fit all the models to that dataset. This works because the reasonable models for a given dataset all produce nearly the same scaling factor.  The rank ordering of models is not altered by the scaling factor. In the Section\ \ref{sec:joint} we lists the rescaling factor for each dataset.  

After deriving new uncertainties, we re-run the minimizer and then run MCMC. If MCMC finds a lower \math{\chi\sp{2}} than the minimizer, we re-run the minimizer starting from the MCMC's best value. The minimizer will find an even better \math{\chi\sp{2}}. We then restart the MCMC from the new minimizer solution.  We ensure that all parameters in four independent MCMC chains converge within 1\% according to the \citet{GelmanRubin1992} test.  We also inspect trace plots for each parameter, parameter histograms, and correlation plots for all parameter pairs.

Our measures of goodness of fit are SDNR, BIC, and AIC values \citep{Liddle2008}:  
\begin{equation}
\label{eqn:BIC}
{\rm BIC} = \chi\sp{2} + k\ln N,
\end{equation}
\begin{equation}
\label{eqn:AIC}
{\rm AIC} = \chi\sp{2} + 2k,
\end{equation}
\noindent where \math{k} is the number of free parameters, N is the number of data points. These criteria penalize additional free parameters in the system, with better fits having lower values. To appropriately compare BIC or AIC values for a given aperture size, and determine the best fit, we use the same uncertainties for each dataset, and model all combinations of ramp models and intrapixel model. SDNR values are used to compare different aperture sizes using the same model. The lowest value defines the best aperture size. 

Equally important is the correlation in the residuals (see Section\ \ref{sec:ch1}). We plot and compare the scaling of binned model residuals versus bin size \citep{pont:2006, winn:2008} with the theoretical \math{1/\sqrt{N}} scaling for the rms of Gaussian residuals.  A significant deviation between those two curves indicates time-correlated variation in the residuals and possible underestimation of uncertainties if only their point-to-point variation is considered.  Note that our uncertainty estimation uses the residuals' global rms, so we already account for a global average of correlated noise.

After MCMC is finished, we study parameter histograms and pairwise correlations plots, as additional indicators of good posterior exploration and convergence.

\section{WASP-14b PHOTOMETRY}
\label{sec:Phot}

For our analyses, we used BCD frames generated in the {\em Spitzer} IRAC pipeline \citep{Fazio2004IRAC}. The pipeline version used for each observation is given in Section\ \ref{sec:obs}. Our data reduction procedure started with applying {\em Spitzer}'s bad pixel masks and with our procedure for flagging additional bad pixels \citep{HarringtonEtal2007natHD149026b}. In each group of 64 frames and at each pixel position, we applied two-iteration outlier rejection, which calculated the frame median and the standard deviation from the median (not mean), and flagged pixels that deviated by more than 4\math{\sigma}. Then we found the stellar centroid for the photometry by using a two-dimensional Gaussian fit to data in an aperture radius of four pixels.
 
After subtracting the mean background (annuli given in Section\ \ref{sec:joint}), light curves were extracted using 5\math{\times}-interpolated aperture photometry \citep{HarringtonEtal2007natHD149026b} for every aperture radius from 2.25 to 4.25 pixels in 0.25 pixel steps.

To calculate the BJD of each exposure we used the mid-exposure time of each frame, based on the UTCS-OBS value in the FITS header and the frame number.  We performed our barycentric light-time correction using our own code and the coordinates of the {\em Spitzer} spacecraft from the Horizons ephemeris system of the JPL.  The times are corrected to BJD\sb{TDB} to remove the effects of leap seconds and light-travel time across the exoplanet's orbit.  

\begin{figure*}[ht!]
\strut\hfill
\includegraphics[width=0.33\textwidth, clip]{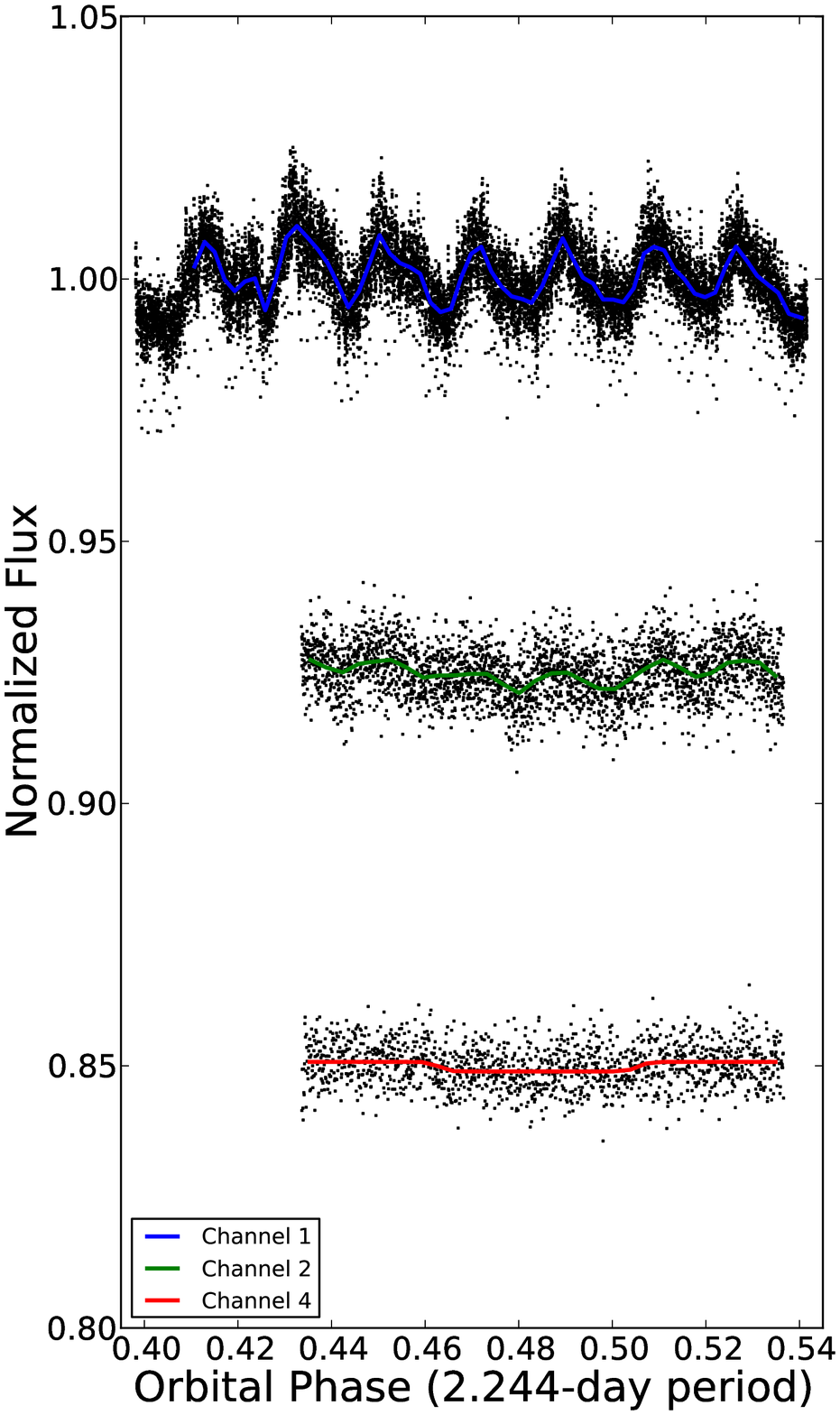}\hfill
\includegraphics[width=0.33\textwidth, clip]{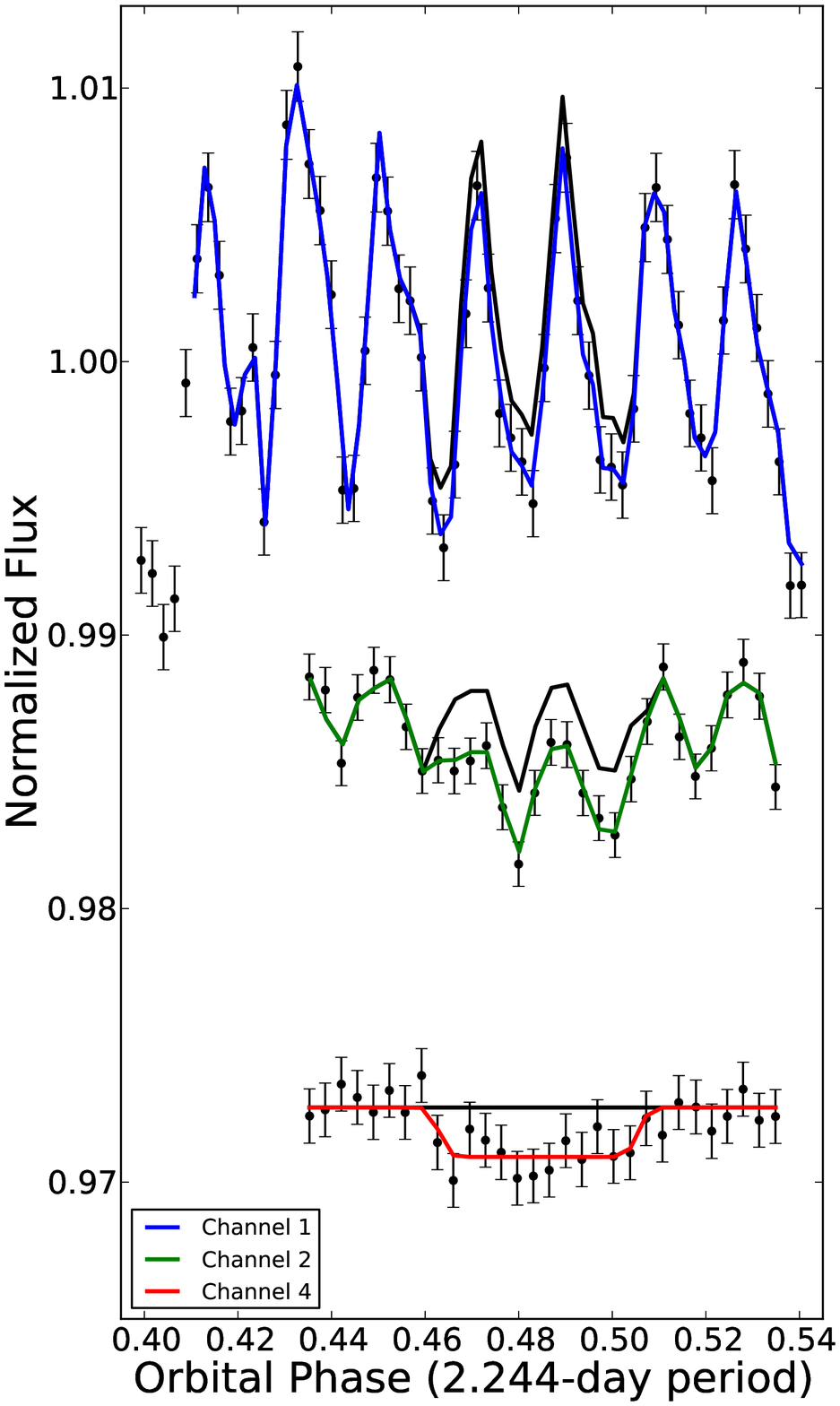}\hfill
\includegraphics[width=0.33\textwidth, clip]{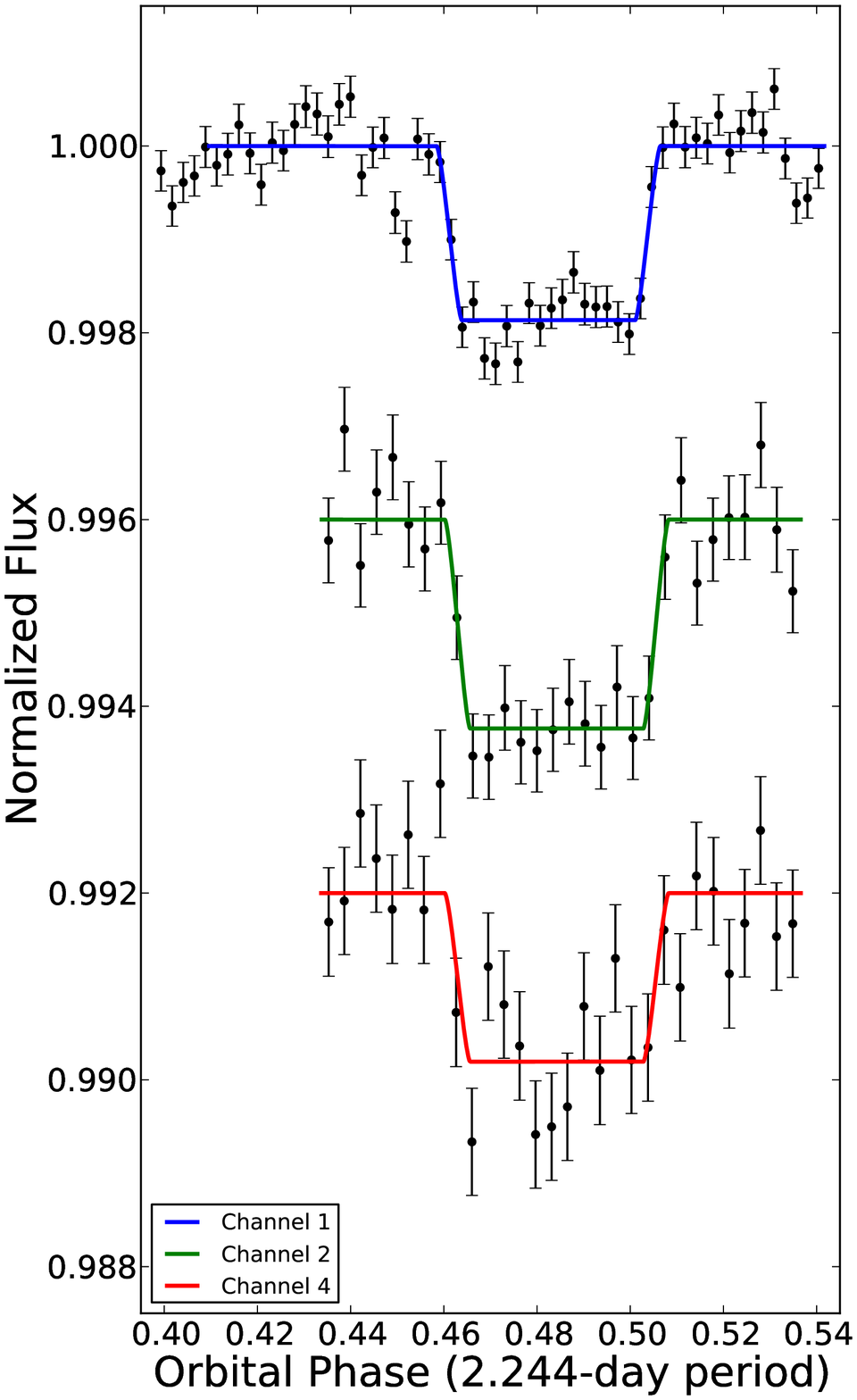}
\hfill\strut
\caption
[Secondary-eclipse light curves of WASP-14b at 3.6, 4.5, and 8.0 {\microns}]
{\label{fig:RawBinNorm1}
Raw (left), binned (center, 60 points per bin), and systematics-corrected (right) secondary-eclipse light curves of WASP-14b at 3.6, 4.5, and 8.0 {\microns}. The results are normalized to the system flux and shifted vertically for clarity.  The colored lines are best-fit models and the error bars are 1\math{\sigma} uncertainties. The black lines in the binned plots are models without an eclipse. As seen in the same plots of channels 2 and 4, a ramp model is not needed to correct for the time-dependent systematic even without clipping any initial data points. The channel 1 model omits early data due to an initial pointing drift (see Section\ \ref{sec:ch1}).\\
(A color version of this figure is available in the online journal.)
}
\end{figure*}

\section{WASP-14b SECONDARY ECLIPSES}
\label{sec:SecEclResults}

Here, we discuss each channel's analysis and model selection in detail, particularly focusing on channel 1, due to the demanding analysis of that data set. In Subsection \ref{sec:ch1} we give our control plots, as an example of how we verify that our results are indeed the best solution for the particular data set. We present each channel separately, followed by a joint fit to all data. Figure \ref{fig:RawBinNorm1} shows our best-fit eclipse light curves. Figure \ref{fig:rms1} shows how the rms of the residuals scales with bin size, a test of correlated noise. In the Appendix we summarize parameters for the WASP-14 system as derived from this analysis and found in the literature.


\begin{figure}[ht!]
    \centering
    \includegraphics[width=1.05\linewidth, clip,  trim=0.75cm 0cm 0cm 0cm]{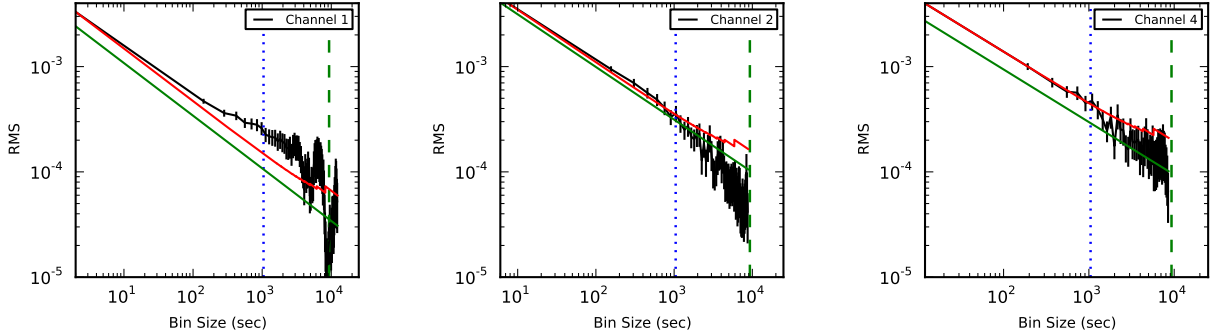}
\caption[Correlations of the residuals for the three secondary eclipse light curves of WASP-14b]
{Correlations of the residuals for the three secondary eclipse light curves of WASP-14b, following \citet{pont:2006}. The black line represents the rms residual flux vs. bin size. The red line shows the predicted standard error scaling for Gaussian noise. The green line shows the Poisson-noise limit. The black vertical lines at each bin size depict 1\math{\sigma} uncertainties on the rms residuals, \math{rms/\sqrt{2N}}, where N is the number of bins (see \citealp[][Section 3.41]{Jeffreys61} and \citealp[][Section 3.3]{Sivia06} for  a derivation including the factor of two, which arises because this is the uncertainty scaling of the rms, not the mean). The dotted vertical blue line indicates the ingress/egress timescale, and the dashed vertical green line indicates the eclipse duration timescale. Large excesses of several \math{\sigma} above the red line would indicate correlated noise at that bin size. Inclusion of 1\math{\sigma} uncertainties shows no noise correlation between the ingress/egress and eclipse duration timescales anywhere except for channel 1 ingress/egress, which hints 3\math{\sigma} at a correlation (adjacent points on this plot are themselves correlated).  Since the relevant timescale for eclipse depths is the duration timescale, we do not scale the uncertainties.  See Section\ \ref{sec:WASP14act} for further discussion.\\
(A color version of this figure is available in the online journal.)
}
\label{fig:rms1}
\end{figure}

\subsection{Channel 1--3.6 {\micron}}
\label{sec:ch1}

The channel-1 observation lasted 7.8 hr, giving ample baseline before and after the secondary eclipse.  The telescope drifted at the start of the observation.   Models with initial data points removed produce better fits with lower values for SDNR.  We therefore ignored some initial data (\sim 36 minutes, 1100 of 13760 points).  Figure \ref{fig:ch1-SDNR1} compares SDNR values for models with different ramps and with and without exclusion of the initial data.

\begin{figure}[ht!]
    \centering
    \includegraphics[width=0.70\linewidth, clip]{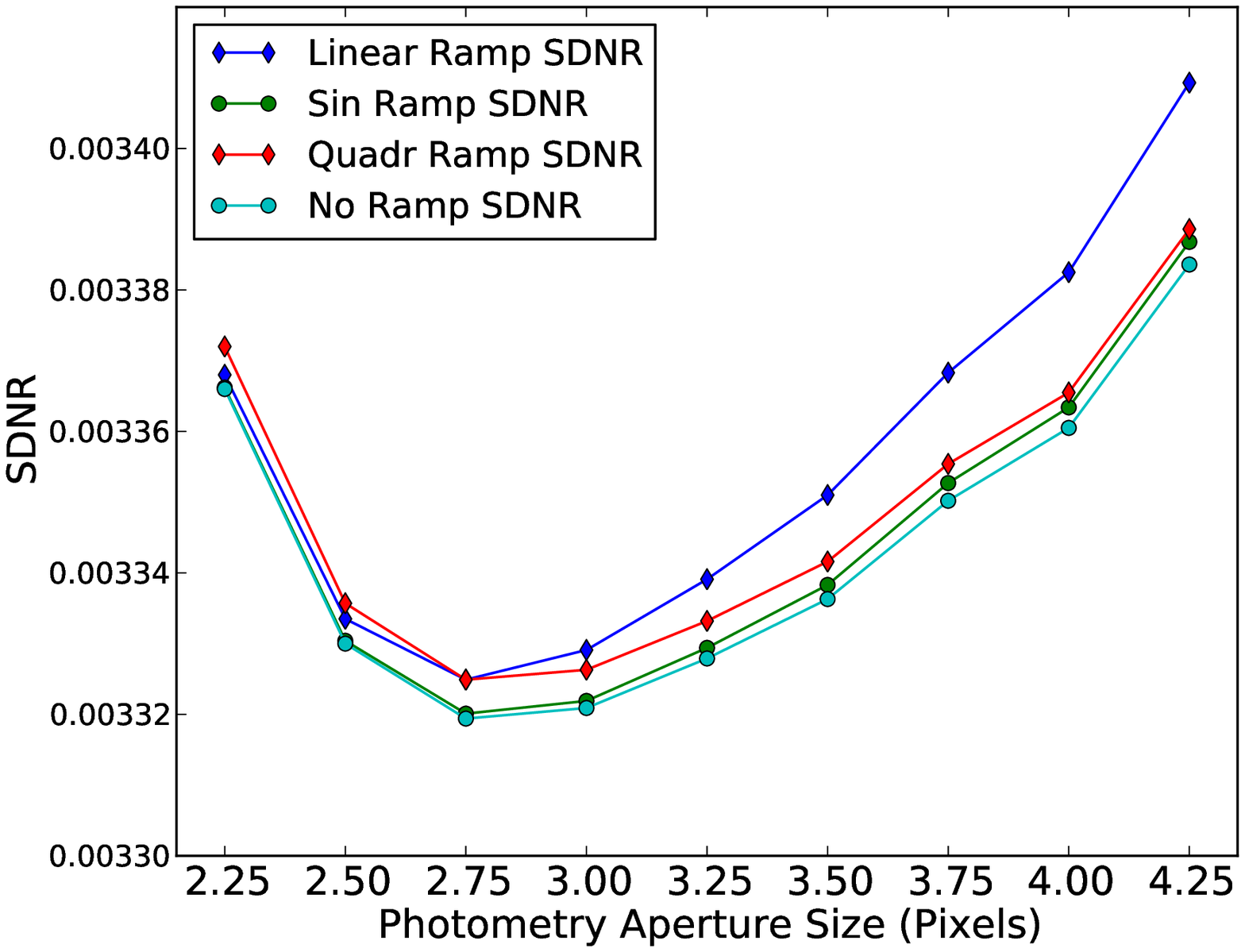}\vspace{-3pt}
    \includegraphics[width=0.70\linewidth, clip]{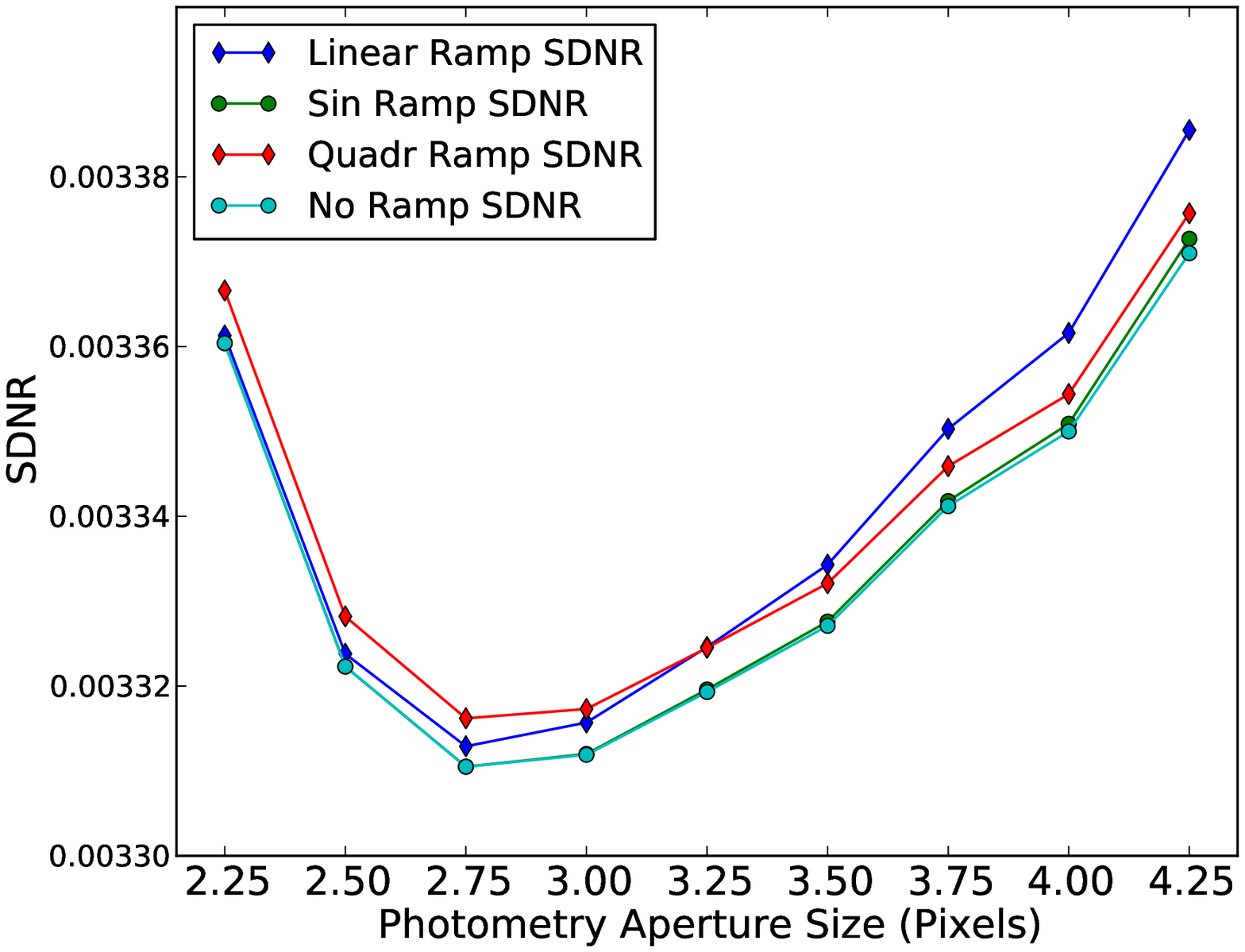}
\caption[Comparison between different ramp models in channel 1]
{SDNR vs. aperture size for different ramp models in channel 1.  A lower value indicates a better model fit. Top: all observational points included ({\em no-preclip}). 
Bottom: same, but with 1100 initial points excluded ({\em preclip}).\\
(A color version of this figure is available in the online journal.)
}
\label{fig:ch1-SDNR1}
\end{figure}

\begin{figure}[ht!]
    \centering
    \includegraphics[width=0.90\linewidth, clip]{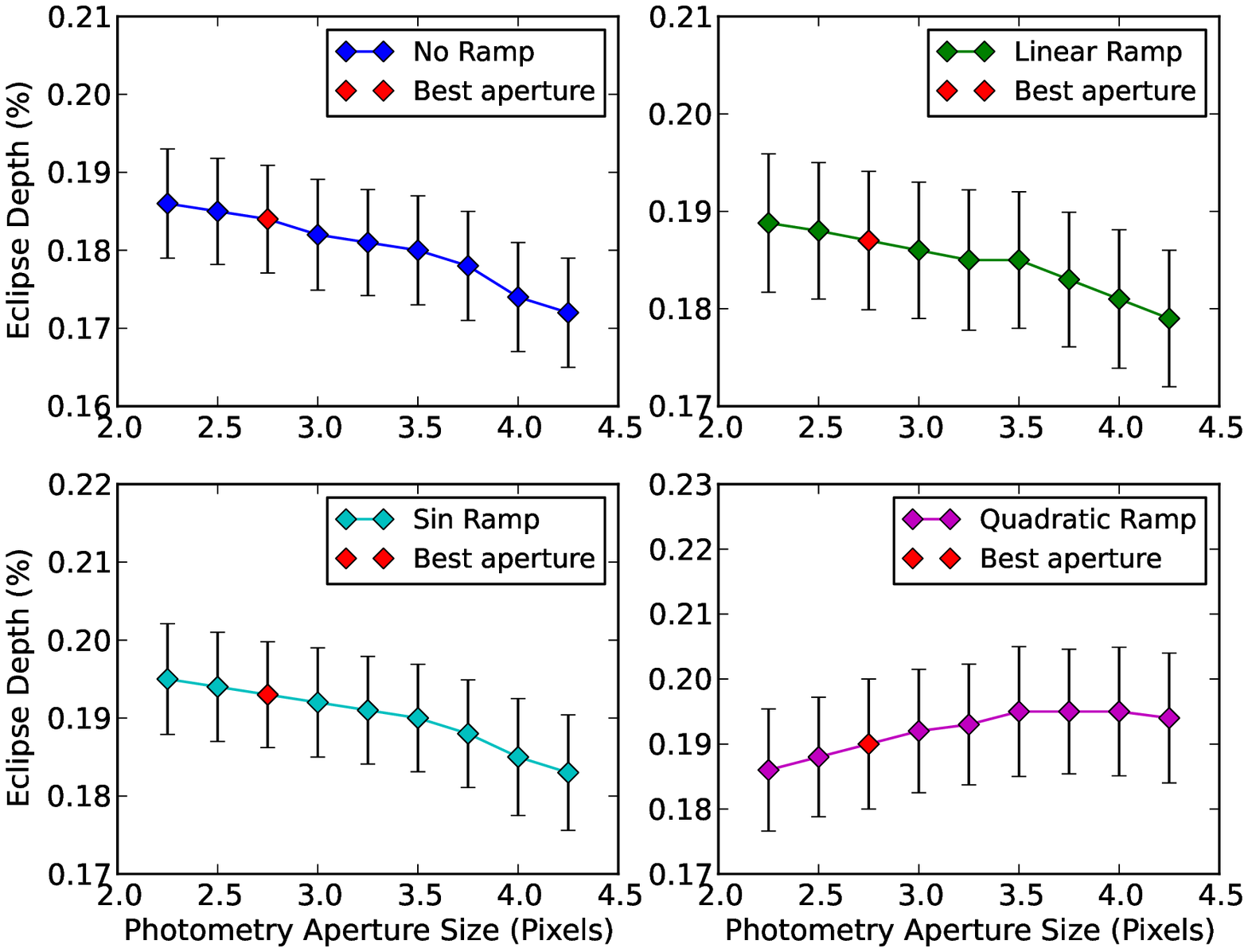}
\caption[Best-fit eclipse depths as a function of photometry aperture size for channel 1]
{Best-fit eclipse depths as a function of photometry aperture size for channel 1. The four best ramp models are plotted (see bellow). The red point indicates the best aperture size for that channel. The eclipse-depth uncertainties are the result of 10\sp5 MCMC iterations. The trend shows insignificant dependence of eclipse depth on aperture size (less than 1\math{\sigma}).\\
(A color version of this figure is available in the online journal.)
}
\label{fig:ch1-depths1}
\end{figure}

Starting from an aperture radius of 2.25 pixels and continuing in increments of 0.25 pixels, we tested all of the ramp models (linear, rising, exponential, sinusoidal, double exponential, logarithmic, etc.). Corresponding equations are listed in \citet{StevensonEtal2012apjHD149026b}. To determine the best solution we consider our best-fit criteria (see Section \ref{sec:LightCurves}) and study the correlation plots. Most of the models produced obvious bad fits, so minimizer and shorter MCMC runs eliminated them. The best aperture radius is 2.75 pixels (see Figure \ref{fig:ch1-SDNR1}, bottom panel). We tested the dependence of eclipse depth on aperture radius \citep{AndersonEtal010ApJ-WASP17b}. The trend in some events may indicate a slightly imperfect background removal (see Figure \ref{fig:ch1-depths1}).  The effect is less than 1\math{\sigma} on the eclipse depth.

\begin{figure}[ht!]
    \centering
    \includegraphics[width=0.68\linewidth, clip, trim=1cm 6cm 0cm 7cm]{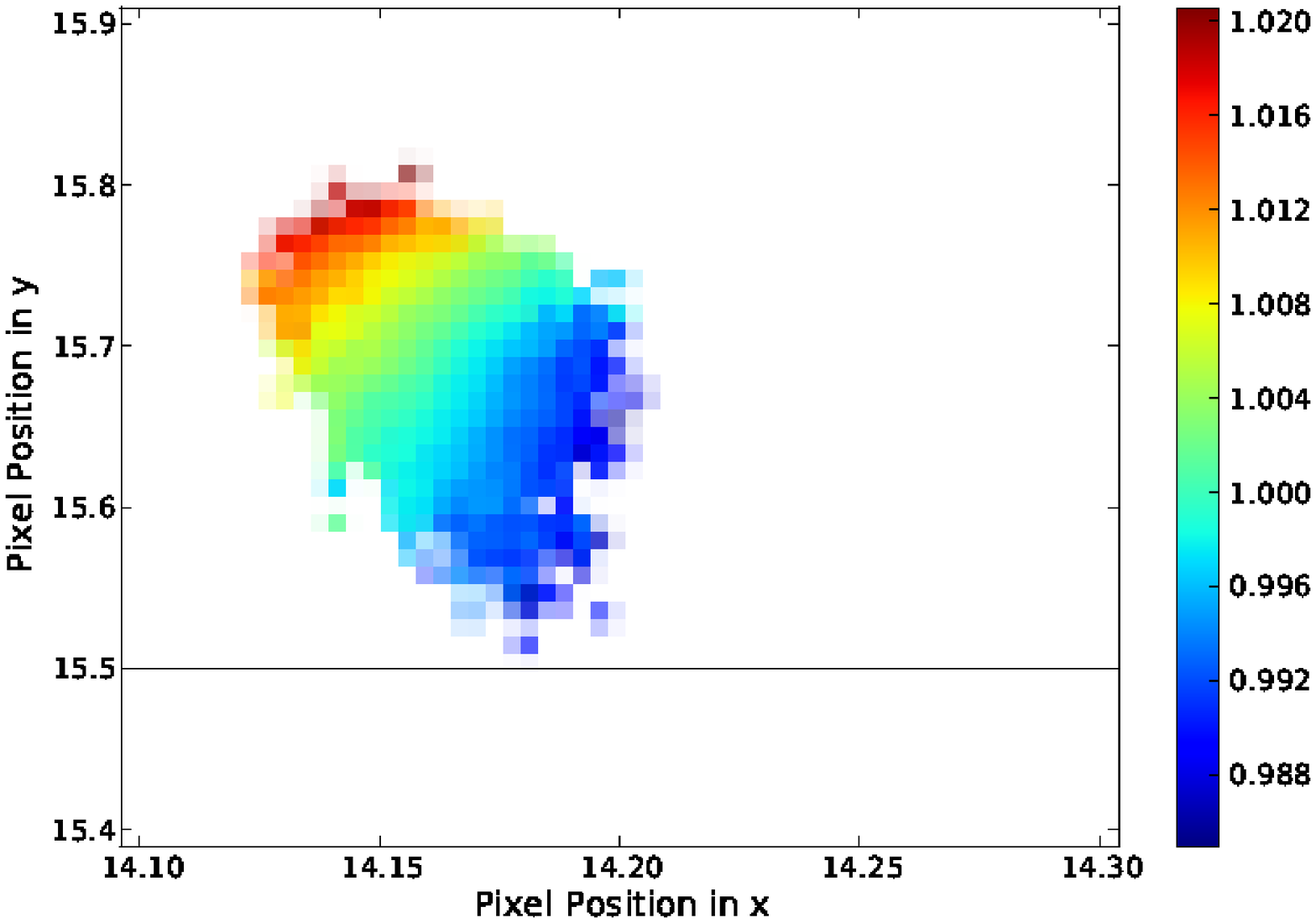}\vspace{-3pt}
    \includegraphics[width=0.70\linewidth, clip, trim=1cm 6cm 0cm 7cm]{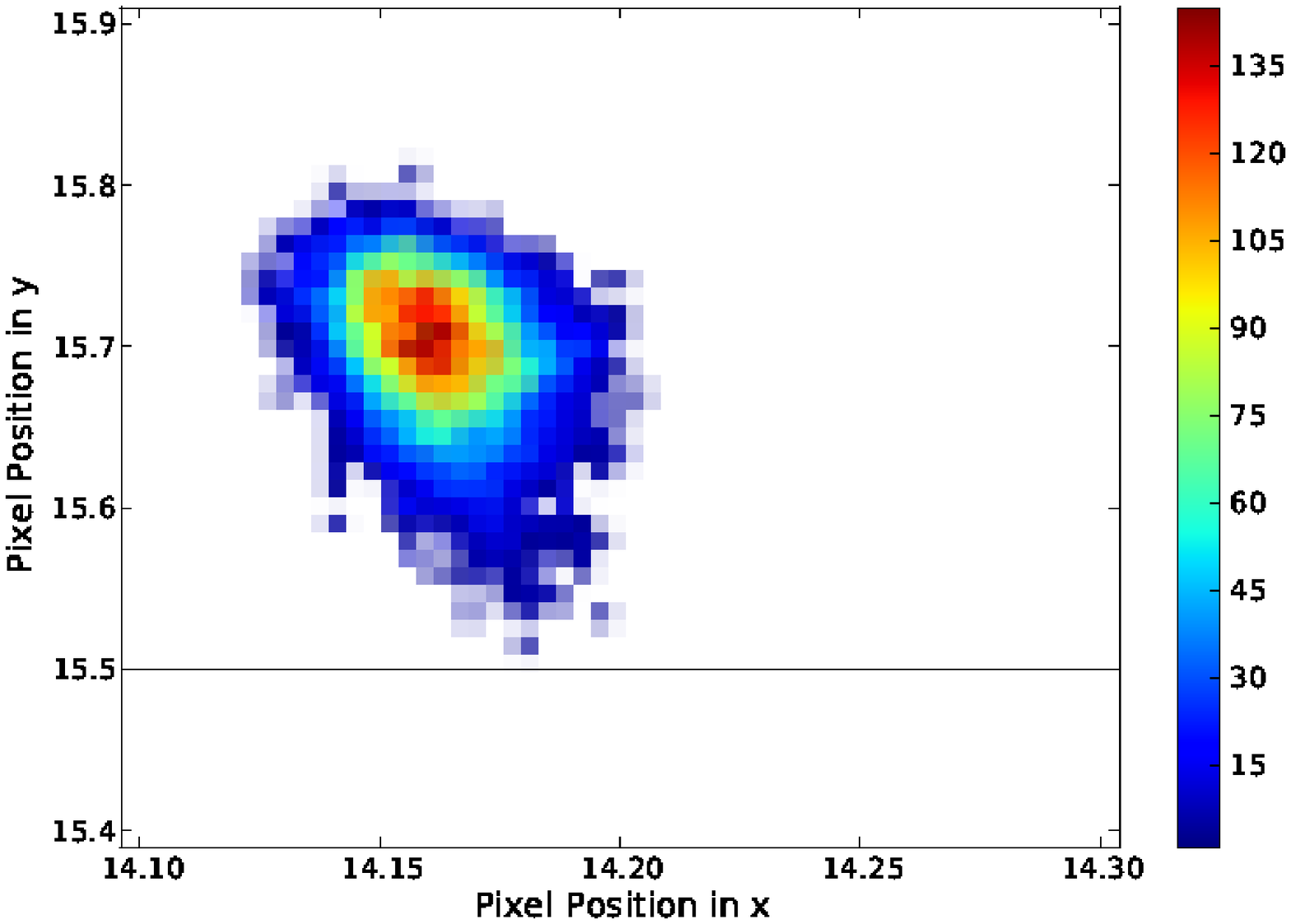}
\caption[BLISS map and pointing histogram of channel 1]
{Top: BiLinearly Interpolated Subpixel Sensitivity (BLISS) map of channel 1.  Redder (bluer) colors indicate higher (lower) subpixel sensitivity.  The horizontal black line defines the lower pixel boundary.
Bottom: Pointing histogram.  Colors indicate the number of points in a given bin.\\
(A color version of this figure is available in the online journal.)
}
\label{fig:SensMap}
\end{figure}

Figure \ref{fig:SensMap} presents the channel-1 BLISS map and Figure \ref{fig:CorrCoef} gives the correlation coefficients between the knot values and the eclipse depth.  As stated in the Section\ \ref{sec:sys}, the most important variable to consider with BLISS is the bin size, which defines the resolution in position space. The position precision for channel 1, measured as the rms of the position difference on consecutive frames, is significantly different for the $x$ and $y$ axes (see Figure \ref{fig:intrapix}). We considered a range of bin sizes for both BLI and NNI around the calculated precision. The best bin size for this data set, determined when NNI outperformed BLI, is 0.004 pixels for $x$ and 0.01 for $y$.

\begin{figure}[ht!]
    \centering
    \includegraphics[width=0.73\linewidth, clip, trim=1cm 6cm 0cm 7cm]{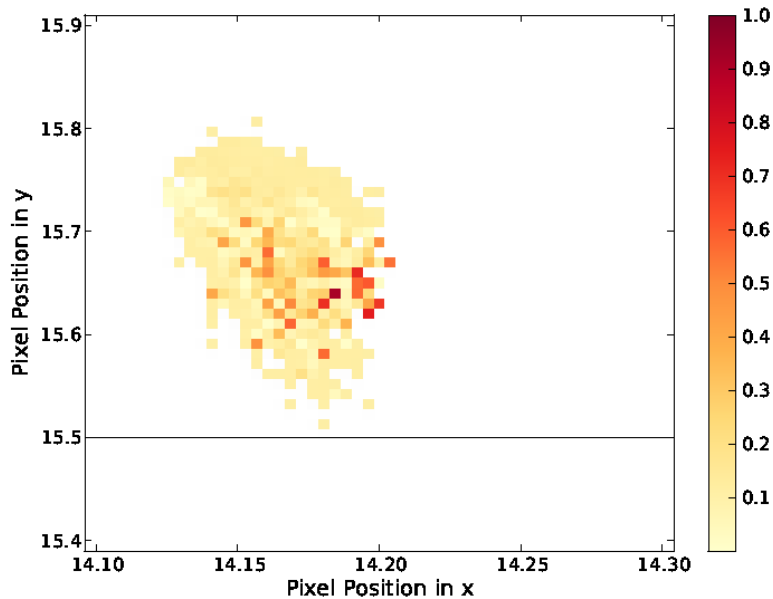}\vspace{-3pt}
\caption[Correlation coefficients between eclipse depth and computed BLISS map knots for channel 1]
{Correlation coefficients between eclipse depth and computed BLISS map knots for channel 1.  The correlation regions (in red) indicate that it is necessary to compute the BLISS map at each MCMC step, to assess the uncertainty on the eclipse depth correctly.\\
(A color version of this figure is available in the online journal.)
}
\label{fig:CorrCoef}
\end{figure}

We also tested two-dimensional polynomial intrapixel models \citep{Knutson08, StevensonEtal2010Natur, CampoEtal2011apjWASP12b}: 
\begin{eqnarray}
\label{eq:vip} V\sb{\rm{IP}}(x,y) = p\sb1y\sp2 + p\sb2x\sp2 + p\sb3xy +
p\sb4y + p\sb5x + 1,
\end{eqnarray} 
\noindent where \math{x} and \math{y} are 
relative to the pixel center nearest the median position and
\math{p\sb{\rm 1}}--\math{p\sb{\rm 5}} are free parameters.  As noted in Section \ref{sec:sys}, we currently lack a quantitative model-selection criterion between polynomial and BLISS intrapixel models, but BIC can apply within a group of BLISS models with the same grid.  BLISS reduces SDNR significantly compared to polynomial models (see Table \ref{table:BLISSPoly}), but so would many models with more free parameters.  We use BLISS because it can handle variations that polynomials cannot follow.  See \citet{StevensonEtal2012apjHD149026b} for other tests that compare polynomial and BLISS intrapixel models.

\begin{figure}[h!]
    \centering
    \includegraphics[width=0.75\linewidth, clip]{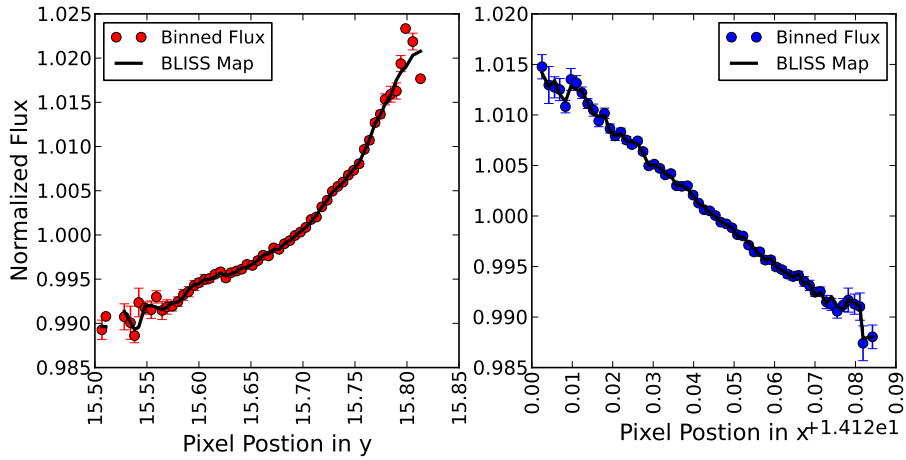}
\caption[BLISS map of channel 1]
{BLISS map and data of channel 1 integrated along the $x$ (right) and $y$ (left) axes.  BLISS effectively fits the position-dependent sensitivity variation.\\
(A color version of this figure is available in the online journal.)
}
\label{fig:intrapix}
\end{figure}

\begin{table}[ht]
\caption{\label{table:BLISSPoly} 
Comparison BLISS And Best Polynomial Model}
\atabon\strut\hfill\begin{tabular}{lccccc}
    \hline
    \hline
    Ramp Model            & \mctc{  BLISS  }                      & \mctc{  Polynomial-Quadratic  }  \\
                  &   SDNR             &    BIC              & SDNR                   & BIC          \\
    \hline
    No ramp       & 0.003313       & 12350.0                 & 0.0033853              & 12593.2      \\ 
    Linear        & 0.003311       & 12342.3                 & 0.0033852              & 12588.5      \\ 
    Sinusoidal    & 0.003316       & 12342.2                 & 0.0033855              & 12590.5      \\
    Quadratic     & 0.003310       & 12351.5                 & 0.0033850              & 12597.3      \\
    \hline
\end{tabular}\hfill\strut\ataboff
\end{table}

To determine the uncertainties in the model parameters, we explored the posterior probability distribution of the model given the data with MCMC. 
We used a Bayesian informative prior for the secondary-eclipse ingress and egress time (\math{t\sb{\rm{2-1}}} = 1046.8 {\pm} 43.9 s), calculated from unpublished WASP photometric and radial-velocity data. All other parameters (eclipse midpoint, eclipse duration, eclipse depth, system flux, and ramp parameters) were left free.


\begin{table}[ht]
\caption{\label{table:Ch1-ramps1} 
Channel 1 Ramp Models}
\atabon\strut\hfill\begin{tabular}{lcccc}
    \hline
    \hline
    Ramp Model    & SDNR            & BIC            & Eclipse Depth (\%)             \\
    \hline
    No ramp       & 0.0033129       & 12350.0        & 0.184 {\pm} 0.007             \\ 
    Linear        & 0.0033105       & 12342.3        & 0.187 {\pm} 0.007             \\ 
    Sinusoidal    & 0.0033162       & 12342.2        & 0.193 {\pm} 0.007             \\
    Quadratic     & 0.0033105       & 12351.5        & 0.190 {\pm} 0.010             \\
    \hline
\end{tabular}\hfill\strut\ataboff
\end{table}

Considering all the above criteria (see also Section \ref{sec:LightCurves}), we selected four ramp models (see Table\ \ref{table:Ch1-ramps1}). The first is without a ramp model, while the other three are: 
\begin{eqnarray}
\vspace{-30pt}
\label{eq:linramp}
R(t) = 1 + r\sb{0}\,(t - 0.5),
\vspace{-30pt}
\end{eqnarray}
\begin{eqnarray}
\vspace{-30pt}
\label{eq:sinramp}
R(t) = 1 + a\,\rm{sin}\,(2\pi(t-t\sb{1})) + b\,\rm{cos}\,(2\pi(t-t\sb{2})),
\vspace{-30pt}
\end{eqnarray}
\begin{eqnarray}
\vspace{-30pt}
\label{eq:quadramp}
R(t) =  1 + r\sb{1}\,(t-0.5) + r\sb{2}\,(t-0.5)\sp{2},
\vspace{-30pt}
\end{eqnarray}
\noindent where \math{t} is orbital phase and \math{a}, \math{b}, \math{r\sb{0},r\sb{1}} and \math{r\sb{2}} are free parameters.

The models produce almost identical SDNR values. However, upon studying the BIC values and the inconsistent trend in the eclipse depths between models with similar BIC values (see Table\ \ref{table:Ch1-ramps1}), we concluded that there is no single best ramp model for this data set.

Therefore, we again use Bayes's theorem and the BIC approximation to the Bayes factor to compare two different models to the data. Following \citet{Raftery1995-BIC} Equations\ (7) and (8), we calculate the posterior odds, i.e., to which extent the data support one model over the other:
\begin{eqnarray}
\label{eq:odds}
\rm{Posterior\,\,\,Odds = Bayes\,\,\,Factor\,\,x\,\, Prior\,\,\,Odds},
\end{eqnarray}
\begin{eqnarray}
\label{eq:prob}
\frac{P(M\sb{2}\,|\,D)}{P(M\sb{1}\,|\,D)} = \frac{P(D\,|\,M\sb{2})}{P(D\,|\,M\sb{1})}\,\,\frac{P(M\sb{2})}{P(M\sb{1})},
\end{eqnarray}
\noindent where \math{M\sb{1}} and \math{M\sb{2}} denote two models, and D denotes the data. \math{P(M\sb{1}\,|\,D)} and \math{P(M\sb{2}\,|\,D)} denote the posterior distributions of the models given the data, \math{P(D\,|\,M\sb{1})} and \math{P(D\,|\,M\sb{2})} denote the marginal probabilities of the data given the model, and \math{P(M\sb{1})} and \math{P(M\sb{2})} denote the prior probabilities of the models.

The first term on the right side of Equation\ (\ref{eq:prob}) is the {\em Bayes factor} for model 2 against model 1, which we will denote as \math{B\sb{21}}. If \math{B\sb{21}} $>$ 1, the data favor model 2 over model 1, and vise versa. 

\citet[see his Equations (20)--(22)]{Raftery1995-BIC} further derives an approximation to the Bayes factor, using BIC, that defines the ratio of marginal probabilities for the two models as:
\begin{eqnarray}
\label{eq:deltaBIC}
B\sb{21} = \frac{P(D\,|\,M\sb{2})}{P(D\,|\,M\sb{1})} \approx e^{-{\Delta {\rm BIC}}/{2}},
\end{eqnarray}
\noindent where \math{\Delta}BIC = BIC\math{(M\sb{2})} - BIC\math{(M\sb{1})}.
We calculate this quantity for each of our ramp models.

\vspace{15pt}
\begin{table}[ht]
\caption{\label{table:deltaBIC} 
Bayes Factor for Model 2 against Model 1}
\atabon\strut\hfill\begin{tabular}{lcccc}
    \hline
    \hline
    Ramp Model    & BIC            & \math{\Delta} BIC     &   \math{B}\sb{21}        & 1\,/\,\math{B}\sb{21}      \\
    \hline
    No ramp       & 12350.0        &     7.8               &     0.02          &  49.4           \\ 
    Linear        & 12342.3        &     0.1               &     0.95          &  1.05           \\ 
    Sinusoidal    & 12342.2        &     0.0               &     ...           &  ...             \\ 
    Quadratic     & 12351.5        &     9.3               &     0.009         &  104.6          \\
    \hline
\end{tabular}\hfill\strut\ataboff
\end{table}

Table\ \ref{table:deltaBIC} gives the probability ratio, or the Bayes factor, for each of our ramp models compared to the model with the smallest BIC value (the sinusoidal model, see Table\ \ref{table:Ch1-ramps1}).  These models are all within the 3\math{\sigma} confidence interval of the best model, indicating an ambiguous situation.  In the atmospheric modeling below, we use the eclipse depth and uncertainty from each of the two extreme models (no-ramp and sinusoidal), and show that the resulting atmospheric models are consistent with each other.  A representative single eclipse depth and uncertainty that spans the two points from the joint fit model (see Section \ref{sec:joint}) is 0.19\% {\pm} 0.01\%, and the corresponding brightness temperature is 2242 {\pm} 55 K.

\subsubsection{On WASP-14 Activity}
\label{sec:WASP14act}

In this channel, we detect time correlation of noise at the 3\math{\sigma} level on time scales of \math{<\ttt{3}} s and \math{\lesssim2\sigma} up to about the 3000 s scale (Figure\ \ref{fig:rms1}, left panel, and Figure\ \ref{fig:resid}). The longest time scale with even a \math{2\sigma} detection of correlation is about 1/7 the eclipse duration, so we do not expect a major effect on the planetary results.  Although not perfect, our ramp and intrapixel models typically remove instrumental effects (e.g., see the middle and right panels of Figure \ref{fig:rms1}), raising the question of stellar activity.

\begin{figure}[ht!]
    \centering
    \includegraphics[width=0.70\linewidth, clip]{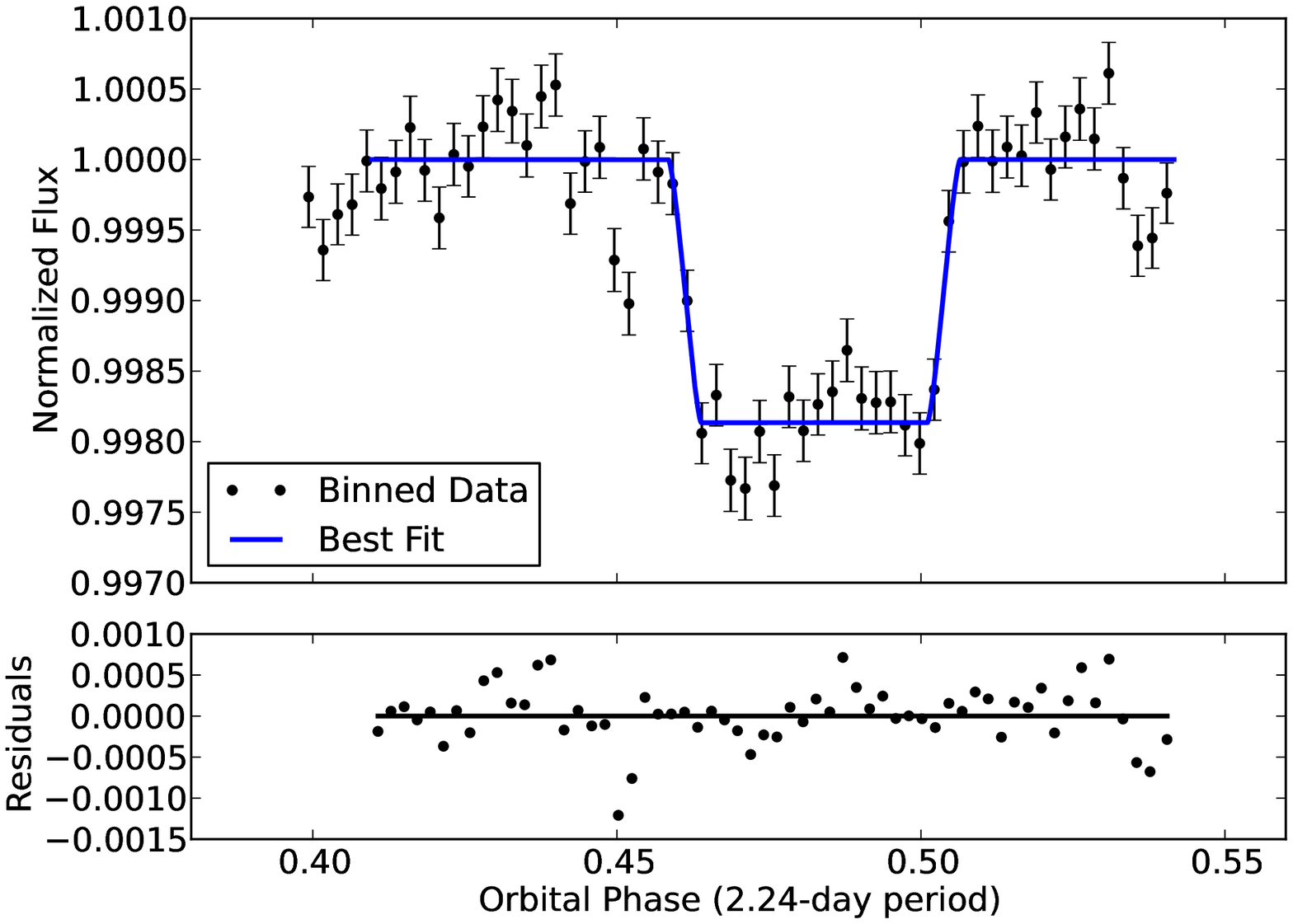}
\caption[Residuals for the channel 1 observations]
{Residuals for the channel 1 observations (lower panel) display some level of correlated noise both in and out of the eclipse.\\
(A color version of this figure is available in the online journal.)
}
\label{fig:resid}
\end{figure}

One would not expect a hot mid-F star (with a small convective zone) like WASP-14 to be active or to show much spot activity even if it were a moderate rotator.  Nonetheless, we analyzed the WASP light curve of WASP-14 to determine whether it shows periodic modulation due to the combination of magnetic activity and stellar rotation.  The stellar rotation values derived by \citet{Joshi2009-WASP14b} together with the estimated stellar radius imply a rotation period of about 12 days or more, assuming that the rotation axis of the star is approximately aligned with the orbital axis of the planet. We used the sine-wave fitting method described by \citet{Maxted2011-WASP41} to calculate a periodogram over 4096 uniformly spaced frequencies from 0 to 1.5 cycles day\sp{-1}. The false-alarm probability (FAP) for the strongest peak in these periodograms was calculated using a bootstrap Monte Carlo method also described by \citet{Maxted2011-WASP41}.

We did not find any significant periodic signals (FAP $<$ 0.05) in the WASP data, apart from frequencies near 1 cycle day\sp{-1}, which are due to instrumental effects. We examined the distribution of amplitudes for the most significant frequency in each Monte Carlo trial and used these results to estimate a 95\% upper confidence limit of 1 milli-magnitude (0.1\%) for the amplitude of any periodic signal in the lightcurve.

In our work on dozens of {\em Spitzer} eclipses, we have often found the same
channel to behave differently at different times, even on the
same star.  Our systematics removal algorithms correct the worst
effects, which are consistent, but there is sometimes still some
significant baseline scatter or oscillation.  While one might expect
certain kinds of stars to be relatively stable, {\em Spitzer} can
reach \math{\sigma\sim 0.01\%} eclipse-depth sensitivity, and
non-periodic stellar oscillations of this scale and at these wavelengths are not
well studied.  So, it is not fully clear whether these effects come
from the observatory or the star.

Since scatter and oscillation often
persist during an eclipse (when the planet is behind the star), and since a change in planetary signal of
the magnitude seen would generally mean an implausibly dramatic
change in the planet, we feel justified in treating the scatter or
oscillation phenomenologically.  In this case, our per-point uncertainties account for a global average of correlated noise.  MCMC accounts for any correlation between eclipse and model
parameters, and the rms versus bin size analysis, now including error bars, determined that the 
time correlation was not significant near the time scale of interest (Figure \ref{fig:rms1}).  Also, a larger uncertainty was assigned to the eclipse depth based on model ambiguity (above), which provides an additional margin of safety.

\subsection{Channel 2--4.5 {\micron}}
\label{sec:ch2}

Channel 2 and 4 were observed at the same time. We first modeled each channel separately, determining the best aperture size, time-variability (ramp) model, and bin size for BLISS. Then we applied a joint fit. For both channels 2 and 4, we again used the Bayesian informative prior for the values of ingress and egress times (\math{t\sb{\rm{2-1}}} = 1046.8 {\pm} 43.9 s), calculated from unpublished WASP photometric and radial-velocity data. All other parameters were left free.  

The observation in channel 2 lasted 5.5 hr. There was no stabilization period observed in the data, so no initial points were removed from the analysis. 

\begin{figure}[ht]
    \centering
    \includegraphics[width=0.70\linewidth, clip]{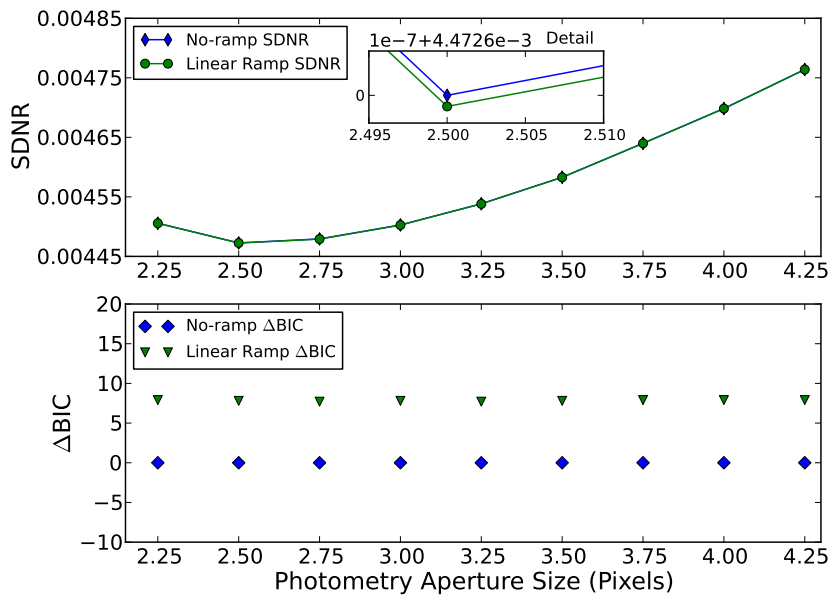}
\caption[Comparison between different ramp models in channel 2]
{Channel 2 comparison between linear and no ramp models. The plots show the SDNR and \math{\Delta}BIC vs. aperture size.  A lower value indicates a better model fit.\\
(A color version of this figure is available in the online journal.)
}
\label{fig:ch2-SDNR1}
\end{figure}

\begin{table}[ht]
\caption{\label{table:Ch2-ramps1} 
Channel 2 Ramp Models}
\atabon\strut\hfill\begin{tabular}{lcccc}
    \hline
    \hline
    Ramp Model    & SDNR           & BIC           & Eclipse Depth (\%)           \\
    \hline
    No Ramp      & 0.0044726       & 2964.2        & 0.224 {\pm} 0.012           \\
    Linear       & 0.0044725       & 2971.9        & 0.224 {\pm} 0.018           \\ 
    Quadratic    & 0.0044723       & 2979.9        & 0.241 {\pm} 0.025           \\ 
    Rising       & 0.0044726       & 2980.1        & 0.224 {\pm} 0.021           \\
    Lin+Log      & 0.0044690       & 2983.9        & 0.228 {\pm} 0.017           \\
    \hline
\end{tabular}\hfill\strut\ataboff
\end{table}

Following the criteria in Section\ \ref{sec:backgr}, we tested each of our ramp models (Table\ \ref{table:Ch2-ramps1}) at each of the aperture radii from 2.25--4.25 pixels in 0.25 pixel increments.  Figure \ref{fig:ch2-SDNR1} shows SDNR and \math{\Delta}BIC versus aperture size for our two best ramp models.  We note insignificantly different SDNR values between the two ramp models, which suggests that the best dataset (aperture radius of 2.50) does not depend on the model being fit.  The BIC favors the no-ramp model. The no-ramp model is 47 times more probable than the linear model.

We also tested the dependence of eclipse depth on aperture radius (see Figure \ref{fig:ch2-depths1}). The eclipse depths are well within 1\math{\sigma}.

\begin{figure}[ht]
    \centering
    \includegraphics[width=0.70\linewidth, clip]{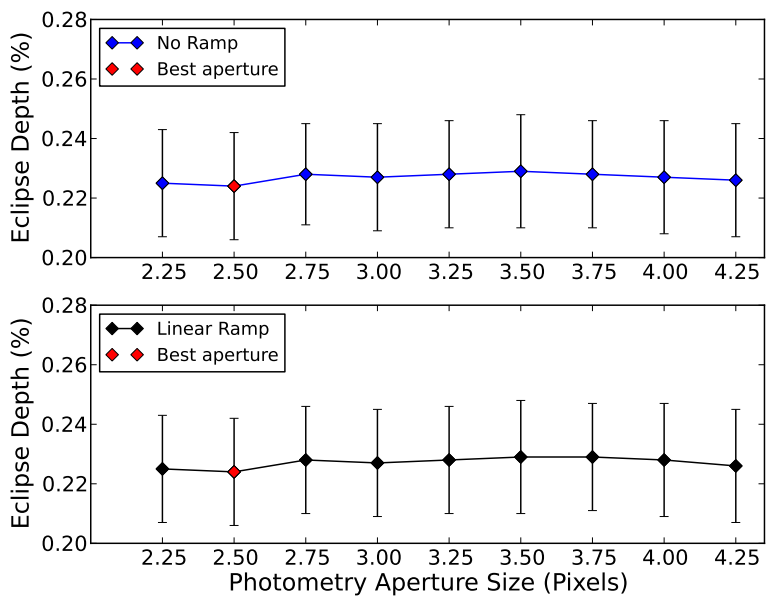}
\caption[Best-fit eclipse depths as a function of photometry aperture size for channel 2]
{Best-fit eclipse depths as a function of photometry aperture size for channel 2. The red point indicates the best aperture size for that channel. The eclipse-depth uncertainties are the result of 10\sp5 MCMC iterations. The trend shows insignificant dependence of eclipse depth on aperture size (much less than 1\math{\sigma}).\\
(A color version of this figure is available in the online journal.)
}
\label{fig:ch2-depths1}
\end{figure}

Prior to the science observations in channels 2 and 4, we observed a 212-frame preflash (see Section\ \ref{sec:obs}) on a diffuse, uniformly bright H\sb{II} emission region centered at $\alpha~=~10,45,02.2$, $\delta~=~-59,41,10.1$. The portion of the array within the aperture of the science observation in each channel was uniformly illuminated. For channel 2, the average flux within the 2.5 pixel aperture is \sim 200 MJy\,sr\sp{-1}, while for the 3.5 pixel aperture of channel 4 it is \sim 1800 MJy\,sr\sp{-1}. 

As expected, channel 2 shows no increase in flux during the preflash observation (see Figure\ \ref{fig:preflash-ch24}, left panel) nor during the main science observation (see Figure \ref{fig:RawBinNorm1}, raw data). The preflash observation in channel 4 saturated within the 30 minutes, eliminating the ramp effect in channel 4.

\begin{figure*}[htb]
\strut\hfill
\includegraphics[width=0.48\textwidth, clip]{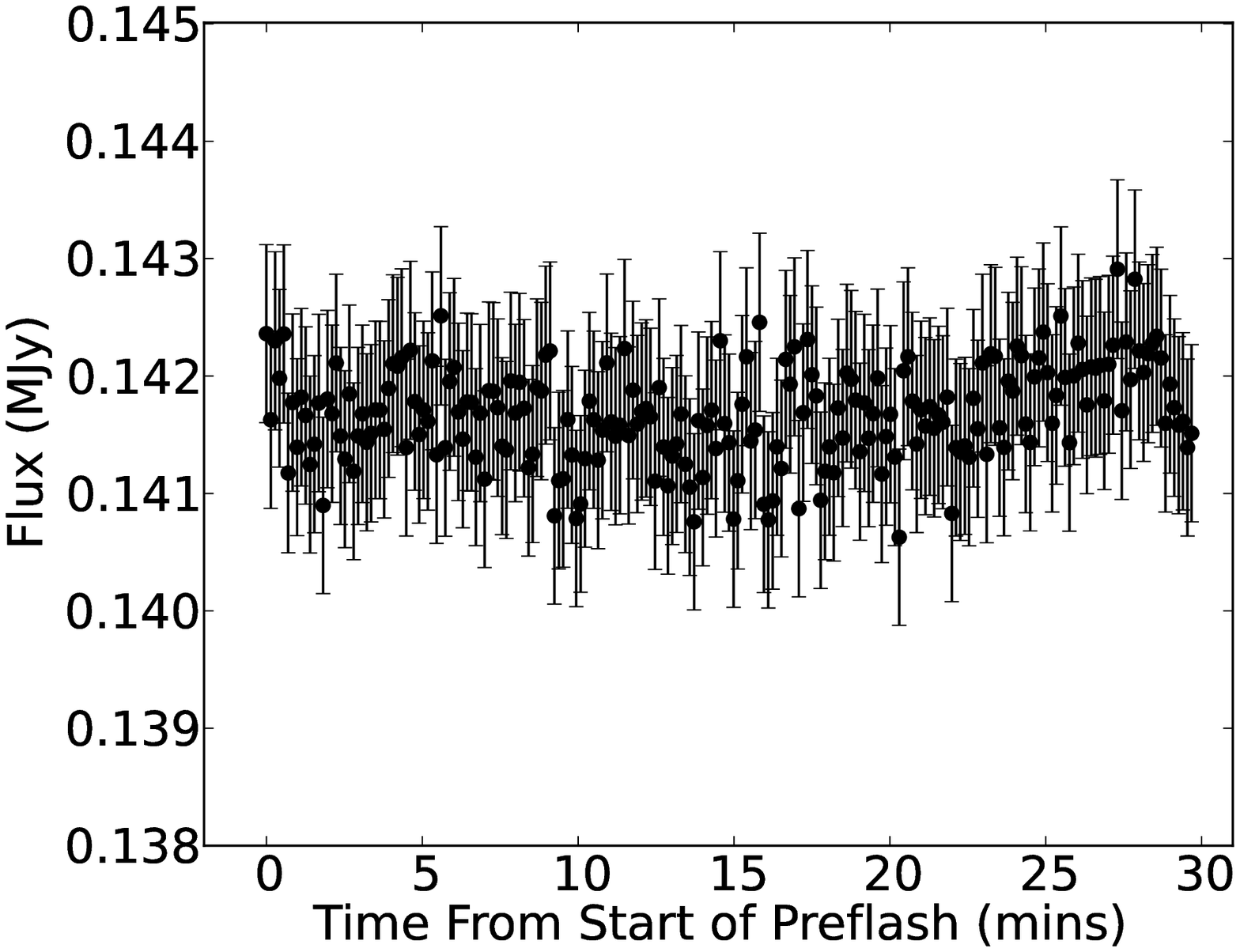}
\includegraphics[width=0.48\textwidth, clip]{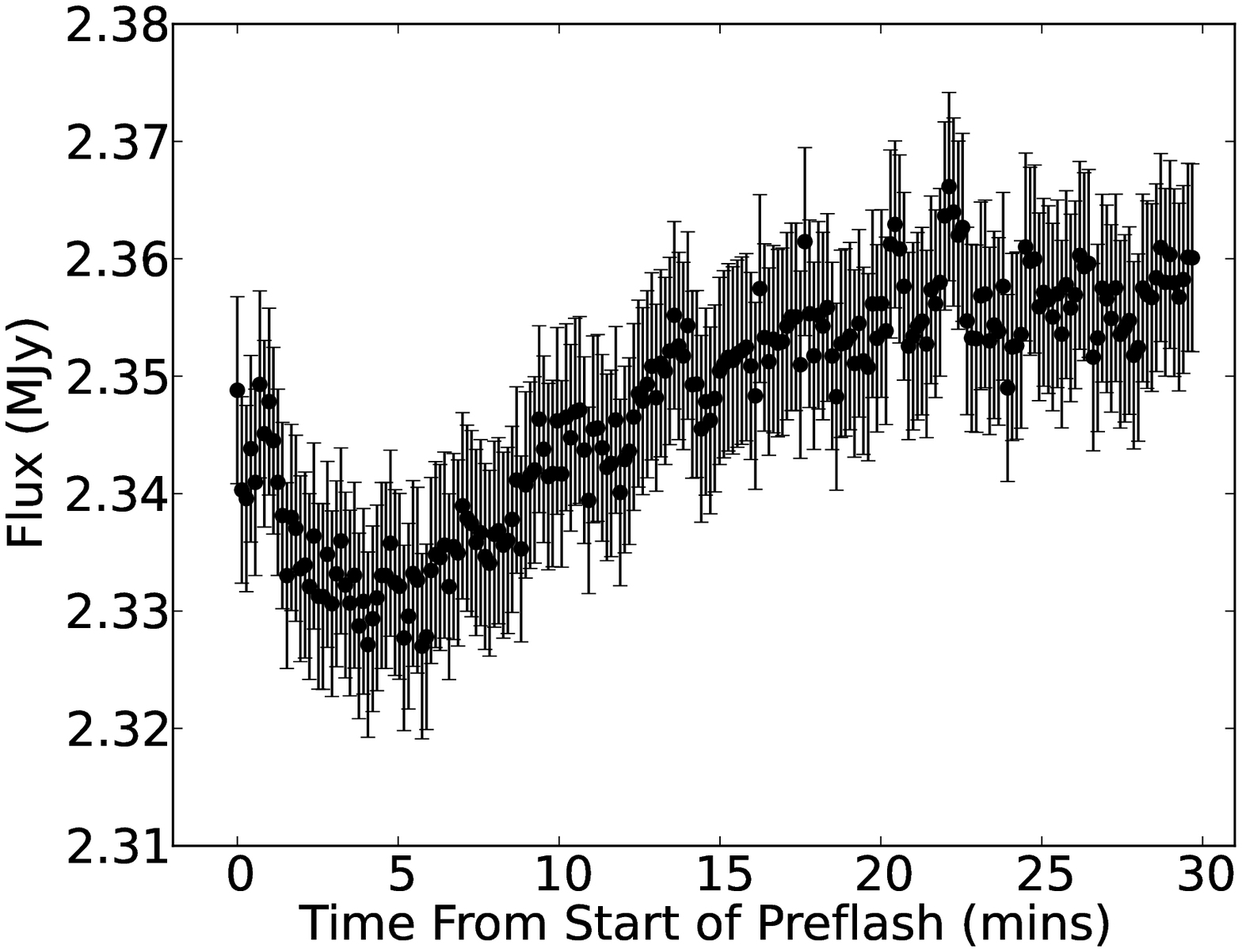}
\caption[Preflash light curves for channel 2 and 4]
{\label{fig:preflash-ch24} Preflash light curves for channel 2 (left) and channel 4 (right). The plots show binned data over 30 minutes of observation. The preflash source is a bright H\sb{II} emission region.  Without a preflash, the science observations would show a similar or possibly longer ramp in channel 4.}
\end{figure*}

Regardless of the preflash observations, we tested the full set of ramp equations and discarded obvious bad fits after shorter runs. Among acceptable fits, the lowest BIC value (see Table\ \ref{table:Ch2-ramps1}) determined that there is no significant ramp effect in the channel 2 dataset.

Each observation ended with a 10-frame, post-eclipse observation of blank sky in the same array position as the science observations to check for warm pixels in the photometric aperture. There were none.

To remove intrapixel variability we again apply our new BLISS technique, and also Equation\ (\ref{eq:vip}).  As with channel 1, the projection plot shows BLISS following significant variations that the polynomial does not fit well. The position precisions in channel 2 are 0.02 pixels for $x$ and 0.014 pixels for $y$. The best bin sizes are 0.028 pixels in $x$ and 0.023 pixels in $y$. The best aperture size, ramp model, and BLISS bin sizes are then used in our joint fit, which gave us the eclipse depths and the brightness temperatures in Section \ref{sec:joint}.

\subsection{Channel 4--8.0 {\micron}}
\label{sec:ch4}

Again, no stabilization period was observed in the 8.0 {\micron} dataset data set, hence no initial data points were removed. The preflash eliminated the ramp entirely, according to BIC (Table\ \ref{table:Ch4-ramps}).

\begin{table}[h]
\caption{\label{table:Ch4-ramps} 
Channel 4 Ramp Models}
\atabon\strut\hfill\begin{tabular}{lcccc}
    \hline
    \hline
    Ramp Model    & SDNR           & BIC           & Eclipse Depth (\%)           \\
    \hline
    No ramp      & 0.0039799       & 1459.2        & 0.181 {\pm} 0.013           \\
    Linear       & 0.0039770       & 1464.3        & 0.182 {\pm} 0.012           \\
    Rising       & 0.0039799       & 1466.4        & 0.198 {\pm} 0.030           \\ 
    Quadratic    & 0.0039763       & 1471.3        & 0.181 {\pm} 0.018           \\ 
    Lin+Log      & 0.0039799       & 1481.0        & 0.181 {\pm} 0.024           \\
    \hline
\end{tabular}\hfill\strut\ataboff
\end{table}

Figure \ref{fig:ch4-SDNR} plots the SDNR and \math{\Delta}BIC values versus aperture size at 8.0 {\micron}. For our two best ramp models (Table\ \ref{table:Ch4-ramps}) the smallest SDNR value is at 3.50 pixels (which determined our best aperture size), and the lowest BIC value at that aperture size is for the model without a ramp. We again test for the dependence of eclipse depth on aperture size (Figure \ref{fig:ch4-depths}).

\begin{figure}[ht]
    \centering
    \includegraphics[width=0.70\linewidth, clip]{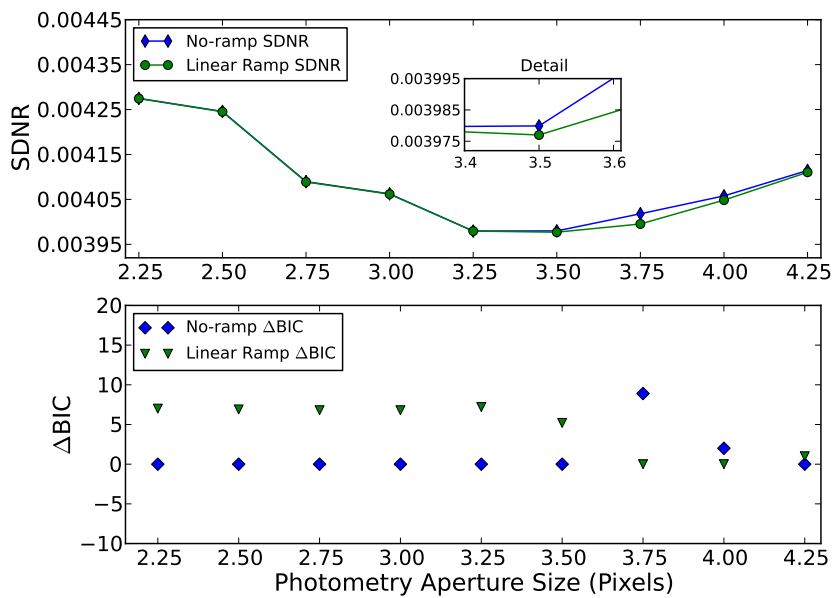}
\caption[Comparison between different ramp models for channel 4]
{Channel 4 comparison between linear and no-ramp models. The plots show SDNR and \math{\Delta}BIC vs. aperture size.  A lower value indicates a better model fit.\\
(A color version of this figure is available in the online journal.)
}
\label{fig:ch4-SDNR}
\end{figure}

\begin{figure}[ht]
    \centering
    \includegraphics[width=0.70\linewidth, clip]{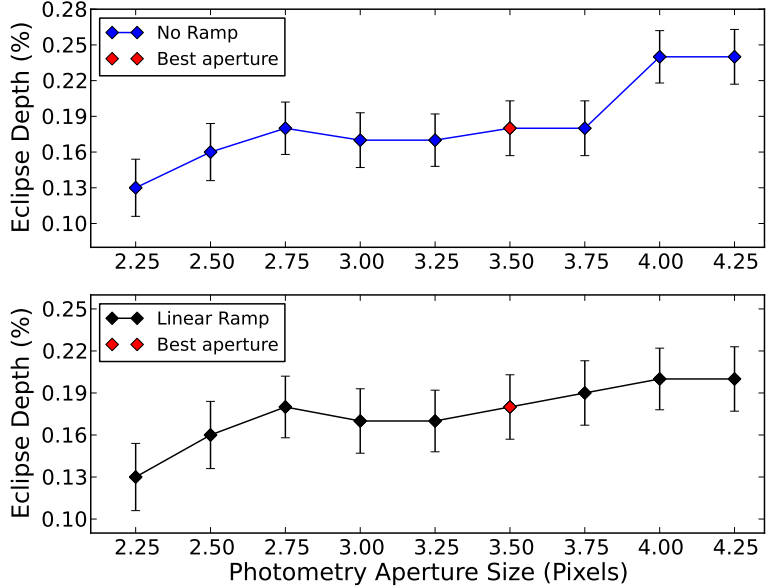}
\caption[Best-fit eclipse depths as a function of photometry aperture size for channel 4]
{Best-fit eclipse depths as a function of photometry aperture size for channel 4. The red point indicates the best aperture size for that channel. The eclipse-depth uncertainties are the result of 10\sp5 MCMC iterations. This channel has the lowest S/N (\sim 8). The aperture size of 2.25 pixels shows excess noise. Excluding it, the trend exhibits insignificant dependence of eclipse depth on aperture size (less than 1\math{\sigma}).\\
(A color version of this figure is available in the online journal.)
}
\label{fig:ch4-depths}
\end{figure}

Even though intrapixel variability is not so strong in channels 3 and 4, pixelation can be significant at any wavelength if the aperture is small (see \citealp{StevensonEtal2012apjHD149026b} and \citealp{Anderson2011-Ch24-WASP17b}). This justifies testing whether BLISS can give a better fit. Upon testing a full set of bin sizes, we concluded that NNI always outperforms BLI, indicating that variability from pixelation is insignificant.

\subsection{Joint Fit}
\label{sec:joint}

Our final models fit all data simultaneously.  The models shared a common eclipse duration for channels 1, 2, and 4 and a common midpoint time for channels 2 and 4, which were observed together.  We used the same priors as above.  The \citet{GelmanRubin1992} convergence diagnostic dropped below 1\% for all free parameters after 50,000 iterations.  Histograms for some interesting parameters for channel 1 appear on the left side of Figure \ref{fig:Corr-Hist}.  The middle plots show the pairwise correlations (marginal distributions) of these parameters.  The histograms on the right are for the joint fit of channels 2 and 4. All other histograms are similarly Gaussian, confirming that the phase space minimum is global and defining the parameter uncertainties.  Tables\ \ref{tab:eclfits-lin} and \ref{tab:eclfits-sin} report two joint-fit results for our two best ramp models in channel 1 (linear and sinusoidal), along with photometric results and modeling choices from the individual fits. Light-curve files including the best-fit models, centering data, photometry, etc., are included as electronic supplements to this article.


\begin{figure*}[h!]
\centerline{
\includegraphics[height=5.9cm, width=5cm ]{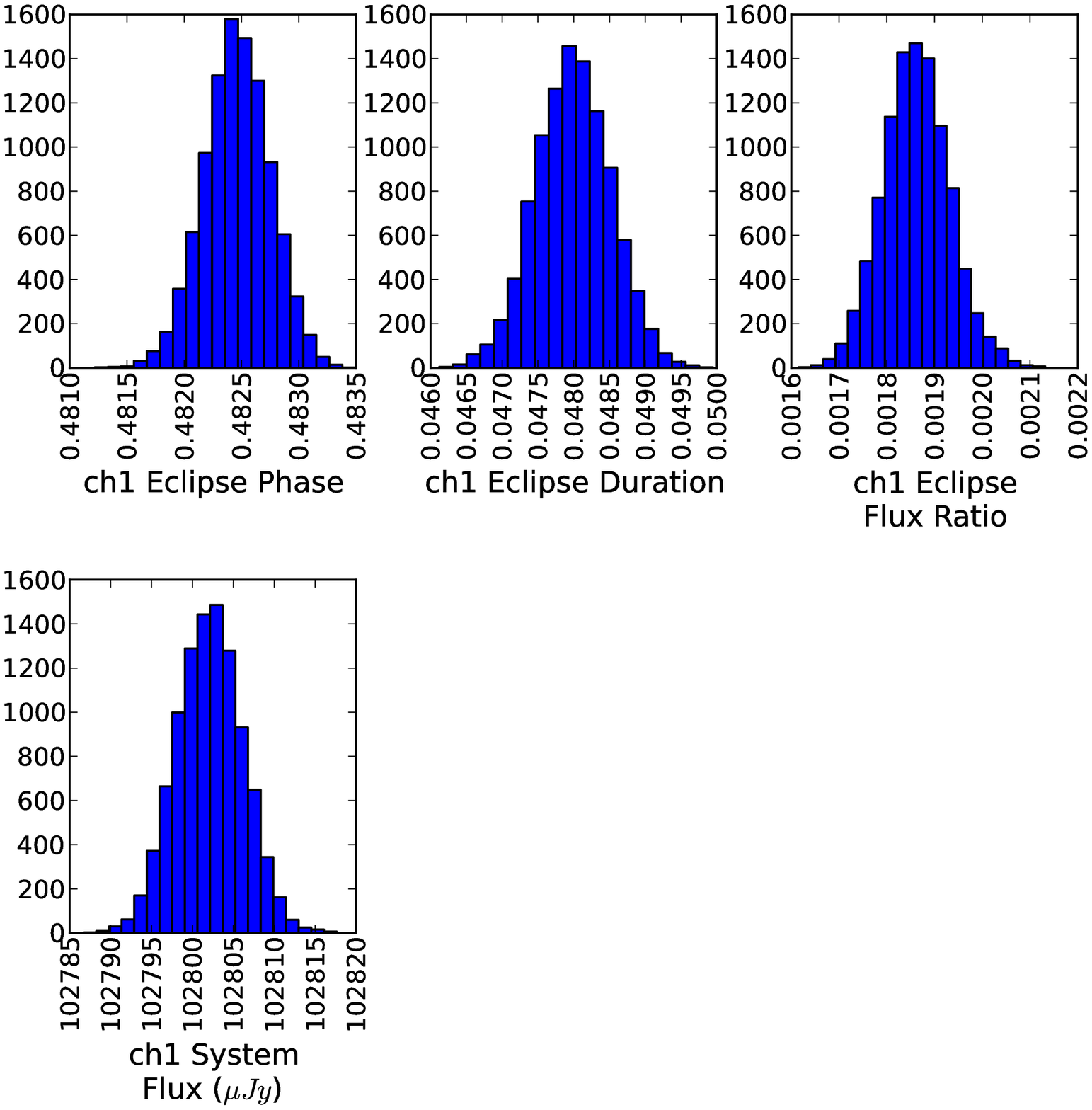}
\includegraphics[height=6cm, clip]{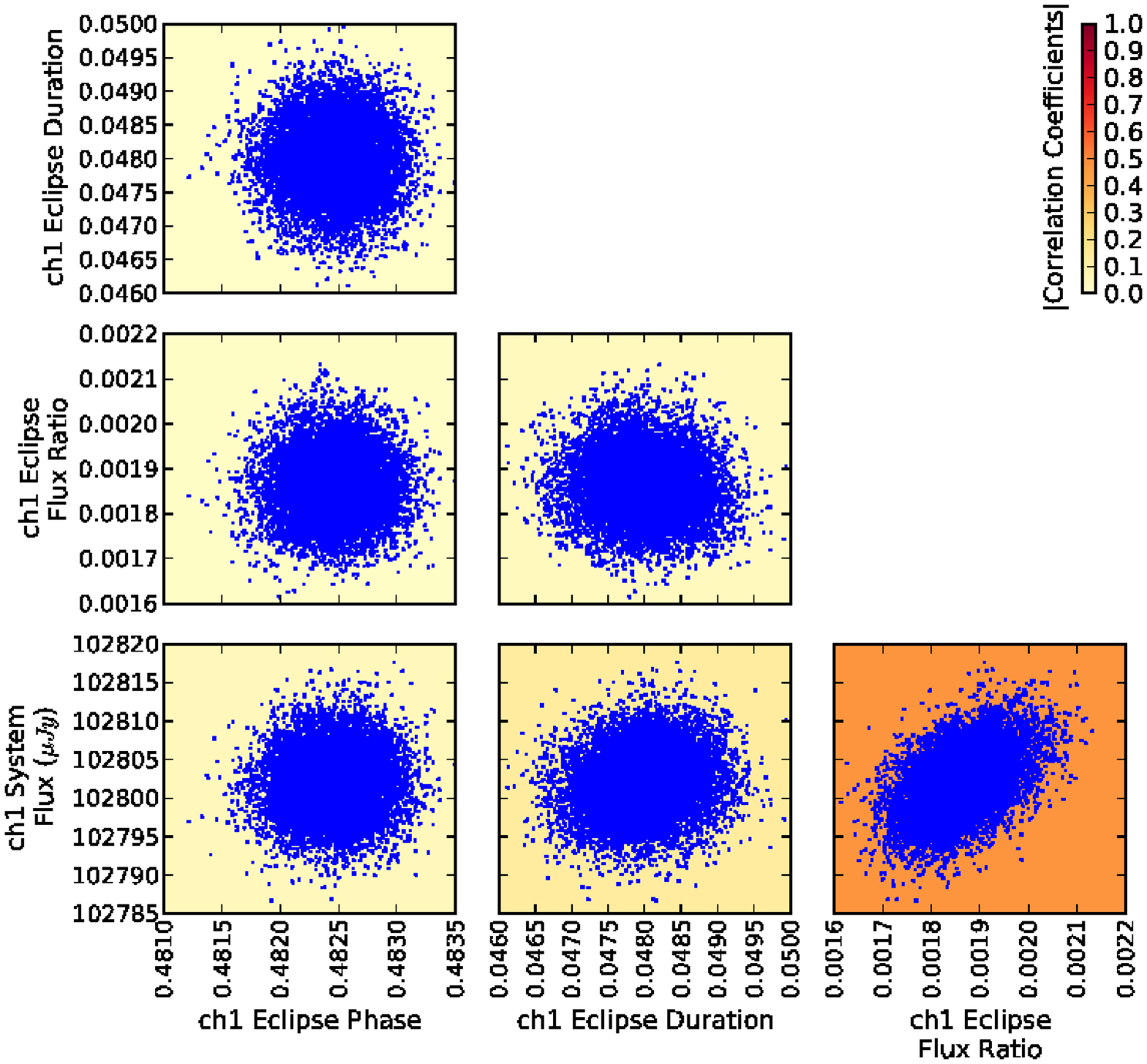}
\includegraphics[height=6cm, width=5cm ]{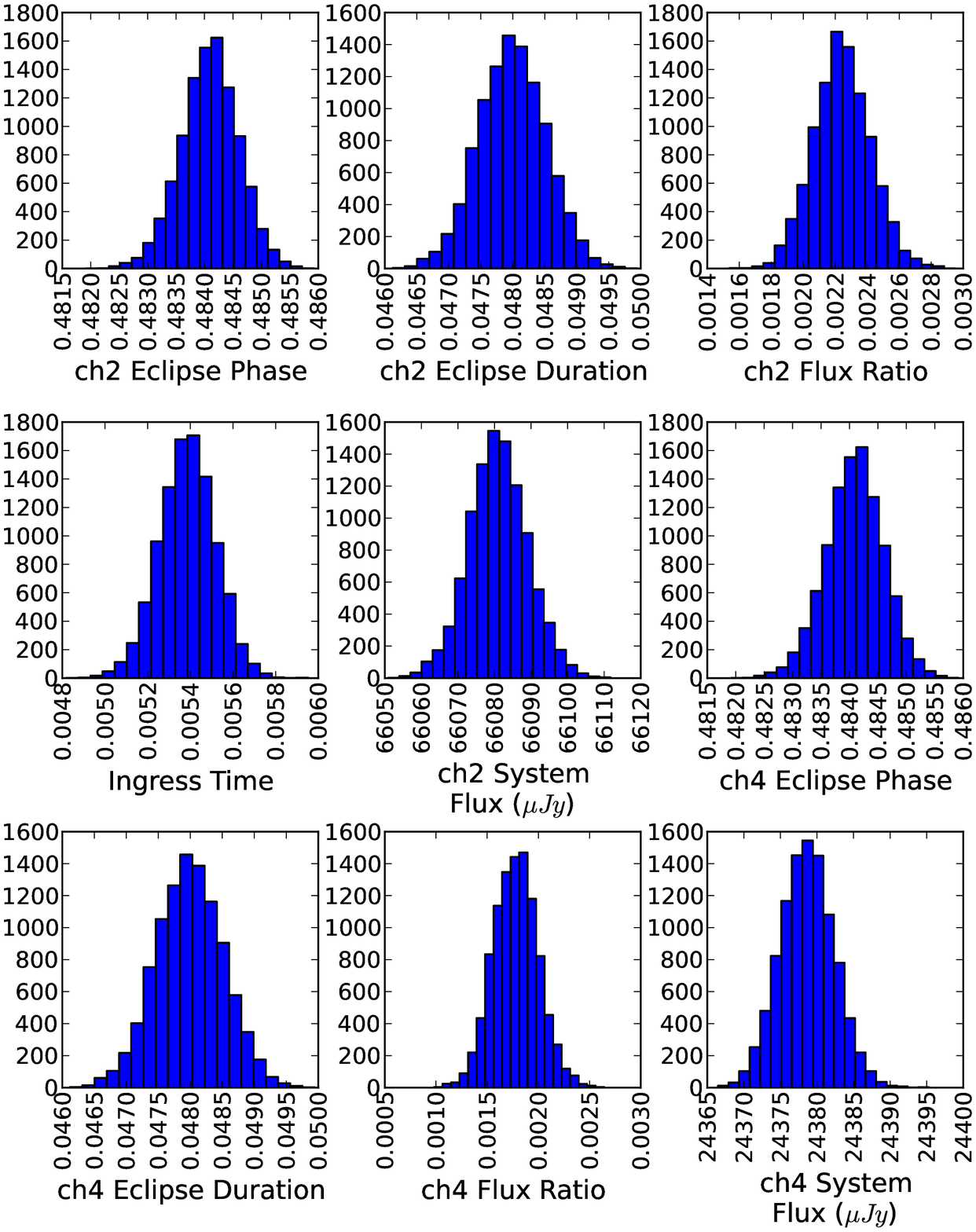}}
\caption[Sample parameter correlations and histograms for channel 1, 2, and 4]
{Left and Center: sample parameter histograms and parameter correlations for channel 1. The background color depicts the absolute value of the correlation coefficient.
Right: sample parameter histograms for channel 2 and channel 4, produced in the joint fit. All other parameter histograms are similarly Gaussian. Every 10th step in the MCMC chain is used to decorrelate consecutive values.\\
(A color version of this figure is available in the online journal.)
}
\label{fig:Corr-Hist}
\end{figure*}

\section{ORBIT}
\label{sec:orbit}

We fit the midpoint times from the {\em Spitzer}\/ lightcurves simultaneously with
the available radial velocity curves and transit photometry in order to
provide updated estimates of system orbital parameters. The timing of secondary eclipse is a strong constraint on the shape and orientation of the
orbit. The two eclipses for the linear and sinusoidal joint fit (Tables\ \ref{tab:eclfits-lin} and \ref{tab:eclfits-sin}) have an insignificant difference in phases (less than 0.5\math{\sigma}), and the linear joint fit has slightly lower BIC value. Hence, we picked the linear joint fit phases for the use in the orbital analysis. The two eclipses occur at phases 0.4825 {\pm} 0.0003 and  0.4841 {\pm} { 0.0005} (using the \citealp{Joshi2009-WASP14b} ephemeris), with a weighted mean after a 37 s eclipse-transit light-time correction of 0.48273 {\pm} 0.00025, indicating that \math{e \cos \omega} = - 0.0271 {\pm} 0.0004. The phases differ from each other by approximately 3\math{\sigma}, but depend strongly on the accuracy of the ephemeris used to compute them.

We fit a Keplerian orbit model to our secondary eclipse times along with
radial velocity data from \cite{Husnoo2011-WASP14b} and
\cite{Joshi2009-WASP14b}, and transit timing data from both amateur
observers and WASP-14b's discovery paper
\citep{Joshi2009-WASP14b}.  The entire data set comprised 38 RV points, six
of which were removed due to the Rossiter-McLaughlin effect, 30 transits,
and two eclipses (see Table\ \ref{tab:ttv1}). All times were adjusted to BJD\sb{TDB}
\citep{EastmanEtal2010apjLeapSec}. The errors were estimated using
our MCMC routine.   This fit gave \math{e = 0.087 \pm
 0.002} and \math{\omega = 107°.1 {\pm} 0°.5}. We did not adjust for any anomalous eccentricity signal from the stellar tidal bulge as described by \citet{ArrasEtal2012MNRAS-RVstar} because the predicted amplitude of this effect is smaller than the uncertainty on the eccentricity, and much smaller than the eccentricity itself. 


\begin{landscape}
\begin{table*}[ht]
\centering
\caption{\label{tab:eclfits-lin} Joint Best-fit Eclipse Light-curve Parameters (Channel 1--Linear Ramp)}
\begin{tabular}{lr@{\,{\pm}\,}lr@{\,{\pm}\,}lr@{\,{\pm}\,}l}
\hline
\hline
Parameter                                                          &   \mctc{  Channel 1       }    &   \mctc{   Channel 2      }    &   \mctc{    Channel 4     }        \\
\hline
Array position (\math{\bar{x}}, pixel)                               &   \mctc{      14.16       }    &   \mctc{      23.82       }    &   \mctc{       24.6       }        \\
Array position (\math{\bar{y}}, pixel)                               &   \mctc{      15.69       }    &   \mctc{      24.11       }    &   \mctc{       21.9       }        \\
Position consistency\tablenotemark{a} (\math{\delta\sb{\rm x}}, pixel)   &   \mctc{      0.005       }    &   \mctc{      0.02        }    &   \mctc{      0.021       }        \\
Position consistency\tablenotemark{a} (\math{\delta\sb{\rm y}}, pixel)   &   \mctc{      0.012       }    &   \mctc{      0.014       }    &   \mctc{      0.025       }        \\
Aperture size (pixel)                                                &   \mctc{      2.75        }    &   \mctc{       2.5        }    &   \mctc{      3.5         }        \\
Sky Annulus inner radius (pixel)                                     &   \mctc{       8.0        }    &   \mctc{      12.0        }    &   \mctc{      12.0        }        \\
Sky Annulus outer radius (pixel)                                     &   \mctc{      20.0        }    &   \mctc{      30.0        }    &   \mctc{      30.0        }        \\
System flux \math{F\sb{\rm s}} (\micro Jy)                         &          102802 & 4            &         66083 & 7              &         24381 & 3              \\
Eclipse depth (\%)                                                 &         0.187 & 0.007          &         0.224 & 0.018          &         0.181 & 0.022              \\
Brightness temperature (K)                                         &         2225 &  39             &         2212  & 94             &        1590 & 116                  \\
Eclipse midpoint (orbits)                                          &        0.4825 & 0.0003         &        0.4842 & 0.0005         &        0.4842 & 0.0005             \\
Eclipse midpoint (BJD\sb{UTC} --2,450,000)                         &      5274.6609 & 0.0006        &      4908.9290 & 0.0011        &      4908.9290 & 0.0011            \\
Eclipse midpoint (BJD\sb{TDB} --2,450,000)                         &      5274.6617 & 0.0006        &      4908.9298 & 0.0011        &      4908.9298 & 0.0011            \\
Eclipse duration  (\math{t\sb{\rm 4-1}}, hrs)                      &          2.59 & 0.03           &         2.59 & 0.03            &         2.59 & 0.03                \\
Ingress/egress time (\math{t\sb{\rm 2-1}}, hrs)                   &         0.290 & 0.007          &         0.290 & 0.007          &          0.290 & 0.007             \\
Ramp name                                                          &   \mctc{     linear       }    &   \mctc{       ...        }    &   \mctc{       ...        }        \\
Ramp, linear term (\math{r\sb{\rm 0}})                             &          0.0044 & 0.0010       &   \mctc{       ...        }    &   \mctc{       ...        }        \\
Intrapixel method                                                  &   \mctc{      BLISS       }    &   \mctc{     BLISS        }    &   \mctc{       ...        }        \\
BLISS bin size in $x$  (pixel)                                       &   \mctc{     0.004        }    &   \mctc{     0.028        }    &   \mctc{       ...        }        \\
BLISS bin size in $y$  (pixel)                                       &   \mctc{     0.01         }    &   \mctc{     0.023        }    &   \mctc{       ...        }        \\
Minimum number of points per bin                                   &   \mctc{        4         }    &   \mctc{        5         }    &   \mctc{       ...        }        \\
Total frames                                                       &   \mctc{      13693       }    &   \mctc{       2972       }    &   \mctc{       1432       }        \\
Rejected frames (\%)                                               &   \mctc{     0.49         }    &   \mctc{     0.34         }    &   \mctc{     3.89         }        \\
Free parameters                                                    &   \mctc{        6         }    &   \mctc{        3         }    &   \mctc{        2         }        \\
AIC value                                                          &   \mctc{  16695.8         }    &   \mctc{   16695.8        }    &   \mctc{     16695.8      }        \\
BIC value                                                          &   \mctc{  16780.7         }    &   \mctc{   16780.7        }    &   \mctc{     16780.7      }        \\
SDNR                                                               &   \mctc{  0.003311        }    &   \mctc{   0.004473       }    &   \mctc{    0.003980      }        \\
Uncertainty scaling factor                                         &   \mctc{  0.031968        }    &   \mctc{   0.294486       }    &   \mctc{     0.342520     }        \\
Photon-limited S/N (\%)                                            &   \mctc{    72.7          }    &   \mctc{    90.4          }    &   \mctc{       68.1       }        \\
\hline\end{tabular}
\begin{minipage}[t]{0.86\linewidth}
\sp{a}{rms frame-to-frame position difference.}
\end{minipage}
\end{table*}
\end{landscape}


\begin{landscape}
\begin{table*}[ht]
\centering
\caption{\label{tab:eclfits-sin} Joint Best-fit Eclipse Light-curve Parameters (Channel 1--Sinusoidal Ramp)}
\begin{tabular}{lr@{\,{\pm}\,}lr@{\,{\pm}\,}lr@{\,{\pm}\,}l}
\hline
\hline
Parameter                                                          &   \mctc{   Channel 1  }        &   \mctc{   Channel 2   }      &   \mctc{    Channel 4     }     \\
\hline
Array position (\math{\bar{x}}, pixel)                               &   \mctc{      14.16       }    &   \mctc{      23.82       }    &   \mctc{       24.6       }      \\
Array position (\math{\bar{y}}, pixel)                               &   \mctc{      15.69       }    &   \mctc{      24.11       }    &   \mctc{       21.9       }      \\
Position consistency\tablenotemark{a} (\math{\delta\sb{\rm x}}, pixel)   &   \mctc{      0.005       }    &   \mctc{       0.02       }    &   \mctc{      0.021       }       \\
Position consistency\tablenotemark{a} (\math{\delta\sb{\rm y}}, pixel)   &   \mctc{      0.012       }    &   \mctc{      0.014       }    &   \mctc{      0.025       }       \\
Aperture size (pixel)                                                &   \mctc{      2.75        }    &   \mctc{      2.5         }    &   \mctc{      3.5         }      \\
Sky Annulus inner radius (pixel)                                     &   \mctc{      8.0         }    &   \mctc{     12.0         }    &   \mctc{     12.0         }       \\
Sky Annulus outer radius (pixel)                                     &   \mctc{     20.0         }    &   \mctc{     30.0         }    &   \mctc{     30.0         }      \\
System flux \math{F\sb{s}} (\micro Jy)                             &       102616 & 7               &         66083 & 7              &       24381 & 3             \\
Eclipse depth (\%)                                                 &         0.193 & 0.007          &         0.224 & 0.017          &         0.181 & 0.021              \\
Brightness temperature (K)                                         &         2258 &  38             &         2212  & 89             &        1590 & 111                  \\
Eclipse midpoint (orbits)                                          &         0.4825 & 0.0003        &          0.4843 & 0.0005       &        0.4843 & 0.0005           \\
Eclipse midpoint (BJD\sb{UTC} --2,450,000)                         &      5274.6609 & 0.0006        &       4908.9291 & 0.0011        &     4908.9291 & 0.0011           \\
Eclipse midpoint (BJD\sb{TDB} --2,450,000)                         &      5274.6617 & 0.0006        &       4908.9298 & 0.0011        &     4908.9298 & 0.0011           \\
Eclipse duration (\math{t\sb{\rm 4-1}}, hrs)                      &           2.59 & 0.03          &            2.59 & 0.03         &          2.59 & 0.03            \\
Ingress/egress time (\math{t\sb{\rm 2-1}}, hrs)                   &          0.290 & 0.007         &           0.290 & 0.007        &         0.290 & 0.007            \\
Ramp name                                                          &   \mctc{     sinusoidal   }    &   \mctc{      ...          }   &   \mctc{       ...         }     \\
Ramp, cosine phase offset (\math{t\sb{\rm 2}})                         &           0.5356 & 0.0016      &   \mctc{      ...           }   &   \mctc{      ...         }      \\
Intrapixel method                                                  &   \mctc{      BLISS       }    &   \mctc{     BLISS         }   &   \mctc{      ...         }      \\
BLISS bin size in $x$  (pixel)                                       &   \mctc{     0.004        }    &   \mctc{     0.028        }    &   \mctc{       ...         }        \\
BLISS bin size in $y$  (pixel)                                       &   \mctc{     0.01         }    &   \mctc{     0.023        }    &   \mctc{      ...         }        \\
Minimum number of points per bin                                   &   \mctc{        4         }    &   \mctc{        5          }   &   \mctc{        ...        }      \\
Total frames                                                       &   \mctc{      13693       }    &   \mctc{       2972        }   &   \mctc{       1432       }      \\
Rejected frames (\%)                                               &   \mctc{    0.49          }    &   \mctc{      0.34         }   &   \mctc{    3.89          }      \\
Free parameters                                                    &   \mctc{        6         }    &   \mctc{        3          }   &   \mctc{        2         }      \\
AIC value                                                          &   \mctc{  16695.9         }    &   \mctc{  16695.9          }   &   \mctc{  16695.9         }      \\
BIC value                                                          &   \mctc{  16780.8         }    &   \mctc{  16780.8          }   &   \mctc{  16780.8         }      \\
SDNR                                                               &   \mctc{ 0.003316         }    &   \mctc{ 0.004473          }   &   \mctc{ 0.003980         }      \\
Uncertainty scaling factor                                         &   \mctc{  0.031968        }    &   \mctc{  0.294485         }   &   \mctc{  0.342520        }      \\
Photon-limited S/N (\%)                                            &   \mctc{       72.6       }    &   \mctc{       90.4        }   &   \mctc{       68.1       }      \\
\hline\end{tabular}
\begin{minipage}[t]{0.86\linewidth}
\sp{a}{rms frame-to-frame position difference.}
\end{minipage}
\end{table*}
\end{landscape}

\begin{table}[hb!]
\vspace{-30pt}
\centering
\caption{\label{tab:ttv1} Transit Timing Data}
\begin{tabular}{lcl}
\hline
\hline
Mid-transit Time (BJD\sb{TDB})  &  Uncertainty          &   Source\tablenotemark{a}                                     \\
\hline
2455695.4082	                & 0.0012	        & V.\ Slesarenkno, E.\ Sokov \tablenotemark{b}                    \\
2455668.4790	                & 0.0011		& Franti\^{s}ek Lomoz						\\
2455652.7744	                & 0.0014		& Stan Shadick, C.\ Shiels\tablenotemark{c}		        \\
2455650.5307	                & 0.0018		& Lubos Br\'{a}t						\\
2455650.52789	                & 0.00076	        & Martin Vra\v{s}t'\'{a}k					\\
2455650.52566	                & 0.00067		& Jaroslav Trnka\tablenotemark{d}				\\
2455632.5807	                & 0.0011		& E.\ Sokov, K.\ N.\ Naumov \tablenotemark{b}                      \\
2455318.45101	                & 0.00085		& Anthony Ayiomamitis						\\
2455302.7464	                & 0.0010		& Stan Shadick\tablenotemark{c}					\\
2455264.6021	                & 0.0012		& Hana Ku\v{c}\'{a}kov\'{a}\tablenotemark{e}			\\
2455264.6017	                & 0.0013		& Radek Koci\'{a}n\tablenotemark{f}				\\
2455219.7290	                & 0.0012		& Lubos Br\'{a}t						\\
2454979.643		        & 0.003			& Wiggins, AXA						        \\
2454968.426		        & 0.001			& Srdoc, AXA						        \\
2454950.4831	                & 0.0021		& Jesionkiewicz, AXA						\\
2454950.4746                	& 0.0014		& Lubos Br\'{a}t						\\
2454950.4745	                & 0.0018		& Hana Ku\v{c}\'{a}kov\'{a}\tablenotemark{e}			\\
2454950.4731	                & 0.0021		& Pavel Marek						        \\
2454950.4728	                & 0.0014		& Wardak, AXA						        \\
2454943.7427	                & 0.0006		& Dvorak, AXA						        \\
2454941.49799	                & 0.00081		& Jaroslav Trnka\tablenotemark{d}				\\
2454941.4916	                & 0.0019		& Franti\v{s}ek Lomoz						\\
2454934.765		        & 0.001			& Brucy Gary, AXA						\\
2454932.5246	                & 0.0014		& Radek D\v{r}ev\v{e}n\'{y}					\\
2454932.5232	                & 0.0011		& Lubos Br\'{a}t						\\
2454932.5222	                & 0.0013		& Jaroslav Trnka\tablenotemark{d}				\\
2454932.5219	                & 0.0015		& T.\ Hynek, K.\ Onderkov\'{a}					\\
2454914.5753	                & 0.0008		& Naves, AXA						        \\
2454887.6457	                & 0.0014		& Georgio, AXA						        \\
\hline
\end{tabular}
\begin{minipage}[t]{0.94\linewidth}
\footnotesize{
\sp{a}{The Amateur Exoplanet Archive (AXA), http://brucegary.net/AXA/x.htm) and Transiting ExoplanetS and Candidates group (TRESCA), http://var2.astro.cz/EN/tresca/index.php) supply their data to the Exoplanet Transit Database (ETD), http://var2.astro.cz/ETD/), which performs the uniform transit analysis described by \citet{Poddany2010}. The ETD Web site provided the AXA and TRESCA numbers in this table, which were converted to BJD\sb{TDB}.\\}
\sp{b}{Sokov E., Naumov K., Slesarenko V.\ et al., Pulkovo Observatory of RAS, Saint-Petersburg, Russia.\\}
\sp{c}{Physics and Engineering Physics Department, University of Saskatchewan, Saskatoon, Saskatchewan, Canada, S7N 5E2.\\}
\sp{d}{Municipal Observatory in Slany Czech Republic.\\}
\sp{e}{Project Eridanus, Observatory and Planetarium of Johann Palisa in Ostrava.\\}
\sp{f}{Koci\'{a}n R., Johann Palisa, Observatory and Planetarium, Technical University Ostrava, 17.\ Listopadu 15, CZ-708 33 Ostrava, Czech Republic.\\}
}
\end{minipage}
\vspace{-30pt}
\end{table}

With our new data, we
refine the ephemeris to \math{T\sb{\rm BJD\sb{\rm TDB}} = 2454827.06666(24) + 
2.2437661(11)\,N}, where \math{T} is the time of transit and \math{N} is the
number of orbits elapsed since the transit time (see Table\ \ref{tab:orbit1}).  We find that the new
ephemeris reduces the difference between the two eclipse phases to less than
1.6\math{\sigma}.  Performing an ephemeris fit to the transit and eclipse data
separately shows that the transit and eclipse periods differ by (1.1 \pm
{ 0.8}) \math{\times} 10\sp{-5} days, a 1.5\math{\sigma} result that limits apsidal motion,
\math{\dot{\omega}}, to less than 0°.0024 day\sp{-1} at the 3\math{\sigma} level
\citep{GimenezBastero1995}. 

\vspace{20pt}
\begin{table*}[ht!]
\centering
\caption{\label{tab:orbit1} Eccentric Orbital Model}
\begin{tabular}{lr@{\,{\pm}\,}lr@{\,{\pm}\,}}
\hline
\hline 
Parameter                                                          &  \mctc{Value}                       \\
\hline
\math{e \sin \omega}\tablenotemark{a}     	                   &  0.0831               & 0.0021      \\
\math{e \cos \omega}\tablenotemark{a}     	                   &  \math{-0.02557}      & 0.00038     \\
\math{e}                          			           &  0.087                & 0.002       \\
\math{\omega} (°)           			           &  \math{-107.1}        & 0.5         \\
\math{P} (days)\tablenotemark{a}      		                   &  2.2437661            & 0.0000011   \\
\math{T\sb{\rm 0}} \tablenotemark{a}\tablenotemark{b}              &  2454827.06666        & 0.00024     \\
\math{K} (m\,s\sp{-1})\tablenotemark{a}\tablenotemark{c}           &  990                  & 3           \\
\math{\gamma} (m\,s\sp{-1})\tablenotemark{a}\tablenotemark{d}      &  \math{-4987.9}      & 1.6         \\
\math{\chi^2}				                           &  \mctc{162}                         \\
\hline                                           
\end{tabular}
\footnotesize{
\begin{minipage}[ht!]{0.53\linewidth}
\hspace{0.7cm} \hangindent=2.4em \sp{a}{Free parameter in MCMC fit.}\\ 
\sp{b}{BJD\sb{TDB}.}\\
\sp{c}{Radial velocity semi-amplitude.}\\
\sp{d}{Radial velocity offset.}
\end{minipage}
}
\end{table*}

The results confirm an eccentric orbit for WASP-14b and improve knowledge of
other orbital parameters.

\section{ATMOSPHERE}
\label{sec:atm}

We explore the model parameter space in search of the best-fitting models for a given data set. The model parameterization is described by \citet{MadhusudhanSeager2009, MadhusudhanSeager2010, Madhusudhan2012}. The sources of opacity in the model include molecular absorption due to H\sb{2}O, CO, CH\sb{4}, CO\sb{2}, TiO, and VO, and collision-induced absorption (CIA) due to H\sb{2}-H\sb{2}. Our molecular line lists are obtained from \citet{Freedman08}, R. S. Freedman (2009, private communication), \citet{Rothman2005-HITRAN, KarkoschkaTomasko2010}, and E. Karkoschka (2011, private communication). Our CIA opacities are obtained from \citet{Borysow1997} and \citet{Borysow2002}. We explore the model parameter space using a MCMC scheme, as described by \citet{MadhusudhanSeager2010}. However, since the number of model parameters (\math{n} = 10) exceed the number of data points (\math{N\sb{\rm{data}}} = 3), our goal is not to find a unique fit to the data but, primarily, to identify regions of model phase space that the data exclude. In order to compute the model planet-star flux ratios to match with the data, we divide the planetary spectrum by a Kurucz model of the stellar spectrum derived from \citet{Castelli2004}. Our models allow constraints on the temperature structure, molecular mixing ratios, and a joint constraint on the albedo and day-night redistribution.

We find that strong constraints can be placed on the presence of a thermal inversion in WASP-14b even with our current small set of observations. At an irradiation of  3 \math{\times} 10\sp{9} erg\,s\sp{-1}\,cm\sp{-2}, WASP-14b falls in the class of extremely irradiated planets that are predicted to host thermal inversions according to the TiO/VO hypothesis of \citet{Fortney2008}. However, the present observations do not show any distinct evidence of a thermal inversion in the dayside atmosphere of WASP-14b. We explored the model parameter space by running \sim 10\sp{6} models with and without thermal inversions, using an MCMC scheme as discussed above. We found that the data could not be explained by a thermal inversion model for any chemical composition. On the other hand, the data are easily fit by models with no thermal inversions. While the brightness temperatures in the 3.6 and 4.5 {\micron} channels are consistent with a blackbody spectrum of the planet at \math{T}\sim 2200 K, the 8 {\micron} flux deviates substantially from the assumption of a blackbody with a brightness temperature of 1668 {\pm} 125 K. In the presence of a thermal inversion, the flux in the 8 {\micron} channel is expected to be much higher than the fluxes in the 3.6 and 4.5 {\micron} channels due to emission features of water vapor and, if present, methane. The low flux observed at 8 {\micron}, therefore, implies strong water vapor and/or methane in absorption, implying the lack of a significant temperature inversion (see \citealp{MadhusudhanSeager2010} for a discussion on inferring thermal inversions). We also note that, as mentioned in Section\ \ref{sec:ch1}, the observations yield different planet--star flux contrasts in the 3.6 {\micron} channel for different choices of ramp models. However, as shown in Figure \ref{fig:spectra}, the two extreme values are still consistent at the 1\math{\sigma} level, and as such, lead to similar model conclusions.


\begin{figure}[hb!]
    \centering
    \includegraphics[width=0.80\textwidth, clip]{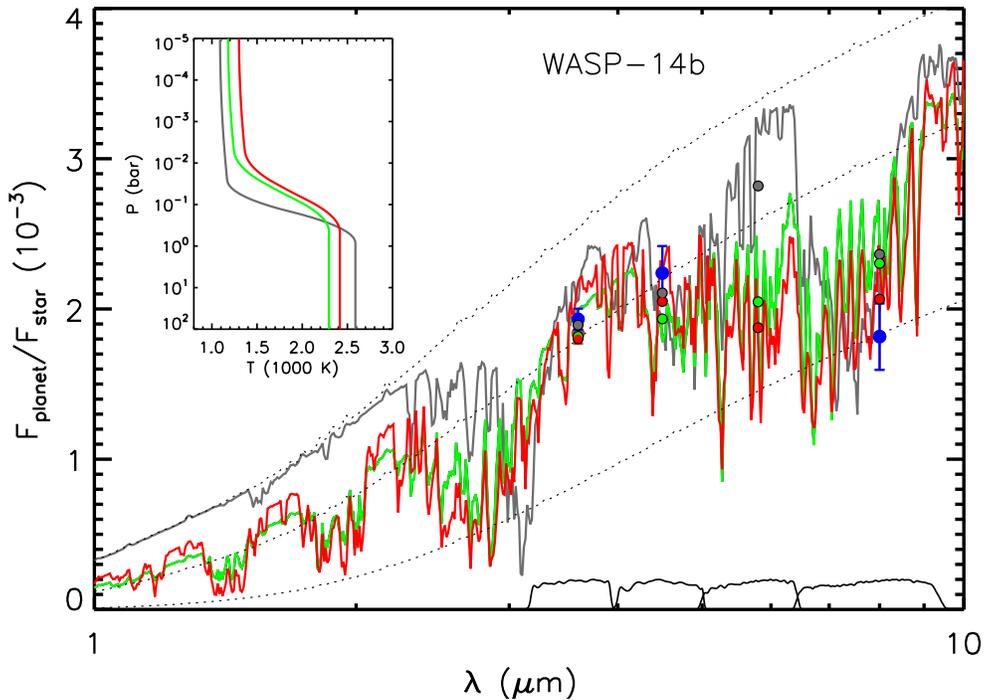}
\vspace{10pt}
\caption[Observations and model spectra for dayside emission from WASP-14b]
{\label{fig:spectra} Observations and model spectra for dayside emission from WASP-14b. The blue filled circles with error bars show our observations in {\em Spitzer} channel 1 (3.6 {\micron}), 2 (4.5 {\micron}), and 4 (8.0 {\micron}). For the 3.6 {\micron} channel, two values are shown, in blue and brown, corresponding to different ramp models used in deriving the eclipse depths (see Section\ \ref{sec:ch1}). The green, red, and gray curves show model spectra with different chemical compositions and without thermal inversions that explain the data; the corresponding pressure--temperature (\math{P-T}) profiles are shown in the inset.  The green model has molecular abundances in thermochemical equilibrium assuming solar elemental abundances. The red model has 10 times lower CO and 6 times higher H\sb{2}O compared to solar abundance chemistry, i.e., more oxygen-rich than solar abundances. The gray model has a carbon-rich chemistry (C/O = 1). The green, red, and gray circles show the model spectra integrated in the {\em Spitzer} IRAC bandpasses. The oxygen-rich (red) model provides a marginally better fit to the data than the solar and carbon-rich models. The black dotted lines show three blackbody planet spectra at 1600 K, 2200 K, and 2600 K.\\
(A color version of this figure is available in the online journal.)
}
\vspace{30pt}
\end{figure}

\newpage
We modeled the dayside atmosphere of WASP-14b using the exoplanetary atmospheric modeling method developed by \citet{MadhusudhanSeager2009,MadhusudhanSeager2010}. We use a one-dimensional line-by-line radiative transfer code to model the planetary atmosphere under the assumption of local thermodynamic equilibrium, hydrostatic equilibrium, and global energy balance at the top of the atmosphere. The latter condition assumes that the integrated emergent planetary flux balances the integrated incident stellar flux, accounting for the Bond albedo (\math{A\sb{B}}) and possible redistribution of energy onto the night side. Our model uses parameterized prescriptions to retrieve the temperature structure and chemical composition from the observations, as opposed to assuming radiative and chemical equilibrium  with fixed elemental abundances \citep{BurrowsEtal2008apjSpectra, Fortney2008}.

We find that the data can be explained by models with a wide range of chemical compositions. Figure \ref{fig:spectra} shows three model spectra with different chemistries, along with the observations: (1) a solar-abundance model (in green in Figure \ref{fig:spectra}) with chemical composition in thermochemical equilibrium assuming solar abundances (TE\sb{\rm solar}), (2) an oxygen-rich model (in red) with 10{\by} lower CO and 8{\by} higher H\sb{2}O, and (3) a carbon-rich model (in gray, e.g., \citealp{MadhusudhanEtal2011natWASP12batm}, \citealp{Madhusudhan2012}). The oxygen-rich model fits the data marginally better than the solar abundance model. A slightly lower CO is favored because of the slightly higher 4.5 {\micron} flux compared to the 3.6 {\micron} flux, which means lower absorption due to CO. Higher absorption due to H\sb{2}O is favored by the low 8 {\micron} point. In principle, a lower CO and a higher H\sb{2}O, compared to TE\sb{\rm solar} values, are both possible by having a C/O ratio less than the solar value of 0.54. However, more data would be required to confirm the low CO requirement, because a blackbody of \sim 2200 K fits the 3.6 and 4.5 {\micron} points just as well. 

Models with high C/O ratios (C/O \math{\geq} 1, i.e., carbon-rich), can lead to strong CH\sb{4}, C\sb{2}H\sb{2}, and HCN absorption in the 3.6 {\micron} and 8 {\micron} channels \citep[e.g.,][]{MadhusudhanEtal2011natWASP12batm, Madhusudhan2011b, Madhusudhan2012}, instead of H\sb{2}O absorption in the low-C/O models. As shown in Figure \ref{fig:spectra}, the C-rich model fits the data as well as the solar-abundance model, but less precisely than the model with low C/O (i.e., enhanced H\sb{2}O and low CO). Although the data marginally favor an oxygen-rich composition in the dayside atmosphere of WASP-14b, new observations are required to provide more stringent constraints on the C/O ratio. Future observations in the near-infrared, from ground and space, can place further constraints on the temperature structure and composition, especially the C/O ratio, of the dayside atmosphere of WASP-14b. In particular, as shown in Figure \ref{fig:spectra}, near-infrared observations in the 1--2.5 {\micron} range probe spectral features of several oxygen- and carbon-bearing molecules such as H\sb{2}O, CO, and CH\sb{4}, mixing ratios of which can provide stringent constraints on the C/O ratio \citep{Madhusudhan2011b}. For example, the oxygen-rich models predict deep absorption features in the H\sb{2}O bands, contrary to the carbon-rich model, which contains no significant water absorption.  {\em Hubble Space Telescope} WFC3 observations in the 1.1--1.7 {\micron} range can test for water absorption. Furthermore, the models with different C/O ratios also predict different continuum fluxes, which can be observed from ground in the \math{J}, \math{H}, and \math{K} bands \citep[see][]{Madhusudhan2012}.


\begin{figure*}[hb!]
    \centering
    \includegraphics[width=0.95\textwidth, clip]{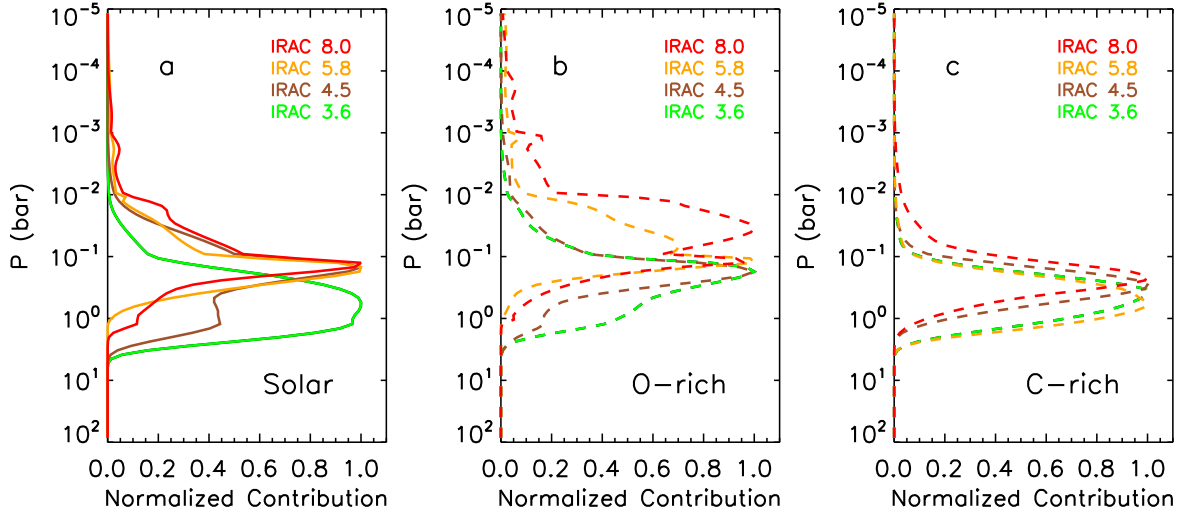}
\caption[Contribution functions for the atmospheric models]
{\label{fig:contr-func} Left: contribution functions in the four {\em Spitzer} channels corresponding to the green model shown in Figure \ref{fig:spectra}. The legend shows the channel center wavelength in {\microns} and the curves are color-coded by the channel. All the contribution functions are normalized to unity.
Middle: contribution functions corresponding to the red model shown in Figure \ref{fig:spectra}.
Right: contribution functions corresponding to the gray model shown in Figure \ref{fig:spectra}.\\
(A color version of this figure is available in the online journal.)
}
\end{figure*}

The models explaining the observations require relatively low day-night energy redistribution in WASP-14b. As shown by the contribution functions in Figure \ref{fig:contr-func}, the 3.6 {\micron} IRAC channel probes the atmosphere between 0.1 and 1 bar.  Consequently, the high brightness temperature in the 3.6 {\micron} channel indicates a hot planetary photosphere. Over the entire model population explored by our retrieval method, we find that the data allow for up to \sim 30\% of the energy incident on the dayside to be redistributed to the night side (i.e., for zero Bond albedo). For the particular best-fitting model (in red) shown in Figure \ref{fig:spectra}, this fraction is \sim 25\%. For non-zero albedos the fraction is even lower; since the quantity we constrain is \math{\eta} = (1-\math{A}\sb{\rm B})(1-\math{f}\sb{\rm r}), where \math{A}\sb{\rm B} is the Bond albedo and \math{f}\sb{\rm r} is the fraction of the dayside incident energy redistributed to the nightside \citet{MadhusudhanSeager2009}. However, the present constraints on the day-night redistribution are only suggestive and new observations are essential to further constrain the energy redistribution in WASP-14b. For example, observations in atmospheric windows at lower wavelengths, e.g., between 1 and 2 {\micron}, where the black-body of the planetary photosphere would peak, would be critical to further constrain the lower atmospheric thermal structure, and hence the energy budget of the planet's dayside atmosphere. More importantly, phase-curve observations are required to constrain the day-night energy redistribution directly \citep[e.g.,][]{Knutson2007, KnutsonEtal2009ApJ-Tres4Inversion}.

\section{DISCUSSION}
\label{sec:discus}

The absence of a thermal inversion in the dayside atmosphere of WASP-14b constrains inversion-causing phenomena in irradiated atmospheres. The canonical argument for such inversions is via absorption in the optical by gaseous TiO and VO \citep{Hubeny2003, Fortney2008}. On the other hand, \citet{Spiegel2009} showed that the high mean molecular masses of TiO and VO would lead to significant gravitational settling of these molecules, thereby depleting them from the upper atmospheres, unless strong vertical mixing keeps them aloft. Additionally, the abundances of inversion-causing molecules might also be influenced by stellar activity and photochemistry \citep{KnutsonHowardIsaacson2010ApJ-CorrStarPlanet}. Consequently, the real cause of thermal inversions in irradiated atmospheres is currently unknown. Nevertheless, models used to infer thermal inversions in the literature have either used parameterized visible opacity sources \citep{BurrowsEtal2008apjSpectra} or parametric temperature profiles (\citealp{MadhusudhanSeager2009}, also used in the present work). To first order, the lack of a thermal inversion in WASP-14b might indicate that the vertical mixing in the dayside atmosphere of WASP-14b is weaker compared to the downward diffusion of TiO and VO.

{\em Spitzer} has observed a number of strongly irradiated hot Jupiters with brightness temperatures in the 1000--2000 K range. The inferences of thermal inversions from emission photometry result from flux excesses in molecular bands where strong absorption is expected \citep{MadhusudhanSeager2010}. In principle, detection of a thermal inversion is possible with just two {\em Warm Spitzer}\/ channels with sufficient S/N if there is a large flux difference between channels 1 and 2 \citep{Knutson08, KnutsonEtal2009ApJ-Tres4Inversion, MadhusudhanSeager2010, MachalekEtal2009ApJ-XO2b, O'DonovanEtal2010ApJ-SpitzerTres2, CristiansenEtal2010ApJ-HATP7b}. Based on the TiO/VO hypothesis described above, \citet{Fortney2008} suggested that depending on the level of irradiation from their parent star, irradiated planets can fall into two categories: the very highly irradiated atmospheres that host thermal inversions and the less-irradiated ones that do not.  However, recent observations have revealed several counterexamples to this hypothesis. \citet{Machalek2008-XO-1b} present evidence for a temperature inversion in XO-1b, despite low irradiation of the planet (\math{T\sb{\rm eq}} = 1209 K), while \citet{FressinEtal2010ApJ-Tres3} show no thermal inversion, although TrES-3b is a highly irradiated planet (\math{T\sb{\rm eq}} = 1643 K). Similarly, WASP-12b, one of the most irradiated hot Jupiters known, has also been reported to lack a significant thermal inversion \citep{MadhusudhanEtal2011natWASP12batm}. In this paper, we present WASP-14b as another counterexample. It is possible that additional parameters (e.g., metallicity, surface gravity, C/O ratio) influence the presence or the absence of a temperature inversion. However, more observations are needed to explain WASP-14b's missing inversion.

\section{CONCLUSIONS}
\label{sec:concl}

During two secondary eclipse events, {\em Spitzer}\/ observed WASP-14b in three IRAC channels: 3.6, 4.5, and 8.0 {\micron}. All eclipses have a high S/N (3.6 {\micron} channel \sim 25, 4.5 {\micron} channel \sim 12, 8.0 {\micron} channel \sim 8), which allowed us to constrain the planetary spectrum and orbital parameters. 

Our observations probe the atmosphere at pressures between 0.01 and 1 bar and indicate the absence of a significant thermal inversion in the dayside atmosphere of WASP-14b. Given WASP-14b's highly irradiated atmosphere, this contradicts predictions that the most-irradiated hot Jupiters should have thermal inversions due to gaseous TiO/VO \citep{Fortney2008}. Additionally, our observations place nominal constraints on the chemical composition and day-night energy redistribution in the atmosphere of WASP-14b. We find that the data can be explained by non-inversion models with nearly solar abundances in chemical equilibrium. A factor of 10 less CO and a factor of 6 higher H\sb{2}O, compared to those obtained with solar abundances, explain the data to within the 1\math{\sigma} uncertainties, on average. Such CO depletion and H\sb{2}O enhancement are, in principle, possible in chemical equilibrium with C/O ratios lower than solar. More data are required to constrain the atmospheric composition of WASP-14b better. 

Because the planet is much brighter than its predicted equilibrium temperature for uniform redistribution (\math{T\sb{\rm eq}} = 1866 K), the best-fitting models limit day-night energy redistribution in WASP-14b to \math{\leq} 30\% for zero Bond albedo.  Thermal phase-curve observations can probe the nightside emission directly and better constrain this quantity.

WASP-14b is one of the most massive transiting planets known, along with CoRoT-3b \citep{Triaud2009-CoRoT3b, Deleuil2008-CoRoT3b}, HAT-P-2b \citep{Bakos2007-HATP-2b, Winn2007-HATP2b, Loeillet2008-HATP-2b}, XO-3b \citep{Hebrard2008-XO3b, Johns-Krull2008-XO-3b, winn:2008}, and WASP-18b \citep{NymeyerEtal2011apjWASP18b}. With the exception of WASP-18b, all of these objects have very eccentric orbits. Classically, closer planets should have more circular orbits due to greater tidal orbital decay. At distances \math{a} $<$ 0.1 AU, circularization should occur in typically a few Myr, compared to common system ages of a few Gyr. However, \citet{Pont2011-mass-period} argue that the time to circularize scales with the planet-star mass ratio, and is also a steep function of the orbital separation scaled to the planet radius (see their Figure 3). For planets with \math{M > M}\sb{\rm J}, the mass-period relation (see their Figure 2) suggests that heavier planets get circularized very close to their parent star, or may not ever reach circularization in their lifetime. A possible explanation is that the planet raises tides on its host star strong enough that the angular momentum of the planet is transferred to the stellar spin, and the planet gets swallowed by the star. This does not oppose the classical tide theory \citep[e.g.,][]{Goldreich1966}, but rather suggests that stopping mechanisms and tidal circularization are related. WASP-14b also has unusually high density for a hot Jupiter, similar to that of some rocky planets (4.6 g\,cm\sp{-3}). The planet's strong signal makes it ideal for further observation to constrain its composition and thus possible formation mechanisms for it and similar objects.

\newpage
\section{SYSTEM PARAMETERS}
\label{sec:app}

Table \ref{table:SystemParams1} lists WASP-14 system parameters derived from our analysis and the literature. The eclipse parameters are listed in Tables \ref{tab:eclfits-lin} and \ref{tab:eclfits-sin}.

\begin{table}[ht!]
\caption{\label{table:SystemParams1} 
System Parameters of WASP-14}
\atabon\strut\hfill\begin{tabular}{llc}
    \hline
    \hline
    Parameter                                                              & Value                   & Reference                 \\
    \hline

    \multicolumn{3}{c}{Orbital parameters}                                                                                       \\ \hline

    Orbital period, \math{P} (days)                                        &    2.2437661 {\pm} 0.0000011    & a                \\
    Semimajor axis, \math{a} (AU)                                          &    0.036 {\pm} 0.001            & b                \\
    Transit time (BJD\sb{TDB})                                             &    2454827.06666 {\pm} 0.00024  & a                \\
    Orbital eccentricity, \math{e}                                         &    0.087 {\pm} 0.002            & a                \\
    Argument of pericenter, \math{\omega} (deg)                            &    \math{-107.1} {\pm} 0.5      & a                \\
    Velocity semiamplitude, \math{K} (m\,s\sp{-1})                         &    990.0 {\pm} 3                & a                \\ 
    Centre-of-mass velocity \math{\gamma} (m\,s\sp{-1})                    &    \math{-4987.9} {\pm} 1.6     & a                \\ \hline
 
    \multicolumn{3}{c}{Stellar parameters}                                                                                       \\ \hline

    Spectral type                                                          &    F5V                           & b                \\
    Mass, \math{M\sb{\rm *}} (\math{M\sb{\odot}})                          &    1.211 $^{+0.127}_{-0.122}$    & b                \\
    Radius, \math{R\sb{\rm *}} (\math{R\sb{\odot}})                        &    1.306 $^{+0.066}_{-0.073}$    & b                \\
    Mean density, \math{\rho\sb{\rm *}} (\math{\rho\sb{\odot}})            &    0.542 $^{+0.079}_{-0.060}$    & b                \\
    Effective temperature, \math{T\sb{\rm eff}} (K)                        &    6475 {\pm} 100                & b                \\
    Surface gravity, log \math{g\sb{\rm *}} (cgs)                          &    4.287 $^{+0.043}_{-0.038}$    & b                \\
    Projected rotation rate, \math{v\sb{\rm *} \sin(i)} (kms\sp{-1})       &    4.9 {\pm} 1.0                 & b                \\
    Metallicity [M/H] (dex)                                                &    0.0 {\pm} 0.2                 & b                \\
    Age (Gyr)                                                              &    \sim 0.5--1.0                 & b                \\
    Distance (pc)                                                          &    160 {\pm} 20                  & b                \\ 
    Lithium abundance, log \math{N}(Li)                                    &    2.84 {\pm} 0.05               & b                \\ \hline

    \multicolumn{3}{c}{Planetary parameters }                                                                                    \\ \hline

    Transit depth, (\math{R\sb{\rm p}}/\math{R\sb{\rm star}})\sp2          &    0.0102 $^{+0.0002}_{-0.0003}$ & b               \\
    Mass, \math{M\sb{\rm p}} (\math{M\sb{\rm J}})                          &    7.341 $^{+0.508}_{-0.496}$    & b               \\
    Radius, \math{R\sb{\rm p}} (\math{R\sb{\rm J}})                        &    1.281 $^{+0.075}_{-0.082}$    & b               \\
    Surface gravity, log \math{g\sb{\rm p}} (cgs)                          &    4.010 $^{+0.049}_{-0.042}$    & b               \\
    Mean density, \math{{\rho}\sb{\rm p}} (g\,cm\sp{-3})                   &    4.6                           & b               \\
    Equilibrium temperature (\math{A}=0), \math{T\sb{\rm eq}} (K)          &    1866.12 $^{+36.74}_{-42.09}$  & b               \\
    \hline
    \multicolumn{2}{l}{\sp{a}  Our analyses (see Section\ \ref{sec:orbit})}\\
    \multicolumn{2}{l}{\sp{b} \citet{Joshi2009-WASP14b}}
\end{tabular}\hfill\strut\ataboff
\end{table}

\section{ACKNOWLEDGMENTS}

We thank Heather Knutson for providing the 3.6 {\micron} {\em Spitzer}\/ data prior to their public release, and Andrew Collier Cameron for useful discussions. We also thank contributors to SciPy, Matplotlib, and the Python Programming Language; other contributors to the free and open-source community; the NASA Astrophysics Data System; and the JPL Solar System Dynamics group for free software and services.  This work is based on observations made with the {\em Spitzer Space Telescope}, which is operated by the Jet Propulsion Laboratory, California Institute of Technology under a contract with NASA. NASA provided support for this work through an award issued by JPL/Caltech and Astrophysics Data Analysis Program grant NNX13AF38G. NM acknowledges support from the Yale Center for Astronomy and Astrophysics through the YCAA postdoctoral Fellowship. JB was partially supported by NASA Earth and Space Sciences Fellowship NNX12AL83H.

{\em Facility:Spitzer}

\newpage
\bibliographystyle{apj}
\bibliography{chap-wasp14b}

\fi

\chapter{SPITZER OBSERVATIONS OF THE THERMAL EMISSION FROM WASP-43b}
\label{chap:WASP43b}

{\singlespacing
\noindent{\bf Jasmina Blecic\sp{1}, Joseph Harrington\sp{1, 2}, Nikku Madhusudhan\sp{3}, Kevin B.\ Stevenson\sp{1}, Ryan A.\ Hardy\sp{1}, Patricio E. Cubillos\sp{1}, Matthew Hardin\sp{1}, M.\ Oliver Bowman, Sarah Nymeyer\sp{1}, David R.\ Anderson\sp{4}, Coel Hellier\sp{4}, Alexis M.S. Smith\sp{4} and Andrew Collier Cameron\sp{5}
}

\vspace{1cm}

\noindent{\em
\sp{1} Planetary Sciences Group, Department of Physics, University of Central Florida, Orlando, FL 32816-2385, USA\\
\sp{2} Max-Planck-Institut f\"{u}r Astronomie, D-69117 Heidelberg, Germany\\
\sp{3} Department of Physics and Department of Astronomy, Yale University, New Haven, CT 06511, USA\\
\sp{4} Astrophysics Group, Keele University, Keele, Staffordshire ST5 5BG, UK\\
\sp{5} SUPA, School of Physics and Astronomy, University of St.\ Andrews, North Haugh, St.\ Andrews, Fife KY16 9SS, UK
}

\vspace{1cm}

\centerline{Received 26 February 2013.}
\centerline{Accepted 7 December 2013.}
\centerline{Published in {\em The Astrophysical Journal} 16 January 2014.}


\vspace{1cm}

\centering{Blecic, J., Harrington, J., Madhusudhan, N., Stevenson, K. B., Hardy, R. A., Cubillos,\\
P. E., Hardin, M., Bowman, O., Nymeyer, S., Anderson, D. R., Hellier, C., Smith,\\
A. M. S., \& Collier Cameron, A. 2014, ApJ, 781, 116}
\vspace{0.2cm} 

\centering{http://adsabs.harvard.edu/abs/2014ApJ...781..116B \\
doi:10.1088/0004-637X/781/2/116}

\vspace{0.5cm}
\centerline{\copyright AAS. Reproduced with permission.}

}

\clearpage

\if \includeSecond y
    
\setcitestyle{authoryear,round}

\section{ABSTRACT}

WASP-43b is one of the closest-orbiting hot Jupiters, with a semimajor axis of \math{a} = 0.01526 {\pm} 0.00018 AU and a period of only 0.81 days. However, it orbits one of the coolest stars with a hot Jupiter (\math{T\sb{\rm{*}}} = 4520 {\pm} 120 K), giving the planet a modest equilibrium temperature of \math{T\sb{\rm{eq}}} = 1440 {\pm} 40 K, assuming zero Bond albedo and uniform planetary energy redistribution. The eclipse depths and brightness temperatures from our jointly fit model are 0.347\% {\pm} 0.013\% and 1670 {\pm} 23 K at 3.6 {\micron} and 0.382\% {\pm} 0.015\% and 1514 {\pm} 25 K at 4.5 {\micron}. The eclipse timings improved the estimate of the orbital period, \math{P}, by a factor of three (\math{P} = 0.81347436 {\pm} 1.4\tttt{-7} days) and put an upper limit on the eccentricity (\math{e = 0.010^{+0.010}_{-0.007}}). We use our {\em Spitzer}\/ eclipse depths along with four previously reported ground-based photometric observations in the near-infrared to constrain the atmospheric properties of WASP-43b. The data rule out a strong thermal inversion in the dayside atmosphere of WASP-43b. Model atmospheres with no thermal inversions and fiducial oxygen-rich compositions are able to explain all the available data. However, a wide range of metallicities and C/O ratios can explain the data. The data suggest low day-night energy redistribution in the planet, consistent with previous studies, with a nominal upper limit of about 35\% for the fraction of energy incident on the dayside that is redistributed to the nightside.

\section{INTRODUCTION}
\label{intro}

Our knowledge of exoplanetary systems is rapidly improving.
Recent {\em Kepler}\/ results \citep{Borucki2011ApKeplerFirstResults,Batalha2012Kepler-secondResults} have shown a striking increase in detections of the smallest candidates, and the planet candidate lists now show that hot Jupiters are much less common than planets smaller than Neptune.  However, nearly all {\em Kepler}\/ candidates are too small, cold, or distant for atmospheric characterization, except the nearby hot Jupiters. Their host stars, bright enough for radial velocity (RV) measurements, subject these planets to a strong irradiating flux, which governs their atmospheric chemistry and dynamics. Their large sizes and large scale heights \citep[e.g.,][]{ShowmanGuillot2002A&A} give the signal-to-noise ratio (S/N) needed for basic atmospheric characterization.

The most common technique for observing hot Jupiters and characterizing their dayside atmospheres is secondary eclipse photometry \citep[e.g., ][]{FraineEtal2013-GJ1214b, CrossfieldEatl2012ApJWASP12b-reavaluation, TodorovEtal2012ApJ-XO4b-HATP6b-HATP8b, DesertEatl2012AAS-Spizter, DemingEtal2011ApJ-CoRoT1-CoRoT2, BeererEtal2011ApJ-WASP4b, Demory2007aaGJ436bspitzer}. During secondary eclipse, when the planet passes behind its star, we see a dip in integrated flux proportional to the planet-to-star flux ratio, or usually 0.02\%--0.5\% in the {\em Spitzer Space Telescope} infrared wavelengths, where the signal is strongest.  This dip is much lower at wavelengths accessible from the ground or from the {\em Hubble Space Telescope}. Techniques such as phase curve measurement \citep{KnutsonEatl2009ApJ-PhaseVariationHD149026b, Knutson2012ApJ-phaseVariation-HD189733b, LewisEtal2013arXiv-PhaseVAr-HATP2b, CowanEtal2012ApJ-ThermalPhaseVar-WASP12b, CowanEtal2012ApJ-ThermalPhases-EarthLikePlanets, CrossfieldEtal2010apjUpsAndb}, transmission spectroscopy \citep{DemingEtal2013arXivHD209458b-Xo1b-WCF3, GibsonEtal2012MNRAS-HD189733b-WFC3, BertaEtal2012ApJ-GJ1214b-WFC3}, and ingress--egress mapping \citep{deWitEtal2012aapFacemap, MajeauEtal2012Facemap} can reveal more than a secondary eclipse but are available for only a small number of high-S/N planets.

A secondary eclipse observed in one bandpass places a weak constraint on an exoplanet's temperature near the average altitude of optical depth unity over that bandpass. Multiple wavelengths constrain the planet's dayside spectrum, potentially yielding insight into the atmospheric composition and temperature structure. Different wavelengths probe different atmospheric levels and can be combined into a broadband spectrum for further atmospheric modeling \citep[e.g.,][]{MadhusudhanSeager2009, StevensonEtal2010Natur}.  Infrared observations are specifically valuable because the most abundant chemical species in planetary atmospheres (aside from H\sb{2} and He), such as H\sb{2}O, CO, CO\sb{2}, and CH\sb{4}, have significant absorption and emission features at these wavelengths \citep[e.g.,][]{MadhusudhanSeager2010}. Constraints on chemical composition and thermal structure are important for both further atmospheric modeling \citep[e.g.,][]{ShowmanEtal2009ApJ-3DCircModel} and studies of the planet's formation. Several recent studies have shown that the atmospheric C/O ratios of giant planets can be significantly different from those of their host stars because of, for example, the formation location of the planet (see, for exoplanets, \citealp{Obergetal2011ApJ-C/O} and \citealp{Madhusudhan2011b}, and for Jupiter, \citealp{Lodders2004ApJ-Jupiter} and \citealp{MousisEtal2012-Jupiter}).

Secondary eclipse observations also provide insight into the exoplanet's orbit. Measuring the time of the secondary eclipse relative to the time of transit can establish an upper limit on orbital eccentricity, \math{e}, and constrain the argument of periapsis, \math{\omega}, independently of RV measurements. Orbital eccentricity is important in dynamical studies and in calculating irradiation levels.  Apsidal precession can also be constrained by eclipse timing and can be used to reveal the degree of central concentration of mass in the planetary interior \citep{ragozzine:2009, CampoEtal2011apjWASP12b, LopezMoralesEtal2010apjWASP12borb}.

WASP-43b was first detected by the Wide-Angle Search for Planets (WASP) team \citep{Hellier2011-WASP43b} in 2009 and 2010 from the WASP-South and WASP-North observatories. The WASP team also performed follow-up measurements with the CORALIE spectrograph, the TRAPPIST telescope, and EulerCAM in 2010 December. These observations revealed a planet with a mass of \math{M\sb{\rm p}} = 1.78 Jupiter masses (\math{M\sb{\rm{J}}}) and a radius of \math{R\sb{\rm p}} = 0.93 Jupiter radii (\math{R\sb{\rm J}}), transiting one of the coldest stars to host a hot Jupiter (type K7V, \math{T\sb{*}} = 4400 {\pm} 200 K). They found the planet to have an exceptionally short orbital period of 0.81 days and a semimajor axis of only 0.0142 AU, assuming the host star has a mass of \math{M\sb{*}} = 0.58 {\pm} 0.05 \math{M\sb{\odot}}. The planet's orbital eccentricity was constrained by the radial velocity and transit data to \math{e} $<$ 0.04 at 3\math{\sigma}.  Spectroscopic measurements of the star revealed a surface gravity of log\,\math{(g)} = 4.5 {\pm} 0.2 (cgs) and a projected stellar rotation velocity of \math{v\sb{*}\sin(i)} = 4.0 {\pm} 0.4 km\,s\sp{-1}, where \math{i} is the inclination of the star's pole to the line of sight. Strong Ca H and K emission indicates that the star is active. The estimated age of the star is \math{400^{+200}_{-100}} Myr.

For low-mass stars like WASP-43, there are notable discrepancies \citep{BergerEtal2006ApJ-LowMassStars} between interferometrically determined radii and radii calculated in evolutionary models \citep[i.e.,][]{ChabrierBaraffe1997-LowMassStars, SiessEtal1997-SyntheticHRDiagram}.  \citet{BoyajianEtal2012-KMStars} presented high-precision interferometric diameter measurements of 33 late-type K and M stars. They found that evolutionary models overpredict temperatures for stars with temperatures below 5000 K by \sim 3\% and underpredict radii for stars with radii below 0.7 \math{R\sb{\odot}} by \sim 5\%. Their Table 11 lists an average temperature and radius for each spectral type in the sample, suggesting that WASP-43, with its measured temperature of 4520 {\pm} 120 K, is likely a K4 star rather than a K7 as reported by \citet{Hellier2011-WASP43b}.

\citet{GillonEtal2012A&A-WASP-43b} analyzed twenty-three transit light curves, seven occultations, and eight new measurements of the star's RV, observed during 2010 and 2011 with TRAPPIST, the Very Large Telescope (VLT), and EulerCAM. They refined eccentricity to \math{e} = 0.0035 {\pm} 0.0043 and placed a 3\math{\sigma} upper limit of 0.0298 using all data simultaneously. They also improved the parameters of the system significantly (\math{M\sb{\rm{p}}} = 2.034 {\pm} 0.052 \math{M\sb{\rm{J}}}, \math{R\sb{\rm{p}}} = 1.036 {\pm} 0.019 \math{R\sb{\rm{J}}}), refined stellar parameters (\math{T\sb{\rm{eff}}} = 4520 {\pm} 120 K, \math{M\sb{*}} = 0.717 {\pm} 0.025 \math{M\sb{\odot}}, \math{R\sb{*}} = 0.667 {\pm} 0.011 \math{R\sb{\odot}}), and constrained stellar density (\math{\rho\sb{*}} = 2.41 {\pm} 0.08 \math{\rho\sb{\odot}}).. They also confirmed that the observed variability of the transit parameters can be attributed to the variability of the star itself (consistent with \citealp{Hellier2011-WASP43b}). In addition, they detected the planet's thermal emission at 1.19 {\micron} and 2.09 {\micron} and used the atmospheric models of \citet{Fortney2005, Fortney2008} to infer poor redistribution of heat to the night side and an atmosphere without a thermal inversion.

In this paper we present two secondary eclipses, observed at 3.6 and 4.5 {\micron} with the Infrared Array Camera \citep[IRAC,][]{Fazio2004IRAC} on the {\em Spitzer Space Telescope}, which further constrain the dayside emission of the planet and improve the orbital parameters of the system. We combine our {\em Spitzer} eclipse depth measurements with on previously reported measurements of thermal emission in the near-infrared from \citet{GillonEtal2012A&A-WASP-43b} and \citet{WangEtal2013-WASP43b} to constrain the atmosphere's energy redistribution and thermal profile by using the retrieval method of \citet{MadhusudhanSeager2009} as subsequently developed.

The following sections present our observations (Section \ref{sec:obs}); discuss photometric analysis (Section \ref{sec:method}); explain specific steps taken to arrive at the fits for each observation and a joint fit (Section \ref{sec:results}); give improved constraints on the orbital parameters based on available RV, eclipse, and transit data (Section \ref{sec:orbit}); discuss implications for the planetary emission spectrum and planetary composition (Section \ref{sec:atm}); state our conclusions (Section \ref{sec:conc}); and, in the Appendix, supply the full set of system parameters from our own work and previous work.  The electronic attachment to this paper includes archival light curve files in FITS ASCII table and IRSA formats.

\vspace{20pt}
\begin{table}[h!]
\caption{\label{table:ObsDates} 
Observation Information}
\atabon\strut\hfill\begin{tabular}{lccc@{ }c@{ }c@{ }l}
    \hline
    \hline 
    Channel & Observation & Start Time    & Duration  & Exposure  & Number of \\
            & Date        & (MJD\sb\rm{UTC})           & (s)       & Time (s)  & Frames    \\
    \hline
    Ch1     & 2011 Jul 30  & 2455772.6845  & 21421     & 2         & 10496     \\
    Ch2     & 2011 Jul 29  & 2455771.8505  & 21421     & 2         & 10496     \\
   \hline  
\end{tabular}\hfill\strut\ataboff
\end{table}

\section{OBSERVATIONS}
\label{sec:obs}

We observed two secondary eclipses of WASP-43b with the {\em Spitzer}\/ IRAC camera in subarray mode (Program ID 70084).  A sufficiently long baseline (Figure \ref{fig:RawBinNorm}) was monitored before the eclipses, providing good sampling of all {\em Spitzer}\/ systematics. To minimize intrapixel variability, each target had fixed pointing. We used the Basic Calibrated Data (BCD) from {\em Spitzer}'s data pipeline, version S.18.18.0. Basic observational information is given in Table \ref{table:ObsDates}.

\section{SECONDARY ECLIPSE ANALYSIS -- METHODOLOGY}
\label{sec:method}

Exoplanet characterization requires high precision, since the planets' inherently weak signals are weaker than the systematics. In addition, {\em Spitzer}'s systematics lack full physical characterizations. We have developed a  modular pipeline, Photometry for Orbits, Eclipses, and Transits (POET), that implements a wide variety of treatments of systematics and uses Bayesian methods to explore the parameter space and information criteria for model choice. The POET pipeline is documented in our previous papers \citep{StevensonEtal2010Natur, CampoEtal2011apjWASP12b, NymeyerEtal2011apjWASP18b, StevensonEtal2012apjHD149026b, BlecicEtal2013ApJ-WASP14b, CubillosEtal2013apjWASP8b}, so we give here just a brief overview of the specific procedures used in this analysis.

The pipeline uses {\em Spitzer}-supplied BCD frames to produce systematics-corrected light curves and parameter and uncertainty estimates, routinely achieving 85\% of the photon S/N limit or better. Initially, POET masks pixels according to {\em Spitzer's} permanent bad pixel masks, and then it additionally flags bad pixels (energetic particle hits, etc.) by grouping sets of 64 frames and performing a two-iteration, 4\math{\sigma} rejection at each pixel location. Image centers with 0.01 pixel accuracy come from testing a variety of centering routines \citep[][Supplementary Information]{StevensonEtal2010Natur}. Subpixel 5\math{\times} interpolated aperture photometry \citep{HarringtonEtal2007natHD149026b} produces the light curves. We omit frames with bad pixels in the photometry aperture. The background, subtracted before photometry, is an average of good pixels within an annulus centered on the star in each frame.


\begin{figure}[t!]
\strut\hfill
\includegraphics[width=0.33\textwidth, clip]{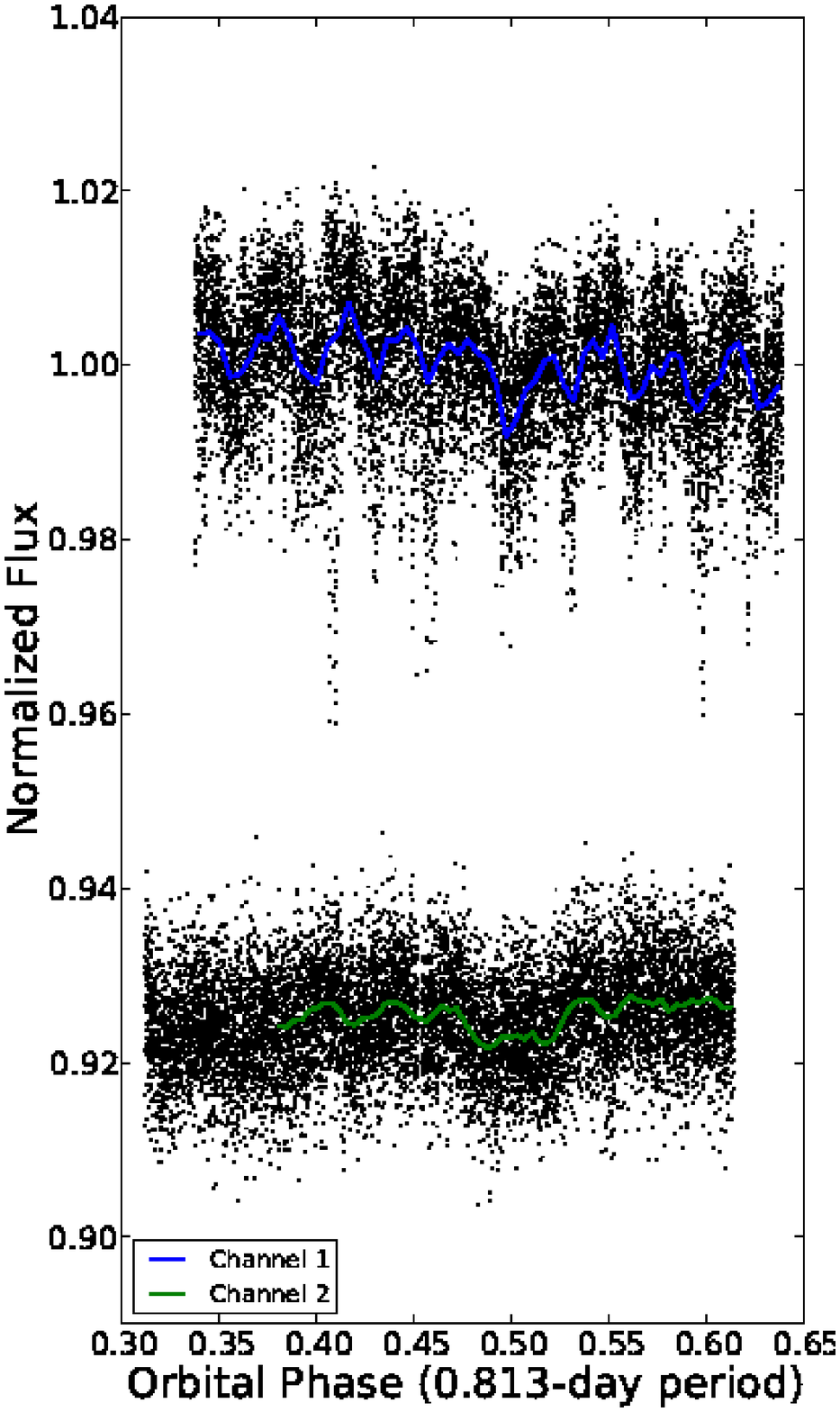}\hfill
\includegraphics[width=0.33\textwidth, clip]{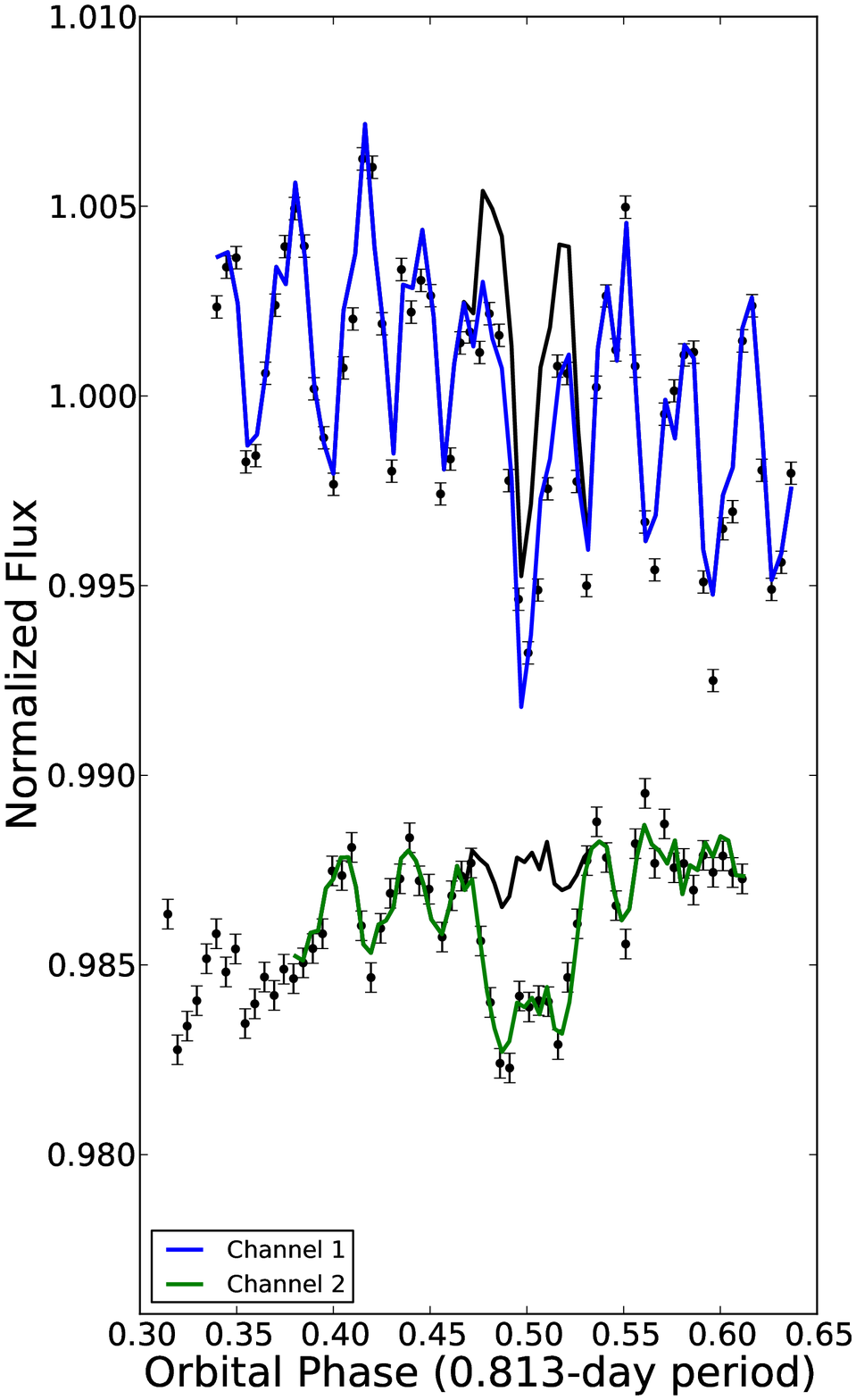}\hfill
\includegraphics[width=0.33\textwidth, clip]{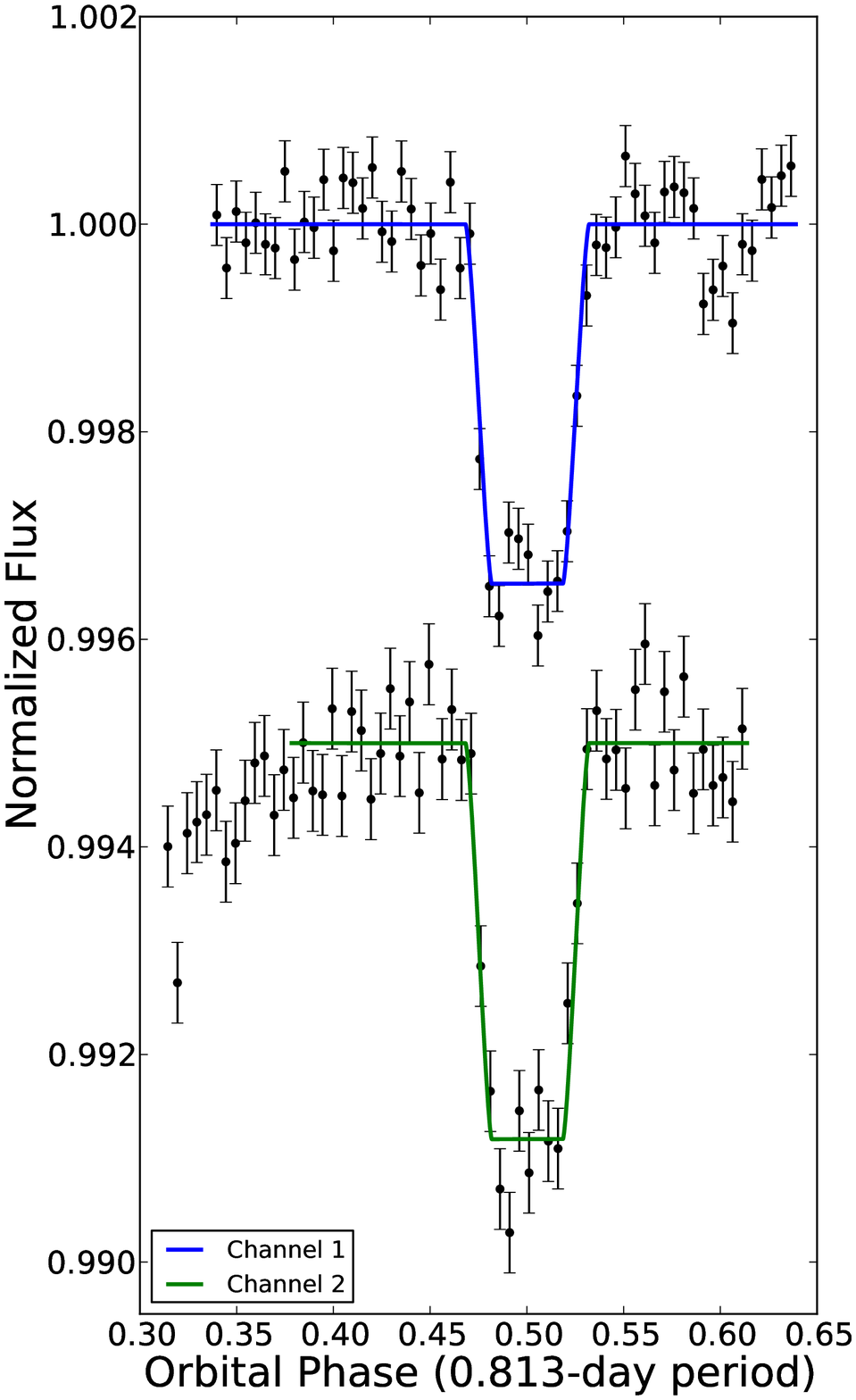}
\hfill\strut
\caption[Secondary-eclipse light curves of WASP-43b at 3.6 and 4.5 {\microns}]
{\label{fig:RawBinNorm}
Raw (left), binned (center, 60 points per bin, with 1\math{\sigma} error bars), and systematics-corrected (right) secondary eclipse light curves of WASP-43b at 3.6 and 4.5 {\microns}. The results are normalized to the system flux and shifted vertically for comparison.  Note the different vertical scales used in each panel. The colored lines are best-fit models. The black curves in panel 2 are models without eclipses. As seen in the binned plots of channel 2, a ramp model is not needed to correct for the time-dependent systematic if initial data points affected by pointing drift are clipped (see Section \ref{sec:method}).\\
(A color version of this figure is available in the online journal.)
}
\end{figure}

Detector systematics vary by channel and can have both temporal (detector ramp) and spatial (intrapixel variability) components. At 3.6 and 4.5 {\micron}, intrapixel sensitivity variation is the dominant effect \citep{CharbonneauEtal2005apjTrES1}, so accurate centering at the 0.01 pixel level is critical. We fit this systematic with a Bilinearly Interpolated Subpixel Sensitivity (BLISS) mapping technique, following \citet{StevensonEtal2012apjHD149026b}, including the method to optimize the bin sizes and the minimum number of data points per bin.

At 8.0, and 16 {\micron}, there is temporal variability, attributed to charge trapping \citep{KnutsonEatl2009ApJ-PhaseVariationHD149026b}. Weak temporal dependencies can also occur at 3.6 and 4.5 {\micron} \citep{Reach2005-IRACCalibration, CharbonneauEtal2005apjTrES1, CampoEtal2011apjWASP12b, DemoryEtal2011apj55Cnc, BlecicEtal2013ApJ-WASP14b}, while weak spatial variability has been seen at 5.8 and 8.0 {\micron} \citep{StevensonEtal2012apjHD149026b, Anderson2011-Ch24-WASP17b}. Thus, we consider both systematics in all channels when determining the best-fit model.

We fit the model components simultaneously using a \citet{MandelAgol2002ApJtransits} eclipse, \math{E(t)}; the time-dependent detector ramp model, \math{R(t)}; and the BLISS map, \math{M(x, y)}:
\begin{eqnarray}
\label{eqn:full}
F(x, y, t) = F\sb{s}\,R(t)\,M(x,y)\,E(t),
\end{eqnarray}
\noindent where \math{F(x,y,t)} is the aperture photometry flux and \math{F\sb{\rm s}} is the constant system flux outside of the eclipse.

To choose the best systematics models, we analyze dozens of model combinations and use goodness-of-fit criteria \citep{CampoEtal2011apjWASP12b}. For a given channel, we first vary the photometric aperture size and the number of initial data points that we exclude because of instrument settling, and then we test different ramp models and bin sizes for the intrapixel model.  To choose the best aperture size and the number of initial points dropped during instrument settling, we minimize the standard deviation of the normalized residuals (SDNR). Ignoring data points from the beginning of the observation is a common procedure \citep{Knutson2011-GJ436b} when searching for the best-fitting ramp. We remove the smallest number of points consistent with the minimal SDNR (see each channel analysis for the number of points discarded).

Once we have found the best dataset in this way, we compare different ramp models by applying the Bayesian Information Criterion:
\begin{equation}
{\rm BIC} = \chi\sp{2} + k\ln N,
\label{BIC}
\end{equation}
\noindent where \math{N} is the number of data points. The best model minimizes the chosen criterion. The level of correlation in the photometric residuals is also considered by plotting root-mean-squared (rms) model residuals versus bin size \citep[time interval,][]{pont:2006, winn:2008, CampoEtal2011apjWASP12b} and comparing this to the theoretical \math{1/\sqrt{2N}} rms scaling (\citealp{BlecicEtal2013ApJ-WASP14b} explains the factor of 2). Sometimes, we prefer less-correlated models with insignificantly poorer BIC values.

We explore the phase space and estimate errors by using a Markov-chain Monte Carlo (MCMC) routine following the Metropolis--Hastings random-walk algorithm, which uses independent Gaussian proposal distributions for each parameter with widths chosen to give an acceptance rate of 30\%--60\%. Each MCMC model fit begins with the Levenberg--Marquardt algorithm (least-squares minimization). We use an informative prior \citep[e.g.,][]{Gelman2002-prior} taken from other work on parameters that are more tightly constrained than what  our fits can achieve. In this work, those are ingress and egress times, which are not well sampled by our observation. All other parameters have flat priors and are free parameters of the MCMC. For each channel, they are listed in Section \ref{sec:joint}. For orbital analysis, they are listed in Section \ref{sec:orbit}. We then run enough MCMC iterations to satisfy the \citet{Gelman1992} convergence test. After every run, we assess convergence by examining plots of the parameter traces, pairwise correlations, autocorrelations, marginal posteriors, best-fitting model, and systematics-corrected best-fitting model. The final fit is obtained from the simultaneous run of all datasets, sharing parameters such as the eclipse midpoint and duration among some or all datasets.

We report the times of our secondary eclipses in both BJD\sb{UTC} (Coordinated Universal Time) and BJD\sb{TT} (BJD\sb{TDB}, Barycentric Dynamical Time), calculated using the Jet Propulsion Laboratory (JPL) Horizons system and following \citet{EastmanEtal2010apjLeapSec}.

\section{SECONDARY ECLIPSE ANALYSIS -- FIT DETAILS}
\label{sec:results}

Light curves for both channels were extracted using every aperture radius from 2.00 to 4.50 pixels, in 0.25 pixel increments.  We tested three centering routines, center of light, two-dimensional (2D) Gaussian fit, and least asymmetry (see Supplementary Information of \citealp{StevensonEtal2010Natur} and \citealp{LustEtal2013apjCentering}). A 2D Gaussian fit found the most consistent stellar centers. We estimated the background flux by using an annulus of 7--15 pixels from the center of the star for both channels. For the secondary eclipse ingress and egress time, we used a Bayesian prior (\math{t\sb{\rm{2-1}}} = 950.5 {\pm} 145.5 s), calculated from unpublished WASP photometric and RV data.

\begin{figure}[hb!]
    \centering
    \includegraphics[width=0.75\linewidth, clip]{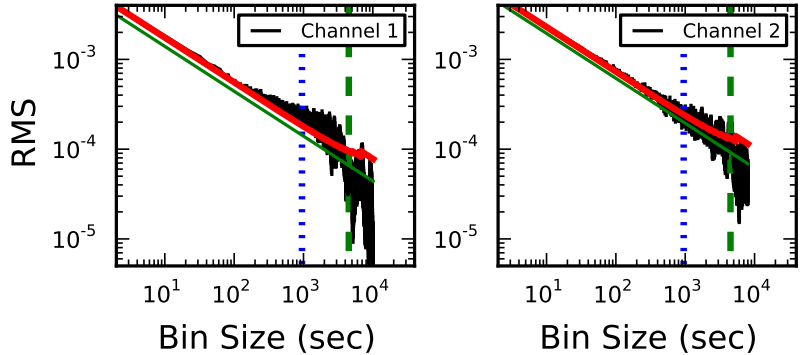}
\caption[Correlations of the residuals for the two secondary eclipses of WASP-43b]
{Correlations of the residuals for the two secondary eclipses of WASP-43b, following \citet{pont:2006}. The black line represents the rms residual flux vs. bin size. The red line shows the predicted standard error scaling for Gaussian noise. The green line shows the theoretical photon noise limit (observed S/N is 80.3\% and 85\% of the photon-limited S/N for channel 1 and channel 2, respectively; see Section \ref{sec:joint}). The black vertical lines at each bin size depict 1\math{\sigma} uncertainties on the rms residuals (\math{rms/\sqrt{2N}}, where N is the number of bins). The dotted vertical blue line indicates the ingress/egress timescale, and the dashed vertical green line indicates the eclipse duration timescale. Large excesses of several \math{\sigma} above the red line would indicate correlated noise at that bin size. Inclusion of 1\math{\sigma} uncertainties shows no noise correlation on the timescales between the ingress/egress duration and the eclipse duration.\\
(A color version of this figure is available in the online journal.)
}
\label{fig:rms}
\end{figure}

Figure \ref{fig:RawBinNorm} shows our systematics-corrected, best-fit light curve models.  Figure \ref{fig:rms} presents the scaling of the rms model residuals versus bin size for both channels, which shows no significant time correlation in the residuals.

\subsection{Channel 1 -- 3.6 {\micron}}
\label{sec:ch1}

The most prominent systematic in this {\em Spitzer}\/ channel is the intrapixel effect. The best BLISS-map bin size is 0.006 pixels when we exclude bins with less than four measurements. The ramp and eclipse models fit without removing initial data points. The smallest value of BIC reveals that the best ramp model is quadratic; this model is 1.2\tttt{30} times more probable than the next-best (linear) model. Table \ref{table:Ch1-ramps} lists the best ramp models, comparing their SDNR, BIC values, and eclipse depths.

\begin{figure}[ht!]
    \centering
    \includegraphics[width=0.73\linewidth, clip]{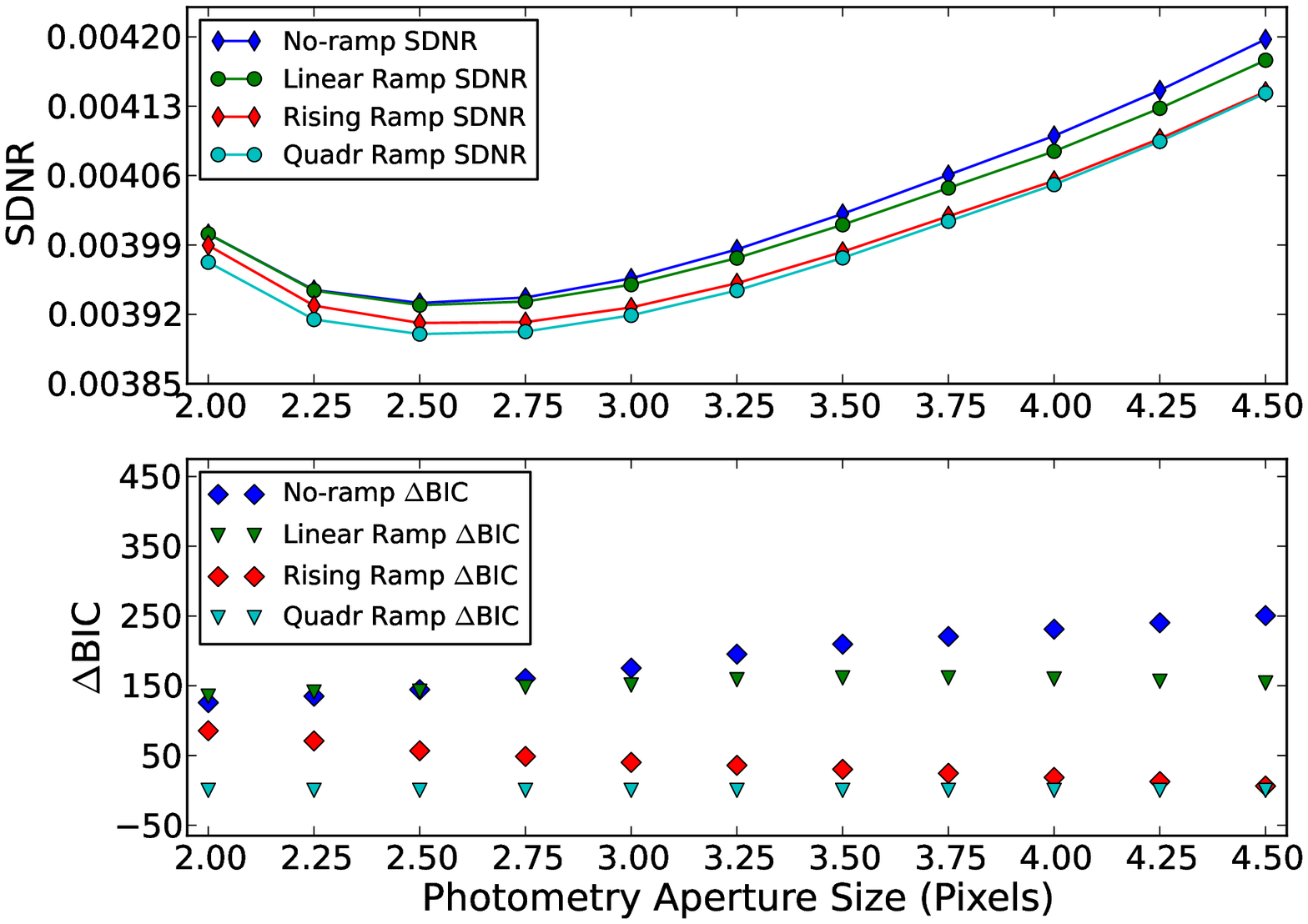}
\caption[Comparison between different ramp models for channel 1]
{Channel 1 comparison between different ramp models. The plots show SDNR vs. aperture size and \math{\Delta}BIC vs. aperture size.  A lower SDNR value indicates a better model fit. The lowest SDNR value marks the best aperture size (2.50 pixels). A lower \math{\Delta}BIC value at the best aperture size indicates which ramp model is the best (quadratic ramp model, green triangles).\\
(A color version of this figure is available in the online journal.)
}
\label{fig:ch1-SDNR}
\end{figure}


\begin{table}[ht!]
\caption{\label{table:Ch1-ramps} 
Channel 1 Ramp Models}
\atabon\strut\hfill\begin{tabular}{lcrcc}
    \hline
    \hline
    Ramp Model    & SDNR           & \math{\Delta}BIC  & Eclipse Depth (\%) \\
    \hline
    Quadratic    & 0.0039001       & 0.0        &  0.344 {\pm} 0.013 \\
    Rising       & 0.0039113       & 56.8       &  0.292 {\pm} 0.012 \\
    No-Ramp      & 0.0039315       & 144.4      &  0.268 {\pm} 0.012 \\
    Linear       & 0.0039293       & 142.1      &  0.270 {\pm} 0.012 \\
    \hline
\end{tabular}\hfill\strut\ataboff
\end{table}

Figure \ref{fig:ch1-SDNR} shows a comparison between the best ramp models and their SDNR and BIC values through all aperture sizes, indicating which aperture size is the best and which model has the lowest BIC value. 

\begin{figure}[ht!]
    \centering
    \includegraphics[width=0.80\linewidth, clip]{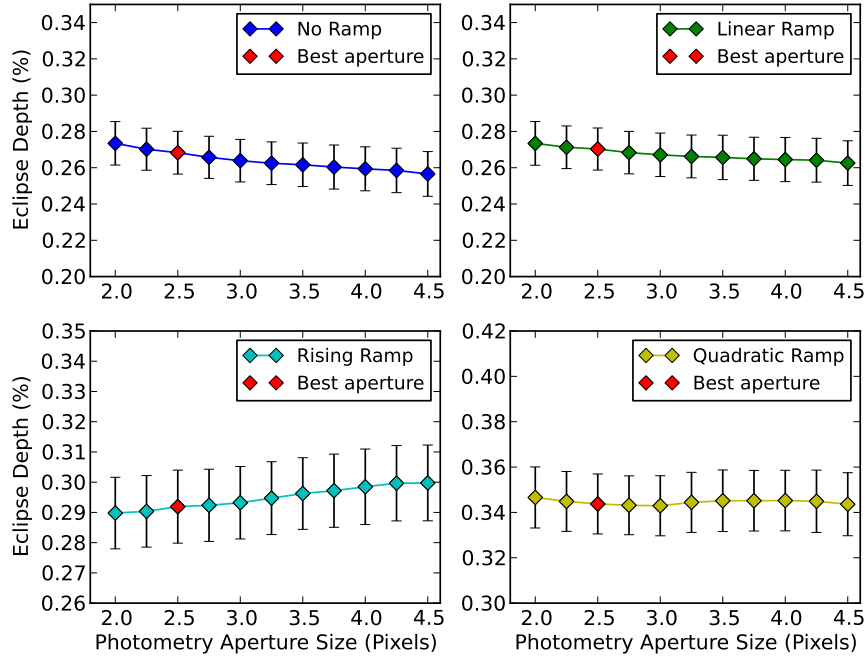}
\caption[Best-fit eclipse depths as a function of photometry aperture size for channel 1]
{Best-fit eclipse depths as a function of photometry aperture size for channel 1. The four best ramp models are plotted. The red point indicates the best aperture size for that channel. The eclipse depth uncertainties are the result of 10\sp{5} MCMC iterations. The trend shows insignificant dependence of eclipse depth on aperture size (much less than 1\math{\sigma}).\\
(A color version of this figure is available in the online journal.)
}
\label{fig:ch1-depths}
\end{figure}

Photometry generates consistent eclipse depths for all tested apertures, with the lowest SDNR at an aperture radius of 2.50 pixels (Figure \ref{fig:ch1-depths}).

\begin{figure}[ht!]
    \centering
    \includegraphics[width=0.75\linewidth, clip]{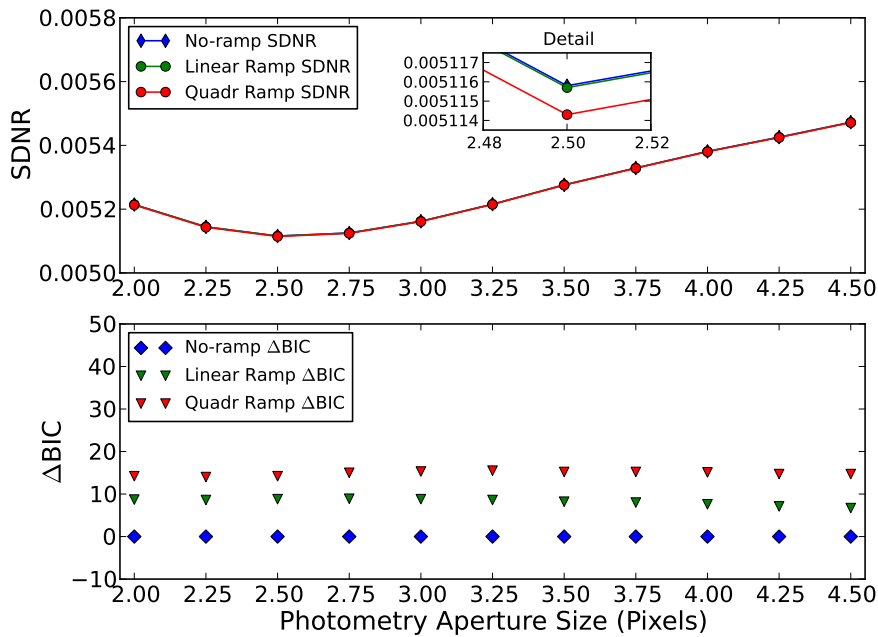}
\caption[Comparison between different ramp models for channel 2]
{Channel 2 comparison between different ramp models. Plot shows SDNR vs. aperture size and \math{\Delta}BIC vs. aperture size.  A lower SDNR value indicates a better model fit. The lowest SDNR value marks the best aperture size (2.50 pixels). Lower BIC values at the best aperture size indicate better models (best: no-ramp model, blue diamonds). The inset shows separation in SDNR for different ramp models at the best aperture size.\\
(A color version of this figure is available in the online journal.) 
}

\label{fig:ch2-SDNR}
\end{figure}

\subsection{Channel 2 -- 4.5 {\micron}}
\label{sec:ch2}

In this channel, we noticed an upward trend in flux at the beginning of the observation, possibly due to telescope settling, which we do not model.  We clipped 2300 initial data points (\sim 38 minutes of observation), the smallest number of points consistent with the minimal SDNR. The 2.50 pixel aperture radius minimizes SDNR (Figure \ref{fig:ch2-SDNR}). 

To remove intrapixel variability, the BLISS bin size is 0.016 pixels, ignoring bins with less than four points. The lowest BIC value corresponds to the model without a ramp (Table \ref{table:Ch2-ramps}), which is 78 times more probable than the linear model.

\begin{figure}[h!]
    \centering
    \includegraphics[width=0.75\linewidth, clip]{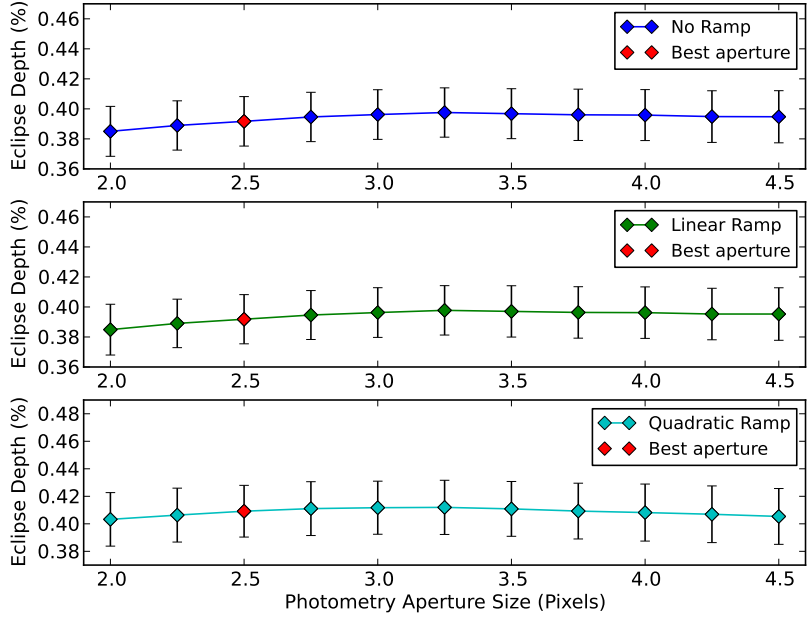}
\caption[Best-fit eclipse depths as a function of photometry aperture size for channel 2]
{Best-fit eclipse depths as a function of photometry aperture size for channel 2. The three best ramp models are plotted. The red point indicates the best aperture size for that channel. The eclipse-depth uncertainties are the result of 10\sp{5} MCMC iterations. The trend shows negligible dependence of eclipse depth on aperture size (much less than 1\math{\sigma}).\\
(A color version of this figure is available in the online journal.)
}
\label{fig:ch2-depths}
\end{figure}

We tested the dependence of eclipse depth on aperture radius, showing that they are all well within 1\math{\sigma} of each other, to validate the consistency of our models (Figure \ref{fig:ch2-depths}).

\begin{table}[ht!]
\caption{\label{table:Ch2-ramps} 
Channel 2 Ramp Models}
\atabon\strut\hfill\begin{tabular}{lcccc}
    \hline
    \hline
    Ramp Model   & SDNR           & \math{\Delta}BIC    & Eclipse Depth (\%)      \\
    \hline
    No-Ramp      & 0.0051158      &  0                   & 0.392 {\pm} 0.016    \\
    Linear       & 0.0051157      &  8.8                 & 0.392 {\pm} 0.016    \\
    Quadratic    & 0.0051143      &  14.2                & 0.409 {\pm} 0.019    \\ 
    \hline
\end{tabular}\hfill\strut\ataboff
\vspace{40pt}
\end{table}

\subsection{Joint Fit}
\label{sec:joint}

To improve accuracy, we share the eclipse width (duration), eclipse midpoint phases, and ingress and egress times in a joint fit of both datasets. Table \ref{tab:eclfits} indicates which parameters are free, shared, or have informative priors. The best ramp models and the best aperture sizes from the separate channel analyses are used in the joint fit. To produce the best joint-model fits, we iterated MCMC until the \citet{Gelman1992} diagnostics for all parameters dropped below 1\%, which happened after \ttt{5} iterations. 

The best joint-model fit parameters are in Table \ref{tab:eclfits}.  Files containing the light curves, best model fits, centering data, photometry, etc., are included as electronic supplements to this article.  The eclipse midpoint time is further used for the subsequent orbital analysis, and the eclipse depths are used for the atmospheric analysis.

\begin{table*}[h!]
\centering
\caption{\label{tab:eclfits} Best-fit Joint Eclipse Light Curve Parameters}
\begin{tabular}{lr@{\,{\pm}\,}lr@{\,{\pm}\,}lr@{\,{\pm}\,}l}
\hline
\hline
Parameter                                                  &   \mctc{    Channel 1     }           &   \mctc{   Channel 2      }     \\
\hline
Array position (\math{\bar{x}}, pixel)                                 &   \mctc{      14.99       }           &   \mctc{       14.8       }     \\
Array position (\math{\bar{y}}, pixel)                                 &   \mctc{      15.01       }           &   \mctc{      15.05       }     \\
Position consistency\tablenotemark{a} (\math{\delta\sb{x}}, pixel)     &   \mctc{      0.009       }           &   \mctc{      0.016       }     \\
Position consistency\tablenotemark{a} (\math{\delta\sb{y}}, pixel)     &   \mctc{      0.012       }           &   \mctc{      0.014       }     \\
Aperture size (pixel)                                                  &   \mctc{      2.5         }           &   \mctc{      2.5         }     \\
Sky annulus inner radius (pixel)                                       &   \mctc{      7.0         }           &   \mctc{      7.0         }     \\
Sky annulus outer radius (pixel)                                       &   \mctc{      15.0        }           &   \mctc{      15.0        }     \\
System flux \math{F\sb{s}} (\micro Jy)\tablenotemark{b}              &         64399.0 & 5.0                 &         37911.0 & 2.0           \\
Eclipse depth (\%)\tablenotemark{b}                                  &           0.347 & 0.013               &         0.382   & 0.015         \\
Brightness temperature (K)                                           &           1670 & 23                &             1514   & 25          \\
Eclipse midpoint (orbits)\tablenotemark{b}                           &          0.4986 & 0.0004              &          0.4986 & 0.0004        \\
Eclipse midpoint (BJD\sb{UTC}--2,450,000)                           &       5773.3172 & 0.0003              &       5772.5037 & 0.0003        \\
Eclipse midpoint (BJD\sb{TDB}--2,450,000)                           &       5773.3179 & 0.0003              &       5772.5045 & 0.0003        \\
Eclipse duration (\math{t\sb\rm{4-1}}, hr)\tablenotemark{b}       &            1.25 & 0.02                &            1.25 & 0.02          \\
Ingress/egress time (\math{t\sb\rm{2-1}}, hr)\tablenotemark{b}    &           0.268 & 0.018               &           0.268 & 0.018         \\
Ramp name                                                            &   \mctc{        quadramp  }           &   \mctc{     no-ramp      }     \\
Ramp, quadratic term\tablenotemark{b}                                &         -0.0827 & 0.0069              &   \mctc{      ...         }      \\
Ramp, linear term\tablenotemark{b}                                   &         -0.0002 & 0.0005              &   \mctc{      ...         }      \\
Intrapixel method                                                    &   \mctc{       BLISS      }           &   \mctc{      BLISS       }     \\
BLISS bin size in $x$  (pixel)                                         &   \mctc{     0.006        }           &   \mctc{     0.016        }     \\
BLISS bin size in $y$  (pixel)                                         &   \mctc{     0.006        }           &   \mctc{     0.016        }     \\
Minimum number of points per bin                                     &   \mctc{        4         }           &   \mctc{        4         }     \\
Total frames                                                         &   \mctc{      10496       }           &   \mctc{      10496       }     \\
Frames used                                                          &   \mctc{      10124       }           &   \mctc{       8004       }     \\
Rejected frames (\%)                                                 &   \mctc{  0.44            }           &   \mctc{  1.12            }     \\
Free parameters                                                      &   \mctc{        7         }           &   \mctc{        2         }     \\
Number of data points in fit                                         &   \mctc{      10124       }           &   \mctc{       8004       }     \\
BIC                                                                  &   \mctc{  18207.3         }           &   \mctc{  18207.3         }     \\
SDNR                                                                 &   \mctc{ 0.0039007        }           &   \mctc{ 0.0051167        }     \\
Photon-limited S/N (\%)                                              &   \mctc{       80.3       }           &   \mctc{       85.0       }     \\
\hline
\vspace{-9pt}
\end{tabular}
\begin{minipage}[t]{0.89\linewidth}
{\bf Notes.}\\
\sp{a}{rms frame-to-frame position difference.}\\
\sp{b}{Free parameter in MCMC fit.  All priors are flat except
\math{t\sb{2-1}}, which uses a Gaussian prior of 950.5 {\pm} 145.5 s, calculated from unpublished WASP photometric and RV data. Parameters
with identical values and uncertainties are fit jointly.}
\end{minipage}
\end{table*}

\newpage
\section{ORBIT}
\label{sec:orbit}

The eclipse midpoint (after a 15.2 s correction for the eclipse transit light-time) has a phase of 0.5001 {\pm} 0.0004, so \math{e \cos \omega} = 0.0001 {\pm} 0.0006, or a \math{3\sigma} upper limit of \math{|e \cos \omega| < 0.0018}, consistent with a circular orbit.

To improve the orbit solution further, we combined data from our observations with data from a variety of sources (see Table \ref{tab:ttv}).  Transit midpoint times were taken from \citet{Hellier2011-WASP43b} and \citet{GillonEtal2012A&A-WASP-43b}, and amateur observations were listed in the Exoplanet Transit Database (see Table \ref{tab:ttv}).  We used {\em CORALIE} RV observations published by \citet{Hellier2011-WASP43b} and \citet{GillonEtal2012A&A-WASP-43b}.  No RV points analyzed were gathered during transit. We subtracted 15.2 s from the eclipse midpoint to correct for light-travel time across the orbit. We corrected all points to TDB if this was not already done \citep{EastmanEtal2010apjLeapSec}.  We converted the amateur data from HJD to BJD, putting all times in a consistent BJD\sb{TDB} format.  There were 49 transit points, 23 RV points, and one effective eclipse observation.  We fit all of these data simultaneously, as described by \citet{CampoEtal2011apjWASP12b}.  The free parameters in this fit were \math{e \sin \omega}; \math{e \cos \omega}; the period, \math{P}; the reference transit midpoint time, \math{T\sb{\rm{0}}}; the RV semi-amplitude, \math{K}; and the RV offset, \math{\gamma}.  The addition of the amateur transit observations improves the uncertainty of \math{P} by a factor of nearly five compared with \citet{GillonEtal2012A&A-WASP-43b}, reducing it to 13 ms.  The fit finds an eccentricity of 0.010$^{+0.010}_{-0.007}$, consistent with a circular orbit and expectations for a close-in planet, where eccentricity should be damped by tidal interactions with the host star \citep{JacksonEtal2008ApJ-TidalEvolution}. Table \ref{tab:orbit} summarizes the fit results.


\begin{table}[ht!]
\vspace{-20pt}
\centering
\footnotesize{
\caption{\label{tab:ttv} Transit Timing Data}
\begin{tabular}{lll@{ }c@{ }}
\hline
\hline
Mid-transit Time  &  Uncertainty          &   Source\tablenotemark{a}  & Quality  \\
 (BJD\sb{TDB})    &                       &                            & Rating   \\
\hline
2456440.77250		& 0.00037		& Phil Evans							 & 1 	\\
2456410.67704	   	& 0.00097		& Robert Majewski				 		 & 5	\\
2456407.42697		& 0.00097		& Enrique D\'{i}ez Alonso			 	         & 3	\\
2456407.41972		& 0.00102		& Ullrich Dittler						 & 3	\\
2456403.34716		& 0.0017		& Jens Jacobsen							 & 5	\\
2456401.72885		& 0.00199		& Alex Chassy							 & 5	\\
2456387.89785		& 0.00046		& Phil Evans							 & 2	\\
2456375.70739		& 0.00158		& Parijat Singh							 & 4	\\
2456368.37413		& 0.00076		& Adam B\"uchner							 & 3	\\	
2456335.83452		& 0.00113		& Phil Evans							 & 3	\\
2456328.51426		& 0.00039		& Juan Lozano de Haro					         & 2	\\
2456328.51296		& 0.00181		& Daniel Staab							 & 4	\\
2456326.88711		& 0.00072		& Phil Evans							 & 3	\\
2456313.87042		& 0.00039		& P. Kehusmaa \& C. Harlingten	                                 & 2	\\
2456288.65334		& 0.00087		& Jordi Lopesino						 & 3	\\
2456283.77139		& 0.00068	        & A. Chapman \& N. D. D\'{i}az		                         & 3	\\
2456250.42005		& 0.00067	        &L. Zhang, Q. Pi \& A. Zhou                                      & 5	\\
2456035.66489       & 0.0005        & George Hall                         & 2         \\
2456015.3273        & 0.00057       & Martin Z\'{i}bar                    & 2 	     \\
2456006.38071       & 0.00101       & Franti\^{s}ek Lomoz                 & 3	     \\
2456006.3781        & 0.00109       & Franti\^{s}ek Lomoz                 & 3	     \\
2456001.49859       & 0.00156	    & Alfonso Carre\~{n}o                     & 3	     \\
2456001.49662       & 0.00035       & Gustavo Muler Schteinman            & 1	     \\
2456001.49531       & 0.00019       & Fernand Emering                     & 1	     \\
2455997.43508       & 0.00086       & Ren\'{e} Roy                        & 3	     \\
2455997.43105       & 0.00051       & Faustino Garcia                     & 3         \\
2455997.43008       & 0.0004        & Nicolas Esseiva             	 & 2	     \\
2455997.42981       & 0.00068       & Juanjo Gonzalez             	 & 3	     \\
2455984.41939       & 0.00064       & Ferran Grau Horta           	 & 3	     \\
2455984.41548       & 0.00126       & Franti\^{s}ek Lomoz            & 3	     \\
2455984.4149        & 0.00047       & Fabio Martinelli            	 & 2         \\
2455984.41472       & 0.00071       & Franti\^{s}ek Lomoz          	 & 3	     \\
2455979.534         & 0.00044       & Nicolas Esseiva            	 & 2	     \\
2455979.5335        & 0.0004        & Juanjo Gonzalez            	 & 2	     \\
2455957.57296       & 0.00122       & Franti\^{s}ek Lomoz          	 & 4	     \\
2455944.55468       & 0.00106       &  Roy Ren\'{e}            		 & 3	     \\
2455940.48744       & 0.0005        & Anthony Ayiomamitis         	 & 2	     \\
2455939.67475       & 0.00052       & Ramon Naves             		 & 2	     \\
2455933.16473       & 0.00025       & Peter Starr            		 & 1	     \\
2455686.68399       & 0.0008        & Stan Shadick           		 & 3	     \\
2455682.61364       & 0.00039       & Tanya Dax, Stacy Irwin       	 & 5	     \\
\hline
\vspace{-9pt}
\end{tabular}
\begin{minipage}[t]{0.70\linewidth}
{\bf Notes.} \sp{a}{The TRansiting ExoplanetS and CAndidates group (TRESCA, http://var2.astro.cz/EN/tresca/index.php) supply their data to the Exoplanet Transit Database (ETD, http://var2.astro.cz/ETD/), which performs the uniform transit analysis described by \citet{Poddany2010}.  The ETD web site provided the numbers in this table, which were converted from HJD (UTC) to BJD (TDB).}
\end{minipage}
}
\end{table}

\begin{landscape}
\begin{table}[ht!]
\centering
\caption{\label{tab:orbit} Eccentric Orbital Model}
\begin{tabular}{lc}
\hline
\hline 
Parameter                                                          &  Value                                  \\
\hline
\math{e \sin \omega}\tablenotemark{a}     	                   &  \math{-0.010}    {\pm} 0.011           \\
\math{e \cos \omega}\tablenotemark{a}     	                   &  \math{-0.0003}   {\pm} 0.0006         \\
\math{e}                          			           &  0.010           $^{+0.010}_{-0.007}$  \\
\math{\omega} (\degree)           			           &  \math{-88}       $^{+5}_{-9}$          \\
\math{P} (days)\tablenotemark{a}      		                   &  \math{P} = 0.81347436 {\pm} 1.4\math{\times 10\sp{-7}} \\
\math{T\sb{\rm 0}} (BJD\sb{TDB}) \tablenotemark{a}                 &  2455528.86857    {\pm} 0.00005         \\
\math{K} (m\,s\sp{-1})\tablenotemark{a}                            &  549              {\pm} 6               \\
\math{\gamma} (m\,s\sp{-1})\tablenotemark{a}                       &  \math{-3595}     {\pm} 4               \\
\math{\chi^2}				                           &  458                                     \\
\hline   
\vspace{-9pt}                                        
\end{tabular}
\begin{minipage}[t]{0.62\linewidth}
\hspace{2.2cm} \hangindent=5.5em {\bf Note.} \sp{a}{Free parameter in MCMC fit.}\\
\comment{
\sp{b}{BJD\sb{TDB}.}\\
\sp{c}{RV semi-amplitude}.\\
\sp{d}{RV offset}}
\end{minipage}
\end{table}

\vspace{1.5cm}

\begin{table}[ht!]
\centering
\caption{\label{tab:ephem} Ephemeris Solutions}
\begin{tabular}{lccc}
\hline
\hline 
                                        & All Data (Transits, Eclipses, RV)	& Transits Only   & Transits Only\\
Parameter                       &  Linear Ephemeris                         & Linear Ephemeris  & Quadratic Ephemeris              \\
\hline
\math{T_0} (BJD\sb{TDB})    	&  2455528.86857    {\pm} 0.00005  &  2455581.74439    {\pm} 0.00004   & 2455581.74437 {\pm} 0.00004    \\
\math{P} (days orbit\sp{-1})    &  0.81347436       {\pm} 1.4\math{\times 10\sp{-7}} & 0.81347450  {\pm} \math{1.5\times 10\sp{-7}}    & 0.81347530   {\pm} 3.8\math{\times 10\sp{-7}}\\
\math{\delta P} (days orbit\sp{-2})\tablenotemark{a}   &  ...  &         ...         &     \math{(-2.5 {\pm} 0.9) \times 10\sp{-9}}\\
\math{\dot{P}\,} (s\,\,yr\sp{-1})\tablenotemark{b}    &  ...  &        ...          &      \math{-0.095 {\pm} 0.036 }      \\
\math{\chi^2}			   &               ...                          &        444.7                  &	   437.9	     \\
BIC                 	           &               ...                  &        452.8                   &    450.0              \\
\hline
\vspace{-9pt}                                         
\end{tabular}
\begin{minipage}[t]{0.95\linewidth}
{\bf Notes.} \\
\sp{a}{\math{\delta P = \dot{P}P}}.\\
\sp{b}{Derived parameter}.
\end{minipage}
\end{table}
\end{landscape}

Because \math{e \sin \omega} is a much larger component of the eccentricity than \math{e \cos \omega},  it is possible that much of this eccentricity signal comes from the effect of the planet raising a tidal bulge on its host star. \citet{ArrasEtal2012MNRAS-RVstar} predict that the RV semi-amplitude of this effect is 8.9 ms\sp{-1}. Since our model shows that \math{eK} = 5$^{+6}_{-3}$ ms\sp{-1}, it is possible that the majority or entirety of the eccentricity signal is due to the tidal bulge interaction, and the true eccentricity is closer to the upper limit derived from the secondary eclipse.

We found that, for a linear ephemeris fit to just the transit timing data, there is considerable scatter in O-C (observed time minus calculated time).  The root-mean-square of the stated transit-time uncertainties is 51 s, while the standard deviation of the residuals is 124 s.  WASP-43b is close enough to its host star that tidal decay is a significant factor in its evolution, so we attempted to estimate the decay rate by adding a quadratic term to our ephemeris model, following \citet{AdamsEtal2010ApJ}.  Our model for the transit ephemeris is now:
\begin{equation}
T\sb{N} = T\sb{0} + PN + \delta P \frac{N(N-1) }{2},
\end{equation}
\noindent where \math{N} is the number of orbits elapsed since the epoch \math{T\sb{\rm{0}}}, \math{P} is the orbital period at \math{T\sb{\rm{0}}}, and \math{\delta P = \dot{P}P}, where \math{\dot{P}} is the short-term rate of change in the orbital period.  Fitting this model to the transit data, we find that \math{\delta P} = (-2.5 {\pm} 0.9)\tttt{-9} days orbit\sp{-2}, or \math{\dot{P}} = -0.095 {\pm} 0.036 s yr\sp{-1}.  This is illustrated in Figure \ref{fig:ttv}.   This is a nondetection, though the best-fit value is comparable to the value of -0.060 {\pm} 0.015 s yr\sp{-1} found by  \citet{AdamsEtal2010ApJ} for OGLE-TR-113b.


\begin{figure}[ht!]
\centerline{
\includegraphics[width=0.80\linewidth, clip, trim=0cm 0cm 0cm 0cm]{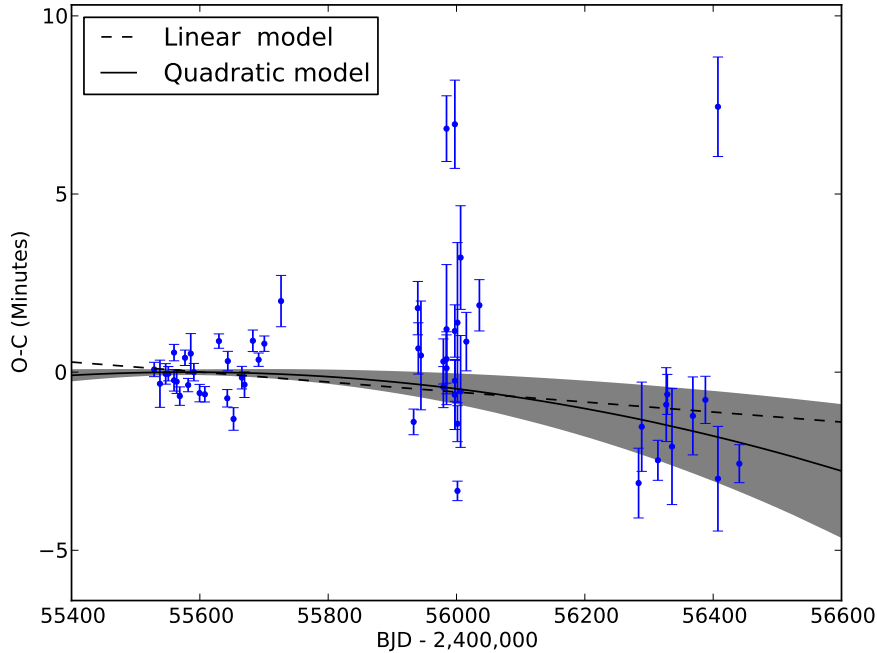}}
\caption[O-C diagram for transit observations of WASP-43b]
{O-C diagram for transit observations of WASP-43b with respect to the linear terms in the best-fit quadratic ephemeris.  The quadratic ephemeris is shown with the 1\math{\sigma} prediction uncertainty in grey.  The best-fit linear ephemeris is also shown as a dashed line.    Only points with an Exoplanet Transit Database quality rating of 3 or better are shown here and used in this analysis.\\
(A color version of this figure is available in the online journal.) 
}
\label{fig:ttv}
\end{figure}

While the suggestion of a quadratic inspiral existed in the data
gathered prior to BJD 2456035, the addition of 17 new amateur
observations from BJD 2456250 to BJD 2456440 (11 of which were of
sufficiently high data quality for inclusion) put the linear fit within
the credible region of the quadratic fit.  

Table \ref{tab:ephem} summarizes all three transit ephemeris models. The first is the linear ephemeris from the fit discussed above. We also fit linear and quadratic models to just the transit data to study any trend in orbital period. These are the second and third fits in Table \ref{tab:ephem}. Their BIC and chi \math{\chi\sp{2}} are comparable to each other but not to the first fit, since the data sets are not identical.

The BIC values still favor the quadratic
(decaying) ephemeris (BIC = 450) over the linear ephemeris (BIC = 452.8)
by a probability ratio of \math{e}\sp{\math{\Delta}BIC/2} = 4.  Two lines of reasoning
favor the linear ephemeris, however, so we consider the linear ephemeris
to be more likely.

First, the inspiral time predicted by the quadratic is extremely short
compared to the planet's lifetime.  We would need to believe that we are
seeing the planet at a very brief and special time in its history to
accept the conclusion of inspiral.

Second, the rms scatter about the quadratic ephemeris is 121 s, much
larger than the 51 s typical transit time uncertainty and not much
different from the linear result. The reduced \math{\chi\sp{2}} for the quadratic fit
is \sim 8, so much of the residual scatter is unexplained by either
model. Possible explanations include stellar activity, transit timing
variations (TTVs), or problems in data processing or reporting. Mistakes
in the time corrections for heterogeneous transit data are unlikely
because the Exoplanet Transit Database indicates that all amateur
observations were submitted in UTC, while the professional data were
unambiguous in their use of TDB.  While it is possible that
uncertainties for certain sets of transit data points may have been
underestimated, the data come from many amateur and professional
sources, and all would have had to make such errors. Since WASP-43 is an
early K-type star, a likely explanation for the scatter is stellar
activity.

Effects like TTVs or tidal infall could still contribute to the scatter,
so we present related calculations below, mainly as motivation and
background for future studies.

From the measured period change of \math{\dot{P}} = -0.095 {\pm} 0.036  s yr\sp{-1}, we adopt a 3\math{\sigma} upper limit, \math{|\dot{P}|} $<$ 0.129 s yr\sp{-1}. For WASP-43b, this translates to a maximum change in the semimajor axis of \math{|\dot{a}|} $<$ 1.9 \tttt{-8} AU yr\sp{-1}.  The three-sigma upper limit on the period decay also suggests an infall timescale of at least 5\tttt{5} ys.  \citet{LevrardEtAl2009ApJ-FallingHotJupiters} give a relation for tidal decay, which for synchronous planetary rotation and negligible eccentricity and obliquity reduces to:
\begin{equation}
\frac{1}{a}\frac{{\rm d} a}{{\rm d} t} = \frac{6}{Q'\sb\star }\frac{M\sb{p}}{M\sb\star}\left(\frac{R\sb\star}{a}\right)\sp5 \left(\omega\sb\star-\frac{2\pi}{P}\right),
\end{equation}
\noindent where \math{Q'\sb{\rm{\star}}} is the ratio of the stellar tidal quality factor to the second-order stellar tidal Love number, \math{k\sb{\rm{2}}}, and \math{\omega\sb{\rm{\star}}} is the stellar rotation rate. The upper limit on the quadratic term implies \math{Q'\sb\star > 12,000}. This is much lower than the values of \ttt{5}--\ttt{10} normally assumed and thus cannot rule out any plausible values.  A small value of \math{Q'\sb{\rm{\star}}} was also found by \cite{AdamsEtal2010ApJ}.

While the quadratic fit failed to produce a useful upper limit on tidal decay, observations with a longer time baseline may yet find secular changes or TTVs.  Until then, the linear ephemeris presented here is the most reliable predictor of future transit times.

\section{ATMOSPHERE}
\label{sec:atm}

We modeled the dayside atmosphere of WASP-43b by using the atmospheric modeling and retrieval method of \citet{MadhusudhanSeager2009, MadhusudhanSeager2010}. The model computes line-by-line radiative transfer in a one-dimensional, plane-parallel atmosphere, with constraints of local thermodynamic equilibrium, hydrostatic equilibrium, and global energy balance. The pressure--temperature profile and components of the molecular composition are free parameters of the model, allowing exploration of models with and without thermal inversions and those with oxygen-rich as well as carbon-rich compositions \citep{Madhusudhan2012}. 

The model includes all the primary sources of opacity expected in hydrogen-dominated giant-planet atmospheres in the temperature regimes of hot Jupiters, such as WASP-43b. The opacity sources include line-by-line absorption due to H\sb{2}O, CO, CH\sb{4}, CO\sb{2}, and NH\sb{3} and collision-induced absorption (CIA) due to H\sb{2}-H\sb{2}. We also include hydrocarbons besides CH\sb{4}, such as HCN and C\sb{2}H\sb{2}, which may be abundant in carbon-rich atmospheres \citep{Madhusudhan2011b, Kopparapu2012ApJ-CarbonRich, Madhusudhan2012}. Since in highly irradiated oxygen-rich atmospheres TiO and VO may be abundant \citep{Fortney2008}, we also include line-by-line absorption due to TiO and VO in regions of the atmosphere where the temperatures exceed the corresponding condensation temperatures. Our molecular line data are from \citet{Freedman08}, R. S. Freedman (2009, private communication), \citet{Rothman2005-HITRAN}, \citet{KarkoschkaTomasko2010}, E. Karkoschka (2011, private communication), and \citet{HarrisEtal2008MNRAS-CarbonStars}. We obtain the H\sb{2}-H\sb{2} CIA opacities from \citet{Borysow1997} and \citet{Borysow2002}. The volume mixing ratios of all the molecules are free parameters in the model. 

We constrain the thermal structure and composition of WASP-43b by combining our {\em Spitzer} photometric observations at 3.6 {\micron} and 4.5 {\micron} combined with previously reported ground-based narrow-band photometric data from \citet{GillonEtal2012A&A-WASP-43b},  obtained using VLT/HAWK-I at 1.19 {\micron} and 2.09 {\micron}, and broadband photometric data from \citet{WangEtal2013-WASP43b} obtained using CFHT/WIRCAMthe in the \math{H} (1.6 {\micron}) and \math{K\sb{s}} (2.1 {\micron}) bands.  
The data also place a joint constraint on the day--night energy redistribution and the Bond albedo by requiring global energy balance, i.e., that the integrated emergent power from the planet does not exceed the incident irradiation.  Given that the number of model parameters is \math{\geq} 10 (depending on the C/O ratio) and the number of available data points is 6, our goal is to find the regions of model space favored by the data, rather than to determine a unique fit. We explore the model parameter space by using an MCMC routine (for details, see \citealp{MadhusudhanSeager2009, MadhusudhanSeager2010, MadhusudhanEtal2011natWASP12batm}).

\begin{figure*}[hb!]
\centerline{
\includegraphics[width = 0.75\textwidth, clip]{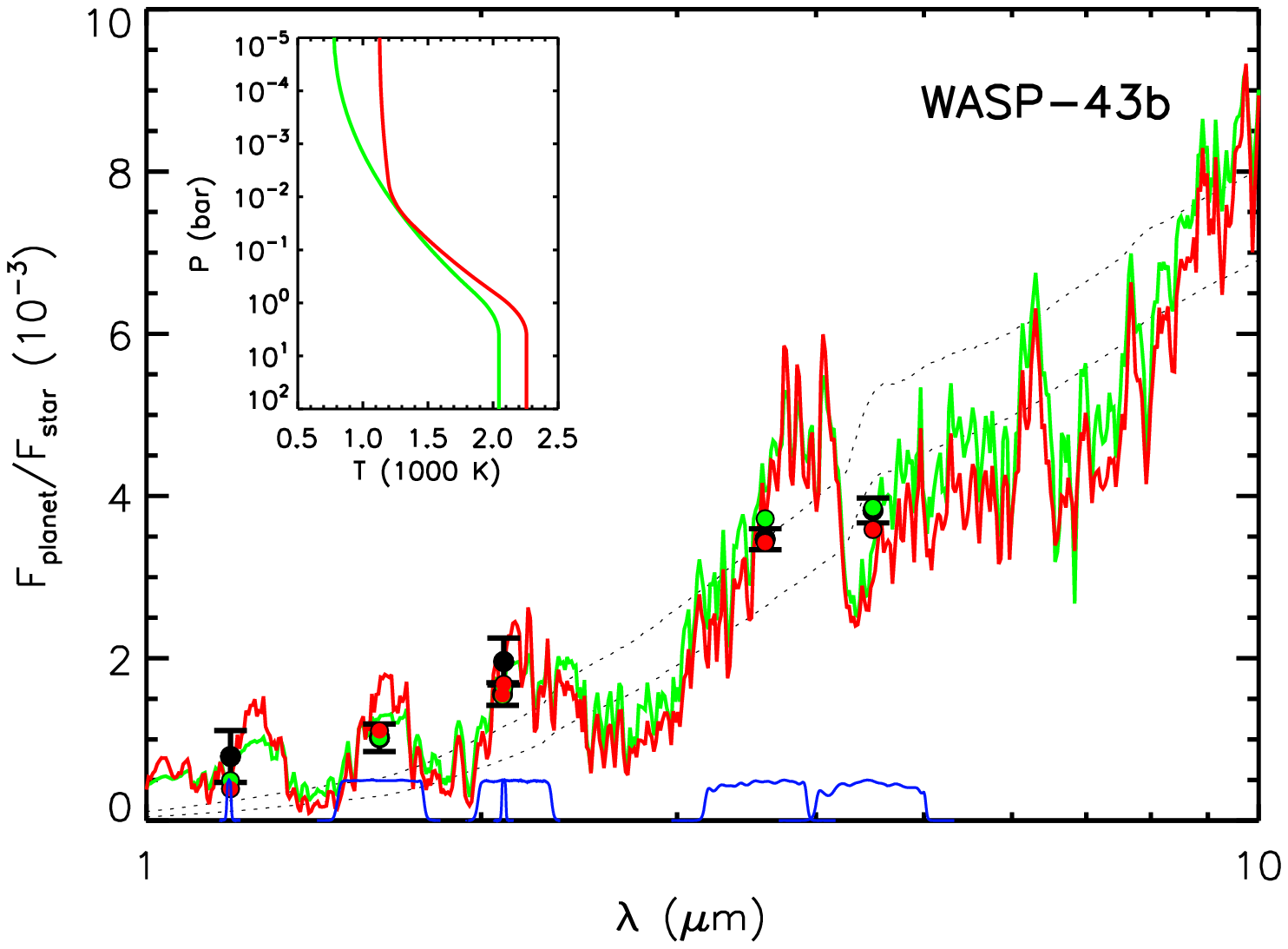}}
\caption[Observations and model spectra for dayside thermal emission from WASP-43b]
{Observations and model spectra for dayside thermal emission from WASP-43b. The black filled circles with error bars show our data in {\em Spitzer} IRAC channels 1 (3.6 {\micron}) and 2 (4.5 {\micron}) and previously published ground-based near-infrared data in narrow-band photometry at 1.19 {\micron} and 2.09 {\micron} \citep{GillonEtal2012A&A-WASP-43b} and in broadband photometry at 1.6 {\micron} and 2.1 {\micron} \citep{WangEtal2013-WASP43b}. The solid curves show the model spectra in the main panel, and the corresponding temperature--pressure profiles, with no thermal inversions, in the inset. The green and red curves correspond to models with compositions of nearly solar and 10 \math{\times} solar metallicity, respectively.  Both models fit the data almost equally well. The dashed curves show blackbody spectra corresponding to planetary brightness temperatures of 1670 K and 1514 K, the observed brightness temperatures in the {\em Spitzer} IRAC channels 1 and 2, respectively.\\
(A color version of this figure is available in the online journal.)
}
\label{fig:atmmodel}
\end{figure*}

The data rule out a strong thermal inversion in the dayside atmosphere of WASP-43b. The data and two model spectra of atmospheres without thermal inversions are shown in Figure \ref{fig:atmmodel}. The ground-based and {\em Spitzer}\/ data provide complementary constraints on the atmospheric properties. The ground-based photometric bandpasses, in narrow bands at 1.19 {\micron} and 2.09 {\micron} \citep{GillonEtal2012A&A-WASP-43b} and in broad bands at 1.6 {\micron} and 2.1 {\micron} \citep{WangEtal2013-WASP43b}, span spectral regions of low molecular opacity probe the deep layers of the atmosphere at pressures of \math{P\sim1} bar, beyond which the atmosphere is optically thick because of collision-induced opacity (the contribution functions are given in Figure\ \ref{fig:contr}). Consequently, the brightness temperatures from such ground-based data constrain the isothermal temperature structure of the deep atmosphere \citep{Madhusudhan2012}.

\begin{figure}[h!]
\centerline{
\includegraphics[width = 0.40\textwidth, clip]{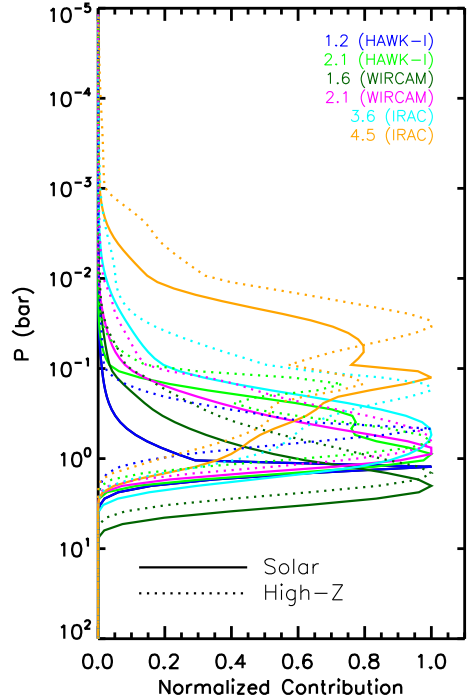}}
\caption[Contribution functions for the atmospheric models]
{Contribution functions for the atmospheric models. The solid (dashed) curves show the contribution functions for the model in Figure ~\ref{fig:atmmodel} with solar (10 \math{\times} solar, "High-\math{Z}") composition. The contribution functions are shown in all the bandpasses corresponding to the data, denoted by the central wavelengths in {\micron} and instruments in parentheses, as shown in the legend. \\
(A color version of this figure is available in the online journal.)
} 
\label{fig:contr}
\end{figure}

On the other hand, the two {\em Spitzer}\/ data sets show lower brightness temperatures at 3.6 {\micron} and 4.5 {\micron} relative to the ground-based data, which is possible only if the temperature structure is decreasing outward in the atmosphere, causing molecular absorption in the {\em Spitzer}\/ bands. The presence of a strong thermal inversion, on the contrary, would have caused molecular emission leading to higher brightness temperatures in the {\em Spitzer} bands relative to the ground-based bands. Consequently, the sum total of {\em Spitzer} and ground-based data rule out a strong thermal inversion in WASP-43b's dayside photosphere.

The molecular composition is less well constrained by the data. Several physically plausible combinations of molecules can explain the absorption in the two {\em Spitzer} bands \citep[e.g.,][]{MadhusudhanSeager2010, Madhusudhan2012}. Figure \ref{fig:atmmodel} shows two oxygen-rich models in chemical equilibrium, with C/O ratios of 0.5 (solar value) but with different metallicities (nearly solar and ten times solar) and thermal profiles, both of which explain the data almost equally well. In both cases, H\sb{2}O absorption in the 3.6 {\micron} band, and H\sb{2}O, CO, and CO\sb{2} absorption in the 4.5 {\micron} band explain the {\em Spitzer} data.

The comparable fits demonstrate the degeneracy between the molecular mixing ratios (via the metallicity) and the temperature gradient. Given the current photometric data, the solar metallicity model with a steep temperature profile (green curve) produces almost as good a fit as the higher metallicity model with a shallower temperature profile (red curve). On the other hand, carbon-rich models with C/O \math{\geq}1 \citep[e.g.,][]{Madhusudhan2012}, with absorption due to CH\sb{4}, CO, C\sb{2}H\sb{2}, and HCN, could also explain the data. As such, the current data are insufficient to discriminate between O-rich and C-rich compositions. Thus, new observations are required to obtain more stringent constraints on the chemical composition of WASP-43b. Observations using the {\em HST}\/ Wide Field Camera 3 in the 1.1--1.8 {\micron} bandpass can help constrain the H\sb{2}O abundance in the atmosphere.  As shown in Figure \ref{fig:atmmodel}, an O-rich composition predicts strong absorption due to H\sb{2}O in the WFC3 bandpass, which would be absent in a carbon-rich atmosphere \citep{MadhusudhanEtal2011natWASP12batm}. Similarly, observations in other molecular bands, such as the CO band at 2.3 {\micron}, can provide constraints on the corresponding molecular mixing ratios.

Our observations provide nominal constraints on the day--night energy redistribution fraction \citep[\math{f\sb{\rm{r}}}; ][]{MadhusudhanSeager2009} in WASP-43b. The models shown in Figure \ref{fig:atmmodel} have \math{f\sb{\rm{r}}} = 16\%--20\%, assuming zero Bond albedo. Our population of model fits to the combined {\em Spitzer} and ground-based data allow for up to \sim 35\% day--night energy redistribution in the planet. Among the acceptable models, those with higher \math{f\sb{\rm{r}}} values require cooler lower atmospheres on the dayside and hence predict lower fluxes in the ground-based channels. While such models produce an acceptable fit to all four data points overall, they predict systematically lower fluxes in the ground-based channels, some fitting the ground-based data points only at the \math{\sigma} lower error bars. Considering the ground-based points alone, without the {\em Spitzer} data, would imply a significantly higher continuum flux and correspondingly a significantly lower day--night redistribution in the planet than \sim 35\%, consistent with the findings of \citet{GillonEtal2012A&A-WASP-43b} and \citet{WangEtal2013-WASP43b}. 

The lack of a strong thermal inversion in WASP-43b is not surprising. At an equilibrium temperature of \math{\sim}1400 K, the dayside atmosphere of WASP-43b is not expected to host gaseous TiO and VO, which have been proposed to cause thermal inversions \citep{Spiegel2009, Hubeny2003, Fortney2008}, though hitherto unknown molecules that could also potentially cause such inversions cannot be ruled out \citep{Zahnle09}. The lack of a thermal inversion is also consistent with the hypothesis of \citet{KnutsonHowardIsaacson2010ApJ-CorrStarPlanet}, since the host star WASP-43 is known to be active \citep{Hellier2011-WASP43b}.

\section{CONCLUSIONS}
\label{sec:conc}

Exoplanet secondary eclipses provide us with a unique way to observe the dayside spectrum of an irradiated planetary atmosphere, where the opacities of the mixture of atmospheric trace molecules determine the thermal structure of the planetary atmosphere.

WASP-43b has a 0.81 day period, making it one of the shortest-period transiting planets. It has a small semimajor axis \citep[0.01526 {\pm} 0.00018 AU,][]{GillonEtal2012A&A-WASP-43b}. WASP-43 is a low-mass star \citep[\math{M\sb{*}} = 0.717 {\pm} 0.025 \math{M\sb{\odot}},][]{GillonEtal2012A&A-WASP-43b} and is also one of the coldest of all stars hosting hot Jupiters. The close proximity of the planet probably induces large tidal bulges on the planet's surface \citep{ragozzine:2009}. The planet's projected lifetime is also unusually short for such a late-type host star, owing to tidal in-spiral. The estimated lifetime for this planet is perhaps 10 Myr--1 Gyr \citep{Hellier2011-WASP43b}.

In this paper we report two {\em Spitzer} secondary eclipse observations, using the IRAC 3.6 and 4.5 {\microns} channels. The S/N of 26 in channel 1 and 24 in channel 2 allowed a nonambiguous analysis. The final eclipse depths from our joint-fit models are 0.347\% {\pm} 0.013\% and 0.382\% {\pm} 0.015\%, in channels 1 and 2, respectively. The corresponding brightness temperatures are 1670 {\pm} 23 K and 1514 {\pm} 25 K.

Our secondary eclipse timings, along with the available RV data and transit photometry from the literature and amateur observations, provide better constraints on the orbital parameters. WASP-43b's orbital period is improved by a factor of three (\math{P} = 0.81347436 {\pm} 1.4\tttt{-7} days). The timing of our secondary eclipse observations is consistent with and suggestive of a circular orbit. 

We combined our {\em Spitzer} eclipse depths with ground-based data in the near-infrared from \citet{GillonEtal2012A&A-WASP-43b} and \citet{WangEtal2013-WASP43b} to constrain the atmospheric properties of WASP-43b. The data rule out a strong thermal inversion in the dayside atmosphere. This is particularly evident because the brightness temperatures in both the {\em Spitzer} channels are lower than those observed in the ground-based channels, suggesting temperatures decreasing outward. The data do not suggest very efficient day--night energy redistribution in the planet, consistent with previous studies, though models with up to \sim 35\% redistribution can explain the data reasonably well. Current data are insufficient to provide stringent constraints on the chemical composition.

WASP-43b is a promising planet for a variety of future observations.
Its high eclipse S/N makes it a prime candidate for dayside mapping
using eclipse ingress and egress data \citep[e.g.,][]{deWitEtal2012aapFacemap, MajeauEtal2012Facemap}. Observations in the HST WFC3 bandpass could break the degeneracy between
O-rich and C-rich atmospheric models.  Finally, the possibility of
measuring orbital decay in the future is exciting because of the unique
constraints this could place on stellar interior parameters.

\newpage
\section{SYSTEM PARAMETERS}
\label{sec:app}

Table \ref{table:SystemParams} lists WASP-43 system parameters derived from our analysis and the literature.

\begin{table}[ht!]
\footnotesize{
\caption{\label{table:SystemParams} 
System Parameters of WASP-43}
\atabon\strut\hfill\begin{tabular}{lcr}
    \hline
    \hline
    Parameter                                                              & Value                   & Reference                 \\
    \hline
    \multicolumn{3}{c}{Eclipse Parameters}                                                                                       \\
    \hline 
    Eclipse midpoint (BJD\sb{TDB}) (2011 Jul 30)                            &    2455773.3179 {\pm} 0.0003    & (1)                \\ 
    Eclipse midpoint (BJD\sb{TDB}) (2011 Jul 29)                            &    2455772.5045 {\pm} 0.0003    & (1)                \\ 
    Eclipse duration  \math{t\sb{\rm 4-1}} (hr)                           &    1.25 {\pm} 0.02              & (1)                \\ 
    Eclipse depth {\em Spitzer} IRAC, 3.6 {\micron} (\%)                   &    0.347 {\pm} 0.013            & (1)                \\ 
    Eclipse depth {\em Spitzer} IRAC, 4.5 {\micron} (\%)                   &    0.382 {\pm} 0.015            & (1)                \\ 
    Eclipse depth VLT HAWK-I, 2.095 {\micron} (ppm)                  &    1560 {\pm} 140               & (2)                \\ 
    Eclipse depth VLT HAWK-I, 1.186 {\micron} (ppm)                  &    790 {\pm} 320                & (2)                \\ 
    Ingress/egress time \math{t\sb{\rm 2-1}} (hr)                        &    0.264 {\pm} 0.018            & (1)                \\\hline  
    \multicolumn{3}{c}{Orbital parameters}                                                                                       \\ \hline
    Orbital period, \math{P} (days)                                        &    \math{P} = 0.81347436 {\pm} 1.4\tttt{-7} & (3) \\  
    Semimajor axis, \math{a} (AU)                                          &    0.01526 {\pm} 0.00018        & (2)                \\ 
    Transit time (BJD\sb{TDB})                                             &    2455726.54336 {\pm} 0.00012  & (2)                \\ 
    Orbital eccentricity, \math{e}                                         &    0.010 $^{+0.010}_{-0.007}$   & (3)                \\ 
    Argument of pericenter, \math{\omega} (deg)                            &    \math{-88} $^{+5}_{-9}$      & (3)                \\ 
    Velocity semiamplitude, \math{K} (m\,s\sp{-1})                         &    549 {\pm} 6                  & (3)                \\ 
    Center-of-mass velocity \math{\gamma} (m\,s\sp{-1})                    &    \math{-3595}     {\pm} 4     & (3)                \\ \hline         
    \multicolumn{3}{c}{Stellar parameters}                                                                                       \\ \hline
    Spectral type                                                          &    see Section \ref{intro}       & (2)                \\ 
    Mass, \math{M\sb{\rm *}} (\math{M\sb{\odot}})                          &    0.717 {\pm} 0.025             & (2)                \\ 
    Radius, \math{R\sb{\rm *}} (\math{R\sb{\odot}})                        &    0.667 $^{+0.011}_{-0.010}$    & (2)                \\ 
    Mean density, \math{\rho\sb{\rm *}} (\math{\rho\sb{\odot}})            &    2.410 $^{+0.079}_{-0.075}$    & (2)                \\ 
    Effective temperature, \math{T\sb{\rm eff}} (K)                        &    4520 {\pm} 120                & (2)                \\ 
    Surface gravity, log \math{g\sb{\rm *}} (cgs)                          &    4.645 $^{+0.011}_{-0.010}$    & (2)                \\ 
    Projected rotation rate, \math{v\sb{\rm *} \sin(i)} (kms\sp{-1})       &    4.0 {\pm} 0.4                 & (2)                \\ 
    Metallicity [Fe/H] (dex)                                               &    -0.01 {\pm} 0.12              & (2)                \\ 
    Distance (pc)                                                          &    80 {\pm} 20                   & (4)                \\ \hline 
    \multicolumn{3}{c}{Planetary parameters }                                                                                    \\ \hline
    Mass, \math{M\sb{\rm p}} (\math{M\sb{\rm J}})                          &    2.034 $^{+0.052}_{-0.051}$    & (2)               \\ 
    Radius, \math{R\sb{\rm p}} (\math{R\sb{\rm J}})                        &    1.036 {\pm} 0.019             & (2)               \\ 
    Surface gravity, log \math{g\sb{\rm p}} (cgs)                          &    3.672 $^{+0.013}_{-0.012}$    & (2)               \\ 
    Mean density, \math{{\rho}\sb{\rm p}} (g\,cm\sp{-3})                   &    1.377 $^{+0.063}_{-0.059}$    & (2)               \\ 
    Equilibrium temperature (A = 0), \math{T\sb{\rm eq}} (K)                 &    1440 $^{+40}_{-39}$           & (2)               \\ 
    \hline\\
    \multicolumn{3}{l}{{\bf Notes. References.} (1) This work (parameters derived using joint fit, see Section \ref{sec:joint});}\\
    \multicolumn{3}{l}{ (2) \citet{GillonEtal2012A&A-WASP-43b}; (3) This work (see Section \ref{sec:orbit}); (4)\citet{Hellier2011-WASP43b}.}
\end{tabular}\hfill\strut\ataboff
}
\end{table}

\section{ACKNOWLEDGMENTS}

We thank the observers listed in Table \ref{tab:ttv} for allowing us to
use their results and the organizers of the Exoplanet Transit
Database for coordinating the collection and uniform analysis
of these data.
We also thank contributors to SciPy, Matplotlib, the Python Programming
Language, the free and open-source community, the
NASA Astrophysics Data System, and the JPL Solar System Dynamics group
for free software and services.
The IRAC data are based on observations made
with the {\em Spitzer Space Telescope}, which is operated by the
Jet Propulsion Laboratory, California Institute of Technology,
under a contract with NASA. Support for this work was
provided by NASA through an award issued by JPL/Caltech; through the
Science Mission Directorate's Planetary Atmospheres Program, grant
NNX12AI69G; Astrophysics Data Analysis Program, grant NNX13AF38G; and by NASA Headquarters under the NASA Earth and Space Science Fellowship Program, grant NNX12AL83H. NM acknowledges support from the Yale Center for Astronomy and Astrophysics through the YCAA postdoctoral Fellowship.

{\em Facility:} Spitzer

\newpage
\bibliographystyle{apj}
\bibliography{chap-wasp43b}

\fi

\chapter{TEA: A CODE FOR CALCULATING THERMOCHEMICAL \\EQUILIBRIUM ABUNDANCES}
\label{chap:TEA}

{\singlespacing
\noindent{\bf Jasmina Blecic\sp{1}, Joseph Harrington\sp{1}, M.\ Oliver Bowman\sp{1}}

\vspace{1cm}

\noindent{\em
\sp{1}Planetary Sciences Group, Department of Physics, University of Central Florida, Orlando, FL 32816-2385, USA}

\vspace{3cm}

\centerline{Submitted to {\em The Astrophysical Journal Supplement Series}} 
\centerline{18 May 2015}

\vspace{5cm}

\centerline{Publication reference:} 
\centerline{Blecic, J., Harrington, J., \& Bowman, M. O. 2015, ArXiv e-prints}
\vspace{0.2cm} 

\centering{http://adsabs.harvard.edu/abs/2015arXiv150506392B}

}
\clearpage

\if \includegTEA y
    
\setcitestyle{authoryear,round}

\section{ABSTRACT}

We present an open-source Thermochemical Equilibrium Abundances (TEA)
code that calculates the abundances of gaseous molecular species.  The
code is based on the methodology of
\citet{WhiteJohnsonDantzig1958JGibbs} and \citet{Eriksson1971}. It
applies Gibbs free-energy minimization using an iterative, Lagrangian
optimization scheme. Given elemental abundances, TEA calculates
molecular abundances for a particular temperature and pressure or a
list of temperature-pressure pairs. We tested the code against the
method of \citet{BurrowsSharp1999apjchemeq}, the free thermochemical
equilibrium code CEA (Chemical Equilibrium with Applications), and the
example given by \citet{WhiteJohnsonDantzig1958JGibbs}. Using their
thermodynamic data, TEA reproduces their final abundances, but with
higher precision. We also applied the TEA abundance calculations to
models of several hot-Jupiter exoplanets, producing expected
results. TEA is written in Python in a modular format. There is a 
start guide, a user manual, and a code document in addition to this
theory paper. TEA is available under a reproducible-research, 
open-source license via {\tt https://github.com/dzesmin/TEA}.

\section{INTRODUCTION}
\label{intro}

There are two methods to calculate chemical equilibrium: using
equilibrium constants and reaction rates, i.e., kinetics, or
minimizing the free energy of a system \citep{bahn1960kinetics,
  ZeleznikGordon:1968}.

The kinetic approach, where the pathway to equilibrium needs to be
determined, is applicable for a wide range of temperatures and
pressures. However, using kinetics for thermochemical equilibrium
calculations can be challenging. Chemical equilibrium can be
calculated almost trivially for several reactions present in the
system, but as the number of reactions increases, the set of numerous
equilibrium constant relations becomes hard to solve
simultaneously. To have an accurate kinetic assessment of the system,
one must collect a large number of reactions and associate them with
the corresponding rates. This is not an issue at lower temperatures,
where reaction rates are well known. However, at high temperatures,
where thermochemical equilibrium should prevail, one needs to know
forward and reverse reactions and corresponding reaction rates, which
are less well known.

The advantage of the free energy minimization method is that each
species present in the system can be treated independently without
specifying complicated sets of reactions a priori, and therefore, a
limited set of equations needs to be solved
\citep{ZeleznikGordon:1960}. In addition, the method requires only
knowledge of the free energies of the system, which are well known,
tabulated, and can be easily interpolated or extrapolated.

Thermochemical equilibrium calculations have been widely used in
chemical engineering to model combustion, shocks, detonations and the
behaviour of rockets and compressors \citep[e.g.,
][]{MillerEtal-annualReview, BelfordStrehlow-annualReview}. In
astrophysics, they have been used to model the solar nebula, the
atmospheres and circumstellar envelopes of cool stars, and the
volcanic gases on Jupiter's satellite Io \citep[e.g.,
][]{LaurettaLoddersFegley997S-nebula, LoddersFegley1993-circum,
  ZolotovFegley1998-Io}. Thermochemistry also governs atmospheric
composition in vast variety of giant planets, brown dwarfs, and
low-mass dwarf stars \citep[][and references therein ]{Lodders02,
  Visscher2006, VisscherEtal2010-chem}.

\subsection{Chemical Models of Exoplanets}
\label{sec:GibbsMinim}

To perform a comprehensive study of a planetary atmosphere, aside from
thermoequilibrium chemistry, one must consider disequilibrium
processes like photochemistry, vertical mixing, horizontal transport,
and transport-induced quenching \citep{MosesEtal2011-diseq,
  MosesEtal2013-COratio, VenotEtal2012-chem, Venot2014-metallicity,
  Agundez2012, Agundez2014, LineEtal2010ApJHD189733b,
  LineEtal2011-kinetics, VisscherMoses2011-quench}. Today, we have 1D
chemical models that integrate thermochemistry, kinetics, vertical
mixing, and photochemistry \citep{LineEtal2011-kinetics,
  MosesEtal2011-diseq, VisscherEtal2010IcarJupiter}. These models have
an ability to smoothly transition from the thermochemical-equilibrium
regime to transport-quenched and photochemical regimes. Specifically,
in giant planets, we can distinguish three chemical layers: deep
within the planetary atmosphere, the temperatures and pressures are so
high that chemical reaction timescales are short, ensuring a chemical
equilibrium composition; at lower temperatures and pressures higher in
the atmosphere, the timescales for chemical reactions slows down,
reaching the vertical transport timescale and smoothing the vertical
mixing-ratio profile by producing quenched abundances; high in the
atmosphere, the host star's ultraviolet radiation destroys stable
molecules, driving photochemical reactions.
  
Photochemical models today face several difficulties. They lack
high-temperature photochemical data, and the list of reactions and
associated rate coefficients are not well defined or are conflicted
\citep{VenotEtal2012-chem, VisscherEtal2010IcarJupiter}. In addition,
the exoplanet photospheres observed with current instruments are
sampled within the region of the atmosphere dominated by vertical
mixing and quenching, but not by photochemistry
\citep{LineYung2013-diseq}.

The majority of early hot-Jupiter atmospheric models assumed chemical
composition consistent with thermochemical equilibrium \citep[e.g.,
][]{BurrowsEtal2007ApJHD209458, FortneyEtal2005apjlhjmodels,
  MarleyEtal2007ApJ, FortneyEtal2010ApJ-transmission,
  BurrowsSharp1999apjchemeq, SharpBurrows2007Apjopacities,
  RogersEtal2009ApJ-groundCoRoT1b}. More recently, a variety of 1D
photochemical models has been used to explore the compositions of hot
Jupiters \citep{MosesEtal2011-diseq, Zahnle09-SulfurPhotoch,
  LineEtal2010ApJHD189733b, KopparapuEtal2012-photoch,
  VenotEtal2012-chem, LineYung2013-diseq, Visscher2006,
  VisscherEtal2010IcarJupiter, VenotEtal2012-chem}. A common
conclusion of these studies is that in hot atmospheres (\math{T}
\math{>} 1200 K), disequilibrium effects are so reduced that
thermochemical equilibrium prevails.

Using secondary eclipse observations as the most fruitful technique
today to assess atmospheric composition \citep[e.g.,
][]{KnutsonEtal2009ApJ-Tres4Inversion,
  KnutsonEtal2009ApJ-redistribution, Machalek2008-XO-1b,
  StevensonEtal2012apjHD149026b, FraineEtal2013-GJ1214b,
  CrossfieldEatl2012ApJWASP12b-reavaluation,
  TodorovEtal2012ApJ-XO4b-HATP6b-HATP8b, DesertEtal2009ApJ-CO,
  DemingEtal2010arXiv-CoRoT12, Demory2007aaGJ436bspitzer,
  MadhusudhanSeager2009ApJ-AbundanceMethod,
  MadhusudhanEtal2011natWASP12batm, BlecicEtal2013ApJ-WASP14b,
  BlecicEtal2014ApJ-WASP43b}, \citet{LineYung2013-diseq} studied the most
spectroscopically active species in the infrared on eight hot planets
(GJ436b, WASP-12b, WASP-19b, WASP-43b, TrES-2b, TrES-3b, HD 189733b,
and HD 149026b), with equilibrium temperatures ranging between 744 K
and 2418 K. They chose to evaluate the presence of disequilibrium
chemistry at 100 mbar, where most secondary-eclipse observations
sample, i.e., where their thermal emission weighting functions usually
peak. They find that all of the models are consistent with
thermochemical equilibrium within 3\math{\sigma} (the work of
\citealp{StevensonEtal2010Natur}, however, questions this conclusion for
GJ436b). They also show that for the hottest planets, (T\sb{100mb}
\math{>} 1200 K ), CH\sb{4}, CO, H\sb{2}O, and H\sb{2} should be in
thermochemical equilibrium even under a wide range of vertical mixing
strengths.

Thermochemical equilibrium calculations are the starting point for
initializing models of any planetary atmosphere. In general,
thermochemical equilibrium governs the composition of the deep
atmospheres of giant planets and brown dwarfs, however, in cooler
atmospheres thermoequilibrium calculations are the necessary baseline
for further disequilibrium assessment. They can also provide a
first-order approximation for species abundances as a function of
pressure, temperature, and metallicity for a variety of atmospheres
\citep[e.g., ][]{VisscherEtal2010IcarJupiter, Lodders02}.

The Gibbs free energy minimization method for calculating
thermochemical equilibrium abundances of complex mixtures was first
introduced by \citet{WhiteJohnsonDantzig1958JGibbs}.  Prior to 1958
all equilibrium calculations were done using equilibrium constants of
the governing reactions. \citet{WhiteJohnsonDantzig1958JGibbs} were
the first to develop a method that makes no distinction among the
constituent species and does not need a list of all possible chemical
reactions and their rates. Rather, it depends only on the chemical
potentials of the species involved. To derive the numerical solution,
they apply two computational techniques: a steepest-descent method
applied to a quadratic fit and the linear programming method.

Following their methodology, \citet{Eriksson1971} developed the SOLGAS
code that calculates equilibrium composition in systems containing
ideal gaseous species and pure condensed phases. Subsequent
modification of this code were made by
\citet{eriksson1973thermodynamic}, \citet{eriksson1975thermodynamic},
and \citet{besmann1977solgasmix}, after which the code was modified
for astrophysical applications and called SOLAGASMIX by
\citet{SharpHuebner90, PetaevWood1998MPSConden},
\citet{BurrowsSharp1999apjchemeq}, and
\citet{SharpBurrows2007Apjopacities}.

The Gibbs free energy minimization approach has been used by many
authors in the exoplanetary field \citep[e.g., ][]{SeagerPhD1999,
  Seager2010-ExoplanetAtmospheres,
  MadhusudhanSeager2009ApJ-AbundanceMethod, MadhusudhanSeager2010,
  SharpHuebner90}. In addition to SOLAGASMIX and other proprietary
codes \citep[e.g., CONDOR by][]{FegleyLodders1994IcarChemiModelSatJup,
  LoddersFegley1994-CONDOR}, and one analytic method to calculate
major gaseous species in planetary atmospheres by
\citet{BurrowsSharp1999apjchemeq}, just one free-software code is
available to the exoplanet community, CEA \citep[Chemical Equilibrium with Applications, {\tt
    http://www.grc.nasa.gov/WWW/CEAWeb}, by][]{GordonMcBride:1994}, .

In this paper, we present a new open-source code, Thermochemical
Equilibrium Abundances (TEA). The TEA code is a part of the
open-source Bayesian Atmospheric Radiative Transfer project ({\tt
  https://github.com/joeharr4/BART}). This project consists of three
major parts: TEA - this code, a radiative-transfer code that 
models planetary spectra, and a statistical module that compares
theoretical models with observations.

TEA calculates the equilibrium abundances of gaseous molecular
species. Given a single \math{T, P} point or a list of \math{T, P}
pairs (the thermal profile of an atmosphere) and elemental abundances,
TEA calculates mole fractions of the desired molecular species. The
code is based on the Gibbs free energy minimization calculation of
\citet{WhiteJohnsonDantzig1958JGibbs} and \citet{Eriksson1971}. TEA
uses 84 elemental species and the thermodynamical data for more then
600 gaseous molecular species available in the provided JANAF (Joint
Army Navy Air Force) tables \citep[{\tt
    http://kinetics.nist.gov/janaf/},][]{ChaseEtal1982JPhJANAFtables,
  ChaseEtal1986bookJANAFtables}. TEA can adopt any initial elemental
abundances. For user convenience a table with solar photospheric
elemental abundances from \citet{AsplundEtal2009-SunAbundances} is
provided.

The code is written in Python in an architecturally modular format. It
is accompanied by detailed documentation, a start guide, the TEA User 
Manual (Bowman and Blecic), the TEA Code Description document 
(Blecic and Bowman), and the TEA Theory document (this paper), so the
user can easily modify it. The code is actively maintained and available
to the scientific community via the open-source development website {\tt
  GitHub.com} ({\tt https://github.com/dzesmin/TEA,
  https://github.com/dzesmin/TEA-Examples}). \\ This paper covers an
initial work on thermochemical calculations of species in gaseous
phases. Implementation of condensates is left for future work.

In this paper, we discuss the theoretical basis for the method applied
in the code. Section \ref{sec:GibbsMinim} explains the Gibbs Free
energy minimization method; Section \ref{sec:Lagrang} describes the
general Lagrangian optimization method and its application in TEA; in
Section \ref{sec:lambda} we introduce the Lambda Correction algorithm
for handling negative abundances that follow from the Lagrangian
method; Section \ref{sec:struct} describes the layout of the TEA code;
Section \ref{sec:applic} explores chemical equilibrium abundance
profiles of several exoplanetary atmospheres; Section
\ref{sec:validity} compares our code to other methods available, and
Section \ref{sec:conc} states our conclusions.

\section{GIBBS FREE ENERGY MINIMIZATION METHOD}
\label{sec:GibbsMinim}

Equilibrium abundances can be obtained by using different combinations
of thermodynamical state functions: temperature and pressure --
(\math{t, p}), enthalpy and pressure -- (\math{H, p}), entropy and
pressure -- (\math{S, p}), temperature and volume -- (\math{t, v}),
internal energy and volume -- (\math{U, v}), etc. Depending on how the
system is described, the condition for equilibrium can be stated in
terms of Gibbs free energy, helmholtz energy, or entropy. If a
thermodynamic state is defined with temperature and pressure, Gibbs
free energy (G) is most easily minimized, since those two states are
its natural, dependent variables.

Gibbs free energy represents a thermodynamic potential that measures
the useful work obtainable by the system at a constant temperature and
pressure. Thus, the Gibbs free energy minimization method minimizes
the total chemical potential of all involved species when the system
reaches equilibrium.


The Gibbs free energy of the system at a certain temperature is the
sum of the Gibbs free energies of its constituents:
\begin{equation}
G_{sys}(T) = \sum_{i}^nG_{i}(T) \, ,
\label{Gibbs-sys}
\end{equation}
\noindent where \math{G\sb{sys}(T)} is the total Gibbs free energy of
the system for \math{n} chemical species, \math{G\sb{i}(T)} is the
Gibbs free energy of a gas species \math{i}, and \math{T} is the
temperature. The total Gibbs free energy of the system is expressed as
the sum of the number of moles \math{x} of the species \math{i},
\math{x\sb{i}}, and their chemical potentials \math{g\sb{i}(T)}:
\begin{equation}
G_{sys}(T) = \sum_{i}^nx_{i}\,g_{i}(T) \, .
\label{Gibbs-pot}
\end{equation}
\noindent The chemical potential \math{g\sb{i}(T)} depends on the
chemical potential at the standard state \math{g\sp{0}\sb{i}(T)} and
the activity \math{a\sb{i}},
\begin{equation}
g_{i}(T) = g_{i}^0 (T)+ RT\ln a_{i} \, ,
\label{pot}
\end{equation}
\noindent where \math{R} is the gas constant, \math{R} =
\math{k\sb{B}N\sb{A}}, and \math{k\sb{B}} and \math{N\sb{A}} are the
Boltzmann constant and Avogadro's number, respectively. Activities
for gaseous species, which are treated as ideal, are equal to the
partial pressures, and for condensates they equal 1:
\begin{equation}
a_{i} = P_{i} = P\,\frac{x_{i}}{N} \, , \, \,\,\,\, {\rm for\,gases}
\label{pot1}
\end{equation}
\begin{equation}
a_{i} = 1 \,, \, \,\,\,\,\,\,\,\,{\rm for \,condensates} \, ,
\label{pot2}
\end{equation}
\noindent where \math{P} is the total pressure of the atmosphere, \math{N}
is the total number of moles of all species involved in the
system. Hence, Equation (\ref{pot}) for gaseous species becomes:
\begin{equation}
g_{i}(T) = g_{i}^0(T) + RT\ln P_i \, .
\label{pot-pot}
\end{equation}
\noindent Combining Equation (\ref{pot-pot}) with Equation
(\ref{Gibbs-pot}), the Gibbs free energy of the system becomes:
\begin{equation}
G_{sys}(T) = \sum_{i}^n x_{i}\Big(g_{i}^0(T) + RT\ln P_i\Big) \, ,\\
\label{Gibbsfree2}
\end{equation}
\noindent or,
\begin{equation}
G_{sys}(T) = \sum_{i}^n\,x_{i}\Big(g_{i}^0(T) + RT\ln P + RT
\ln\,\frac{x_{i}}{N}\Big) \, ,
\label{GibbsFull}
\end{equation}
\noindent For our purposes, it is more convenient to write Equation
(\ref{GibbsFull}) in unitless terms:
\begin{eqnarray}
\label{eq:eqmin}
\frac{G_{sys}(T)}{RT} = \sum_{i=1}^n x_{i} \Big[\frac{g_{i}^0(T)}{RT}
  + \ln P + \ln\frac{x_{i}}{N}\Big]\, .
\end{eqnarray}
Equation (\ref{eq:eqmin}) requires a knowledge of the free energy of
each species as a function of temperature. These can be obtained from
the JANAF tables \citep[{\tt http://kinetics.nist.gov/janaf/},
][]{ChaseEtal1982JPhJANAFtables, ChaseEtal1986bookJANAFtables,
  BurrowsSharp1999apjchemeq}, or easily derived from other tabulated
functions.

To extract free energies, \math{g\sb{i}\sp{0}(T)/RT}, from the JANAF
tables, we used the expression given in \citet[][]{Eriksson1971},
Equation (2):
\begin{eqnarray}
\label{eq:JANAFconv}
\frac{g_{i}^0(T)}{RT} = 1/R\Big[\frac{G_{i}^0 - H_{298}^0}{T}\Big] +
\frac{\Delta_f H_{298}^0}{RT}\, ,
\end{eqnarray}
\noindent where \math{g\sb{i}\sp{0}(T)} is given in J/mol, \math{R} =
8.3144621 J/K/mol, \math{H\sb{298}\sp{0}} is the enthalpy (heat
content) in the thermodynamical standard state at a reference
temperature of 25\sp{o}C = 298.15 K, \math{G\sb{i}\sp{0}} is the Gibbs
free energy in J/mol, (\math{G\sb{i}^0 - H\sb{298}\sp{0}/T}) is the
free-energy function in J/K/mol, and \math{\Delta\sb{f}
  H\sb{298}\sp{0}} is the heat of formation at 298.15 K in
kJ/mol. Thus, our conversion equation becomes:
\begin{eqnarray}
\label{eq:JANAFconvFinal}
\frac{g_{i}^0(T)}{RT} = 1/R\Big[\frac{G_{i}^0 - H_{298}^0}{T}\Big] +
\frac{\Delta_f H_{298}^0 1000}{RT}\, ,
\end{eqnarray}
\noindent \math{G\sb{i}^0 - H\sb{298}\sp{0}/T} is the fourth term in
the JANAF tables and \math{\Delta\sb{f} H\sb{298}\sp{0}} is the
sixth. The free energy function of a species corresponding to a
temperature other than those provided in the JANAF tables is
calculated using spline interpolation.

Alternatively, the free energies can be calculated using the eighth
term in the JANAF tables, following Equation (3) from
\citet[][]{Eriksson1971}:
\begin{eqnarray}
\label{eq:JANAFconvFinal2}
\frac{g_{i}^0(T)}{RT} = -\,ln\,(10)\,log_{10}\,(K_{f})\, ,
\end{eqnarray}
\noindent where \math{K\sb{f}} is the equilibrium constant of
formation.

To determine the equilibrium composition, we need to find a
non-negative set of values \math{x\sb{i}} that minimizes Equation
(\ref{eq:eqmin}) and satisfies the mass balance constraint:
\begin{eqnarray}
\sum_{i=1}^n a_{ij}\, x_{i} = b_{j}\,, \,\,\,\,(j = 1, 2, ...,m)\, ,
\label{masbal}
\end{eqnarray}
\noindent where the stoichiometric coefficient \math{a\sb{ij}}
indicates the number of atoms of element \math{j} in species \math{i}
(e.g., for CH\sb{4} the stoichiometric coefficient of C is 1 and the
stoichiometric coefficient of H is 4), and \math{b\sb{j}} is the total
number of moles of element \math{j} originally present in the mixture.

We use the reference table containing elemental solar abundances given
in \citet{AsplundEtal2009-SunAbundances} Table 1 for \math{b}
values. \citet{AsplundEtal2009-SunAbundances} adopt the customary
astronomical scale for logarithmic abundances, where hydrogen is
defined as log \math{\epsilon{H}} = 12.00, and log \math{\epsilon{X}}
= log(\math{N\sb{X}/N\sb{H}})+12, where \math{N\sb{X}} and
\math{N\sb{H}} are the number densities of element \math{X} and
\math{H}, respectively. Thus, their values are given in {\em dex}
(decimal exponent) units. We transform these values into elemental
fractions by number, i.e., ratio of number densities. We convert each
species dex elemental abundance into number density and divide it by
the hydrogen number density \citep[Section
  3]{AsplundEtal2009-SunAbundances}. The final output are fractional
abundances (mole mixing fractions), i.e., the ratio of each species'
number of moles to the number of moles in the mixture.

\section{LAGRANGIAN METHOD OF STEEPEST DESCENT}
\label{sec:Lagrang}

To find equilibrium abundances of the desired molecular species at a
given temperature and pressure, we need to minimize Equation
(\ref{eq:eqmin}). To do so, we have to apply a technique that
minimizes a multi-variate function under constraint. There are many
optimization techniques used to find the minima of a function subject
to equality constraints (e.g., line search method, Dantzig-simplex
method for linear programming, Newton-Raphson method,
Hessian-conjugate gradient method, Lagrangian steepest-descent
method). The main advantage of the Lagrangian steepest-descent method
is that the number of equations to solve scales with the number of
different types of atoms present in the mixture, which is usually a
much smaller number than the possible number of molecular
constituents. This allows the code to be executed much faster than in
other methods.

Gradient descent, also known as steepest descent, is an algorithm for
finding a local minimum of a function. At each iteration, the method
takes steps towards the minimum, where each step is proportional to
the negative gradient of the function at the current point. If a
function \math{f(x)} is defined and differentiable in the neighborhood
of a point \math{a}, then \math{f(x)} decreases most rapidly in the
direction of the negative gradient, \math{ - \nabla f(\math{a})}. From
this, it follows that if \math{b} = \math{a - \lambda\nabla
  f(\mathbf{a})}, then \math{f(a) > f(b)} if \math{\lambda} is small
enough. Starting with a guess \math{x\sb{0}} for a local minimum of
\math{f}, and considering a sequence \math{x\sb{0}, x\sb{1},
  x\sb{2},...} such that \math{x\sb{n+1} = x\sb{n} - \lambda\nabla
  f(x\sb{n}),\, n \geq 0}, one gets \math{f(x\sb{0})} \math{\geq}
\math{f(x\sb{1})} \math{\geq} \math{f(x\sb{2})} \math{\geq}... . This
sequence of \math{x\sb{n}} converges to a desired local minimum if the
correct \math{\lambda} value is assigned. The value of \math{\lambda}
can vary at each iteration. If the function \math{f} is convex, the
local minimum is also the global minimum.

\begin{figure*}[hb!]
\centerline{
\includegraphics[width=0.51\textwidth]{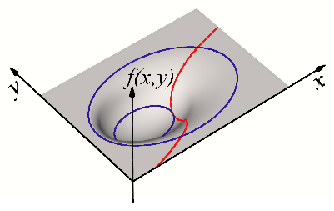}\hspace{0.5cm}
\includegraphics[width=0.45\textwidth]{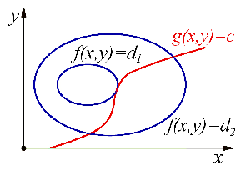}}
\caption[Example of the Lagrangian minimization approach]
  {Example of the Lagrangian minimization approach.  {\bf
    Left:} The 3D illustration of the minimization problem. Blue lines
  indicate starting and ending values of \math{f(x,y)} during
  minimization. An (\math{x}, \math{y}) pair is found that minimizes
  \math{f(x,y)} (bottom blue line) subject to a constraint
  \math{g(x,y)=C} (red line).  {\bf Right:} Contour map of the left
  figure. The point where the red line (constraint) tangentially
  touches a blue contour is the solution. Since \math{d_1<d_2}, the
  solution is the minimum of \math{f(x,y)}.}
\label{fig:Dz-slike}
\end{figure*}

Our code implements a more complex version of the method outlined
above. The problem consists of some function \math{f(x,y)} subject to
a constraint \math{g(x, y)=C}. In this case, we need both \math{f} and
\math{g} to have continuous first partial derivatives. Thus, we
introduce a new variable called the {\em Lagrangian multiplier},
\math{\pi}, where:
\begin{eqnarray}
\Lambda (x, y, \lambda) = f(x, y) \pm \pi\, (g\,(x, y) - C)\, ,
\label{Lagr-mult}
\end{eqnarray}
\noindent which allows us to find where the contour of \math{g(x,
  y)=C} tangentially touches \math{f(x,y)} (Figure
\ref{fig:Dz-slike}). The point of contact is where their gradients
are parallel:
\begin{eqnarray}
\nabla_{xy}\,f(x, y) = - \pi\nabla_{xy}\,g(x, y) \, .
\label{Lagr-cond}
\end{eqnarray}
\noindent The constant \math{\pi} allows these gradients to have
different magnitudes. To find the minimum, we need to calculate all
partial derivatives of the function \math{\Lambda}, equate them with
zero,
\begin{eqnarray}
\nabla_{x, y, \pi}\,\Lambda(x, y, \pi) = 0 \, ,
\label{Lagr-final-cond}
\end{eqnarray}
\noindent and follow the same iteration procedure as explained above.

\subsection{Lagrangian Method in TEA}
\label{sec:LagrTEA}

To implement this in our code, we followed the methodology derived in
\citet{WhiteJohnsonDantzig1958JGibbs}. We applied an iterative
solution to the energy minimization problem, where the mole numbers of
the desired molecular species are recomputed at each step and the new
direction of steepest descent is calculated. This produces improved
mole number values, which however, could be negative. Thus, two short
procedures are required in each iteration cycle: solving a set of
simultaneous linear equations for an improved direction of descent
(described in this Section) and approximately minimizing a convex
function of one variable, \math{\lambda}, to ensure that all improved
mole number values are positive (Section \ref{sec:lambda}).

To calculate the direction of steepest descent (following the
methodology derived in Section \ref{sec:Lagrang}) and initiate the
first iteration cycle, we first need to solve the mass balance
equation, Equation (\ref{masbal}). We start from any positive set of
values for the initial mole numbers, \math{y = (y\sb{1}, y\sb{2}, ...,
  y\sb{n})}, as our initial guess:
\begin{eqnarray}
\sum_{i=1}^n a_{ij}\, y_{i} = b_{j}\, \,\,\,\,(j = 1, 2, ...,m) \, .
\label{masbal2}
\end{eqnarray}
\noindent To satisfy the mass balance Equation (\ref{masbal2}), some
\math{y\sb{i}} variables must remain as free parameters. In solving
these equations, we leave as many free parameters as we have elements
in the system, thus ensuring that the mass balance equation can be
solved for any number of input elements and output species the user
chooses. We set all other \math{y\sb{i}} to a known, arbitrary
number. Initially, the starting values for the known species are set
to 0.1 moles, and the mass balance equation is calculated. If that
does not produce all positive mole numbers, the code automatically
sets known parameters to 10 times smaller and tries again. The initial
iteration input is set when all mole numbers are positive, and the
mass balance equation is satisfied.

To follow with the Lagrangian method, we denote two terms in Equation
 (\ref{eq:eqmin}) as:
\begin{eqnarray}
\label{eq:c}
c_i = \frac{g_{i}^0(T)}{RT} + \ln P \, ,
\end{eqnarray}
\noindent where \math{P} is the pressure in bar. Using \math{c\sb{i}},
we denote the right side of Equation (\ref{eq:eqmin}) as the variable
\math{f\sb{i}(Y)}:
\begin{eqnarray}
f_i(Y) = y_i\Big[c_i + \ln\frac{y_{i}}{\bar{y}}\Big]\, ,
\label{f}
\end{eqnarray}
\noindent where \math{Y} = (\math{y\sb{1}, y\sb{2}, ..., y\sb{n}}) and
\math{\bar{y}} is the total initial number of moles.  The left 
side of Equation (\ref{eq:eqmin}), \math{G\sb{sys}(T)/RT}, we denote
as function \math{F(Y)}:
\begin{eqnarray}
F(Y) = \sum_{i=1}^n\,y_i\Big[c_i + \ln\frac{y_{i}}{\bar{y}}\Big]\, .
\label{sumf}
\end{eqnarray}
Then, we do a Taylor series expansion of the function \math{F} about
\math{Y}. This yields a quadratic approximation \math{Q(X)}:
\begin{eqnarray}
Q(X) = F(X)\Big|_{X=Y} + \sum_i \frac{\partial F}{\partial
  x_i}\Big|_{X=Y} \, \Delta_i + \\ \nonumber
\frac{1}{2}\sum_i\sum_k\frac{\partial^2\,F}{\partial x_i\partial
  x_k}\Big|_{X=Y}\, \Delta_i\, \Delta_k\, .
\label{quadrApp}
\end{eqnarray}
\noindent where \math{\Delta\sb{i} = x\sb{i} - y\sb{i}}, and
\math{x\sb{i}} are the improved mole numbers. This function is
minimized using the Lagrangian principle. We now introduce Lagrangian
multipliers as \math{\pi\sb{j}}:
\begin{eqnarray}
G(X) = Q(X) + \sum_j \pi_j (- \sum_i a_{ij}\,x_i +b_j)\, ,
\label{minim}
\end{eqnarray}
\noindent and calculate the first derivatives, \math {\partial G/
  \partial x\sb{i}}, of the new function. We equate them to zero to
find the minima, \math {\partial G/ \partial x\sb{i}} = 0.

We solve for \math{x\sb{i}} from Equation (\ref{minim}) by combining
Equation (\ref{masbal2}) and (\ref{f}) with the fact that
\math{\bar{x}} is the sum of the improved mole numbers, \math{\bar{x}
  = \sum_{i=1}^n x_i}. The improved number of moles, \math{x\sb{i}},
are given as:
\begin{eqnarray}
x_i = - f_i(Y) + (\frac{y_{i}}{\bar{y}})\,\bar{x} + (\sum_{j=1}^m
\pi_j\,a_{ij})\,y_i\, ,
\label{x_i2}
\end{eqnarray}
\noindent while the Lagrangian multipliers, \math{\pi\sb{j}}, are
expressed as:
\begin{eqnarray}
\sum_{j=1}^m \pi_j\,\sum_{i=1}^n\,a_{ij}\,y_i = \sum_{i=1}^n y_i
\Big[\frac{g_{i}^0(T)}{RT} + \ln P +\ln {\frac{y_{i}}{\bar{y}}}\Big]
\, ,
\label{pi}
\end{eqnarray}
\noindent where \math{j} iterates over the \math{m} elements and
\math{i} iterates over the \math{n} species. \math{\bar{x}} and
\math{\bar{y}} are the sums of improved and initial number of
moles, respectively. Using Equation (\ref{f}), we can now rewrite Equation
(\ref{pi}) as:
\begin{eqnarray}
\sum_{j=1}^m \pi_j\,b_j = \sum_{i=1}^n f_i(Y) \, .
\label{pi2}
\end{eqnarray}
If we further denote the constants with:
\begin{eqnarray}
r_{jk} = r_{kj} = \sum_{i=1}^n (a_{ij}\,a_{ik})\,y_{i}, 
\label{r}
\end{eqnarray}
\noindent combining Equations (\ref{x_i2}), (\ref{pi2}), and (\ref{r}),
we get the following system of \math{m+1} equations that can easily be
solved:
\begin{align}
  r_{11}\pi_1 + r_{12}\pi_2 + ... + r_{1m}\pi_m + b_1\,u =
  \sum_{i=1}^n a_{i1}\,f_i(Y)\, , \nonumber \\ r_{21}\pi_1 +
  r_{22}\pi_2 + ... + r_{2m}\pi_m + b_2\,u = \sum_{i=1}^n
  a_{i2}\,f_i(Y)\, , \nonumber \\ . \hspace{120pt} , \nonumber
  \\ . \hspace{120pt} , \label{final-set} \\ . \hspace{120pt} ,
  \nonumber \\ r_{m1}\pi_1 + r_{m2}\pi_2 + ... + r_{mm}\pi_m + b_m\,u
  = \sum_{i=1}^n a_{im}\,f_i(Y)\, , \nonumber \\ b_1\pi_1 + b_2\pi_2 +
  ... + b_m\pi_m + 0\,u = \sum_{i=1}^n f_i(Y)\, , \nonumber
\end{align}
\noindent where:
\begin{eqnarray}
u = -1 + \bar{x}/\bar{y}\, .
\label{u}
\end{eqnarray}
The solutions to Equations (\ref{final-set}) and (\ref{u}) will give
\math{\pi\sb{j}} and \math{u}, and from them using Equation
(\ref{x_i2}) we can calculate the next set of improved mole numbers,
i.e., an improved direction of descent, \math{\Delta\sb{i} = x\sb{i} -
  y\sb{i}}.

\section{LAMBDA CORRECTION ALGORITHM}
\label{sec:lambda}

Solving a system of linear equations (i.e.,
performing the Lagrangian calculation) can also lead to negative mole
numbers for some species, so a short
additional step is needed to eliminate this possibility and guarantee
a valid result.

To do so, the difference between the initial and final values given by
the Lagrangian calculation, \math{\Delta\sb{i} = x\sb{i} - y\sb{i}},
we will call the total distance for each species. To ensure that all
improved mole numbers are positive, we introduce a new value,
\math{\lambda}, that defines the fraction of the total distance as
\math{\lambda\Delta\sb{i}} (see Figure \ref{fig:lambda}).

\begin{figure}[!h]
\centering
\includegraphics[height=6.0cm, clip, trim=0cm 0cm 0cm 0cm]{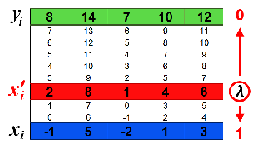}
\caption[Simplified illustration of the lambda correction algorithm]
  {Simplified illustration of the lambda correction algorithm.
  Initial values for one hypothetical Lagrangian iteration cycle,
  \math{y\sb{i}}, are given in green. These values are all positive
  and satisfy the mass balance equation, Equation \ref{masbal2}. The
  {x\sb{i}} values, given in blue, are the values produced by the
  Lagrangian calculation. These values can be negative, but they also
  satisfy the mass balance equation.  The \math{x\sp{'}\sb{i}} values,
  given in red, are produced by choosing the maximum value of lambda
  that ensures all positive and non-zero \math{x\sp{'}\sb{i}}. These
  values become the new initial values of \math{y\sb{i}} for the next
  iteration cycle.}
\label{fig:lambda}
\end{figure}

The computed changes, \math{\lambda\Delta\sb{i}}, are considered to be
{\em directional numbers} indicating the preferred direction of
descent the system moves to. Other than providing all positive mole
numbers, we determine the value \math{\lambda} so that the Gibbs
energy of the system must decrease, i.e., the minimum point is not
passed (see Equation \ref{direc-deriv}).

At each Lagrangian iteration cycle we start with the initial positive
values, \math{y\sb{i}} and we get the next set of improved values
\math{x\sb{i}} given as:
\begin{eqnarray}
x_i = y_i + \Delta_i \, .
\label{xi-lambda}
\end{eqnarray}
\noindent Since we do not want any \math{x\sb{i}} to be negative, the
variable \math{\lambda} performs a small correction:
\begin{eqnarray}
x_i^{'} = y_i + \lambda\Delta_i \, .
\label{xi_corr}
\end{eqnarray}
\noindent \math{\lambda} takes values between 0 and 1, where value
of {\em zero} implies no step is taken from the iteration's original
input, \math{y\sb{i}}, and {\em one} implies that the full Lagrangian
distance is travelled, \math{\Delta_i}. We now rewrite Equation
(\ref{f}) using Equation (\ref{xi_corr}) as:
\begin{eqnarray}
f_{i}(X^{'}) = x_{i}^{'}\Big(\frac{g_{i}^{0}(T)}{RT} + \ln P +\ln
\frac{x_{i}^{'}}{\bar{x}^{'}}\Big) ,
\label{f_X}
\end{eqnarray}
\noindent which can be written in the form:
\begin{eqnarray}
f_i(\lambda) = (y_i+\lambda\Delta_i)\Big(\frac{g_{i}^0(T)}{RT} + \ln P
+\ln \frac{y_{i}+
  \lambda\Delta_i}{{\bar{y}}+\lambda{\bar{\Delta}}}\Big) ,
\label{f_i_lambda}
\end{eqnarray}
\noindent where \math{\bar{\Delta}} = \math{\bar{y} -
  \bar{x}}. Summing over \math{i}, we get a new function,
\math{F(\lambda)}:
\begin{eqnarray}
F(\lambda) = \sum_i\,(y_i+\lambda\Delta_i)\Big(\frac{g_{i}^0(T)}{RT} +
\ln P +\ln \frac{y_{i}+
  \lambda\Delta_i}{{\bar{y}}+\lambda{\bar{\Delta}}}\Big) .
\label{Flambda}
\end{eqnarray}

\comment{ 
We choose \math{\lambda} values using sampling that will ensure
effective exploration of the [0, 1] range. Half of the range is
sampled exponentially, and the other half linearly, totalling 150
points. The exponential sampling is chosen to prevent steps that can
cross the negative threshold for \math{\Delta\sb{i}} (where mole
numbers become negative). For lower temperatures, this threshold has
proven to be very low, thus it requires the smallest step in
\math{\lambda} space. Increasing the number of points in
\math{\lambda} space in certain cases (very low temperatures) can
provide better numerical precision.
}

Thus, to ensure that the new corrected values \math{x\sb{i}\sp{'}} are
all positive, the distance travelled will be limited to fractional
amounts defined by \math{\lambda\Delta\sb{i}}, using the largest
possible value of \math{\lambda} that satisfies the conditions:

{
\begin{enumerate}
\setlength\itemsep{0ex}
\setlength\topsep{0ex}
\setlength\partopsep{0ex}
\setlength\parsep{0ex}

\item The function called the {\em directional derivative} is defined
  and exists:
\begin{eqnarray}
\frac{dF(\lambda)}{d\lambda} = \sum_{i=1}^n \Delta_i
\Big[\frac{g_{i}^0(T)}{RT} + \ln P + \ln \frac{y_{i} +
    \lambda\Delta_i}{\bar{y} + \lambda\bar{\Delta}}\Big] \, .
\label{direc-deriv}
\end{eqnarray}
\item The directional derivative does not become positive (the minimum
  point is not passed).
\end{enumerate}
}

Every new iteration starts with a different set of \math{y\sb{i}},
thus changing the convex function \math{F(\lambda)}, Equation
\ref{Flambda}, and producing a new minimum. This yields to a new
\math{\lambda} value.  \math{\lambda} will be found to approach unity
after some number of iterations. Unity in \math{\lambda} indicates the
solution is near.

We repeat the Lagrangian method and the lambda correction until a
pre-defined maximum number of iterations is met. The final abundances
are given as fractional abundances (mole mixing fractions), i.e., the
ratio of each species' mole numbers to the total sum of mole numbers
of all species in the mixture.

\section{CODE STRUCTURE}
\label{sec:struct}

The TEA code is written entirely in Python and uses the Python
packages NumPy ({\tt http://numpy.org/}) and ({\tt
  http://www.scipy.org/}) along with SymPy, an external linear
equation solver ({\tt http://sympy.org/}).

The code is divided into two parts: the pre-pipeline that makes the
thermochemical data library and stoichiometric tables, and the
pipeline that performs abundance calculations.  Given elemental
abundances, TEA calculates molecular abundances for a particular
temperature and pressure or a list of temperature-pressure
pairs. Documentation is provided in the TEA User Manual (Bowman and
Blecic) and the TEA Code Description (Blecic and Bowman) that
accompany the code. Figure \ref{fig:TEAflow} shows the layout of the
TEA program's flow.  Its modules are:

\begin{figure*}[ht!]
\centering
\includegraphics[width=13cm, trim=22 100 27 110,
clip=true]{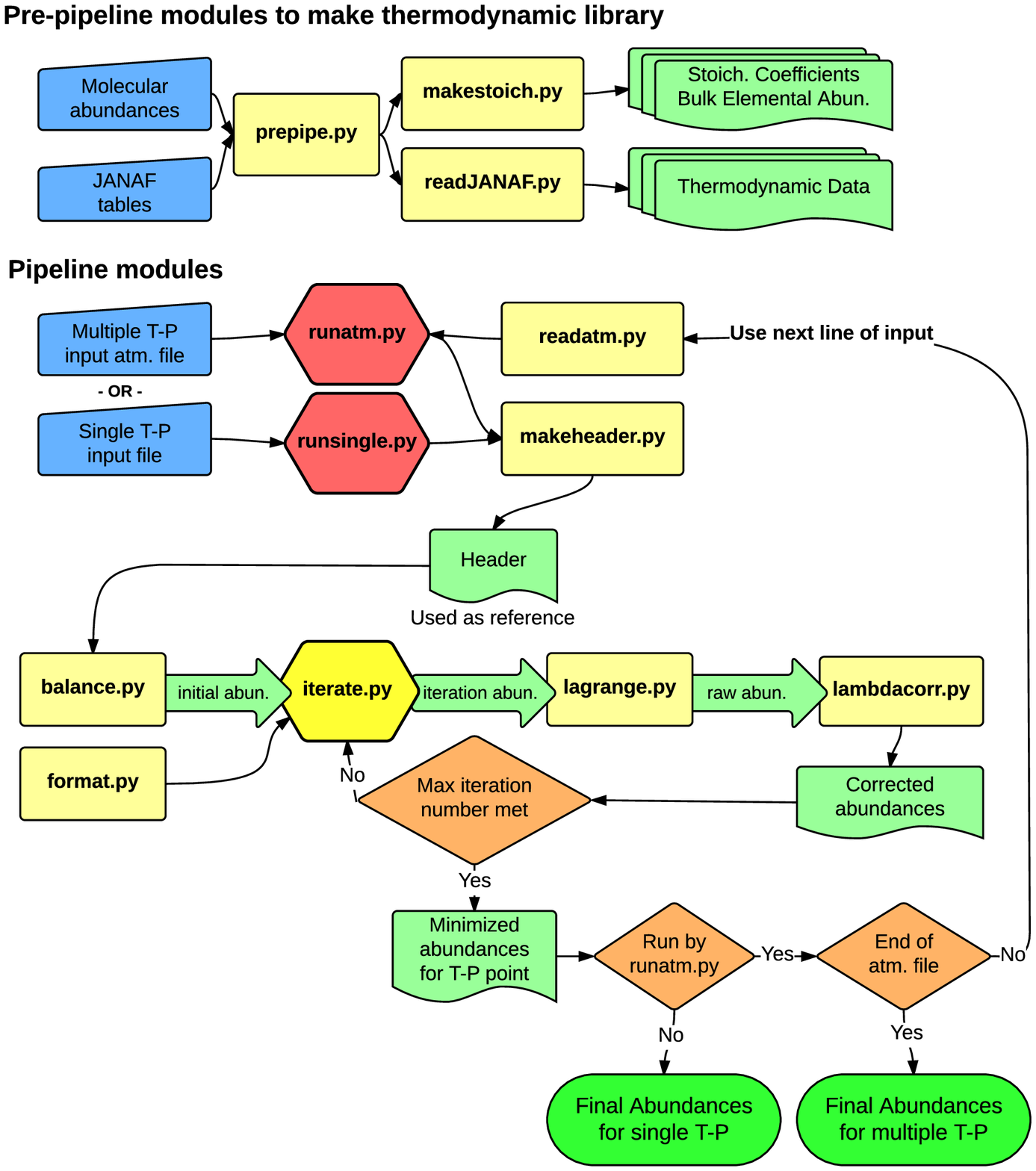}
\includegraphics[width=12.2cm, trim=22 340 27 320,
clip=true]{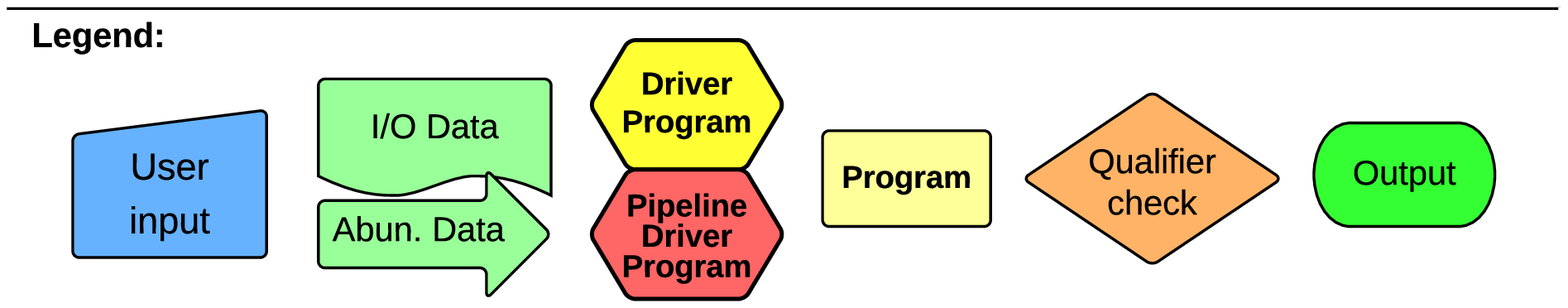}
\caption[Layout of the TEA pre-pipeline and pipeline modules]
 {Layout of the TEA pre-pipeline and pipeline modules.  The
  modules have one of three roles: scientific calculation, file or
  data structure support, or execution of the calculation programs
  over temperature and pressure points in an iterative manner. In
  addition to the modules shown, TEA has three supporting modules:
  readconfig.py, makeatm.py, and plotTEA.py. All modules are described
  in the text.}
\label{fig:TEAflow}
\end{figure*}

{
\begin{enumerate}
\setlength\itemsep{0ex}
\setlength\topsep{0ex}
\setlength\partopsep{0ex}
\setlength\parsep{0ex}

   \item {\bf prepipe.py}: Runs the {\tt readJANAF.py} and {\tt
     makestoich.py} modules and provides their common setup.
   \item {\bf readJANAF.py}: Extracts relevant from all available
     NIST-JANAF Thermochemical Tables and writes ASCII files.
   \item {\bf makestoich.py}: Reads the chemical formula to obtain
     species names and their stoichiometric coefficients from each
     JANAF file, and elemental solar abundances from an ASCII file
     based on \citet{AsplundEtal2009-SunAbundances} Table 1. The code
     produces an output file containing species, stoichiometric
     coefficients, and abundances.
   \item {\bf runsingle.py}: Runs TEA for a single \math{T, P} pair.
   \item {\bf runatm.py}: Runs TEA over a pre-atmosphere file
     containing a list of \math{T, P} pairs.
   \item {\bf readatm.py}: Reads the pre-atmospheric file with
     multiple \math{T, P} pairs.
   \item {\bf makeheader.py}: Combines the stoichiometric information,
     Gibbs free energy per species at specific temperatures, and the
     user input to create a single file with relevant chemical
     informations further used by the pipeline. 
   \item {\bf balance.py}: Uses species and stoichiometric
     information to establish viable, mass-balanced, initial mole
     numbers.
   \item {\bf format.py}: Auxiliary program that manages input/output
     operations in each piece of the pipeline.
   \item {\bf lagrange.py}: Uses data from the most recent
     iteration's corrected mole numbers and implements the Lagrangian
     method for minimization.  Produces output with raw, non-corrected
     mole numbers for each species (values are temporarily allowed to
     be negative).
   \item {\bf lambdacorr.py}: Takes non-corrected mole numbers and
     implements lambda correction to obtain only valid, positive
     numbers of moles.  Output is the corrected mole
     numbers for each species.
   \item {\bf iterate.py}: Driver program that repeats {\tt
     lagrange.py} and {\tt lambdacorr.py} until a pre-defined maximum
     number of iterations is met.
   \item {\bf readconfig.py}: Reads TEA configuration file.
   \item {\bf makeatm.py}: Makes pre-atmospheric file for a multiple
     \math{T, P} run.
   \item {\bf plotTEA.py}: Plots TEA output, the atmospheric file with
     final mole-fraction abundances.
\end{enumerate}
}

\section{APPLICATION TO HOT-JUPITER ATMOSPHERES}
\label{sec:applic}
In this section, we illustrate several applications of the TEA
code. We produced molecular abundances profiles for models of
hot-Jupiter planetary atmospheres, given their temperature-pressure
profiles.

The temperature and pressure (\math{T-P}) profiles adopted for our
thermochemical calculations are shown in Figure
\ref{fig:profiles}. The left and middle panel show the \math{T-P}
profiles of WASP-12b from \citet{StevensonEtal2014-WASP12b} with the
C/O ratio of 0.5 and 1.2, respectively.  The right panel shows the
thermal profile of WASP-43b from
\citet{StevensonEtal2014-PhaseCurve-WASP43b} with solar
metallicity. These profiles are chosen for their relevance to
atmospheric conditions at secondary eclipses.

\begin{figure*}[hb!]
{\centering
\includegraphics[width=0.70\textwidth]{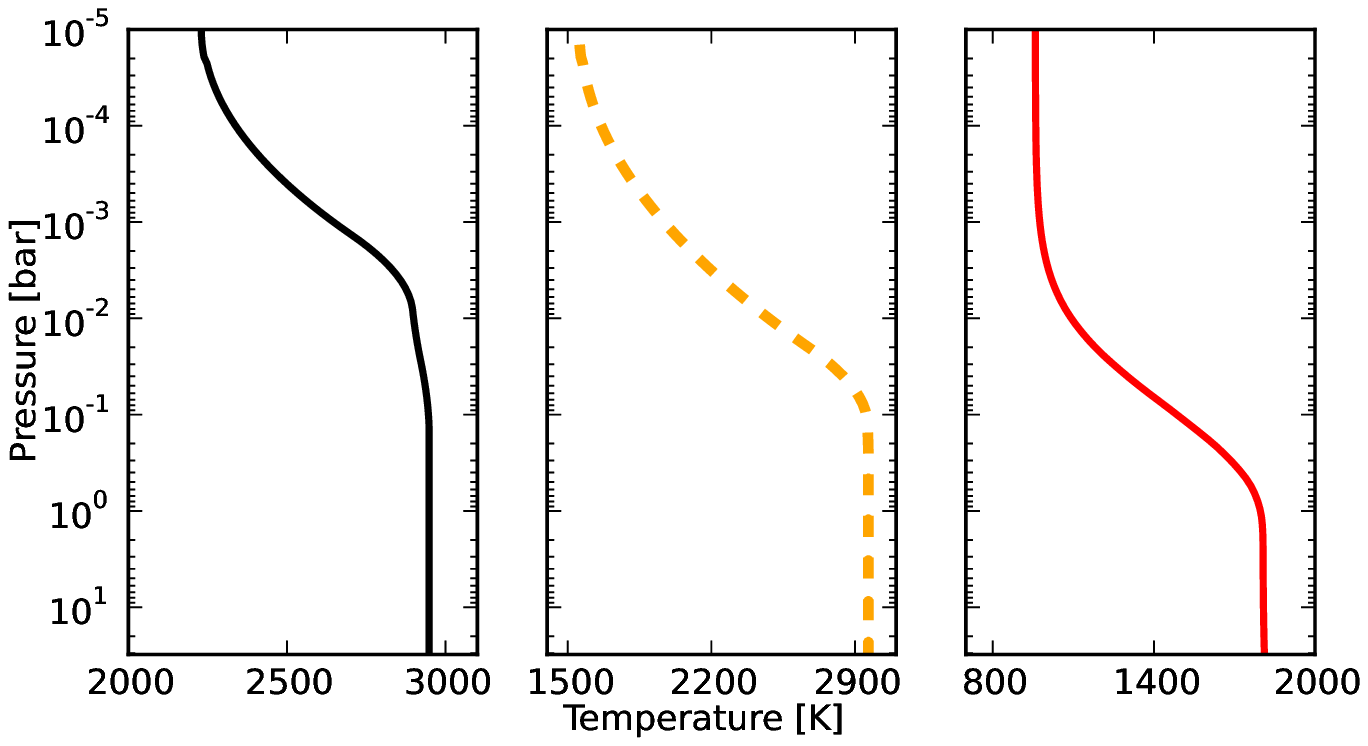}
\caption[Pressure and temperature profiles of WASP-12b and WASP-43b]
  {\label{fig:profiles} The left and middle panels show the
  O-rich and C-rich temperature and pressure (\math{T-P}) profile of
  WASP-12b from \citet{StevensonEtal2014-WASP12b} with C/O = 0.5 and
  C/O = 1.2 respectively. The right panel shows the \math{T-P} profile
  of WASP-43b from \citet{StevensonEtal2014-PhaseCurve-WASP43b} with
  solar metallicity.}  
}

\includegraphics[width=0.53\linewidth,
  clip]{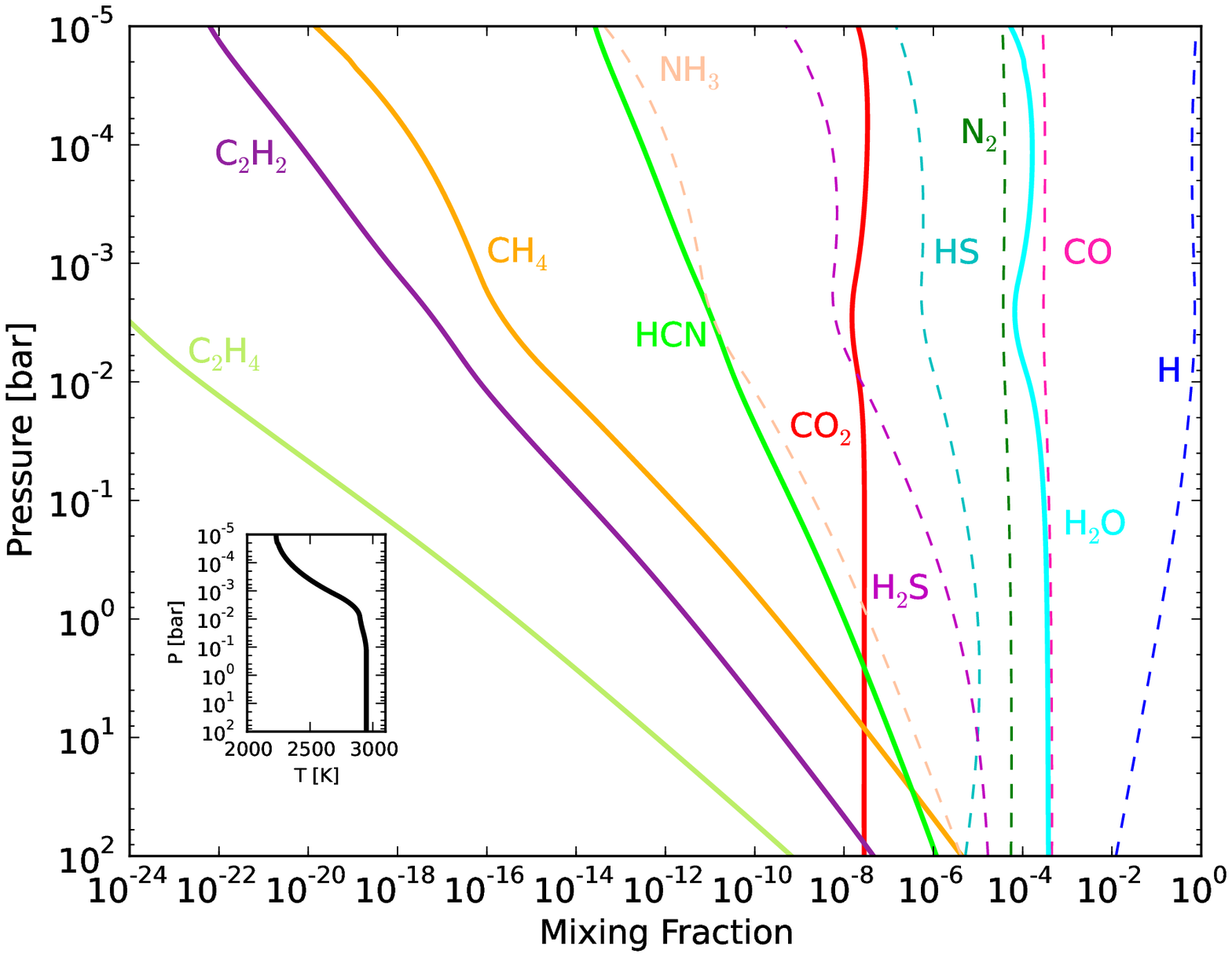}\hspace{-15pt}
\includegraphics[width=0.53\linewidth,  clip]{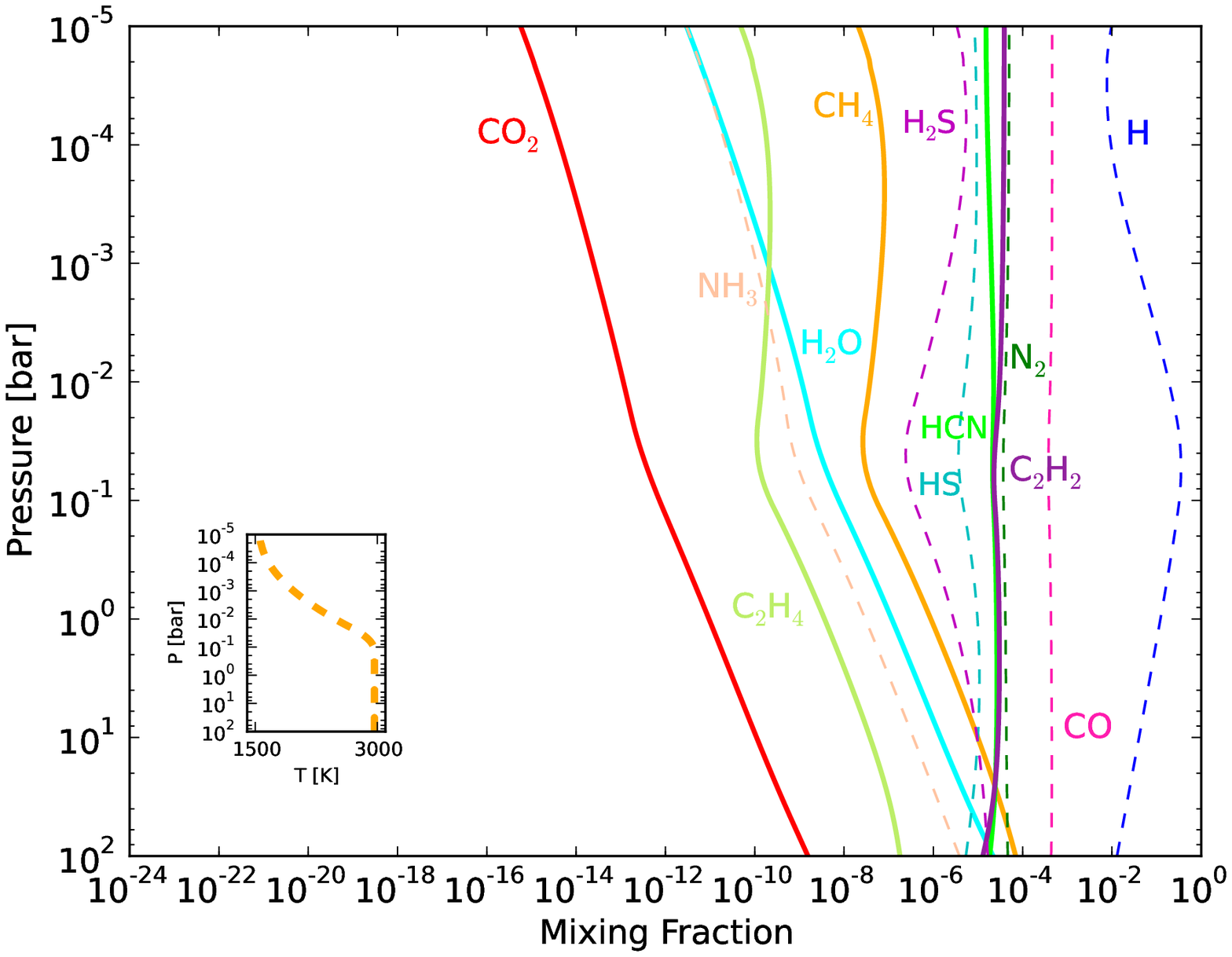} 
\caption[Comparison between vertical thermochemical equilibrium distributions
  for WASP-12b O-rich and WASP-12b C-rich elemental abundance profiles]
  {\label{fig:WASP-12b}
  Comparison between vertical thermochemical equilibrium distributions
  for WASP-12b O-rich, left panel, and WASP-12b C-rich, right panel,
  elemental abundance profile. The inset plots are the \math{T-P}
  profiles shown in Figure \ref{fig:profiles}, left and middle
  panels.}
\end{figure*}

We chose elemental-abundance profiles with C/O \math{>} 1 and C/O
\math{<} 1 and three profiles with solar, 10 times solar, and 50 times
solar elemental abundances to show the influence of the C/O ratio and
metallicity on the chemistry and composition of extrasolar giant
planets.

We adopt \citet{AsplundEtal2009-SunAbundances} photospheric solar
abundances as our baseline. To change the elemental abundance profile,
set them to a certain C/O ratio, or enhance metallicity, we use our
Python routine, {\tt makeAbun.py}. This routines is the part of the
BART project and it is available to the community via {\tt Github.com}
under an open-source licence ({\tt
  https://github.com/joeharr4/BART}). For different metallicities, the
routine multiples the elemental abundances of all species except for
hydrogen and helium, preserving the ratio of major atomic species like
C, N, and O.  \comment{ To get a certain C/O ratio \math{>} 1, we
  swapped the elemental abundances of carbon and oxygen and then
  decrease the carbon elemental abundance to get the desired ratio.  }

We chose to run the models for all plausible, spectroscopically active
species in the infrared relevant for hot-Jupiter atmospheres: H\sb{2},
CO, CO\sb{2}, CH\sb{4}, H\sb{2}O, HCN, C\sb{2}H\sb{2}, C\sb{2}H\sb{4},
N\sb{2}, NH\sb{3}, HS, and H\sb{2}S. Our input species are: H, He, C,
N, O, S.

Figure \ref{fig:WASP-12b} shows results for WASP-12b. Each \math{T-P}
profile is sampled 100 times uniformly in log-pressure space. Figure
\ref{fig:WASP-43b-metal} shows the TEA runs for WASP-43b with
different metallicities. This \math{T-P} profile is sampled 90 times in
uniformly log-pressure space.

As expected, Figure \ref{fig:WASP-12b} shows that H\sb{2}O, CH\sb{4},
CO, CO\sb{2}, C\sb{2}H\sb{2}, C\sb{2}H\sb{4}, and HCN are under the
strong influence of the atmospheric C/O ratio in hot Jupiters
\citep[e.g., ][]{Lodders02, Seager05, FortneyEtal2005apjlhjmodels,
  MadhusudhanEtal2011natWASP12batm, MadhusudhanEtal2011-Cplanets,
  Madhusudhan2012-COratio, MadhusudhanSeager2011-GJ436b,
  MosesEtal2013-COratio}. These species are plotted in solid lines,
while species with only small influence from the C/O ratio are plotted
as dashed lines.

The results also show, as expected, that CO is a major atmospheric
species on hot Jupiters for all C/O ratios and metallicities (Figures
\ref{fig:WASP-12b} and \ref{fig:WASP-43b-metal}), because CO is
chemically favored over H\sb{2}O. Other oxygen-bearing molecules like
H\sb{2}O and CO\sb{2} are more abundant when C/O\math{<}1, while
CH\sb{4}, C\sb{2}H\sb{2}, and C\sb{2}H\sb{4} become significant
species when C/O\math{>}1. Species like N\sb{2} and NH\sb{3} that do
not contain carbon or oxygen are much less affected by the C/O ratio.


\begin{figure}[h!]
\vspace{-50pt}
    \centering \includegraphics[width=0.53\linewidth,
      clip]{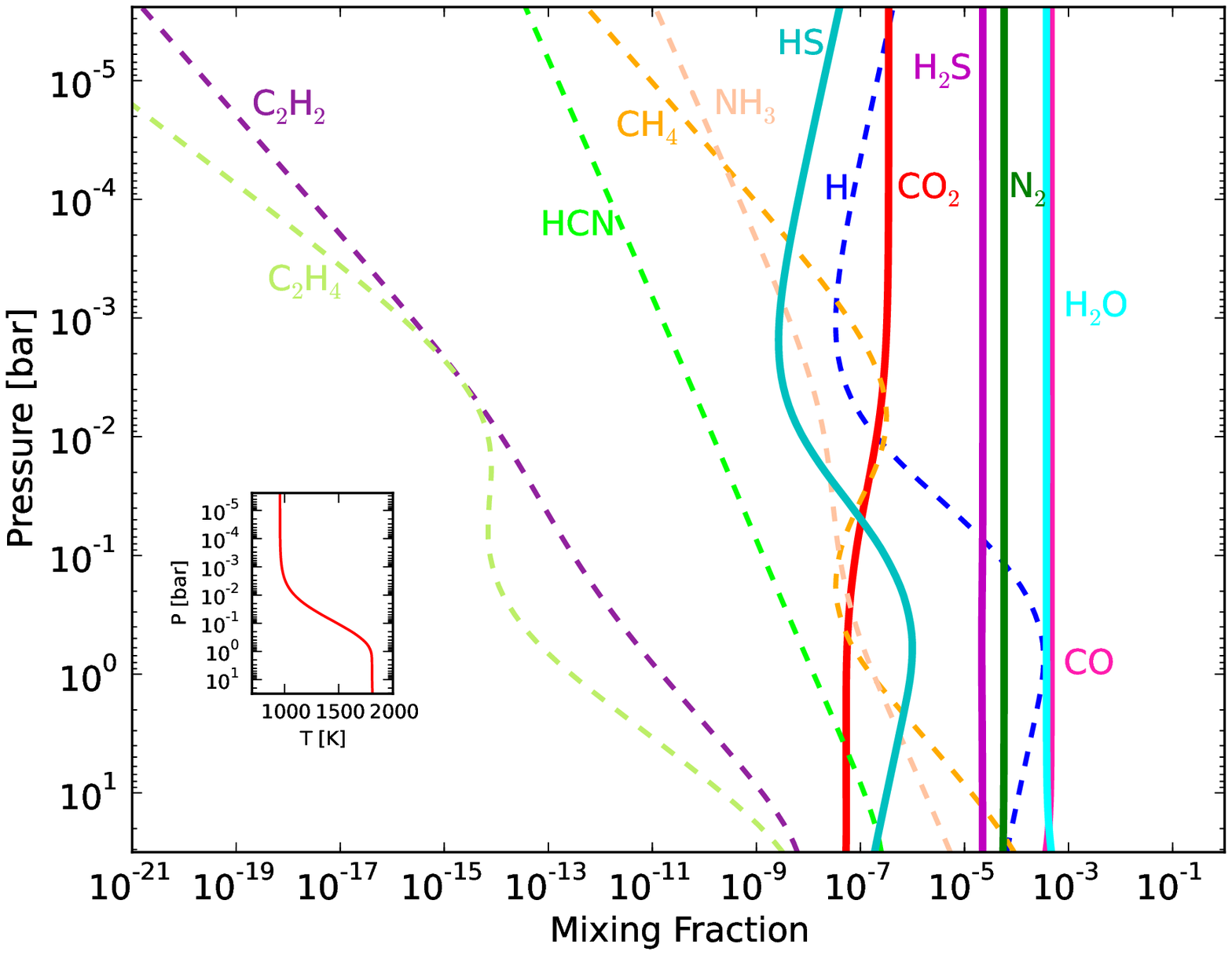}\hspace{10pt}
    \includegraphics[width=0.53\linewidth,
      clip]{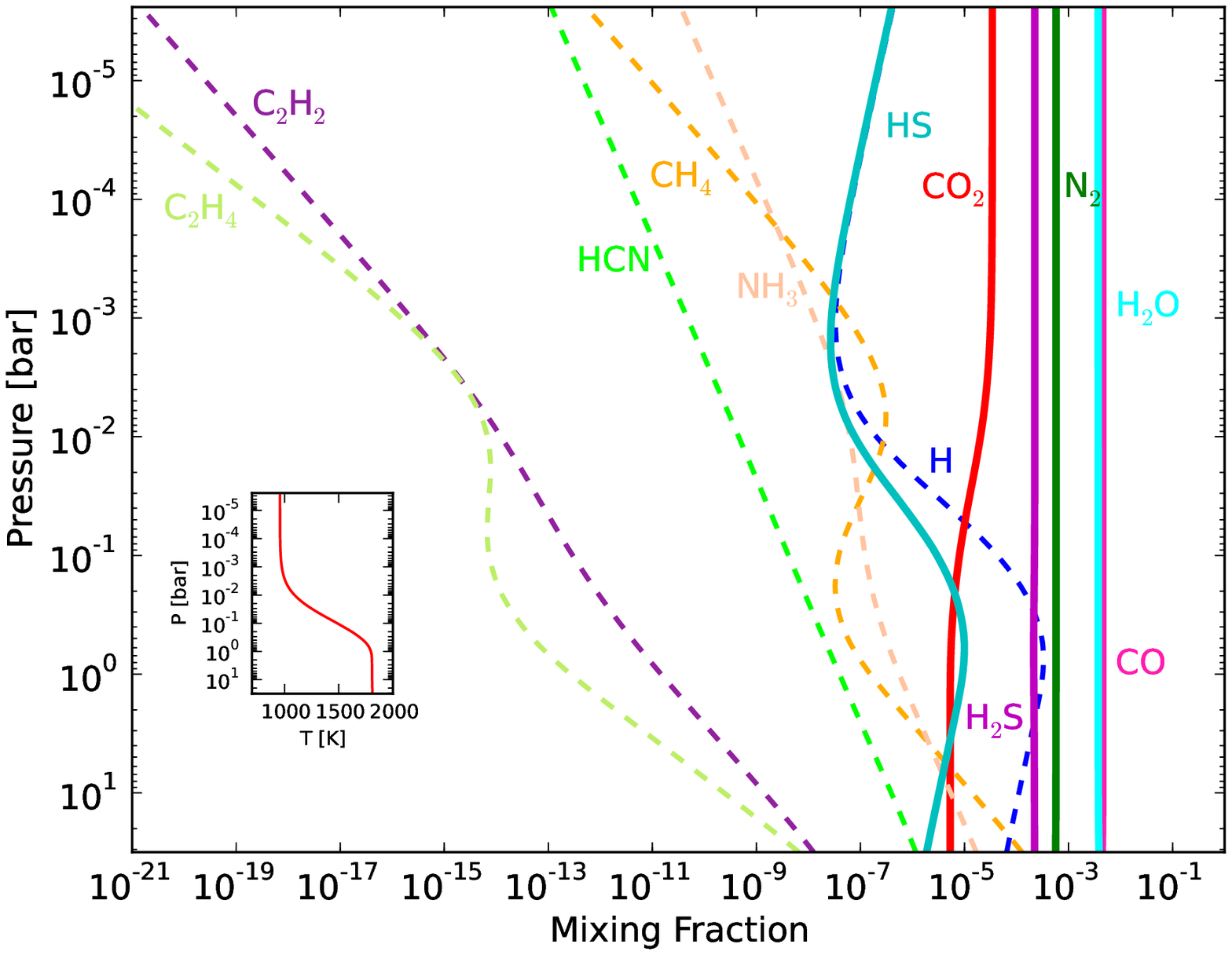}\hspace{10pt}
    \includegraphics[width=0.53\linewidth,
      clip]{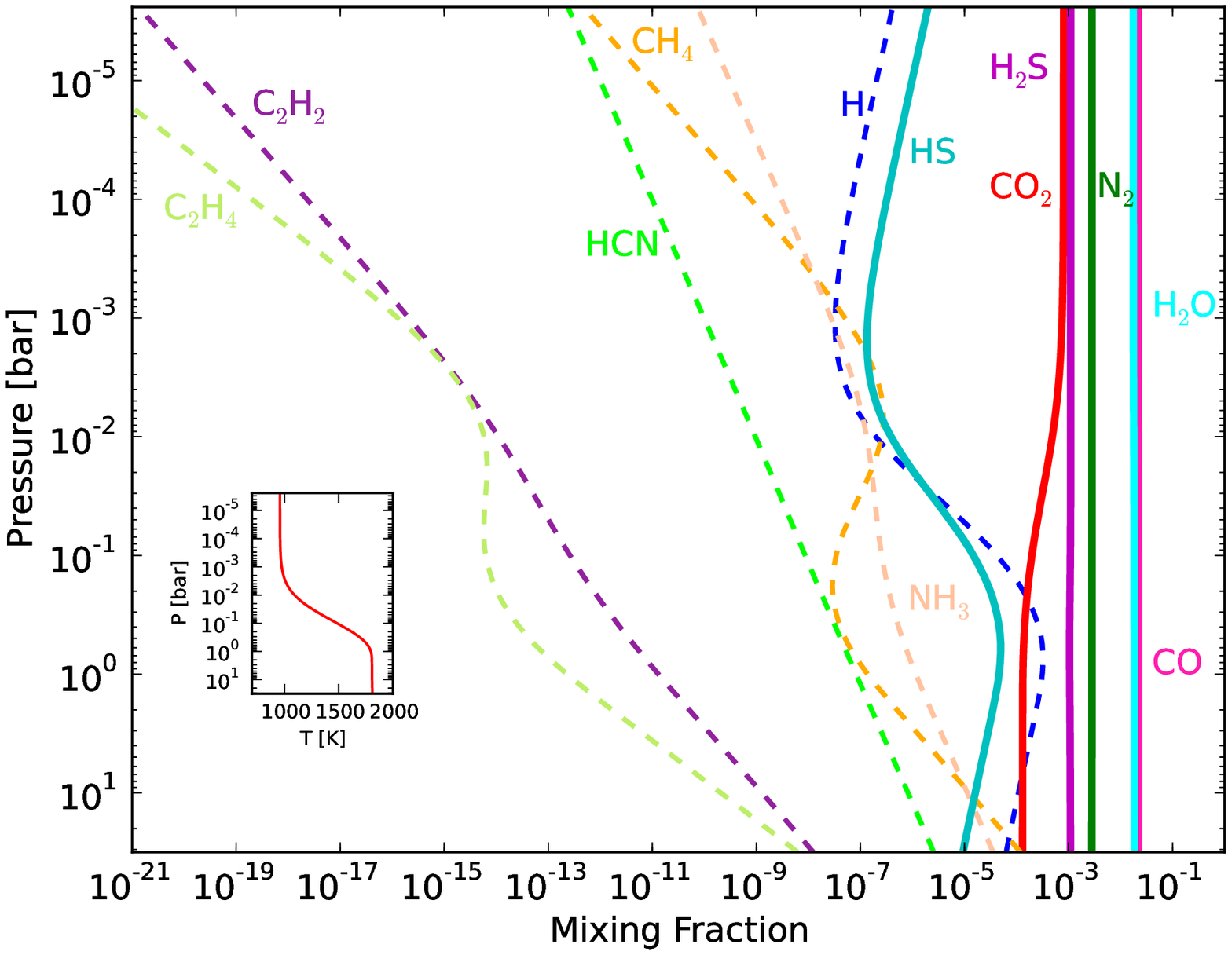}
    \caption[Thermochemical equilibrium vertical distributions for different metallicities of WASP-43b]
      {\label{fig:WASP-43b-metal} Thermochemical equilibrium
      vertical distributions for different metallicities of WASP-43b
      assuming the \math{T-P} profile in Figure \ref{fig:profiles},
      right panel (profile given in inset). Three metallicity cases
      with \math{\zeta}= 1, 10, and 50 are shown from the top to the
      bottom.}
\end{figure}

\newpage
H\sb{2}O is abundant in hot-Jupiter atmospheres \citep[e.g.,
][]{BurrowsSharp1999apjchemeq, Lodders02, HubenyBurrows2007,
  SharpBurrows2007Apjopacities} due to the large solar abundances of
oxygen and hydrogen. Even disequilibrium processes like photochemistry
cannot deplete its abundance. Photochemical models by
\citet{MosesEtal2011-diseq} and \citet{LineEtal2010ApJHD189733b,
  LineEtal2011-kinetics} predict that water will be recycled in
hot-Jupiter atmospheres, keeping H\sb{2}O abundances close to
thermochemical equilibrium values. A low water abundance seems to
occur only in atmospheres with a C/O\math{>}1.

CO\sb{2}, although present in hot-Jupiter atmospheres and
spectroscopically important, is not a major constituent, and it
becomes even less abundant when C/O\math{>}1. Although photochemistry
can greatly enhance the HCN, C\sb{2}H\sb{2}, and C\sb{2}H\sb{4}
abundances \citep{MosesEtal2013-COratio}, we also see that with
C/O\math{>}1, they are the most abundant constituents.

In Figure \ref{fig:WASP-43b-metal}, the species strongly influenced by
metallicity are again plotted as solid lines. In general, we see, as
expected \citep[e.g., ][]{LineEtal2011-kinetics, Lodders02,
  Venot2014-metallicity}, that the shapes of the vertical
distributions are mostly preserved for all metallicities. However, the
thermochemical mixing ratio of CO\sb{2}, CO, H\sb{2}O, N\sb{2}, HS,
and H\sb{2}S vary by several orders of magnitude over the range of
metallicities, while CH\sb{4} and hydrocarbons change very little.

When the metallicity changes from 1 to 50, the abundance of CO\sb{2}
experiences the most dramatic change. It increases by a factor of
1000, confirming it as the best probe of planetary metallicity
\citep{Lodders02, Zahnle09-SulfurPhotoch}.  CO\sb{2} abundance is the
quadratic function of metallicity \citep{Venot2014-metallicity}, while
CO, H\sb{2}O, HS, H\sb{2}S, and N\sb{2} abundances, for species that
either contain one metal atom or are the major reservoirs of carbon
and nitrogen, increase linearly with metallicity \citep{
  Visscher2006}. For this metallicity range, the CO, H\sb{2}O, HS,
H\sb{2}S, and N\sb{2} abundances change by a factor of 100, while
NH\sb{3}, CH\sb{4}, C\sb{2}H\sb{2}, C\sb{2}H\sb{4}, and HCN change by
a factor of 10 or less.

\section{COMPARISON TO OTHER METHODS}
\label{sec:validity}

To test the validity of our code, we performed 4 different tests.  We
compared the output of TEA with the example from
\citet{WhiteJohnsonDantzig1958JGibbs} using their thermodynamic data.
We also compared the TEA output with the output of our TEBS
(Thermochemical Equilibrium by \citeauthor{BurrowsSharp1999apjchemeq})
code that implements the \citet{BurrowsSharp1999apjchemeq} analytical
method for calculating the abundances of five major molecular species
present in hot-Jupiter atmospheres (CO, CH4, H2O, N2, NH3). As another
comparison, we used the free thermochemical equilibrium code CEA
(Chemical Equilibrium with Applications, available from NASA Glenn
Research Center at {\tt http://www.grc.nasa.gov/WWW/CEAWeb/}). This
code uses the Newton-Raphson descent method within the Lagrange
optimization scheme to solve for chemical abundances. Their approach
is described by \citet{GordonMcBride:1994, McBrideGordon:1996}, and
\citet{ZeleznikGordon:1960, ZeleznikGordon:1968}. The thermodynamic
data included in the CEA code are partially from the JANAF tables
\citep{ChaseEtal1986bookJANAFtables} that we used in our TEA code, but
also from numerous other sources \citep[e.g.,][]{Cox1982-CEAdata,
  Gurvich1989-CEAdata, McBrideGordonReno:1993a}. Lastly, we derived
CEA free energies and used them as input to TEA, to compare the CEA
and TEA outputs.

Our first comparison was done using the example from
\citet{WhiteJohnsonDantzig1958JGibbs}. We determined the composition
of the gaseous species arising from the combustion of a mixture of
hydrazine, N\sb{2}H\sb{4}, and oxygen, O\sb{2}, at \math{T} = 3500 K
and the pressure of 750 psi = 51.034 atm. We used the free-energy
functions and \math{b\sb{j}} values (total number of moles of element
j originally present in the mixture) from their Table 1. We reproduced
their abundances, Table \ref{table:White-TEA}, with slightly higher
precision probably due to our use of double precision.

\vspace{15pt}
\begin{table}[!h]
\caption{\label{table:White-TEA} Comparison
  \citeauthor{WhiteJohnsonDantzig1958JGibbs} {\em vs.} TEA}
\atabon\strut\hfill\begin{tabular}{lcccc} \hline \hline Species &
\math{\frac{g\sb{i}\sp{0}(T)}{RT}} &
\citeauthor{WhiteJohnsonDantzig1958JGibbs} & TEA & Difference \\ & &
abundances & abundances & \\ \hline H & -10.021 & 0.040668
& 0.04065477 & -0.00001323 \\ H\sb{2} & -21.096 & 0.147730 &
0.14771009 & -0.00001991 \\ H\sb{2}O & -37.986 & 0.783153 & 0.78318741
& 0.00003441 \\ N & -9.846 & 0.001414 & 0.00141385 & -0.00000015
\\ N\sb{2} & -28.653 & 0.485247 & 0.48524791 & 0.00000091 \\ NH &
-18.918 & 0.000693 & 0.00069312 & 0.00000012 \\ NO & -28.032 &
0.027399 & 0.02739720 & -0.00000180 \\ O & -14.640 & 0.017947 &
0.01794123 & -0.00000577 \\ O\sb{2} & -30.594 & 0.037314 & 0.03730853
& -0.00000547 \\ OH & -26.111 & 0.096872 & 0.09685710 & 0.00001490
\\ \hline
\end{tabular}\hfill\strut\ataboff
\end{table}

Figure \ref{fig:CEATEATEBS}, left panel, shows the CEA, TEA, and TEBS
runs for the temperatures between 600 and 3000 K, pressure of 1 bar,
and solar abundances. The runs were performed with the input and
output species that all codes contain (H, C, O, N, H\sb{2}, CO,
CH\sb{4}, H\sb{2}O, N\sb{2}, NH\sb{3}). We also run the comparison
just between CEA and TEA, Figure \ref{fig:CEATEATEBS}, right panel,
for the WASP-43b model atmosphere that we described in Section
\ref{sec:applic}.  We used the pressure and temperature profile shown
in Figure \ref{fig:profiles}, right panel, and solar elemental
abundances. The temperatures and pressures range from 958.48 to
1811.89 K and 1.5\math{\times}10\sp{-5} to
3.1623\math{\times}10\sp{1} bar, respectively. We included the same
species as in Section \ref{sec:applic} with the exclusion of the
C\sb{2}H\sb{2} and HS species, because CEA does not carry the
thermodynamical parameters for them.

\begin{figure*}[h!]
\centering \includegraphics[height=6.5cm, clip = True, trim=0.75cm
  0.1cm 0.1cm 0cm]{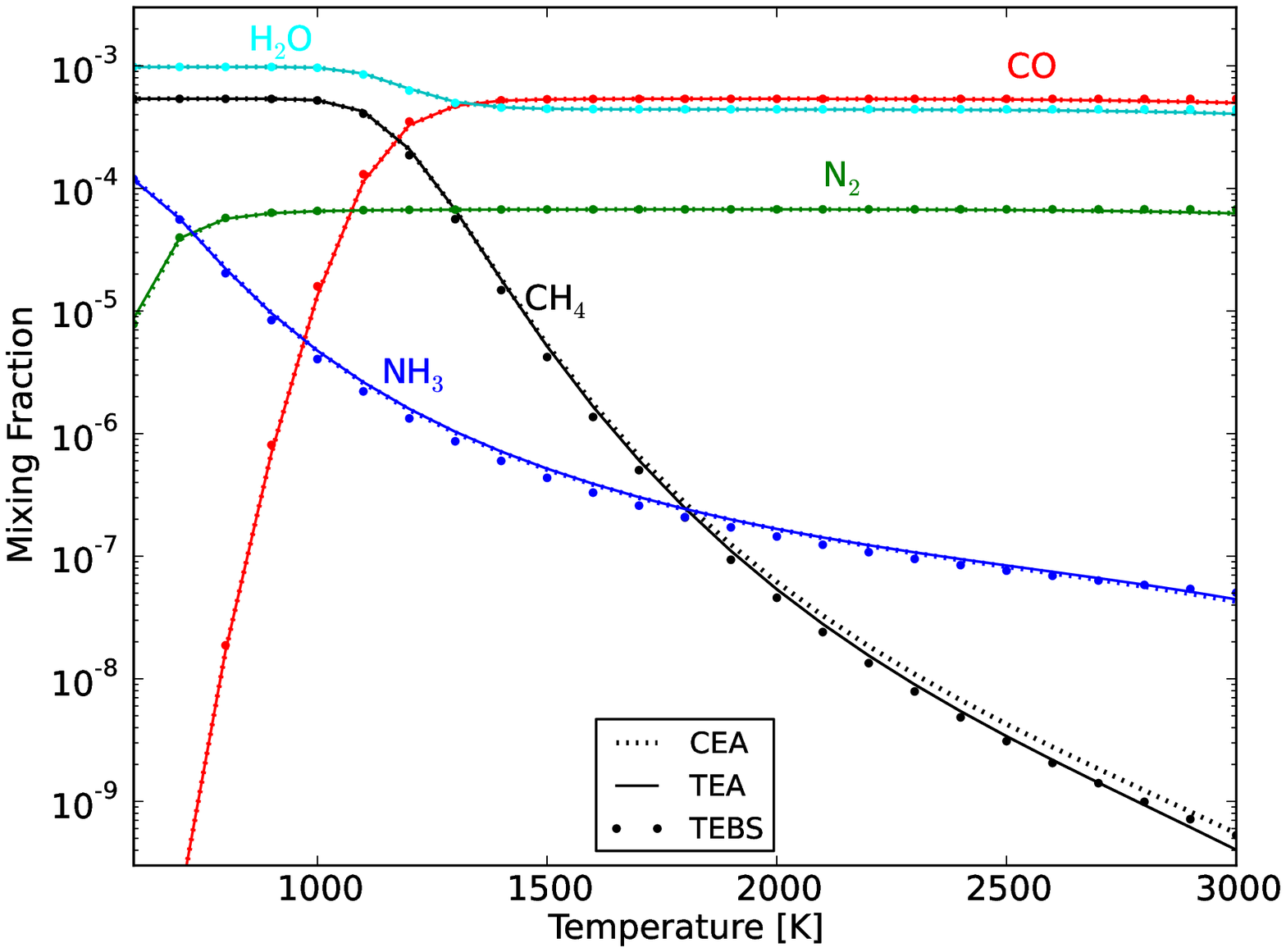}\hspace{-10pt}
\includegraphics[height=6.5cm, clip = True, trim=0.75cm 0.1cm 0.1cm
  0cm]{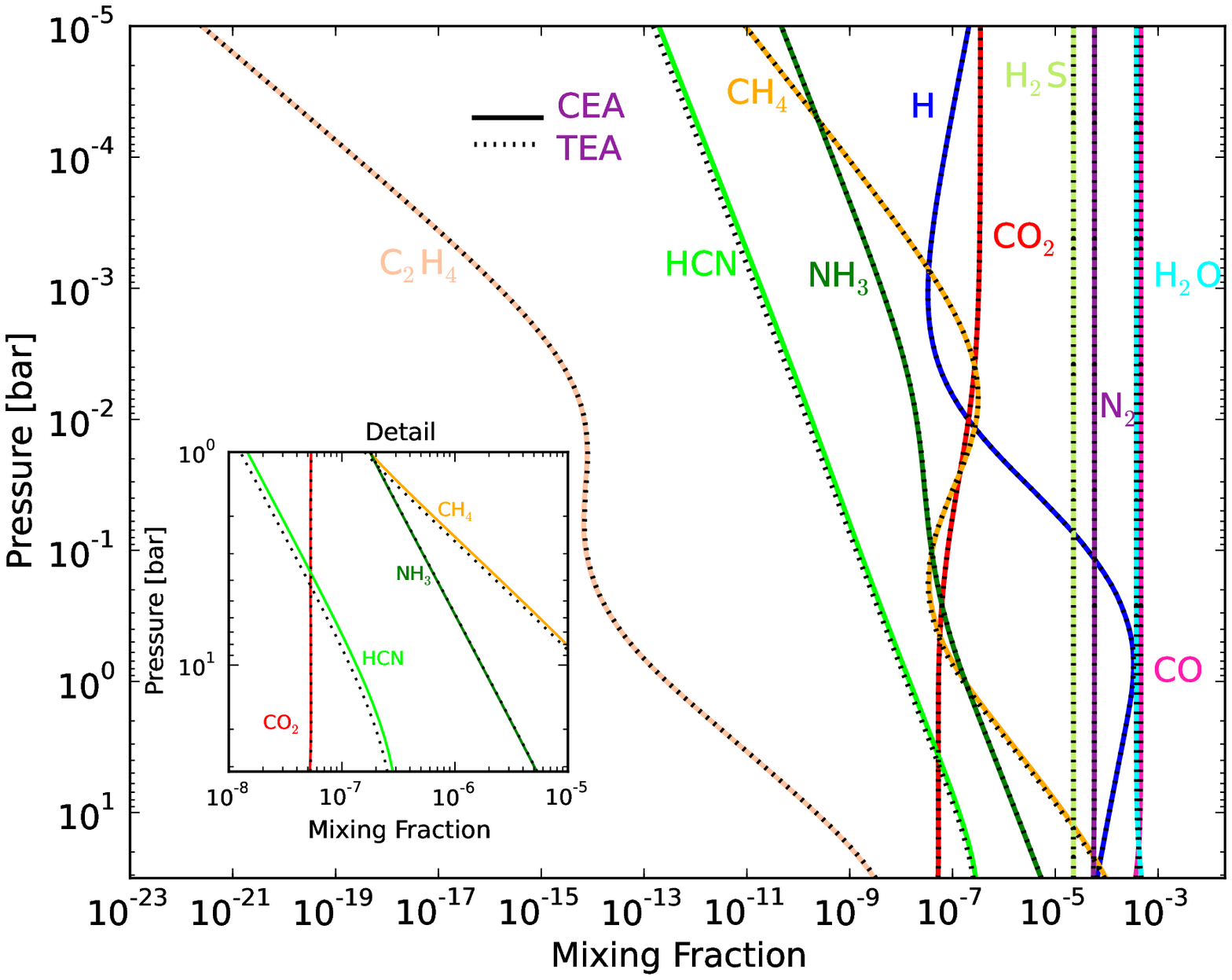}
\caption[Comparison between TEA, CEA, and TEBS]
  {{\bf Left:}\label{fig:CEATEATEBS} Comparison TEA, CEA and
  TEBS. TEBS is an analytic method, while CEA and TEA are numerical
  methods. We show the major spectroscopically-active species in the
  infrared that can be produced by all three methods. We run the codes
  for the same range of temperatures and the pressure of \math{P} = 1
  bar. Each species in each method is plotted with a different line
  style, but with the same color. The TEBS final abundances are
  plotted as dots, CEA as dashed lines, while TEA is plotted as solid
  lines. {\bf Right:} Comparison of the TEA results with the results
  from CEA. CEA and TEA are both numerical methods that use Gibbs free
  energy minimization method with similar optimization scheme. We show
  the most plausible and most abundant spectroscopically-active
  species in the infrared expected to be present in hot-Jupiter
  atmospheres, that all codes can cover. In the inset plot, we show a
  detail (zoom-in part), pointing out species lines
  that do not overlap.  The \math{T-P} profile used for this run is
  given in the right panel of Figure \ref{fig:profiles}. Tables
  \ref{table:CEA-TEA-left} and \ref{table:CEA-TEA-right} list
  differences between the final abundances for random three \math{T,
    P} points chosen from each run.}
\end{figure*}

In the left panel of Figure \ref{fig:CEATEATEBS}, we see that for the
most species and temperatures CEA and TEA lines overlap (CEA result is
plotted in dashed and TEA in solid lines).  However, CH\sb{4} species
abundances above \math{T} \sim1700 K do not overlap.  TEBS colored
dots do not overplot either CEA or TEA curves, but follow them
closely.  This method is derived for only five major molecular species
and is based on a few simple analytic expressions.

In Figure \ref{fig:CEATEATEBS}, right panel, we again see that most
species overlap, except HCN and CH\sb{4}.  The HCN curves (for CEA and
TEA runs) differ for the full temperature range (see the inset
figure), while, as before, CH\sb{4} curve differs slightly only for
pressures above \sim0.1 bar and temperatures above \sim1700 K (see Figure 
\ref{fig:profiles} for the \math{T-P} profile used for this run). 

The differences seen in Figure \ref{fig:CEATEATEBS} come from the
different sources of thermodynamic data used for CEA and TEA 
(see Tables \ref{table:CEA-TEA-left} and \ref{table:CEA-TEA-right}). 
When the CEA thermodynamic data are used as input to TEA, all species final
abundances match, see Figure \ref{fig:CEATEA}. Section
\ref{sec:freeEner}, below, elaborates on this and investigate the difference
in free energy input values used for CEA and TEA.

\begin{figure*}[hb!]
\centering \includegraphics[height=6.5cm, clip = True, trim=0.75cm
  0.1cm 0.1cm 0cm]{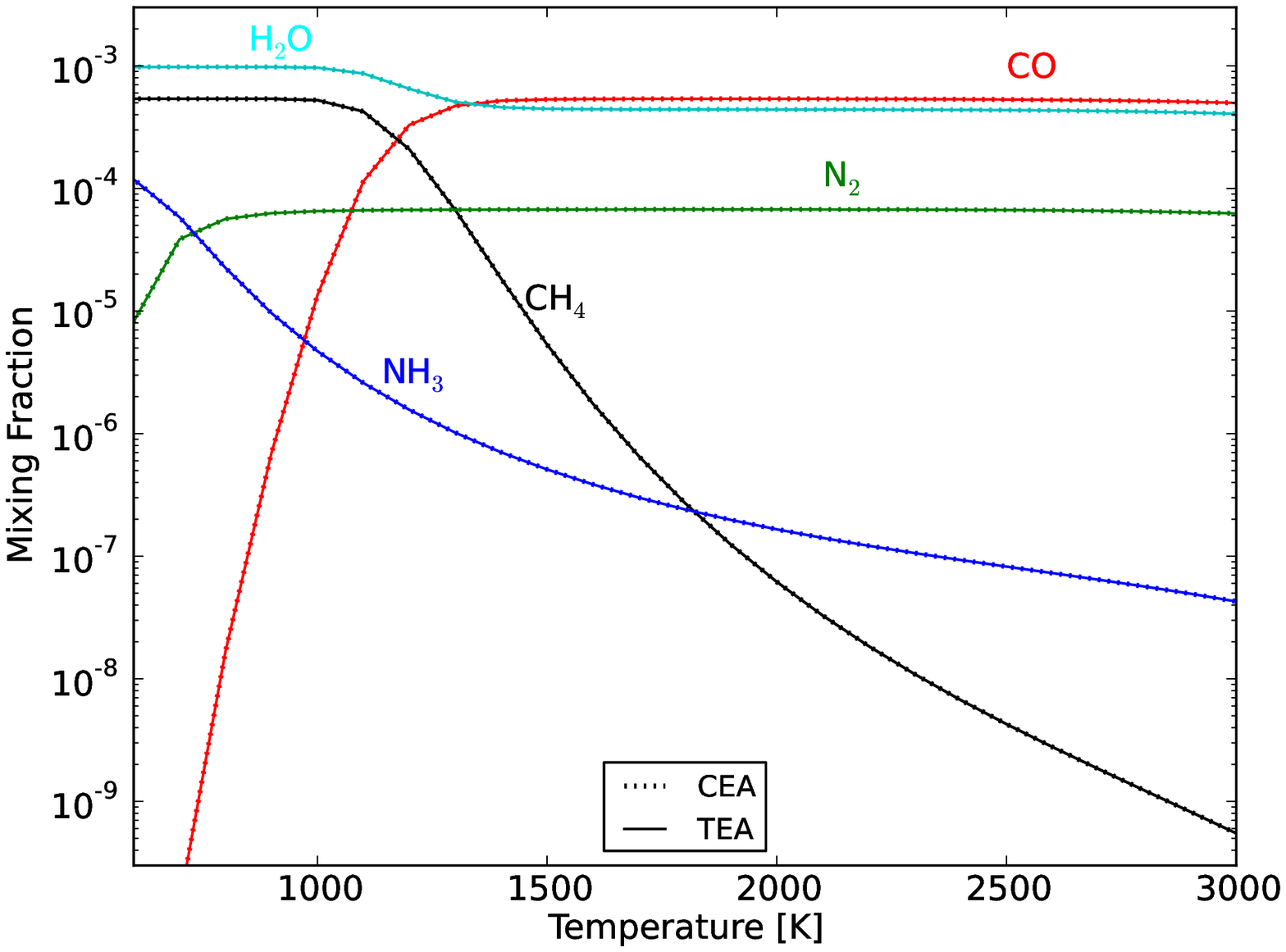}\hspace{-10pt}
\includegraphics[height=6.5cm, clip = True, trim=0.75cm 0.1cm 0.1cm
  0cm]{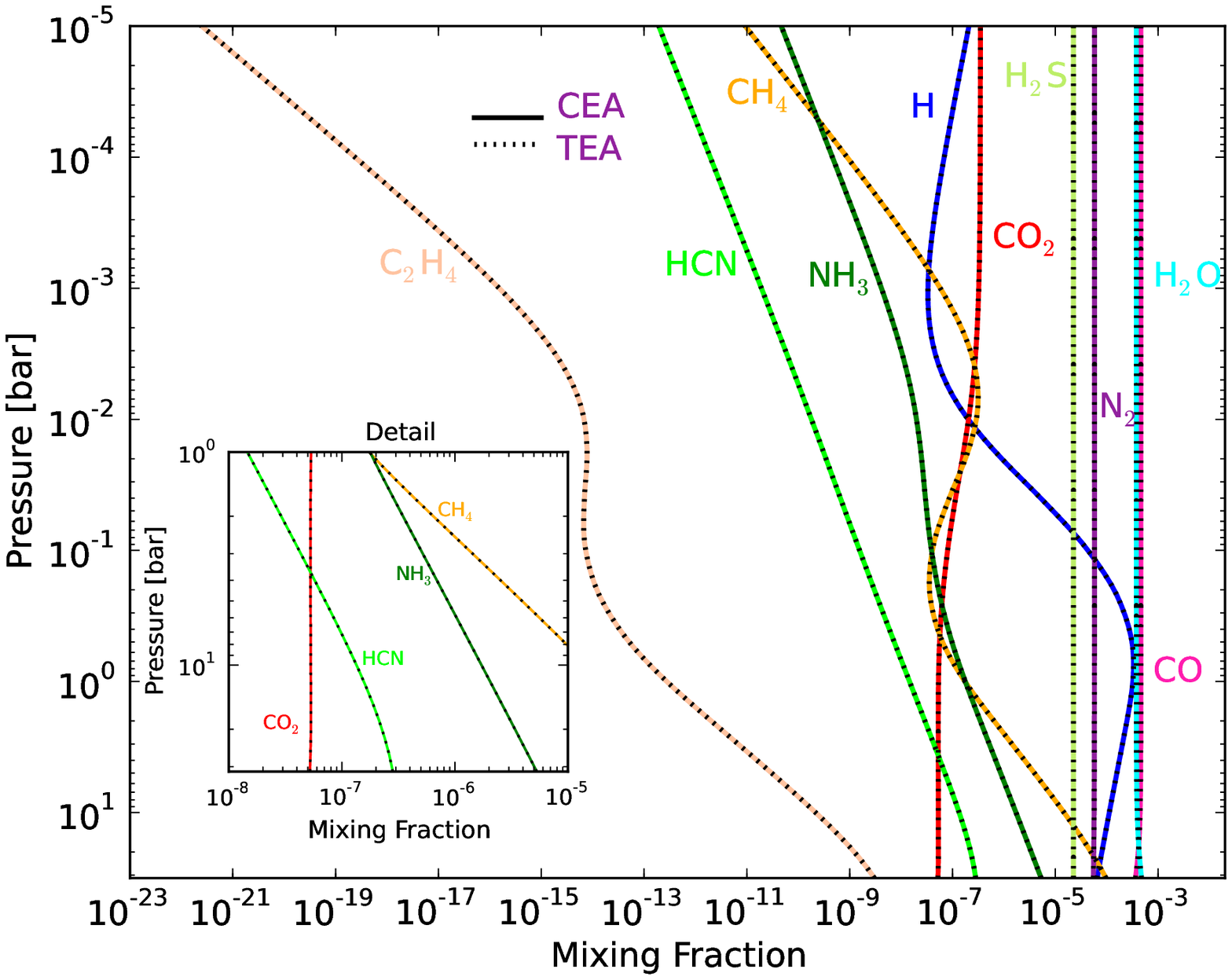}
\caption[Comparison between TEA and CEA using CEA thermodynamic data]
  {{\bf Left:}\label{fig:CEATEA} Comparison TEA and CEA using
  CEA thermodynamic data provided in their {\tt thermo.inp} file. The
  comparison is done for the same conditions as in Figure
  \ref{fig:CEATEATEBS}. Tables \ref{table:CEA-TEA-left} and
  \ref{table:CEA-TEA-right} list differences between the final
  abundances for random three \math{T, P} points chosen from each
  run.}
\end{figure*}

\subsection{Comparison of Free Energy Values in CEA and TEA}
\label{sec:freeEner}

The thermodynamic data used for CEA are in the form of polynomial
coefficients, and are listed in the {\tt termo.inp} file provided with
the CEA code. The format of this library is explained in Appendix A of
\citet{McBrideGordon:1996}. For each species, the file lists, among
other data, the reference sources of the thermodynamic data, the
values of the standard enthalpy of formation, \math{\Delta\sb{f}
  H\sb{298}\sp{0}}, at the reference temperature of 298.15 K and
pressure of 1 bar, and coefficients of specific heat, \math{C\sb{p}
  \sp{0}}, with integration constants for enthalpy, \math{H\sp{o}},
and entropy, \math{S\sp{o}}, for temperature intervals of 200 to 1000
K, 1000 to 6000 K, and 6000 to 20000 K.

The JANAF tables list the reference sources of their thermodynamic
parameters in \citet{ChaseEtal1986bookJANAFtables}. The data are also
available at {\tt http://kinetics.nist.gov/janaf/}. 

The difference in thermodynamic parameters between CEA and TEA is
noticeable even in their \math{\Delta\sb{f} H\sb{298}\sp{0}} values.
The source of standard enthalpies of formation in CEA for, e.g., HCN
and CH\sb{4} is \citet{gurvich1991criteria}, page 226 and 36,
respectively, and their respective values are 133.08 and -74.60
kJ/mol. The source of standard enthalpies of formation in the JANAF
tables is listed in \citet{ChaseEtal1986bookJANAFtables} on page 600
and 615, respectively, and their respective values are 135.14 and
-74.873 kJ/mol.

TEA uses JANAF tables to calculate the values of free energies for
each species following Equation \ref{eq:JANAFconv}.  To calculate the
values of free energies used in CEA, we started from Chapter 4 in
\citet{GordonMcBride:1994}. Our goal is to plug CEA free energies
into TEA and test whether TEA will produce the same final abundances
as CEA does.

As explained in Section 4.2, the thermodynamic functions specific
heat, enthalpy, and entropy as function of temperatures are given as:
\begin{eqnarray}
\frac {C_p^{o}}{R} = \sum\,a_i\,T^{q_{i}} \, ,
\label{Cp}
\end{eqnarray}
\begin{eqnarray}
\frac {H^{o}}{RT} = \frac{\int C_p^{o}\, dT}{RT} \, ,
\label{H}
\end{eqnarray}
\begin{eqnarray}
\frac {S^{o}}{R} = \int \frac{C_p^{o}}{RT}\, dT \, .
\label{S}
\end{eqnarray}
\noindent These functions are given in a form of seven polynomial
coefficients for specific heat, \math{C\sb{p}\sp{o}/R}, and two
integrations constants (\math{a\sb{8}} and \math{a\sb{9}}) for
enthalpy, \math{H\sp{o}/RT}, and entropy, \math{S\sp{o}/R}:
\begin{eqnarray}
\label{CpPoly}
\frac {C_p^{o}}{R} = a_1T^{-2} + a_2T^{-1} + a_3 + a_4T + a_5T^{2} +
a_6T^{3} + \\ \nonumber a_7T^{4} \, ,
\end{eqnarray}
\begin{eqnarray}
\label{HPoly}
\frac {H^{o}}{RT} = -\,\,a_1T^{-2} + a_2T^{-1}\,lnT + a_3 +
a_4\frac{T}{2} + a_5\frac{T^{2}}{3} + \\ \nonumber a_6\frac{T^{3}}{4}
+ a_7\frac{T^{4}}{5} + \frac{a_8}{T} \, ,
\end{eqnarray}
\begin{eqnarray}
\label{SPoly}
\frac {S^{o}}{R} = -\,\,a_1\frac{T^{-2}}{2} -\,a_2T^{-1} + a_3\,lnT +
a_4T + a_5\frac{T^{2}}{2} + \\ \nonumber a_6\frac{T^{3}}{3} +
a_7\frac{T^{4}}{4} + a_9 \, .
\end{eqnarray}
To derive free energies in the form that TEA uses them, we rewrite
Equation \ref{eq:JANAFconv} for one species as:
\begin{eqnarray}
\label{eq:JANAF}
\frac{g^0(T)}{RT} = 1/R\Big[\frac{G_{T}^0 - H_{298}^0}{T}\Big] +
\frac{\Delta_f H_{298}^0}{RT}\, ,
\end{eqnarray}
\noindent The first term on the right side can be expressed in the
following format \citep[][Page 3]{Chase1974janaf}:
\begin{eqnarray}
\label{eq:Chase1974}
\frac{G_{T}^0(T) - H_{298}^0}{T} = -\,\, S_T^{o} + \frac{(H_T^{o} -
  H_{298}^0)}{T} \, .
\end{eqnarray}
\noindent Thus, we rewrite Equation \ref{eq:JANAF} as:
\begin{eqnarray}
\label{eq:JANAF2}
\frac{g^0(T)}{RT} = 1/R\Big[\-\,\, S_T^{o} + \frac{(H_T^{o} -
    H_{298}^0)}{T}\Big] + \frac{\Delta_f H_{298}^0}{RT}\, ,
\end{eqnarray}
\begin{eqnarray}
\label{eq:JANAF2}
\frac{g^0(T)}{RT} = 1/R\Big[\-\,\, S_T^{o} + \frac{H_T^{o}}{T} -
  \frac{H_{298}^0}{T}\Big] + \frac{\Delta_f H_{298}^0}{RT}\, ,
\end{eqnarray}
\noindent To see Equations \ref{HPoly} and \ref{SPoly} inside Equation
\ref{eq:JANAF2}, we multiply and divide the first and second term on
the right with \math{R} and get:
 \begin{eqnarray}
\label{eq:JANAF4}
\frac{g^0(T)}{RT} = \frac{S_T^{o}}{R} + \frac{H_T^{o}}{RT} -
\frac{H_{298}^0}{RT} + \frac{\Delta_f H_{298}^0}{RT}\, ,
\end{eqnarray}
In the CEA analysis paper, Section 4.1, \citet{GordonMcBride:1994}
state that they have arbitrary assumed \math{H\sp{o}(298.15) =
  \Delta\sb{f}H\sp{o}(298.15)}. Adopting this assumption leads to:
\begin{eqnarray}
\label{eq:JANAF4}
\frac{g^0(T)}{RT} = \frac{S_T^{o}}{R} + \frac{H_T^{o}}{RT} -
\frac{\Delta_f H_{298}^0}{RT} + \frac{\Delta_f H_{298}^0}{RT}\, .
\end{eqnarray}
\noindent The last two terms cancel leading to a simple expression for
free energies:
\begin{eqnarray}
\label{eq:JANAF4}
\frac{g^0(T)}{RT} = \frac{S_T^{o}}{R}  + \frac{H_T^{o}}{RT}\, ,
\end{eqnarray}
\noindent The first term on the right side is Equation \ref{SPoly}, while
the second term is Equation \ref{HPoly}; expressions with polynomial
coefficients that are given in the CEA {\tt thermo.inp} file.

Following the last conclusion, we calculated the free energies for each
species of interest and used them as input to TEA.  Figure
\ref{fig:CEATEA} shows the comparison between CEA and TEA using CEA
free energies. We see that all species overlap. Tables
\ref{table:CEA-TEA-left} and \ref{table:CEA-TEA-right} give the exact
values of free energies used and the final abundances
for several (\math{T, P}) points that showed the largest differences
between CEA and TEA runs in Figure \ref{fig:CEATEATEBS}. It also lists
the free energies calculated using JANAF tables and the final
abundances produced by TEA using JANAF thermodynamic data.

As seen in Figure \ref{fig:CEATEA}, although CEA uses Newton-Raphson
and TEA the Lagrangian method of steepest descent, both approaches,
using the same inputs (free energies), find the same final
abundances. Table \ref{table:CEA-TEA-left}, (groups {\em CEA final
  abundances} and {\em TEA final abundances using CEA free energies}),
shows values identical for most species between the two tests.  A few
cases show that abundance ratios are inconsistent at the 10\sp{-5}
level. Table \ref{table:CEA-TEA-right} displays the same trend. The
differences in the fifth decimal place may indicate that, somewhere in
CEA, a calculation is carried out in 32-bit precision, possibly due to
a literal single-precision number in the source code.  Python floating
literals are in 64-bit precision by default.

\begin{landscape}
\begin{table*}[!t]
\caption{\label{table:CEA-TEA-left} Difference between CEA {\em vs.} TEA
 for Figures \ref{fig:CEATEATEBS} and \ref{fig:CEATEA}, Left Panels}
\atabon\strut\hfill\begin{tabular}{lcccccc} \hline \hline Pressure &
Temp & \multicolumn{5}{c}{Species} \\ (bar) & (K) & CO & CH4 & H2O &
N2 & NH3 \\ \hline \multicolumn{7}{c}{\bf CEA free energies} \\ \hline
1.0000e+00 & 2500.00 & -33.80559930 & -34.91655970 & -40.12912409 &
-27.71757996 & -32.70878374 \\ 1.0000e+00 & 2700.00 & -33.69182758 &
-35.33414843 & -39.64186320 & -27.99507610 & -33.05270703
\\ 1.0000e+00 & 2900.00 & -33.61649725 & -35.76252669 & -39.25604592 &
-28.25692510 & -33.39666924 \\ \hline \multicolumn{7}{c}{\bf TEA
  (JANAF) free energies} \\ \hline 1.0000e+00 & 2500.00 & -33.80793700
& -34.70780992 & -40.12426098 & -27.72037451 & -32.73695542
\\ 1.0000e+00 & 2700.00 & -33.69214791 & -35.08662806 & -39.63255004 &
-27.99496246 & -33.08302621 \\ 1.0000e+00 & 2900.00 & -33.61466712 &
-35.47533692 & -39.24191944 & -28.25439783 & -33.42896561 \\ \hline
\multicolumn{7}{c}{\bf CEA final abundances} \\ \hline 1.0000e+00 &
2500.00 & 5.3129e-04 & 4.2666e-09 & 4.3546e-04 & 6.6686e-05 &
8.2252e-08 \\ 1.0000e+00 & 2700.00 & 5.2311e-04 & 1.8387e-09 &
4.2855e-04 & 6.5661e-05 & 6.4332e-08 \\ 1.0000e+00 & 2900.00 &
5.0876e-04 & 8.2340e-10 & 4.1586e-04 & 6.3844e-05 & 4.9466e-08
\\ \hline \multicolumn{7}{c}{\bf TEA final abundances using CEA free
  energies} \\ \hline 1.0000e+00 & 2500.00 & 5.3129e-04 & 4.2665e-09 &
4.3547e-04 & 6.6685e-05 & 8.2251e-08 \\ 1.0000e+00 & 2700.00 &
5.2311e-04 & 1.8387e-09 & 4.2856e-04 & 6.5661e-05 & 6.4332e-08
\\ 1.0000e+00 & 2900.00 & 5.0876e-04 & 8.2339e-10 & 4.1586e-04 &
6.3844e-05 & 4.9466e-08 \\ \hline \multicolumn{7}{c}{\bf TEA final
  abundances using JANAF free energies} \\ \hline 1.0000e+00 & 2500.00
& 5.3129e-04 & 3.3976e-09 & 4.3547e-04 & 6.6685e-05 & 8.3987e-08
\\ 1.0000e+00 & 2700.00 & 5.2312e-04 & 1.4194e-09 & 4.2856e-04 &
6.5661e-05 & 6.6260e-08 \\ 1.0000e+00 & 2900.00 & 5.0878e-04 &
6.1471e-10 & 4.1586e-04 & 6.3845e-05 & 5.1339e-08 \\ \hline
\end{tabular}\hfill\strut\ataboff
\end{table*}
\end{landscape}

\begin{landscape}
\begin{table*}[!t]
\caption{\label{table:CEA-TEA-right} Difference between CEA {\em vs.} TEA
 for Figures \ref{fig:CEATEATEBS} and \ref{fig:CEATEA}, Right Panels}
\atabon\strut\hfill\begin{tabular}{lcccccccc} \hline \hline Pressure &
Temp & \multicolumn{7}{c}{Species} \\ (bar) & (K) & CO & CO2 & CH4 &
H2O & HCN & NH3 & H2S \\ \hline \multicolumn{9}{c}{\bf CEA free
  energies} \\ \hline 3.8019e-01 & 1719.64 & -34.9307244 & -58.4124145
& -33.595356 & -43.7404858 & -19.8124391 & -31.4721776 & -30.6054338
\\ 1.6596e+00 & 1805.28 & -34.7237371 & -57.3627896 & -33.695587 &
-43.1403205 & -20.4938031 & -31.5886558 & -30.7577826 \\ 2.1878e+01 &
1810.15 & -34.7128505 & -57.3065771 & -33.701803 & -43.1083097 &
-20.5310828 & -31.5955222 & -30.7664570 \\ \hline
\multicolumn{9}{c}{\bf TEA (JANAF) free energies} \\ \hline 3.8019e-01
& 1719.64 & -34.9288052 & -58.4130468 & -33.5180429 & -43.7386156 &
-19.6678405 & -31.4830322 & -30.5760402 \\ 1.6596e+00 & 1805.28 &
-34.7231685 & -57.3648328 & -33.6061366 & -43.1392058 & -20.3573682 &
-31.6020990 & -30.7285988 \\ 2.1878e+01 & 1810.15 & -34.7123566 &
-57.3086978 & -33.6116396 & -43.1072342 & -20.3950903 & -31.6091097 &
-30.7372812 \\ \hline \multicolumn{9}{c}{\bf CEA final abundances}
\\ \hline 3.8019e-01 & 1719.64 & 4.5960e-04 & 5.8035e-08 & 4.9221e-08
& 3.7681e-04 & 5.6243e-09 & 7.9716e-08 & 2.2498e-05 \\ 1.6596e+00 &
1805.28 & 4.5918e-04 & 5.2864e-08 & 4.4665e-07 & 3.7724e-04 &
2.4131e-08 & 2.8864e-07 & 2.2504e-05 \\ 2.1878e+01 & 1810.15 &
4.0264e-04 & 5.3052e-08 & 5.6873e-05 & 4.3396e-04 & 2.3851e-07 &
3.7082e-06 & 2.2520e-05 \\ \hline \multicolumn{9}{c}{\bf TEA final
  abundances using CEA free energies} \\ \hline 3.8019e-01 & 1719.64 &
4.5959e-04 & 5.8035e-08 & 4.9219e-08 & 3.7682e-04 & 5.6240e-09 &
7.9714e-08 & 2.2498e-05 \\ 1.6596e+00 & 1805.28 & 4.5918e-04 &
5.2865e-08 & 4.4667e-07 & 3.7724e-04 & 2.4131e-08 & 2.8864e-07 &
2.2504e-05 \\ 2.1878e+01 & 1810.15 & 4.0263e-04 & 5.3053e-08 &
5.6875e-05 & 4.3396e-04 & 2.3851e-07 & 3.7083e-06 & 2.2519e-05
\\ \hline \multicolumn{9}{c}{\bf TEA final abundances using JANAF free
  energies} \\ \hline 3.8019e-01 & 1719.64 & 4.5959e-04 & 5.8326e-08 &
4.5480e-08 & 3.7681e-04 & 4.8604e-09 & 8.0472e-08 & 2.2497e-05
\\ 1.6596e+00 & 1805.28 & 4.5922e-04 & 5.3200e-08 & 4.0512e-07 &
3.7719e-04 & 2.0937e-08 & 2.9102e-07 & 2.2504e-05 \\ 2.1878e+01 &
1810.15 & 4.0694e-04 & 5.3429e-08 & 5.2592e-05 & 4.2965e-04 &
2.1124e-07 & 3.7386e-06 & 2.2519e-05 \\ \hline
\end{tabular}\hfill\strut\ataboff
\end{table*}
\end{landscape}

\section{REPRODUCIBLE RESEARCH LICENSE}
\label{sec:RR}

Reproducing a lengthy computation, such as that implemented in TEA,
can be prohibitively time consuming \citep{stodden2009legal}.  We have released TEA under an
open-source license, but this is not enough, as even the most
stringent of those licenses (e.g., the GNU General Public License)
does not require disclosure of modifications if the researcher does
not distribute the code.  So that the process of science can proceed
efficiently, there are several terms in our license to ensure
reproducibility of all TEA results, including those from derivative
codes. A key term requires that any reviewed scientific publication using TEA
or a derived code must publish that code, the code output used in the
paper (such as data in tables and figures, and data summarized in the
text), and all the information used to initialize the code to produce
those outputs in a reproducible research compendium (RRC).  The RRC must
be published with the paper, preferably as an electronic supplement, or
else in a permanent, free-of-charge, public internet archive, such as
github.com.  A permanent link to the archive must be published in the
paper, and the archive must never be closed, altered, or charged for.
Details and examples of how to do this appear in the license and
documents accompanying the code, along with additional discussion.  The
RRC for this paper, including the TEA package and documentation, is
included as an electronic supplement, and is also available via
{\tt https://github.com/dzesmin/ RRC-BlecicEtal-2015a-ApJS-TEA/}.

\section{CONCLUSIONS}
\label{sec:conc}

We have developed an open-source Thermochemical Equilibrium Abundances
code for gaseous molecular species. Given elemental abundances and one
or more temperature-pressure pairs, TEA produces final mixing
fractions using the Gibbs-free-energy minimization method with an
iterative Lagrangian optimization scheme.

We applied the TEA calculations to several hot-Jupiter \math{T-P}
models, with expected results. The code is tested against the
original method developed by \citet{WhiteJohnsonDantzig1958JGibbs},
the analytic method developed by \citet{BurrowsSharp1999apjchemeq},
and the Newton-Raphson method implemented in the free Chemical
Equilibrium with Applications code. Using the free energies listed in 
\citet{WhiteJohnsonDantzig1958JGibbs}, their example, and derived free
energies based on the thermodynamic data provided in CEA's {\tt
  thermo.inp} file, TEA produces the same final
abundances, but with higher precision.

Currently, TEA is specialized for gaseous species, with the
implementation of condensates left for future work. In opacity
calculations at low temperatures (below 1000 K), the inclusion of
condensates is necessary as it reduces the gas phase
contribution to opacity \citep[e.g., ][]{SharpHuebner90, Lodders02,
  BurrowsSharp1999apjchemeq}.

The thermochemical equilibrium abundances obtained with TEA can be
used in all static atmospheres, atmospheres with vertical transport
and temperatures above 1200 K (except when ions are present), and as a
starting point in models of gaseous chemical kinetics and abundance
retrievals run on spectroscopic data. TEA is currently used to
initialize the atmospheric retrial calculations in the open-source
BART project (available at {\tt https://github.com/joeharr4/BART}).

TEA is written in a modular way using the Python programming
language. It is documented (the Start Guide, the User Manual, 
the Code Document, and this theory paper are provided with the code), 
actively maintained, and available to the community via the open-source
development sites {\tt https://github.com/dzesmin/TEA} and \\
{\tt https://github.com/dzesmin/TEA-Examples}.

\section{ACKNOWLEDGMENTS}

This project was completed with the support of the
NASA Earth and Space Science Fellowship Program, grant NNX12AL83H,
held by Jasmina Blecic, PI Joseph Harrington, and through the Science
Mission Directorate's Planetary Atmospheres Program, grant
NNX12AI69G. We would like to thank Julianne Moses for useful
discussions, and Kevin B. Stevenson and Michael R. Line for
temperature and pressure profiles. We also thank contributors to
SciPy, NumPy, Matplotlib, and the Python Programming Language; the
open-source development website GitHub.com; and other contributors to
the free and open-source community. \\

\newpage
\bibliographystyle{apj}
\bibliography{chap-TEA}

\fi

\chapter{THE BAYESIAN ATMOSPHERIC RADIATIVE TRANSFER CODE AND APPLICATION TO WASP-43b }
\label{chap:BART}

{\singlespacing
\noindent{\bf Jasmina Blecic\sp{1}, Joseph Harrington\sp{1}, P. Cubillos\sp{1}, M. Oliver Bowman\sp{1}, Patricio Rojo\sp{2}, M. Stemm \sp{1}, Nathaniel B.\ Lust\sp{1}, Ryan C.\ Challener\sp{1}, Austin J.\ Foster\sp{1}, Andrew S.\ D.\ Foster\sp{1}, Sarah D.\ Blumenthal\sp{1}, Dylan Bruce\sp{1}, Thomas J. Loredo\sp{3}}

\vspace{1cm}

\noindent{\em
\sp{1}Planetary Sciences Group, Department of Physics, University of Central Florida, Orlando, FL 32816-2385, USA \\
\sp{2}Department of Astronomy, Universidad de Chile, Santiago de Chile, Chile \\
\sp{3}Center for Radiophysics and Space Research, Space Sciences Building, Cornell University, Ithaca, NY 14853-6801
}

\vspace{5cm}



}
\clearpage

\if \includeBART y
    \setcitestyle{authoryear,round}

\section{ABSTRACT}

This paper is a part of three contributed papers \citep{HarringtonEtal2015-BART, CubillosEtal2015-BART} that describe a new open-source retrieval framework, Bayesian Atmospheric Radiative Transfer (BART). BART is a line-by-line radiative-transfer code initialized by a thermochemical equilibrium abundances code and driven through the parameter phase space by a differential-evolution Markov-chain Monte Carlo sampler. The three major parts of BART, the Thermochemical Equilibrium Abundances module \citep[\href{https://github.com/dzesmin/TEA}{{\tt TEA}}, ][]{BlecicEtal2015apjsTEA}, the radiative-transfer module (\href{https://github.com/exosports/transit}{{\tt Transit}}), and the Multi-core Markov-chain Monte Carlo statistical module \citep[\href{https://github.com/pcubillos/MCcubed}{{\tt MC\sp{3}}}, ][]{CubillosEtal2015apjRednoise} are self-sufficient modules that can be used for other scientific purposes. BART and its submodules are available to the community under the reproducible-research license via \href{https://github.com/exosports/BART}{https://github.com/exosports/BART}. In this paper, we describe the implementation of the initialization routines, the atmospheric profile generator, the eclipse module, the best-fit routines, and the contribution function module. Other modules and routines are presented in the collaborative paper by \citet{CubillosEtal2015-BART}. We also present a comprehensive atmospheric analysis of all WASP-43b secondary eclipse data obtained from the space- and ground-based observations using BART.

\section{INTRODUCTION}
\label{intro}

The rapid increase of detected extrasolar planets in the last decade (1593 confirmed and 3751 candidates as of August 5\sp{th}, 2015; \href{exoplanets.org}{ exoplanets.org}) and the number of novel techniques employed to analyze their data \citep[e.g.,][]{SwainEtal2008ApJhd209458bspec, CarterWinn2010, KnutsonEatl2009ApJ-PhaseVariationHD149026b, StevensonEtal2012apjHD149026b, DemingEtal2013arXivHD209458b-Xo1b-WCF3, deWitEtal2012aapFacemap, DemingEtal2015-PixelDecor}, have prompted theorists to develop different methods to model their atmospheres. To get an insight into their thermal structure and chemical composition, initial methods started from the first principles in a one-dimensional (1D) scheme \citep[e.g.,][]{FortneyEtal2005apjlhjmodels} and later developed into complex three-dimensional (3D) models that study the atmospheric dynamics and circulations \citep[e.g.,][]{ShowmanEtal2009-3Dcirc, Dobbs-DixonAgol2013-3DRT}.

Today, we have two major approaches to atmospheric modeling. One applies a direct, forward modeling technique that provides a set of parameters to generate the observed spectra, and the other uses the observations to determine the model's best-fit parameters and their uncertainties. In the direct approach, a set of physically motivated models is generated and the comparison is done by eye. Usually, in these models the molecular abundances are assumed to be in the thermochemical equilibrium \citep[e.g.,][]{Lodders09, FortneyEtal2005apjlhjmodels, Fortney2008, BurrowsEtal2008apjSpectra}. The inverse, essentially Bayesian based retrieval technique, determines the properties of the planetary atmosphere based on the available observations. It uses a statistical algorithm (usually a Markov-chain Monte Carlo method, MCMC) to explore the posterior distribution of the model given the data. By searching for the regions of space that provide the best fit to the data, this approach determines the uncertainties in the model parameters. 

The retrieval approach in characterizing exoplanetary atmospheres was first introduced by \citet{MadhusudhanSeager2009ApJ-AbundanceMethod}. Providing a set of free parameters, six for the temperature and pressure profile, \math{T(p)}, and four for species mixing ratios (abundances), they used a multidimensional grid optimization scheme in a line-by-line radiative-transfer model to explore the parameter phase space. Soon after, \citet{MadhusudhanSeager2010} utilized the first application of the MCMC algorithm in the retrieval to explore the likelihood space of models with millions of samples.

In 2012, \citet{LeeEtal2012-CF} introduced an optimal estimation retrieval algorithm \citep{irwin2008nemesis, Rodgers2000-Retrieval} that applies the correlated-K technique in an iterative scheme \citep{LacisOinas1991-correlatedK}. They used K-distribution tables pre-calculated from line-list databases, allowing rapid integration of the model spectra and an order of magnitude faster exploration of the phase space than line-by-line algorithms. This method requires fewer model evaluations attributed to the assumption that the parameter error distributions are Gaussians, and offers a single best-fit solution calculated using the Levenberg-Marquardt algorithm \citep{Levenberg1944, Marquardt1963}.  \citet{LineEtal2012-Retrieval-Intro} applies this approach to the line-by-line radiative transfer.


The same year, \citet{BennekeSeager2012-Retrieval} introduced another Bayesian-based retrieval algorithm, which considers non-Gaussian uncertainties of the atmospheric parameters and returns to computationally intensive exploration of the phase space. In addition, they account for the presence of a cloud deck or solid surface.

In 2014, \citet{LineEtal2014-Retrieval-I} tested three recently proposed retrieval approaches on a synthetic water-dominated hot Jupiter: optimal estimation, differential evolution Markov chain Monte Carlo (DEMC), and bootstrap Monte Carlo. They found a good agreement between the three methods only when the observations are of good quality (a high spectral resolution and high signal-to-noise data), and there are more observed spectral channels than free parameters. For weak and noisy exoplanetary signals, the best approach is to proceed with DEMC \citep{LineEtal2013-Retrieval-II, LineEtal2014-Retrieval-III}.

In 2015, \citet{Waldmann2015-TAU} introduced their novel inverse retrieval code, \math{\tau-REx}, Tau Retrieval for Exoplanets. It utilizes molecular line lists from the {\em ExoMol} project, a custom built software that identifies likely absorbers/emitters in the spectra, a Bayesian partition function (also called the Bayesian Evidence) that allows appropriate selection of the best-fit model, and two independent algorithms for sampling the parameter space: the nested sampling and the classical MCMC algorithm. This approach allows them to thoroughly explore the likelihood space of models with a large number of parameters through the multi-core processor scalability.

Very recently, \citet{benneke2015strict} introduced his new Self-Consistent Atmospheric RetrievaL framework for ExoplaneTs (SCARLET). In addition to retrieving molecular abundances, this code determines the full range of self-consistent scenarios by accounting for the uncertainties resulting from our limited knowledge of the chemical, dynamical, and cloud formation processes in planetary atmospheres. 

\begin{figure*}[t!]
\centerline{
\includegraphics[width=0.99\textwidth, clip=True]{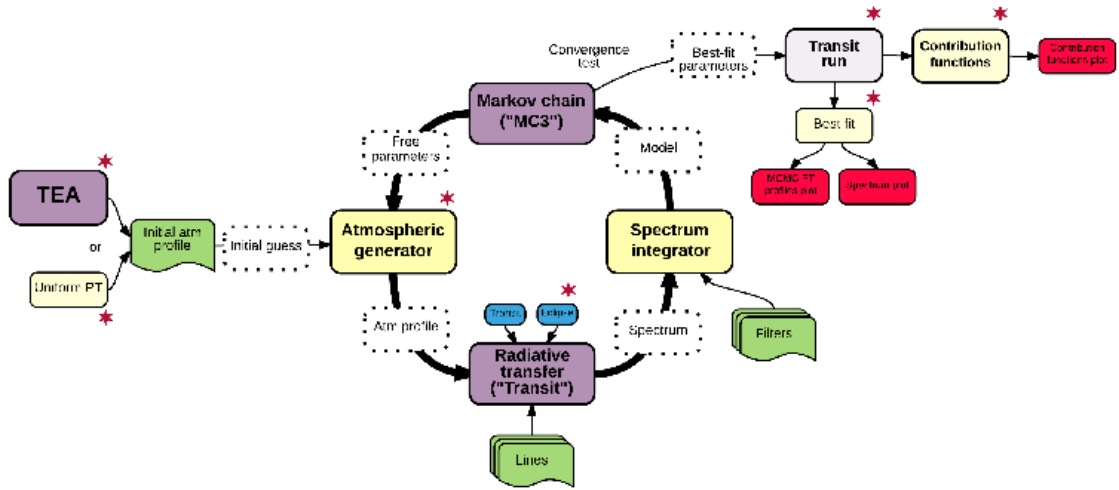}}
\caption[Simplified BART flow chart]{\footnotesize
Simplified BART flow chart. In purple are the three major BART modules. {\tt TEA} and {\tt MC\sp{3}} are written in Python, while {\tt Transit} is written in the C language. Other supporting modules are written in Python and given in yellow. In white are the inputs/outputs of each module. In green are the input files. In blue are transit and eclipse raypath-solution submodules. Stars denote modules and packages that will be described in more detail in this paper.}
\label{fig:BARTChart}
\end{figure*}

Here, we present a new Bayesian line-by-line radiative-transfer framework. This paper is a part of three contributed papers that describe the architecture, individual modules and packages, and theory implemented in the open-source Bayesian Atmospheric Radiative Transfer (BART) code. Figure \ref{fig:BARTChart} shows a simplified flow of the BART code. The underlying theory and the code structure of the modules and packages marked by the stars will be covered in more detail in this paper. Other modules are described in the collaborative paper by \citet{CubillosEtal2015-BART}. In addition, each paper presents an atmospheric analysis of a hot-Jupiter planet using BART to show its diverse features.

In Section \ref{sec:BART}, we present BART and describe the implementation of the initialization routines, atmospheric profile generator, eclipse module, best-fit routines, and contribution function module. In Section \ref{sec:analWASP43b}, we present an atmospheric analysis of all available WASP-43b secondary eclipse data using BART. In Section \ref{sec:conc}, we state our conclusions.

\section{THE BAYESIAN ATMOSPHERIC RADIATIVE  TRANSFER CODE}
\label{sec:BART}

Bayesian Atmospheric Radiative Transfer (BART) is an open-source Bayesian, thermochemical, radiative-transfer code written in Python and C, and available to the community under the reproducible-research license via \href{https://github.com/exosports/BART}{https://github.com/exosports/BART}. The code initializes a model for the atmospheric retrieval calculation, generates thousands of theoretical model spectra using parametrized pressure and temperature profiles and line-by-line radiative-transfer calculation, and employs a statistical package to compare the models with the observations. Given transit or eclipse observations at multiple wavelengths, BART retrieves the thermal profile and chemical abundances of the selected atmospheric species.

BART consists of three major parts (Figure \ref{fig:BARTChart}, purple boxes): the Thermochemical Equilibrium Abundances ({\tt TEA}) module, a radiative-transfer module ({\tt Transit}), and the Multi-core Markov-chain Monte Carlo statistical module ({\tt MC\sp{3}}). Each of the modules works independently and can be used for other scientific purposes. {\tt TEA} is an open-source thermochemical equilibrium abundances code that calculates the mixing fraction of gaseous molecular species \citep{BlecicEtal2015apjsTEA}. The code is written in Python and available to the community via \href{https://github.com/dzesmin/TEA}{https://github.com/dzesmin/TEA}. {\tt Transit} is an open-source  radiative-transfer code originally developed by Patricio Rojo. {\tt Transit} applies 1D line-by-line radiative-transfer computation in local thermodynamic equilibrium to generate model spectra. This code, written in C, was built specifically to attempt to detect water in the extrasolar planet HD 209458b using transit spectroscopy. Since then, {\tt Transit} has been significantly improved to handle eclipse geometry, multiple line-list and CIA opacity sources, and to perform opacity grid and Voigt profile calculation (see the collaborative paper by \citealt{CubillosEtal2015-BART}). The user's and programmer's documentation are added to the package. The code is available via \href{https://github.com/exosports/transit}{https://github.com/exosports/transit}. {\tt MC\sp{3}} is an open-source model-fitting tool \citep{CubillosEtal2015apjRednoise}. It uses Bayesian statistics to estimate the best-fit values and the credible region of the model parameters. The code is written in Python and C and available at \href{https://github.com/pcubillos/MCcubed}{https://github.com/pcubillos/MCcubed}.

To start off the retrieval (\ref{fig:BARTChart}), BART generates an initial atmospheric model (see Section \ref{sec:Init}). Given a pressure array, BART evaluates the temperature profile by using one of the two parametrization schemes (see Section \ref{sec:atmGenerator}). Then, for given elemental and molecular species, the species mixing ratios are calculated using the TEA module or the routine that produces a vertically uniform abundances profile. The \math{T(p)} profile and species scaling factors are free parameters of the models. Given an initial guess, the atmospheric generator produces the atmospheric model, passes it to {\tt Transit} to calculate an emergent spectrum for the desired geometry (transit or eclipse), and integrates the spectrum over the observational bandpasses. The band-integrated values are sent to {\tt MC\sp{3}} to compare them against the observations and calculate \math{\chi\sp{2}}. Then, {\tt MC\sp{3}} generates a new set of free parameters, repeating the process until the phase space of parameters is fully explored and the Gelman and Rubin convergence test is satisfied \citep{GelmanRubin1992}. Upon finishing the loop, the {\tt MC\sp{3}} returns the sampled parameter posterior distribution, best-fit values, and 68\% credible region. For details about the spectrum integrator and the {\tt MC\sp{3}} loop see collaborative paper by \citet{CubillosEtal2015-BART}.

The best-fit parameters are then used to run the {\tt Transit} module once more to reproduce the spectra of the best-fit model atmosphere (see Section \ref{sec:bestFit}). In this run, {\tt Transit} generates a file with the optical depth values for each atmospheric layer and wavelength, which is used to calculate the contribution functions of each observation (see Section \ref{sec:cf}). Finally, BART plots the best-fit spectrum with data and model band-integrated values, the \math{T(p)} profile, and the contribution functions. In addition, {\tt MC\sp{3}} plots the parameters' traces, pairwise posterior distributions, and marginal posterior histograms.

In the following sections, we describe in more detail the implementation of the initialization routines, Section \ref{sec:Init}; the atmospheric profile generator, Section \ref{sec:atmGenerator}; the eclipse module, Section \ref{sec:eclipse}; the best-fit routines, Section \ref{sec:bestFit}; and the contribution function module, Section \ref{sec:cf}. The {\tt TEA} and {\tt MC\sp{3}} modules, given in short in this paper (Sections \ref{sec:TEA} and \ref{sec:MC3}), are described in detail in \citet{BlecicEtal2015apjsTEA} and \citet{CubillosEtal2015apjRednoise} respectively.

\begin{figure}[ht!]
\centerline{
\includegraphics[width=7.0cm, trim=0 0 0 20, clip=True]{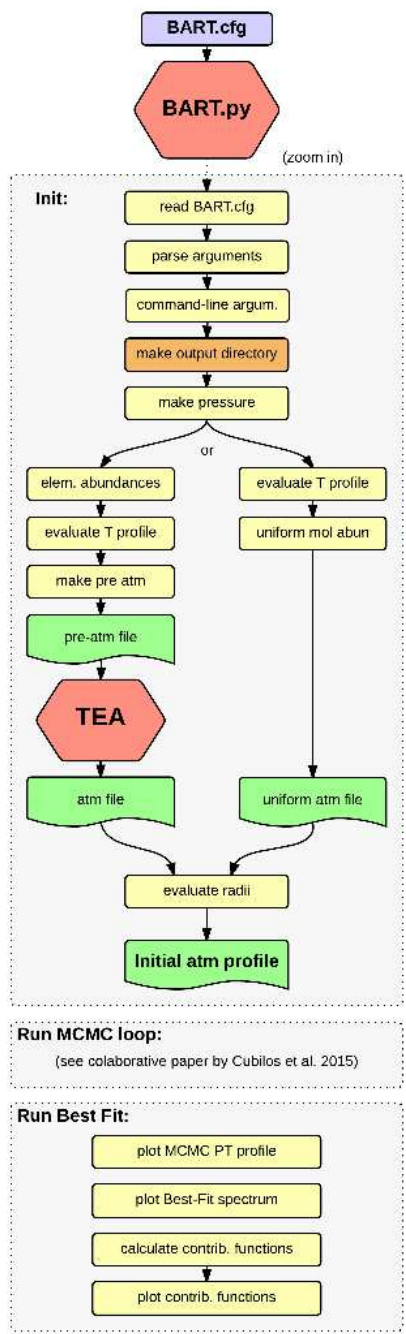}}
\vspace{-20pt}
\caption[Flow chart of the BART driver, BART.py.]{\footnotesize
Flow chart of the BART driver, BART.py. The gray sections show the initialization routines, the MCMC loop (described in the collaborative paper by \citealt{CubillosEtal2015-BART}), and the best-fit routines. All routines are called inside {\tt BART.py}.}
\label{fig:BARTpy}
\end{figure}

\subsection{BART Initialization}
\label{sec:Init}

BART is configured with a single ASCII file, {\tt BART.cfg}. {\tt BART.cfg} is the main input file for the BART driver, {\tt BART.py}, that executes all other modules and subroutines. Figure \ref{fig:BARTpy} shows a simplified execution order of the routines called in {\tt BART.py}.

In this section, we describe in more detail the routines listed in the {\tt Init} section of Figure \ref{fig:BARTpy}. Section {\tt Run MCMC Loop} is described in the collaborative paper by \citet{CubillosEtal2015-BART}, and Section {\tt Run BEST Fit} is described in Section \ref{sec:bestFit} of this paper.

The initialization starts with parsing the arguments from {\tt BART.cfg} and any additional arguments provided on the command line. {\tt BART.cfg} gathers information from the user about the output directory, initial atmospheric model, data and uncertainties, and arguments' values needed to run {\tt TEA}, {\tt Transit}, and {\tt MC\sp{3}}. These inputs constrain the final atmospheric model.

\newpage
The user chooses the pressure grid in logarithmic or linear space, the desired elemental and molecular species, the elemental mixing ratios, a C/O ratio and/or solar metallicity factor, and a parametrization scheme for the temperature profile. This generates a pre-atmospheric file for the {\tt TEA} module. {\tt TEA} is executed to calculate thermochemical equilibrium abundances for the species of interest. Then, based on the hydrostatic balance equation and the referenced planetary radius and pressure, the radius array is calculated and added to the final {\tt TEA} output. This produces the initial atmospheric profile for the first {\tt MC\sp{3}} iteration. 

In addition to using {\tt TEA} to calculate the species' equilibrium abundances, the user has an option to  choose a vertically-uniform-abundances profile. Upon setting the pressure array and evaluating the temperature profile, the user sets the desired mixing fractions for the input and output species and calls a routine to generate a uniform atmospheric model.

\subsection{Atmospheric Profile Generator}
\label{sec:atmGenerator}

BART provides two parametrized temperature-profile schemes. One originally developed by \citet{Guillot2010A-LinePTprofile}, and the other based on the parametrization described in \citet{MadhusudhanSeager2009ApJ-AbundanceMethod}.

\subsubsection{Parametrization Scheme I}
\label{sec:Line}


This parametrization scheme was originally formulated by \citet{Guillot2010A-LinePTprofile} and subsequently modified by \citet{ParmentierGuillot2014-LinesPTprofile, LineEtal2013-Retrieval-II}, and \citet{HengEtal2012LinePTprofile} to include  more freedom for the case when temperature inversion is present in a planetary atmosphere. This approach is usually denoted as the three-channel approximation, where the planet's temperature is given as:
\begin{equation}
T^4(\tau) = \frac{3T_{\rm{int}}^4}{4}\big(\frac{2}{3} + \tau\big) + \frac{3T_{\rm{irr}}^4}{4}(1 -\alpha)\,\xi_{\gamma_{\rm 1}}(\tau) + \frac{3T_{\rm{irr}}^4}{4}\,\alpha\,\xi_{\gamma_{\rm 2}}(\tau)\, ,
\end{equation}
\noindent with \math{\xi\sb{\gamma_{\rm i}}} defined as:
\begin{equation}
\xi_{\gamma_{\rm i}} = \frac{2}{3} + \frac{2}{3\gamma_{\rm i}}\big[1 + \big(\frac{\gamma_{\rm i}\tau}{2} - 1\big)\,e^{-\gamma_{\rm i}\tau}\big] + \frac{2\gamma_{\rm i}}{3} \big(1 - \frac{\tau^2}{2}\big)\,E_2(\gamma_{\rm i}\tau)\, .
\end{equation}
\noindent \math{\gamma\sb{\rm 1}} and \math{\gamma\sb{\rm 2}} are ratios of the mean opacities in the visible to the infrared, given as \math{\gamma\sb{\rm 1} = \kappa\sb{\upsilon\sb{1}}/\kappa\sb{IR}} and \math{\gamma\sb{\rm 2} = \kappa\sb{\upsilon\sb{2}}/\kappa\sb{IR}}. The parameter \math{\alpha} ranges between 0 and 1 and describes the separation of the two visible streams, \math{\kappa\sb{\rm{\upsilon{1}}}} and \math{\kappa\sb{\rm{\upsilon{2}}}}. \math{E\sb{2}(\gamma\tau)} is the second-order exponential integral function. The planet internal flux  is given as:
\begin{equation}
T\sb{\rm{irr}} = \beta\,\big(\frac{R_{*}}{2a})^{1/2}\, T_{*} \, 
\end{equation}
\noindent where \math{R\sb{*}} and \math{T\sb{*}} are the stellar radius and temperature, and \math{a} is the semimajor axis. The incident solar flux is denoted as \math{T\sb{\rm{irr}}}. Both \math{T\sb{\rm{int}}} and \math{T\sb{\rm{irr}}} variables have fixed values. The parameter \math{\beta} has a value around 1 and accounts for albedo, emissivity, and day-night redistribution.  The parameter \math{\tau} is the infrared optical depth calculated using the mean infrared opacity, \math{\kappa\sb{\rm{IR}}}, pressure \math{P}, and the planet surface gravity \math{g} at one 1 bar level:
\begin{equation}
\tau = \frac{\kappa_{\rm{IR}}\,P}{g} \, ,
\end{equation}
This parametrized approach has five free parameters: \math{\kappa\sb{\rm{IR}}}, \math{\kappa\sb{\rm{\upsilon{1}}}}, \math{\kappa\sb{\rm{\upsilon{2}}}}, \math{\alpha}, and \math{\beta}. The energy balance at the top of the atmosphere is accounted through the parameter \math{\beta}. The existence of a temperature inversion is allowed through the parameters \math{\kappa\sb{\rm{\upsilon{1}}}} and \math{\kappa\sb{\rm{\upsilon{2}}}}. For boundaries imposed on these parameters see \citet{LineEtal2014-Retrieval-I}, Section 3.2.

\subsubsection{Parametrization Scheme II}
\label{sec:Madhu}

Our second choice for the temperature-profile generator is the parametrization scheme similar to the one developed by \citet{MadhusudhanSeager2009ApJ-AbundanceMethod}. We made some minor changes to this method, which are described below.

In this scheme, the profiles are generated for an inverted and non-inverted atmosphere separately. The atmosphere is divided into three layers based on the physical constraints expected in hot Jupiters, Figure \ref{fig:MadhuPT}.

\begin{figure}[h]
    \centering
    \includegraphics[clip,width=0.63\textwidth]{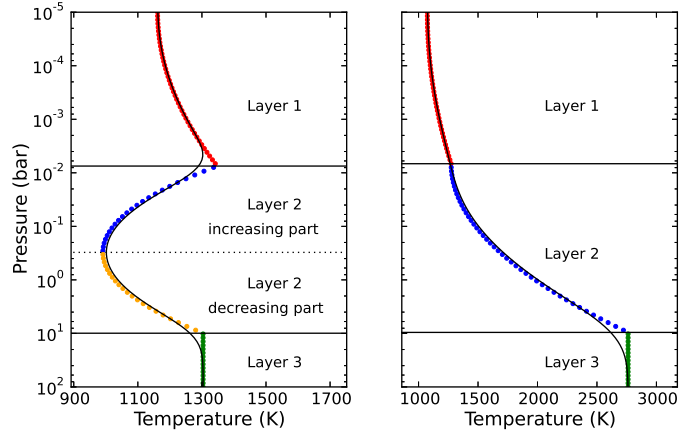}
\caption[Parametric T(p) profiles for inverted and non-inverted atmosphere based on \citet{MadhusudhanSeager2009ApJ-AbundanceMethod}]{\footnotesize Parametric T(p) profiles for inverted and non-inverted atmosphere based on \citet{MadhusudhanSeager2009ApJ-AbundanceMethod}. The profiles include three layers: 1. {\em Mesopheric layer, Layer 1} - layer below 10\sp{-5} bars,  important for atmospheric escape and photochemistry. The atmosphere is transparent to the outgoing radiation in infrared and optical. This layer is heated from the lower layers and cooled with increasing altitude; 2. {\em Stratospheric layer, Layer 2} - in this layer a temperature inversion can occur depending on the level of irradiation from the host star and the presence of strong absorbers. Radiation is the dominant transport mechanism. Most spectral features come from this layer; 3. {\em Isothermal layer, Layer 3} - due to the strong radiation from the parent star the radiative-convective boundary of hot-Jupiter planets is shifted deep in the planetary atmosphere. Strong irradiation is absorbed higher in the atmosphere and does not reach this layer, leading to an isothermal temperature profile. Radiation that comes from this layer behaves like a black body spectra.
{\bf Left:} For the inverted atmosphere, Layer 2 consists of two parts: one where temperature decreases with height, and the other where temperature increases with height (thermal inversion occurs). Different colors depict partial profiles for each layer generated using our Equation (\ref{invert}). The dots display the number of levels in the atmosphere, which is equally spaced in log-pressure space. The profile has six free parameters: ${P}_{1}$, ${P}_{2}$,  ${P}_{3}$, ${T}_{3}$, ${\alpha_{1}}$, and ${\alpha_{2}}$. The thin black line shows the smoothed profile using a 1D Gaussian filter. {\bf Right:} Non-inverted atmospheric profile. The different colors depict partial profiles for each layer generated using our Equation (\ref{NONinvert}). The profile has five free parameters: ${P}_{1}$, ${P}_{3}$, ${T}_{3}$, ${\alpha_{1}}$, and ${\alpha_{2}}$. The thin black line shows the smoothed profile using a 1D Gaussian filter.}
\label{fig:MadhuPT}
\end{figure}

The following set of equations, as given by \citet{MadhusudhanSeager2009ApJ-AbundanceMethod}, mimics the behaviour in each atmospheric layer (see the caption of Figure \ref{fig:MadhuPT} for more details):
\begin{align}
\label{MadhuPT}
  P_0 &< P < P_1& &P = P_0e^{\alpha_1(T - T_0)^{\beta_1}} &\hspace{0.2cm} \rm{layer\,\,1} \nonumber \\
  P_1 &< P < P_3& &P = P_2e^{\alpha_2(T - T_2)^{\beta_2}} &\hspace{0.2cm} \rm{layer\,\,2}  \\
  P~&> P_3& &T = T_3 &\hspace{0.2cm} \rm{layer\,\,3}  \nonumber
\end{align}
The set contains 12 variables: \math{P}\sb{0}, \math{P}\sb{1}, \math{P}\sb{2},  \math{P}\sb{3}, \math{T}\sb{0}, \math{T}\sb{1}, {T}\sb{2}, {T}\sb{3}, \math{\alpha\sb{1}}, \math{\alpha\sb{2}}, \math{\beta\sb{1}}, and \math{\beta\sb{2}}. 

To decrease the number of free parameters, we first set \math{P}\sb{0} to the pressure at the top of the atmosphere. The parameters \math{\beta\sb{1}} and \math{\beta\sb{2}} are empirically determined to be \math{\beta\sb{1}} = \math{\beta}\sb{2} = 0.5 \citep{MadhusudhanSeager2009ApJ-AbundanceMethod}. 
Two of the parameters can be eliminated based on the two constraints of continuity at the two layer boundaries. 

The temperature \math{T\sb{3}} is estimated based on the effective (surface) temperature of the planet. When a planet total emissivity in the observed wavelength band is less than one, due to the presence of an atmosphere, its emissivity is less than that of a black body and the actual temperature of the object is higher than the effective temperature. Thus, to account for the presence of an atmosphere and spectral features, we use a scaling factor of 1 to 1.5 to constrain the maximum range of the \math{T\sb{3}} temperature.

The effective temperature of the planet is calculated based on the energy balance equation as:
\begin{equation}
\label{Teff}
T_{\rm eff}^4 = f\,\, T_{\rm eff}^{*\,\,\,4}\, \Big(\frac{R}{a}\Big)^2\,(1-A)\, ,
\end{equation}
\noindent where factor \math{f} describes the energy redistribution from the day to the night side. \math{f} = 1/4 defines the uniform redistribution of energy between the day and the night side of the planet. Since we are observing the planet dayside during secondary eclipse, we are interested in the case when the energy received is uniformly redistributed on the planet dayside, and none of the energy is transferred to the night side. In that case, the factor \math{f} is 1/2. For zero albedo, Equation (\ref{Teff}) becomes:
\begin{equation}
\label{PlanetTeff}
T_{\rm eff}^4 = \frac{1}{2}\,T_{\rm eff}^{*\,\,\,4}\, \Big(\frac{R}{a}\Big)^2\, .
\end{equation}
To remove one more free parameter for a non-inversion case, we rewrite Equation (\ref{MadhuPT}) to distinguish the increasing and decreasing part of the Layer 2 curve:
\begin{align}
\label{eqn:JB-PT}
  P_0 &< P < P_1& &P = P_0e^{\alpha_1(T - T_0)^{1/2}} &\hspace{0.2cm} \nonumber \\
  P_1 &< P < P_2& &P = P_2e^{\alpha_2(T - T_2)^{1/2}} &\hspace{0.2cm} \\
  P_2 &< P < P_3& &P = P_2e^{-\alpha_2(T - T_2)^{1/2}} &\hspace{0.2cm} \nonumber \\
  P~&> P_3& &T = T_3 &\hspace{0.2cm}\nonumber
\end{align}

\vspace{8pt}
\noindent {\bf 5.3.2.2.1 \hspace{2pt} Inverted \math{T(p)} Profile} 
\vspace{6pt}

The parametric profile for the inverted atmosphere has six free parameters: \math{P}\sb{1}, \math{P}\sb{2},  \math{P}\sb{3}, \math{T}\sb{3}, \math{\alpha\sb{1}}, and \math{\alpha\sb{2}}. We calculate the \math{T\sb{0}}, \math{T\sb{1}}, and \math{T\sb{2}} temperatures as:
\begin{align}
\label{invert}
  T_2 = T_3 - \big(\frac{log(P_3/P_2)}{\alpha_2}\big)^2   \nonumber \\
  T_0 = T_2 - \big(\frac{log(P_1/P_0) }{\alpha_1}\big)^2 + \big(\frac{log(P_1/P_2)}{-\alpha_2}\big)^2  &\hspace{0.4cm} \\
  T_1 = T_0 + \big(\frac{log(P_1/P_0)}{\alpha_1}\big)^2 &\hspace{0.2cm} \nonumber
\end{align}
An example of an inverted \math{T(p)} profile is shown in Figure \ref{fig:MadhuPT}, left panel. To remove sharp kinks on the layer boundaries, we use a Gaussian filter.

\vspace{8pt}
\noindent {\bf 5.3.2.2.2 \hspace{2pt} Non-Inverted \math{T(p)} Profile}
\vspace{6pt}

For the non-inverted atmosphere, we assume that the Layer 2 follows an adiabatic temperature profile and exclude \math{P\sb{2}} as a free parameter. Thus, the parametric profile for the inverted atmosphere has five free parameters: \math{P}\sb{1}, \math{P}\sb{3}, \math{T}\sb{3}, \math{\alpha\sb{1}}, and \math{\alpha\sb{2}}.  We calculate \math{T\sb{0}} and \math{T\sb{1}} as:
\begin{align}
\label{NONinvert}
  T_1 = T_3 - \big(\frac{log(P_3/P_1)}{\alpha_2}\big)^2   \\
  T_0 = T_1 - \big(\frac{log(P_1/P_0)}{\alpha_1}\big)^2  &\hspace{0.1cm} \nonumber
\end{align}
An example of a non-inverted \math{T(p)} profile is shown in Figure \ref{fig:MadhuPT}, right panel. To remove sharp kinks on the layer boundaries, we againuse a Gaussian filter.

\subsubsection{Species Factors}
\label{sec:specsFac}

In addition to the temperature profile parameters, the atmospheric generator accepts free parameters for the molecular species scaling factors.  This way, we allow for non-equilibrium conditions in the planetary atmosphere to occur. We have as many free parameters as species we want to fit in our model. Our scaling factors are constant with altitude, and we are retrieving log of the species abundances to prevent negative and physically un-plausible mixing ratios (and to allow them to vary over several orders of magnitude).

\subsection{TEA Module}
\label{sec:TEA}

To initialize the retrieval and calculate the equilibrium abundances for the species of interest, we use the Thermochemical Equilibrium Abundances code \citep[TEA][]{BlecicEtal2015apjsTEA}. TEA calculates the mixing fractions of gaseous molecular species following the methodology by \citet{WhiteJohnsonDantzig1958JGibbs} and \citet{Eriksson1971}. Given a \math{T(p)} profile and elemental abundances, TEA determines the mixing fractions of the desired molecular species by minimizing the total Gibbs free energy of the system:
\begin{eqnarray}
\label{eq:eqmin}
\frac{G_{sys}(T)}{RT} = \sum_{i=1}^n x_{i} \Big[\frac{g_{i}^0(T)}{RT}
  + \ln P + \ln\frac{x_{i}}{N}\Big]\, ,
\end{eqnarray}
\noindent under the mass balance constraint:
\begin{eqnarray}
\sum_{i=1}^n a_{ij}\, x_{i} = b_{j}\,, \,\,\,\,(j = 1, 2, ...,m)\, ,
\label{masbal}
\end{eqnarray}
\noindent where \math{G\sb{sys}(T)} is the total Gibbs free energy of
the system, \math{n} is the number of chemical species, \math{T} is the
temperature, \math{R} is the gas constant, \math{P} is the pressure, \math{x\sb{i}} is the number of moles of species \math{i}, \math{a\sb{ij}} is the stoichiometric coefficient that indicates the number of atoms of element \math{j} in species \math{i}, \math{m} is the number of elemental species originally present in the mixture, \math{b\sb{j}} is the total number of moles of element \math{j}, and \math{g\sb{i}\sp{0}(T)} is the free energy of the species. 

The minimization is done using an iterative Lagrangian steepest-descent method that minimizes a multi-variate function under constraint. In addition, to guarantee physically plausable results, i.e., positive mixing fractions, TEA implements the lambda correction algorithm.

Equation (\ref{eq:eqmin}) requires a knowledge of the free energy of
species as a function of temperature. These are obtained from
the JANAF (Joint Army Navy Air Force) tables \citep[\href{http://kinetics.nist.gov/janaf/}{http://kinetics.nist.gov/janaf/},][]{ChaseEtal1982JPhJANAFtables, ChaseEtal1986bookJANAFtables}. Thus, TEA has an access to 84 elemental species and the thermodynamical data for more than
600 gaseous molecular species. We use the reference table containing elemental solar abundances given in \citet[][]{AsplundEtal2009-SunAbundances}, their Table 1. 

TEA is tested against the original method developed by \citet{WhiteJohnsonDantzig1958JGibbs}, the analytic method developed by \citet{BurrowsSharp1999apjchemeq}, and the Newton-Raphson method implemented in the free Chemical Equilibrium with Applications code \\(CEA, \href{http://www.grc.nasa.gov/WWW/CEAWeb/}{http://www.grc.nasa.gov/WWW/CEAWeb/}). Using the free energies listed in \citet[][]{WhiteJohnsonDantzig1958JGibbs}, their Table 1, and derived free energies based on the thermodynamic data provided in CEA's {\tt
  thermo.inp} file, TEA produces the identical final
abundances for both approaches, but with a higher numerical precision.

The thermochemical equilibrium abundances obtained with TEA can be
used in static atmospheres, and as a starting point in models of gaseous chemical kinetics and abundance retrievals. TEA is written in Python in a modular way, documented (the start guide, the user manual, the code document, and the theory paper are provided with the code), actively maintained, and available to the community via the open-source
development sites \href{https://github.com/dzesmin/TEA}{https://github.com/dzesmin/TEA} and \href{https://github.com/dzesmin/TEA-Examples}{https://github.com/dzesmin/TEA-Examples}.

\subsection{Statistical Module}
\label{sec:MC3}

BART explores the parameter space of thermal profiles and species abundances using the Multi-core Markov-chain Monte Carlo module \citep[MC\sp{3},][]{CubillosEtal2015-BART, CubillosEtal2015apjRednoise}. MC\sp{3} is an open-source general-purpose statistical package for model fitting. Using Bayesian Inference through a MCMC algorithm, MC\sp{3} provides two routines to sample the parameters' posterior distributions: Differential-Evolution \citep[DEMC, ][]{Braak2006DifferentialEvolution} or Metropolis Random Walk (using multivariate Gaussian proposals). It handles Bayesian priors (uniform, Jeffrey's, or informative), and implements the Gelman-Rubin convergence test \citep{GelmanRubin1992}. It utilizes single-CPU and multi-core computation, supported through Messaging Passing Interface, MPI. The code, written in Python with several C-routines, is documented and available to the community via \href{https://github.com/pcubillos/}{https://github.com/pcubillos/MCcubed}.

\subsection{Eclipse Geometry}
\label{sec:eclipse}

The radiative-transfer equation allows us to generate and study planetary spectra. A planetary emergent spectrum depends on where and how much incoming and outgoing radiation is absorbed, scattered, and re-radiated. It carries information about the atmospheric temperature, pressure, and chemical composition. To develop the radiative-transfer equation, we need to know the observing geometry and the energy transitions of the atoms and molecules present in the planetary atmosphere. 

During transit, the planet passes in front of the host star and the incoming radiation travels through the planetary limb probing upper layers of the planetary atmosphere. During eclipse, the planet is passing behind the star, allowing us to measure the planet dayside thermal emission. 

The raypath solution for the transit geometry was part of the original {\tt Transit} module (see collaborative paper by \citealt{CubillosEtal2015-BART}). Below, we describe the implementation of the eclipse module that calculates the raypath solution for the eclipse geometry.

The eclipse module sets eclipse geometry, calculates optical depth, and returns the emergent flux at the top of the atmosphere. The workflow of the eclipse module is given in Figure \ref{fig:eclipseFlow}.
\begin{figure}[b!]
\vspace{-15pt}
\centerline{
\includegraphics[width=0.5\textwidth, clip=True]{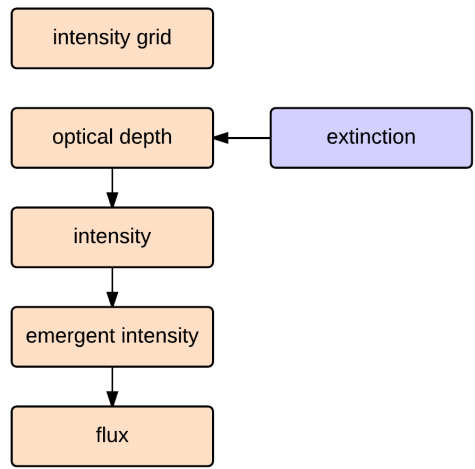}}
\caption[Work flow of the eclipse module]{\footnotesize
Work flow of the eclipse module. In orange are the routines written for the eclipse geometry. In blue is the routine from the original {\tt Transit} code that calculates extinction.}
\label{fig:eclipseFlow}
\end{figure}
{\tt Transit} treats an atmosphere as a plane-parallel radiation field under the local thermodynamic equilibrium conditions. Irradiation from the host star reaches the top of the atmosphere and propagates through it in 1D (see Figure \ref{fig:eclgeom}). The interaction of photons travelling through a planetary atmosphere is expressed in terms of optical depth. The change in the monochromatic optical depth along a path (\math{ds}) is given as:
\begin{equation}
\label{tau}
d\tau_\nu = - e\,\,ds\, ,
\end{equation}
\noindent where \math{e} is the extinction and \math{s} is the distance travelled from a certain layer to the top layer of the planetary atmosphere. {\tt  Transit} calculates extinction either through a line-by-line calculation or by interpolating the pre-calculated opacity grid. The negative sign comes from the fact that we are measuring optical depth from the top of the atmosphere, where it equals zero.

A ray, emerging at the top of the atmosphere at an angle in respect to the normal, travels a distance:
\begin{equation}
\label{ds}
ds = - \frac{dz}{cos\theta} .
\end{equation}

\begin{figure}[h!]
\centerline{
\includegraphics[width=0.40\textwidth, clip=True, trim=0 -20 0 0]{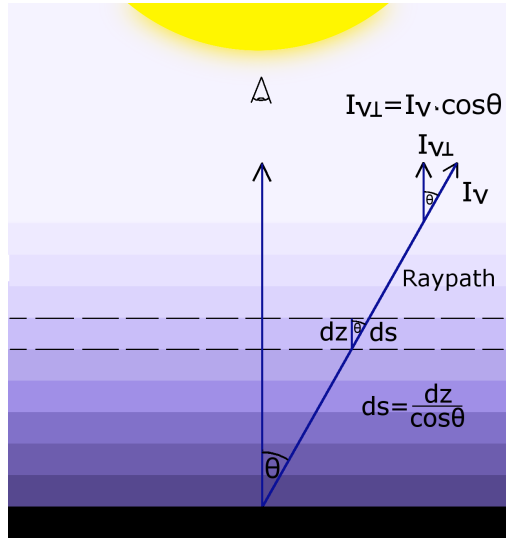}}
\caption[Eclipse geometry]{\footnotesize
Monochromatic beam of radiation travelling through a planetary atmosphere and emerging at the top of the atmosphere at an angle ${\theta}$. ${ds}$ is the distance along the path, and ${I_{\nu \perp}}$ is the intensity in the observer's direction.}
\label{fig:eclgeom}
\end{figure}

\noindent If we denote cos\,\math{\theta} = \math{\mu}, the radiative-transfer equation as a function of the optical depth and incident angle becomes:
\begin{equation}
\label{eq:RT-tau-final}
\mu \frac{dI_\nu}{d\tau_\nu}=I_\nu - S_\nu\, .
\end{equation}
\noindent where \math{I_\nu} is the intensity of radiation and \math{S_\nu} is the source function. Under the LTE and no scattering, Kirchoff's Law is valid, and the source function is equal to the black body function leading to:
\begin{equation}
\label{solve}
-\mu\frac{d I_\nu}{d \tau} = - I_\nu  +  B_\nu.
\end{equation}
\noindent By multiplying Equation (\ref{solve}) by \math{\exp\,(-\tau/\mu)}, recognizing derivative terms, and integrating from the bottom to the top of the atmosphere, we get the intensity at the top of the atmosphere:
\begin{eqnarray}
I_\nu\,(\tau=0) =  \int_{0}^{\tau}\frac{B_\nu}{\mu}\,e^{-\tau/\mu} d\tau\, ,
\label{final}
\end{eqnarray}
\noindent In {\tt Transit}, intensity is given in erg/s/cm/sr.

\subsubsection{Intensity Grid and Observed Flux}
\label{sec:IntensFlux}

We are interested in the flux coming from the dayside of the planet at the observer's location. Thus, we need to account for the distance to the observer.

\begin{figure}[h]
\centerline{
\includegraphics[width=0.45\textwidth, height=6.5cm, trim= 0 0 0 200]{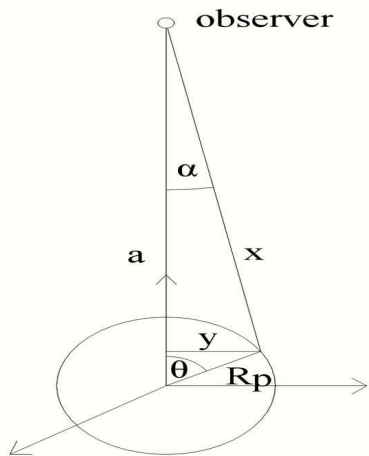}}
\caption[Observing geometry]{\footnotesize
Observing geometry. \math{a} is the shortest distance between the planet and the observer. \math{x} is the  distance between the observer and the point at the top of the atmosphere, where the intensity emerges at the angle \math{\theta}. \math{R\sb{p}} is the radius of the planet.}
\label{fig:obserDist}
\end{figure}

To get the surface flux, we would integrate over the solid angle for half of the sphere (dayside). If we assume the uniform \math{T(p)} profile and opacities on the planet dayside, the intensities at different angles (\math{\theta}) emerging at the top of the atmosphere are the same, and we can take them outside of the integral.

This gives us the well-known expression for the surface flux:
\begin{eqnarray}
\label{Sur-flux}
F_\nu  & = & \int_\Omega I_\nu\,cos \theta \,d\Omega\, , \nonumber \\
F_\nu  & = & \pi\,I_\nu\, ,
\end{eqnarray}
\noindent where cos\,\math{\theta} describes the angle between the intensity vector (where the ray is coming out of the atmosphere) and the direction of the solid angle (\math{I_\nu\,cos \theta} is denoted as \math{I\sb{\nu\,\perp}} in Figure \ref{fig:eclgeom}).

However, the flux at the observer's location is different than the surface flux. In addition to angle \math{\theta}, it also depends on the angle \math{\alpha} (see Figure \ref{fig:obserDist}):
\begin{eqnarray}
\label{ObsFlux}
F_\nu & = & \int_\Omega I_{\nu\perp}\,cos \alpha \,d\Omega\, ,
\end{eqnarray}
\noindent where \math{\Omega} is the solid angle,  \math{d\Omega = \sin\alpha\,d\alpha\,d\phi}, giving: 
\begin{eqnarray}
F_\nu  & = & 2\pi \int_0^{\pi/2} I_{\nu\perp} \cos\alpha\sin\alpha\,d\alpha\, .
\label{fluxGen2}
\end{eqnarray}
\noindent To solve this integral, we calculate intensities at several angles across the planet's dayside and perform hemispheric integration to get the flux at the observer's location.

From the geometry in Figure \ref{fig:obserDist} we see that \math{\sin \alpha = \frac{R_p}{x}\,\approx\,\frac{R_p}{a}}, where \math{R\sb{p}} is the radius of the planet, \math{x} is the hypotenuse, and \math{a} is the distance to the observer. Thus, our integral becomes:
\begin{eqnarray}
\label{limits}
F_\nu  & = & 2\pi\,\int_0^{R_p/a} I_{\nu\perp} \cos\alpha\sin\alpha\,d\alpha\, .
\end{eqnarray}
\noindent By expressing angle \math{\alpha} through angle \math{\theta}, and assuming that at very large distances \math{x} \math{\approx} \math{a}, \math{\sin\alpha \approx \alpha}, and \math{\cos\alpha \approx 1}, we get:
\begin{eqnarray}
a \, \alpha & = & R_p\,sin\theta\, . 
\label{y-approx2}
\end{eqnarray}
\noindent If we substitute this in Equation (\ref{limits}) and change limits, we get:
\begin{eqnarray}
F_\nu  & = & 2\pi\,\Big(\frac{R_p}{a}\Big)^2\int_0^{\pi/2} I_{\nu\perp}\,\sin\theta\, d(\sin\theta)\, .
\label{obsFlux2}
\end{eqnarray}
\noindent We choose several angles across the planet dayside and calculate the intensity in the observers direction, and then integrate along the donut-shaped segments that account for the calculated intensity, so that the full solid angle between [0, \math{\frac{\pi}{2}}] is covered (Figure \ref{fig:donuts}).

\begin{figure}[h]
\centerline{
\includegraphics[width=0.45\textwidth, clip=True, trim=0 0 0 420]{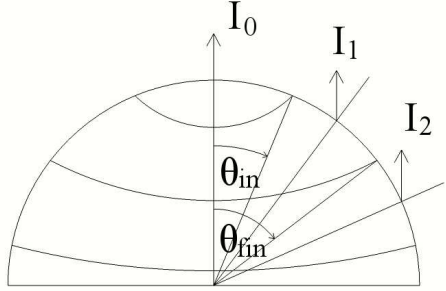}}
\caption[Intensity grid]{\footnotesize
Intensity grid. The flux at the observer's location is integrated along the donut-shaped segments for each intensity at angle \math{\theta} along the whole hemisphere.}
\label{fig:donuts}
\end{figure}

If we denote the intensities at different angles \math{\theta}, \math{I\sb{\nu\perp}(\theta)}, as \math{I\sb{i}}, Equation (\ref{obsFlux2}) becomes:
\begin{eqnarray}
F_\nu  & = & 2\pi\,\Big(\frac{R_p}{a}\Big)^2\,\sum_{(0, \frac{\pi}{2})} I_i\,\,\frac{(\sin\theta)^2}{2}\Big|_{\theta_{in}}^{\theta_{fin}} \, ,\\
F_\nu  & = & \pi\,\Big(\frac{R_p}{a}\Big)^2\,\sum_{(0, \frac{\pi}{2})} I_i\,(\sin\theta)^2\Big|_{\theta_{in}}^{\theta_{fin}} \, .
\label{obsFluxFinal}
\end{eqnarray}
\noindent {\tt Transit} calculates and outputs the following flux:
\begin{eqnarray}
F_\nu  & = & \pi\,\sum_{(0, \frac{\pi}{2})} I_i\,(\sin\theta)^2\Big|_{\theta_{in}}^{\theta_{fin}} \, .
\label{Final}
\end{eqnarray}
\noindent The distance factor \math{(\frac{R\sb{p}}{a})^2} is dropped and included in the Spectrum integrator (see collaborative paper by \citealt{CubillosEtal2015-BART}).

\subsection{Best Fit}
\label{sec:bestFit}

Upon running the desired number of {\tt MC\sp{3}} iterations and checking for the Gelman and Rubin convergence, {\tt MC\sp{3}} generates the best-fit parameters file that we use to run the {\tt Transit} module one more time. This generates the final BART outputs: the best-fit atmospheric file, a file that carries the optical depths for each layer in the atmosphere, and the best-fit flux values. Using these, we calculate the flux ratio, plot the best-fit spectrum and the PT-profile posteriors, and calculate and plot the contribution functions (see Figures \ref{fig:BARTChart} and \ref{fig:BARTpy}, right and bottom parts, respectively).

\subsection{Contribution Functions}
\label{sec:cf}

The contribution functions provide the information of how well the observer (telescope) sees a particular layer of the atmosphere, i.e., where the emission measured by the telescope originates \citep{Chamberlain1878-PlanAtmosp, Orton1977-CF, KnutsonEtal2009ApJ-redistribution, Griffith1998-CF, LeeEtal2012-CF}. 

The intensity measured at each wavelength comes from different atmospheric layers. To assess the contribution from a certain layer to the observed intensity, we calculate two quantities: the transmission weighting function and the Planck function at a given temperature. The weighting function, which describes how transmission is changing with altitude, is the kernel of the radiative-transfer integral. It 'weights' the contribution to the intensity from the Planck functions at different log-pressure altitudes. The contribution function, the product of the weighting function and the Plank function, is the integrand of the radiative-transfer integral.

The weighting function is defined as a partial derivative of the transmission function:
\begin{eqnarray}
\label{transm}
\mathcal{T} = e^{-\tau/\mu}\, ,
\end{eqnarray}
\begin{eqnarray}
\label{weight}
W = \frac{d\mathcal{T}}{d\,z}\, ,
\end{eqnarray}
\noindent where \math{z} is the altitude.

The altitude where the peak of the weighting function is found depends on the opacity at that wavelength. In other words, it is sensitive to the atmospheric thermal structure and composition (mixing fractions). The contribution function assesses vertical sensitivity of the emission spectrum by giving the pressure level at which thermal emission from the atmosphere contributes most to the intensity observed at the top of the atmosphere in each wavelength.

To investigate the contribution from a certain atmospheric layer to the observed intensity, we start with Equation (\ref{final}) that defines the intensity at the top of the atmosphere:
\begin{eqnarray}
I_\nu\,(\tau=0) & = & \int_{0}^{\tau}\frac{B_\nu}{\mu}\,e^{-\tau/\mu} d\tau\, , 
\label{final0}
\end{eqnarray}
\noindent and rewrite it as:
\begin{eqnarray}
I_\nu\,(\tau=0) & = &  \int_{0}^{\tau}\frac{B_\nu}{\mu}\,d(e^{-\tau/\mu})\,(-{\mu}) \, .
\label{final1}
\end{eqnarray}
\noindent Using the fact that:
\begin{eqnarray}
d(e^{-\tau/\mu}) = e^{-\tau/\mu}\, (-\frac{1}{\mu})\, d{\tau}\, ,
\label{der}
\end{eqnarray}
\noindent we get:
\begin{eqnarray}
I_\nu\,(\tau=0) =  \int_{\tau}^{0}{B_\nu}\,d(e^{-\tau/\mu})\, .
\label{final2}
\end{eqnarray}
\noindent The above equation can be rewritten in terms of altitude:
\begin{eqnarray}
I_\nu\,(z) =  \int_{0}^{\infty}{B_\nu}\,\frac{d(e^-\frac{\tau(z)}{\mu})}{d\,z}\, d\,z \,.
\label{final3}
\end{eqnarray}
\noindent By using Equation (\ref{transm}), Equation (\ref{final3}) simplifies to:
\begin{eqnarray}
I_\nu\,(z) =  \int_{0}^{\infty}{B_\nu}\,\frac{d\,\mathcal{T}}{d\,z}\, d\,z \,,
\label{final4}
\end{eqnarray}
\noindent i.e., 
\begin{eqnarray}
I_\nu\,(z) =  \int_{0}^{\infty}{B_\nu}\,W(z) \, d\,z \,,
\label{final5}
\end{eqnarray}
\noindent where \math{W(z)} is the weighting function. Since pressure is closely related to altitude, Equation (\ref{final2}) can be
rewritten in terms of pressure and the weighting and the contribution functions redefined accordingly. Starting with the hydrostatic balance equation:
\begin{eqnarray}
\frac{d\,p}{d\,z} = -g\, \rho\,,
\label{hydrobal}
\end{eqnarray}
\noindent where \math{g} is gravity and \math{\rho} is density, and the equation of state, \math{p = \rho\,R\,T}, Equation (\ref{hydrobal}) can be rewritten as:
\begin{eqnarray}
\frac{d\,p}{p} = -\frac{d\,z}{H} \,,
\label{hydrobal1}
\end{eqnarray}
\noindent where \math{H} is the scale height given as \math{H = R\,T/g} (\math{R} is the gas constant and \math{T} is the surface temperature). Using the fact that \math{log\,p} = \math{\frac{d\,p}{p}}, Equation (\ref{hydrobal1}) can be expressed as:
\begin{eqnarray}
d(logp) = -\frac{d\,z}{H} \,,
\label{hydrobal2}
\end{eqnarray}
\noindent and Equation (\ref{final4}), given in terms of altitude, can be rewritten in terms of pressure:
\begin{eqnarray}
I_\nu\,(p) =  \int_{p_{b}}^{p_{t}}{B_\nu}\,\frac{d\,\mathcal{T}}{d(log\,p)}\, d(log\,p) \,,
\label{final5}
\end{eqnarray}
\noindent where \math{p\sb{b}} and \math{p\sb{t}} are pressures at the top and the bottom of the atmosphere. The weighting function is now given as:
\begin{eqnarray}
\label{weightP}
W(p) = \frac{d\mathcal{T}}{d(log\,p)}\, ,
\end{eqnarray}
\noindent and the contribution function as:
\begin{eqnarray}
\label{CF}
CF(p) = {B_\nu}\,\frac{d\mathcal{T}}{d(log\,p)}\, ,
\end{eqnarray}
\noindent i.e.,
\begin{eqnarray}
\label{CF2}
CF(p) = {B_\nu}\,\frac{d(e^-\frac{\tau(p)}{\mu})}{d(log\,p)} \,.
\end{eqnarray}

Since the weighting function is the convolution of the rising transmission and falling density, it is bell-shaped (Gaussian-shaped). The weighting function will have a Gaussian shape for all wavelengths that do not sense the surface. When the atmosphere is more transparent, the peak of the weighting function moves towards larger pressures (lower altitudes). For wavelengths where the atmosphere is completely transparent, the weighting function is below the surface and the shape becomes exponential. Above the peak, the telescope does not sense the atmosphere that well due to the low atmospheric density and few emitting molecules. Below the peak, the emitting radiation is mostly absorbed by the atmosphere above.

The weighting and contribution functions are the key quantities in temperature retrievals; thus, they strongly depend on the best-fitting models. BART calculates the contribution functions after {\tt MC\sp{3}} has determined the best-fit parameters. Using them, we run the {\tt Transit} module again and reproduce the optical-depth array of the best-fit model. The optical depths are used to calculate contribution functions at each wavelength across the planet's spectrum. The band-averaged contribution functions are obtained by integrating the calculated contribution functions across the filter bandpasses of our observations (the transmission response functions) at every pressure layer.

\newpage
\section{ATMOSPHERIC ANALYSIS OF WASP-43\lowercase{b}}
\label{sec:analWASP43b}

We model the dayside atmosphere of WASP-43b using BART. In the following Section \ref{sec:WASP43b}, we give a comprehensive overview of the current studies done on WASP-43b and outline the atmospheric analysis done in this paper. In Section \ref{sec:data}, we list the data used in the analysis. In Section \ref{sec:retMod}, we present our modeling strategy. In Section \ref{sec:BARTsetup}, we describe the BART setup used in this analysis. In Sections \ref{sec:four} and \ref{sec:seven}, we present our results. In Section \ref{sec:CO}, we discuss constraints on the WASP-43b C/O ratio. In Section \ref{sec:cfWASP43b}, we analize WASP-43b contribution functions. In Section \ref{sec:diss}, we discuss our results, and in Section \ref{sec:conc}, we state our conclusions.

\subsection{WASP-43b Previous Studies}
\label{sec:WASP43b}


WASP-43b is one of the closest-orbiting hot Jupiters, rotating around one of the coldest stars that hosts hot Jupiters.  Attributed to the small radius of the host star, its cool temperature, and small semi-major axis, the system produces deep eclipses both in transit and occultation. This makes WASP-43b one of the most favorable objects today for space- and ground-based observations and a perfect target for atmospheric characterization.

WASP-43b was discovered by \citet{HellierEtal2011aaWASP43bdisc} using the WASP-South camera in conjunction with radial-velocity measurements from the Euler/CORALIE spectrograph. They determined that WASP-43 is a K7V main-sequence star with an effective temperature of 4400 {\pm} 200 K and a surface gravity of \math{4.65^{+0.06}_{-0.04}} (cgs). The star is active, has slightly above-solar metal abundances, and a rotational period of 15.6 {\pm} 0.4 days.  In 2013, \citet{czesla2013x} reported the detection of WASP-43b in the X-ray band using XMM-Newton. They derived WASP-43b's bolometric luminosity and estimated its mass-loss rate. Two years later, \citet{salz2015high} observed the planet again in the X-ray band, confirming that WASP-43b has one of the highest mass-loss rates among hot-Jupiters.

\citet{GillonEtal2012AA-WASP-43b} presented twenty-three transit light curves and seven occultation light curves with eight new radial-velocity measurements of the star. This large data set allowed them to significantly improve the parameters of the system and break the degeneracy of the stellar solution from \citet{HellierEtal2011aaWASP43bdisc}. They confirmed the planet's circular orbit, \math{e} = \math{0.0035^{+0.0060}_{-0.0025}}, and constrained stellar mass and radius to 0.717 {\pm} 0.025 \math{M\sb{\rm sun}} and 0.667 {\pm} 0.011 \math{R\sb{\rm sun}}, respectively. The planet's mass and radius are deduced to 2.034 {\pm} 0.052 \math{M\sb{\rm j}} and 1.036 {\pm} 0.019 \math{R\sb{\rm j}}, respectively. WASP-43b's high density of 2.41 {\pm} 0.08 \math{\rho\sb{sun}} favors an old age and a massive core. They detected the emission of the planet at 2.09 and 1.19 {\microns}. and modeled the planetary atmosphere using the methods described in \citet{FortneyEtal2005apjlhjmodels, Fortney2008} and \citet{FortneyEtal2007apjH2OVapor}. The results showed poor redistribution of the heat to the night side, favoring a model with no thermal inversion.

\citet{WangEtal2013-WASP43b} reported the ground-based detection of WASP-43b thermal emission in the H (1.6 {\micron}) and Ks (2.1 {\micron}) band observed using a WIRCam instrument on the Canada-France-Hawaii Telescope. They combined their observations with the narrow-band observations at 1.19 and 2.09 {\micron} from \citet{GillonEtal2012AA-WASP-43b}. The presence of thermal inversion could not be constrained by the data and the eclipse depths are consistent with a single black body curve at the temperature of \sim1850 K. 

\citet{BlecicEtal2014-WASP43b} presented the first observations of WASP-43b using the {\em Spitzer Space Telescope} at 3.6 and 4.5 {\micron}. The eclipse timings improved the orbital period by a factor of three compared to \citet{GillonEtal2012AA-WASP-43b}, P = 0.81347436 {\pm} 0.00000014 days, and put an upper limit on eccentricity, \math{e} = \math{0.010^{+0.010}_{-0.007}}. Combining their observations with the previous measurements from \citet{GillonEtal2012AA-WASP-43b} and \citet{WangEtal2013-WASP43b}, they confirmed a weak day-night redistribution, and ruled out a strong thermal inversion. The data were insufficient to break the degeneracy between the O-rich and C-rich, and the solar and super-solar metallicity models.

\citet{ChenEtal2014-WASP43b} observed one transit and one eclipse of WASP-43b simultaneously in the g', r', i', z', J, H, K bands using the GROND instrument on the MPG/ESO 2.2 m telescope at La Silla in Chile. The planetary dayside emission was detected in the K-band and marginally in the i' band. In their models, they investigated a Rayleigh scattering caused by reflective hazes.

\citet{LineEtal2013-Retrieval-II} performed a retrieval analysis using the eclipse depths from \citet{BlecicEtal2014-WASP43b}, \citet{GillonEtal2012AA-WASP-43b}, and \citet{WangEtal2013-WASP43b}. They ruled out a C to O ratio larger than one, confirmed the absence of a temperature inversion, and determined that the WASP-43b atmospheric composition is consistent with thermochemical equilibrium within 3\math{\sigma}.

\citet{ZhouEtal2014-WASP-43b} reported on the Ks-band secondary eclipse observation of WASP-43b using the IRIS2 infrared camera on the Anglo-Austrian Telescope.  Using published P-T profiles, they found that WASP-43b is marginally consistent with the carbon-rich composition, with a hazy layer at the top of the atmosphere.

\citet{StevensonEtal2014-WASP43b} reported the spectroscopic phase-curve observations of WASP-43b using the {\em Hubble Space Telescope}. They constructed a longitudinally-resolved brightness temperature map as a function of the optical depth. Using the same observations, \citet{KreidbergEtal2014-WASP43b} performed a precise measurement of the water abundance and determined the metallicity of WASP-43b to be 0.4 to 3.5 times solar, assuming the solar C/O ratio. With the same data, \citet{kataria2014atmospheric} explored the atmosphere of WASP-43b using a 3D atmospheric circulation model coupled with a non-gray radiative-transfer code \citep{FortneyEtal2006apjAtmDynamics, ShowmanEtal2009-3Dcirc}. They found that the 5\math{\times}solar model (without TiO/VO) is the best match to the data, and unlike the results from previous work, the inclusion of TiO and VO revealed a localized temperature inversions on the dayside at pressures below \sim100 mbar. \citet{benneke2015strict} presents the analysis of WASP-43b transmission spectra using the same HST/WFC3 observations. He finds the water abundance consistent with the solar composition and a C/O \math{<} 0.9 at the 3\math{\sigma} level.

In this paper, we performed a comprehensive atmospheric analysis of all WASP-43b secondary eclipse data from the space- and ground-based observations using BART. Our goals were to put additional constraints on the WASP-43b atmospheric model, to compare our results with the results from the literature, and to investigate cases previously unexplored on WASP-43b.

\begin{table}[ht!]
\caption{\label{table:eclDep} Eclipse Depths}
\atabon\strut\hfill\begin{tabular}{lcc}
    \hline
    \hline
    Source                                            & Wavelength      & Eclipse Depth               \\
                                                      & \micron        &     (\%)                     \\
    \hline
    \citet{GillonEtal2012AA-WASP-43b}, {\em VLT/HAWK I}   & 0.90            & 0.021 {\pm} 0.019             \\ 
                                                      & 1.19            & 0.079 {\pm} 0.032             \\ 
                                                      & 2.09            & 0.156 {\pm} 0.014             \\
    \hline
    \citet{WangEtal2013-WASP43b}, {\em WIRCam}       & 1.65            & 0.103 {\pm} 0.017             \\
                                                      & 2.19            & 0.194 {\pm} 0.029             \\
    \hline  
    \citet{ChenEtal2014-WASP43b}, {\em GROND}        & 0.806           & 0.037 {\pm} 0.022             \\
                                                      & 2.19            & 0.197 {\pm} 0.042             \\
    \hline
    \citet{BlecicEtal2014-WASP43b}, {\em Spitzer}    & 3.6             & 0.347 {\pm} 0.013             \\
                                                      & 4.5             & 0.382 {\pm} 0.015             \\
    \hline
    \citet{ZhouEtal2014-WASP-43b}, {\em IRIS2/AAT}   & 2.15            & 0.181 {\pm} 0.027             \\
    \hline
    \citet{StevensonEtal2014-WASP43b}, {\em HST}     & 1.1425          & 0.0365 {\pm} 0.0045           \\
                                                      & 1.1775          & 0.0431 {\pm} 0.0039           \\
                                                      & 1.2125          & 0.0414 {\pm} 0.0038           \\
                                                      & 1.2475          & 0.0482 {\pm} 0.0036           \\
                                                      & 1.2825          & 0.0460 {\pm} 0.0037           \\
                                                      & 1.3175          & 0.0473 {\pm} 0.0033           \\
                                                      & 1.3525          & 0.0353 {\pm} 0.0034           \\
                                                      & 1.3875          & 0.0313 {\pm} 0.0030           \\
                                                      & 1.4225          & 0.0320 {\pm} 0.0036           \\
                                                      & 1.4575          & 0.0394 {\pm} 0.0036           \\
                                                      & 1.4925          & 0.0439 {\pm} 0.0033           \\
                                                      & 1.5275          & 0.0458 {\pm} 0.0035           \\
                                                      & 1.5625          & 0.0595 {\pm} 0.0036           \\
                                                      & 1.5975          & 0.0614 {\pm} 0.0037           \\
                                                      & 1.6325          & 0.0732 {\pm} 0.0042           \\
    \hline
    \citet{StevensonEtal2015-newSpitzer}, {\em Spitzer} & 3.6             & 0.3300 {\pm} 0.0089           \\
                                                      & 4.5             & 0.3827 {\pm} 0.0084           \\
    \hline
\end{tabular}\hfill\strut\ataboff
\end{table}

\subsection{Data}
\label{sec:data}

Table \ref{table:eclDep} lists the source, the observing instrument and bandpass, the center wavelength, and the measured eclipse depth used in our analysis.

\subsection{Retrieved models}
\label{sec:retMod}

In an attempt to test the conclusions from recent studies \citep[e.g.,][]{hansen2014features, Swain2013-WASP12b} on how the inclusion of additional opacity sources influences the best-fit model, we generated multiple atmospheric cases and compared them using statistical factors (Table \ref{tab:cases}). We constructed four cases where we fit four major molecular species, H\sb{2}O, CO\sb{2}, CO, and CH\sb{4}, with a different number of opacity sources, and three cases where we fit seven molecular species, H\sb{2}O, CO\sb{2}, CO, CH\sb{4}, NH\sb{3}, HCN, and C\sb{2}H\sb{2}, with a different number of opacity sources. In addition, we generated our initial atmospheric model using the thermochemical equilibrium abundances profile and compared it to a commonly used vertically-uniform abundances profile \citep[e.g., ][]{MadhusudhanSeager2010, LineEtal2013-Retrieval-II}. We also generated and compared models with C/O ratios of 0.5, 0.7, and 1.

\begin{table}[h!]
\centering
\caption{\label{tab:cases} Atmospheric Cases}
\begin{tabular}{l@{\hspace{25pt}}l@{\hspace{25pt}}l@{\hspace{25pt}}l@{\hspace{7pt}}}
    \hline
    \hline
Case        & Fitted                    & Opacity                    & Initial    \\
            & Species                   & Sources                    & Abund. Profile      \\
    \hline
Case 1      & 4\tablenotemark{a}   & 4\tablenotemark{a}    & Uniform      \\
Case 2      & 4\tablenotemark{a}   & 4\tablenotemark{a}    & Equilibrium  \\
Case 3      & 4\tablenotemark{a}   & 7\tablenotemark{b}   & Equilibrium  \\
Case 4      & 4\tablenotemark{a}   & 11\tablenotemark{c}  & Equilibrium  \\
Case 5      & 7\tablenotemark{b}  & 7\tablenotemark{b}   & Uniform      \\
Case 6      & 7\tablenotemark{b}  & 7\tablenotemark{b}   & Equilibrium  \\
Case 7      & 7\tablenotemark{b}  & 11\tablenotemark{c}  & Equilibrium  \\
    \hline

\end{tabular}
\begin{minipage}[t]{0.58\linewidth}
\footnotesize
\sp{a}{H\sb{2}O, CO\sb{2}, CO, CH\sb{4}.\\}
\sp{b}{H\sb{2}O, CO\sb{2}, CO, CH\sb{4}, NH\sb{3}, HCN, C\sb{2}H\sb{2}.\\}
\sp{c}{H\sb{2}O, CO\sb{2}, CO, CH\sb{4}, NH\sb{3}, HCN, C\sb{2}H\sb{2}, C\sb{2}H\sb{4}, H\sb{2}S, TiO, VO.}
\end{minipage}
\end{table}

To assess different models quantitatively, we used three factors: the reduced \math{\chi\sp{2}}, \math{\chi\sp{2}\sb{\rm red}} = \math{\frac{\chi\sp{2}}{N -k}}, where \math{N} is the number of data points, and \math{k} is the number of free parameters; the Bayesian Information Criterion, BIC, \math{\chi\sp{2} + k\,ln(N)}; and standard deviation of the residuals, SDR. BIC allows us to compare goodness-of-fit for the models generated on the same dataset. Although in general, models with more free parameters improve the fit, BIC adds a penalty for any additional parameters in the system by  increasing its value. Lower BIC value indicate a better fit.

\subsection{BART Setup}
\label{sec:BARTsetup}

The pressure range for all of our models was constrained between 10\sp{2} and 10\sp{-5} bars, and sampled 100 times uniformly in log space. We used the parametrization from Section \ref{sec:Line} as our \math{T(p)} profile generator. We performed several trial runs including all five free parameters of the \math{T(p)} profile and concluded that we can fix \math{\gamma\sb{2}} and \math{\alpha} to zero; thus, the free parameters of the \math{T(p)} profile were: \math{\kappa\sb{\rm IR}}, \math{\gamma\sb{1}}, and \math{\beta}. In addition, we constrained the temperature range between 300 and 3000 K to prevent {\tt MC\sp{3}}'s random walk from stepping outside of the plausible temperature range (this range is bounded by the HITRAN/HITEMP databases' partition functions). The remaining free parameters were scaling factors of the species abundances fit in our models. 

For an equilibrium abundances profile case, we chose the temperature profile's initial parameters based on several trial runs, and calculated the mixing ratios using {\tt TEA}. For a vertically uniform abundances case, we used the mixing ratios typical for the expected temperature regimes of WASP-43b as our initial guess.  The mixing ratios of all input atomic species were taken from \citet{AsplundEtal2009-SunAbundances}.

When we wanted to directly compare our results with the results from \citet{LineEtal2013-Retrieval-II} and \citet{KreidbergEtal2014-WASP43b}, we included only He, H\sb{2}, CO, CO\sb{2}, CH\sb{4}, and H\sb{2}O in the atmospheric models. In our other runs, we included other expected and spectroscopically active species relevant for hot-Jupiter atmospheres: HCN, C\sb{2}H\sb{2}, C\sb{2}H\sb{4}, NH\sb{3}, HS, H\sb{2}S, TiO, and VO, in addition to the four major molecular species. We neglected the effect of clouds and scattering in all of our models.

We used the HITRAN database as our main source for the molecular line-list data (\href{https://www.cfa.harvard.edu/hitran}{https://www.cfa.harvard.edu/hitran}). For H\sb{2}O, CO, and CO\sb{2}, we used the molecular line-list data from the HITEMP database, \citet{rothman2010-hitemp}; for CH\sb{4}, NH\sb{3}, C\sb{2}H\sb{2}, and H\sb{2}S, from \citet{rothman2013-hitran2012}; for HCN and C\sb{2}H\sb{4}, from \citet{rothman2009-hitran2008}. The opacity source for HS, although potentially significant in hot-Jupiter planetary atmospheres \citep{Zahnle09-SulfurPhotoch}, could not be found in the HITRAN database, so they were not included in our analysis. The partition functions for the HITRAN opacity sources were calculated based on \citet{LaraiaEtal2011TIPS}. We also included the line-list data for TiO \citep{Schwenke1998TiO} and VO (B. Plez 1999, private communication). In addition to the molecular line lists, we included the H\sb{2}-H\sb{2} collision induced opacities from \citet{BorysowEtal2001-H2H2highT} and \citet{Borysow2002-H2H2lowT}, and H\sb{2}-He collision induced opacities from \citet{RichardEtal2012-HITRAN-CIA}.

\begin{figure*}[ht!]
\vspace{-20pt}
\centering
\includegraphics[width=.25\textwidth, height=3.5cm]{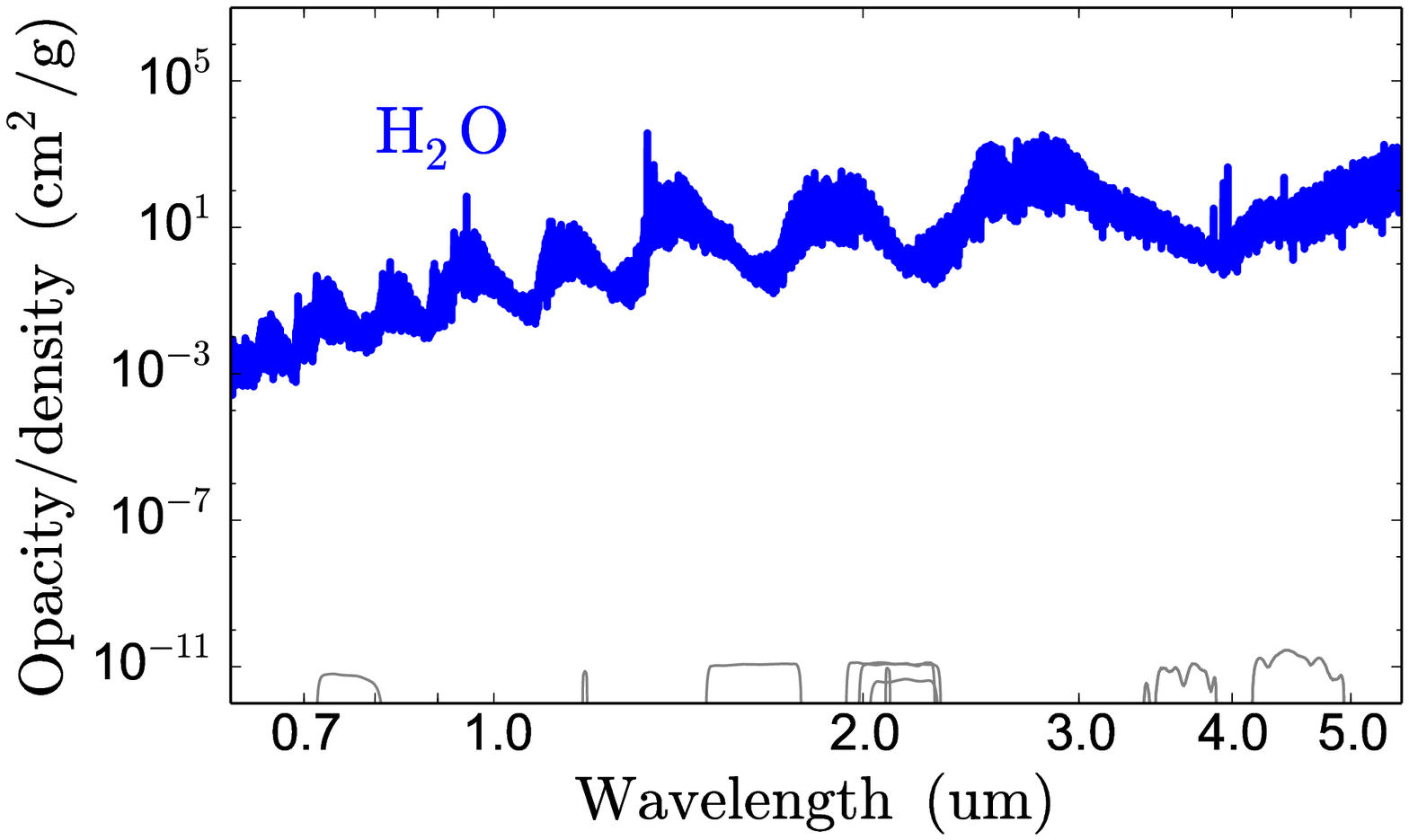}\hspace{-13pt}
\includegraphics[width=.25\textwidth, height=3.5cm]{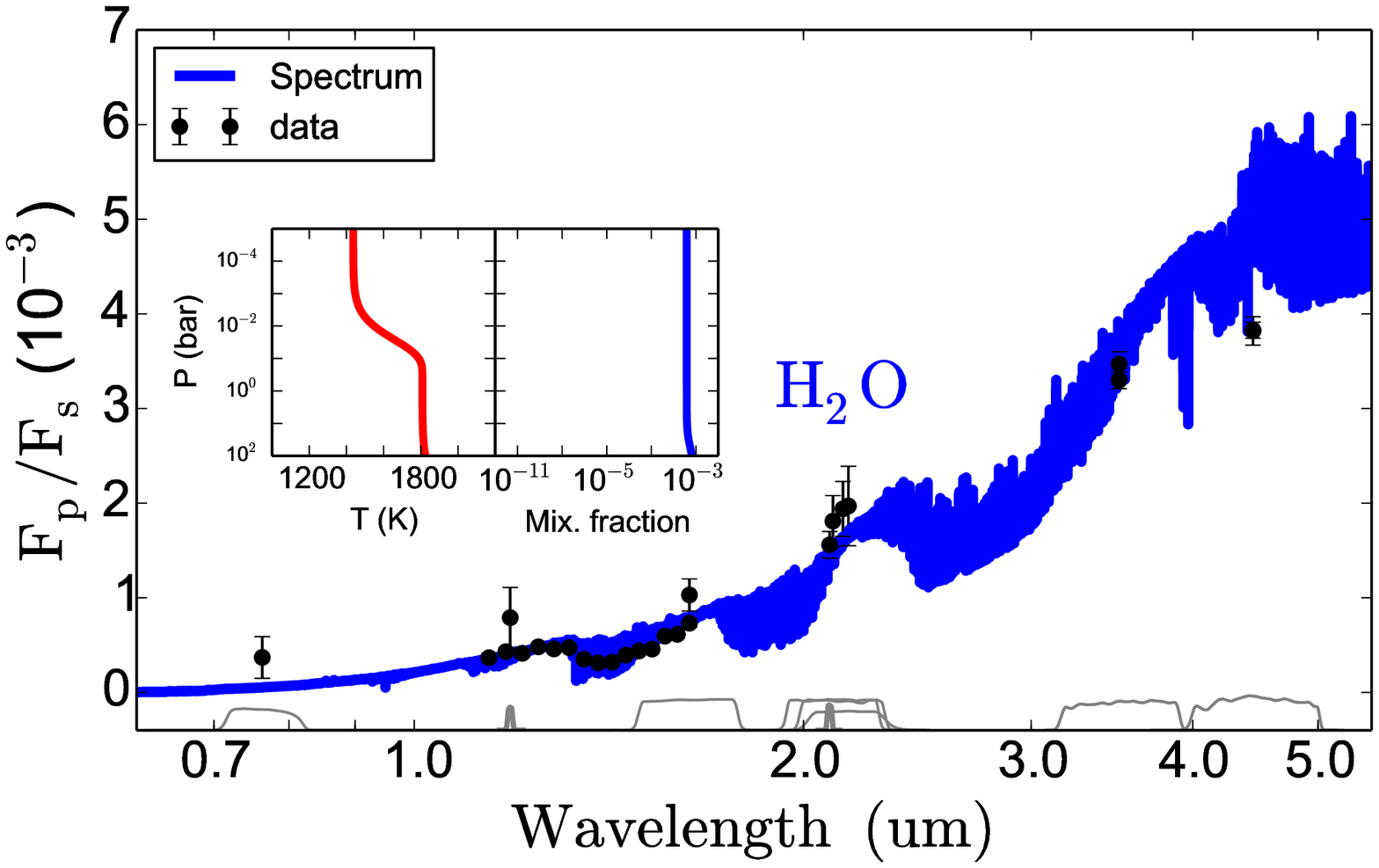}\hspace{-13pt}
\includegraphics[width=.25\textwidth, height=3.5cm]{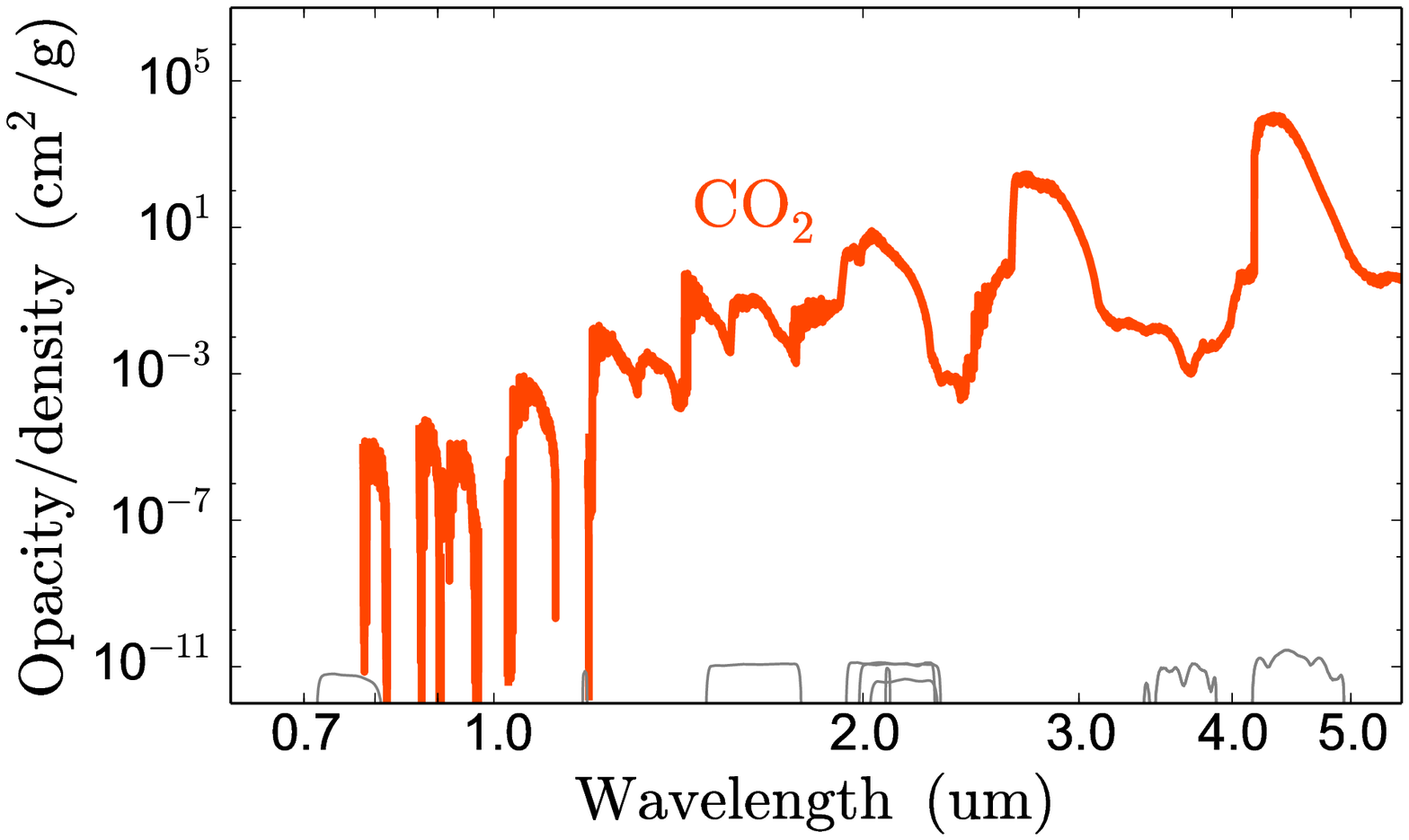}\hspace{-13pt}
\includegraphics[width=.25\textwidth, height=3.5cm]{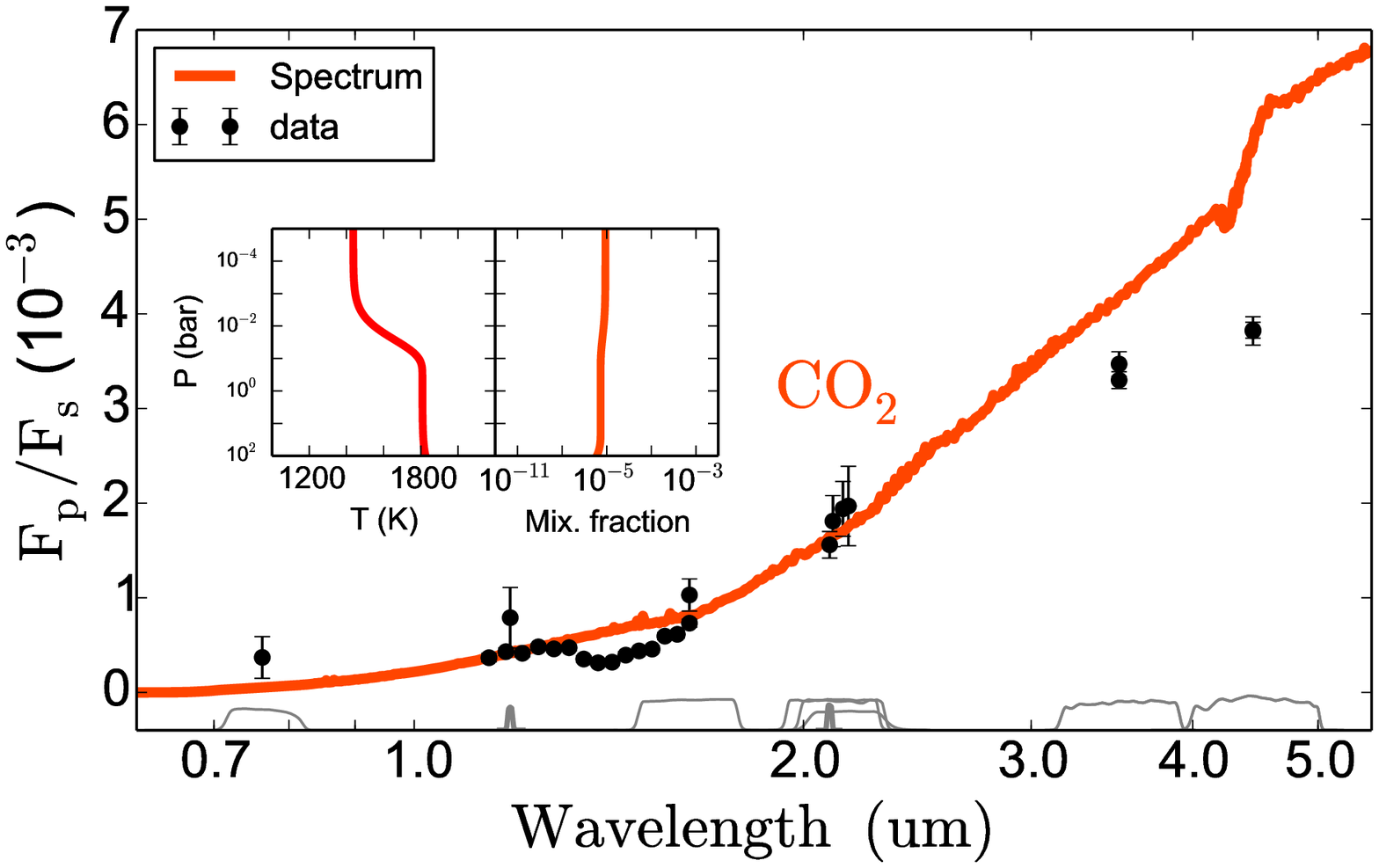}
\vspace{-2pt}

\includegraphics[width=.25\textwidth, height=3.5cm]{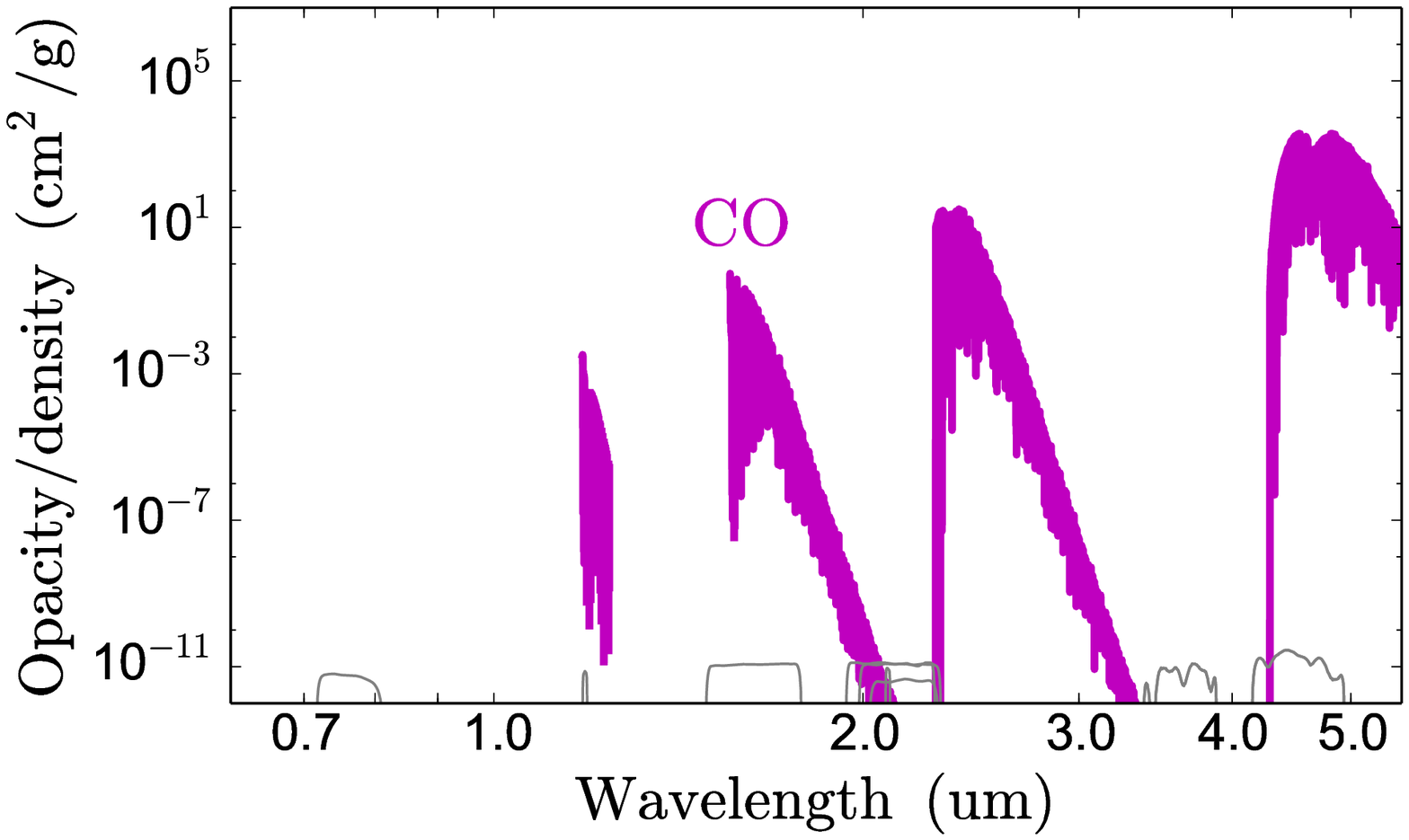}\hspace{-13pt}
\includegraphics[width=.25\textwidth, height=3.5cm]{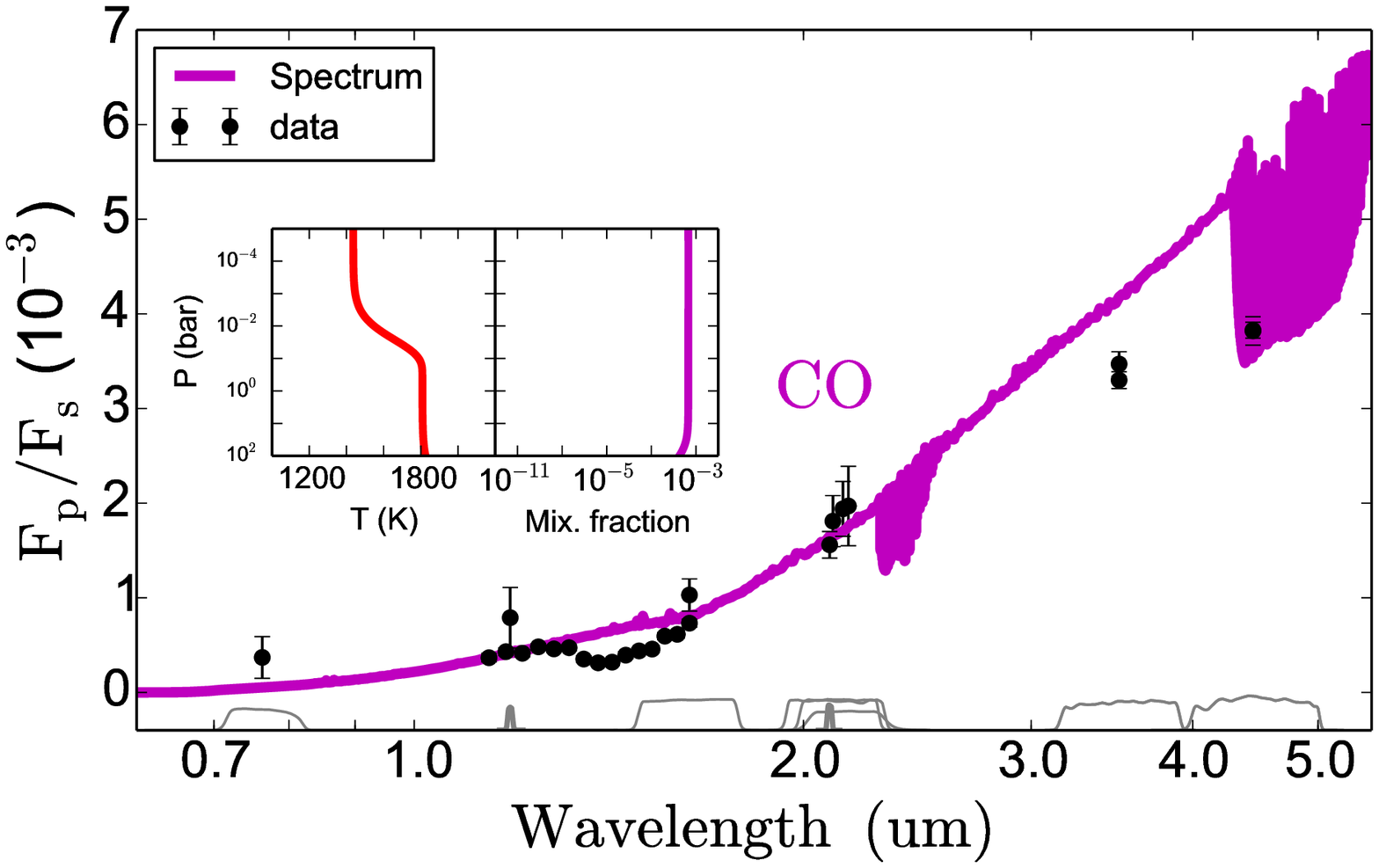}\hspace{-13pt}
\includegraphics[width=.25\textwidth, height=3.5cm]{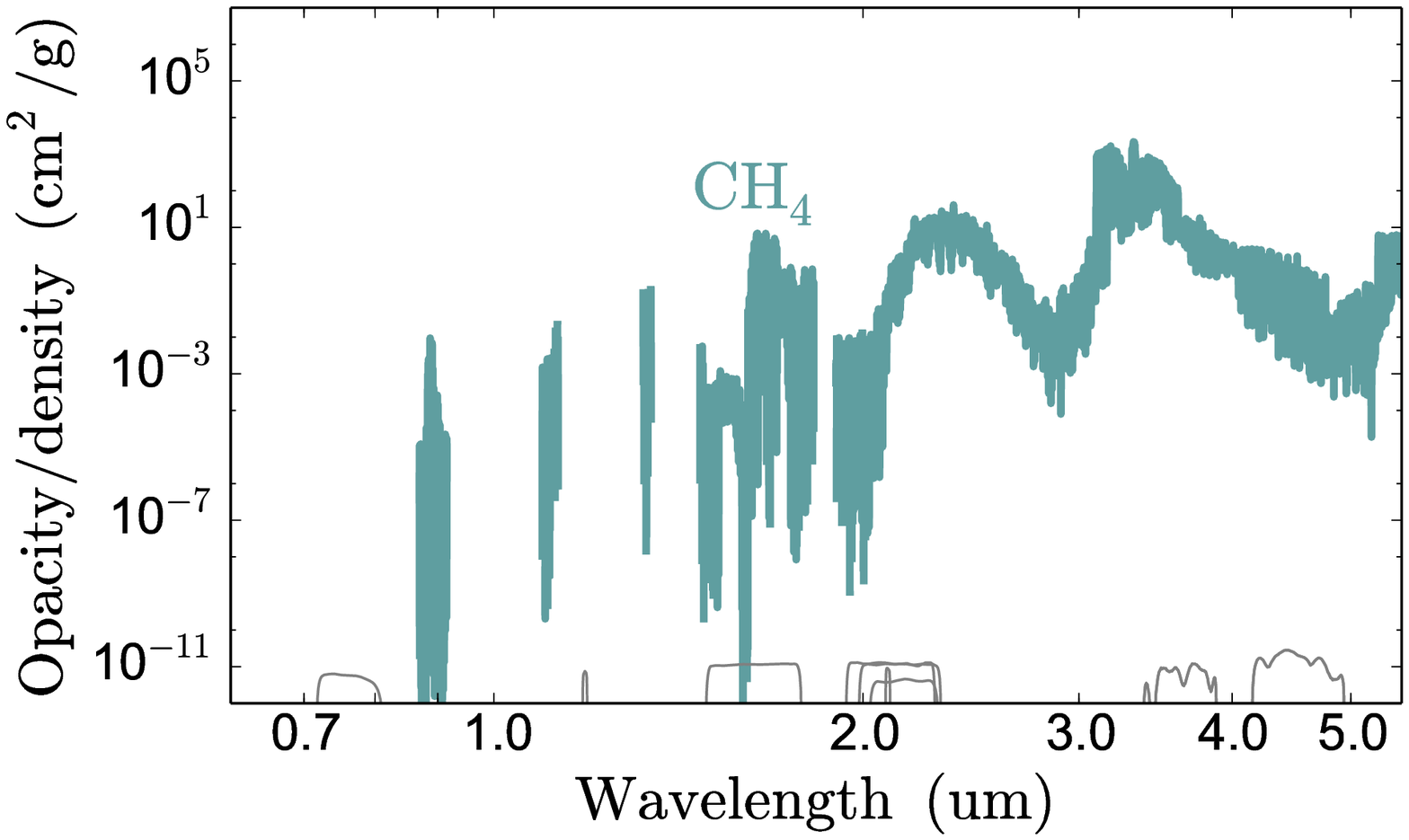}\hspace{-13pt}
\includegraphics[width=.25\textwidth, height=3.5cm]{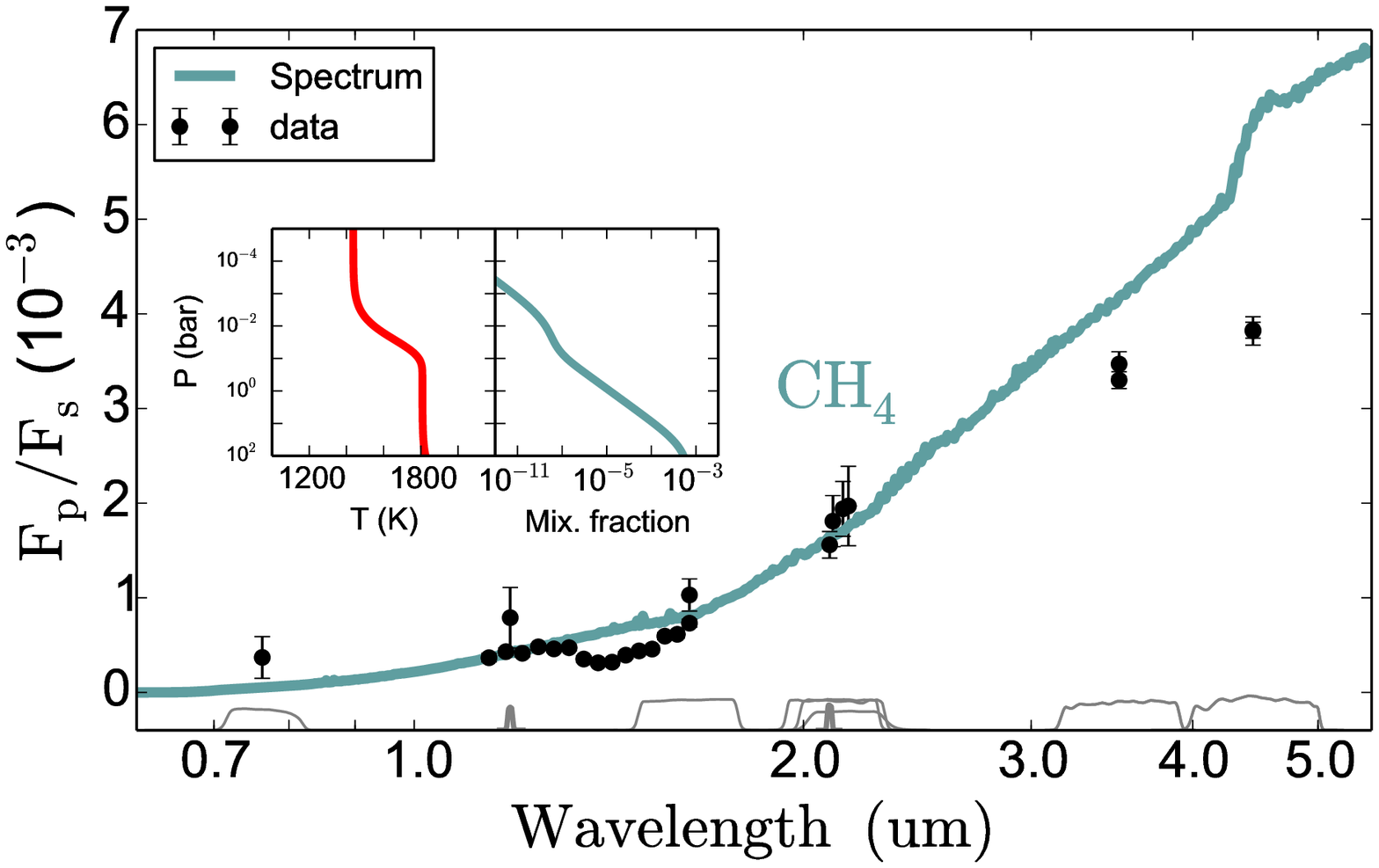}
\vspace{-2pt}

\includegraphics[width=.25\textwidth, height=3.5cm]{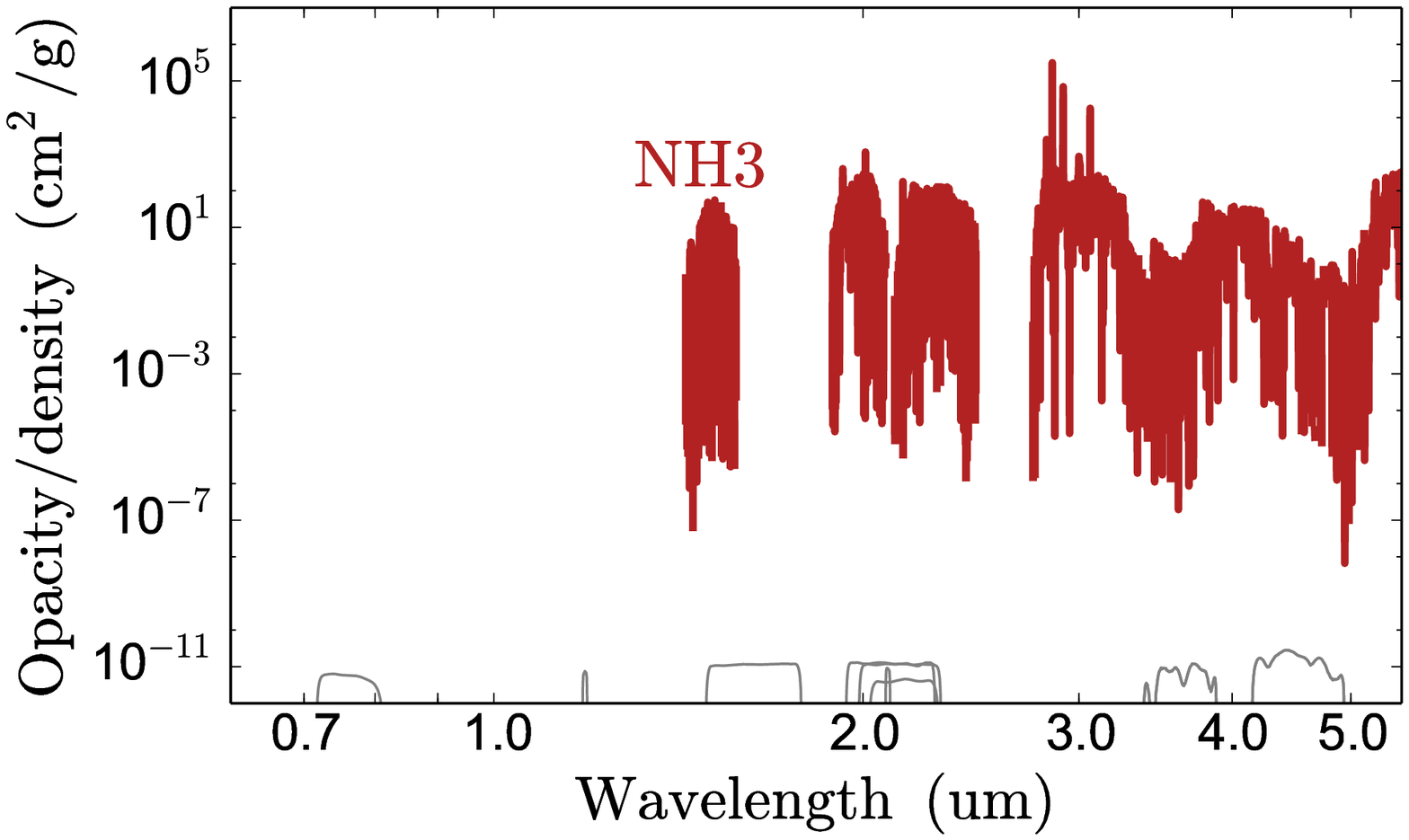}\hspace{-13pt}
\includegraphics[width=.25\textwidth, height=3.5cm]{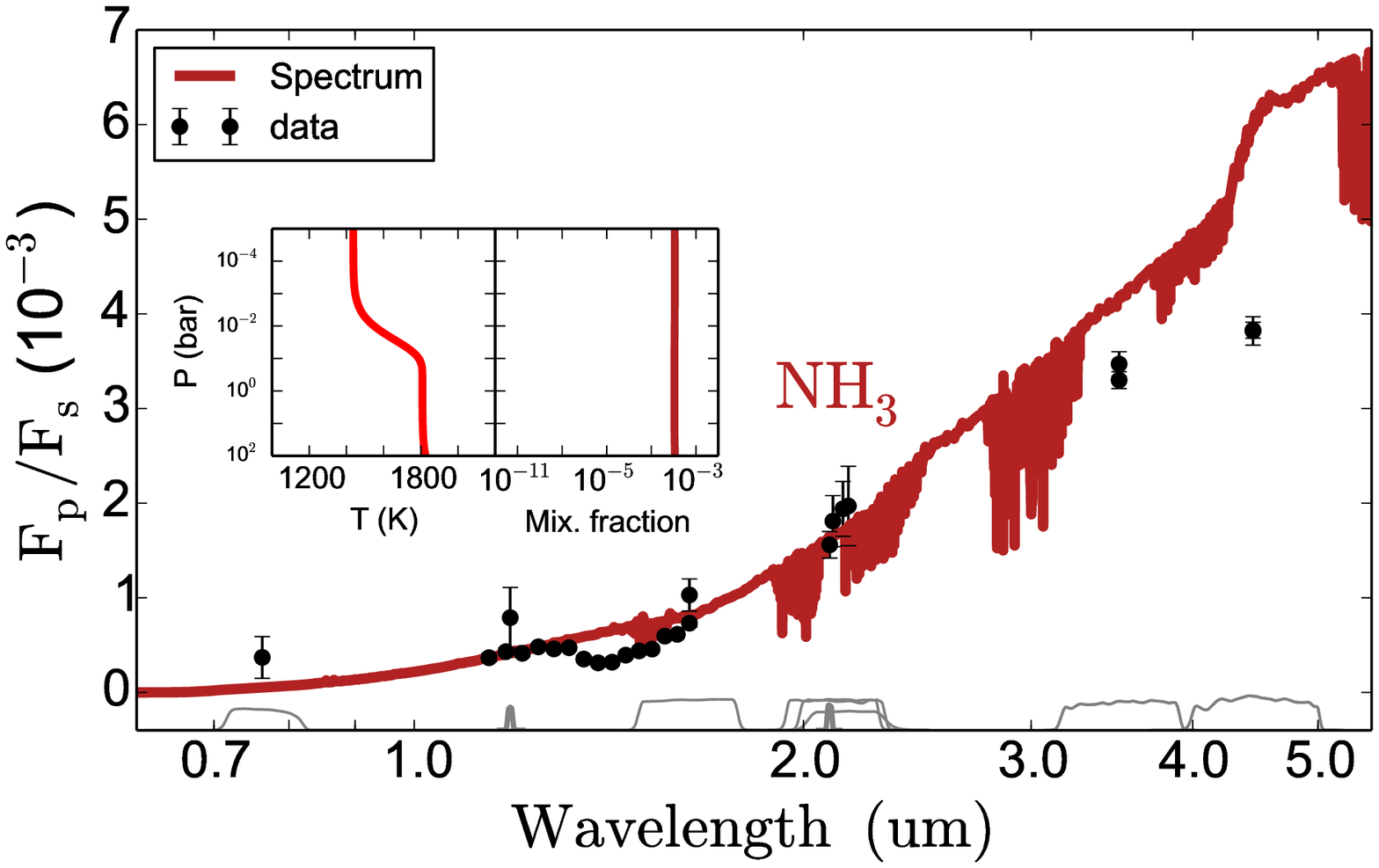}\hspace{-13pt}
\includegraphics[width=.25\textwidth, height=3.5cm]{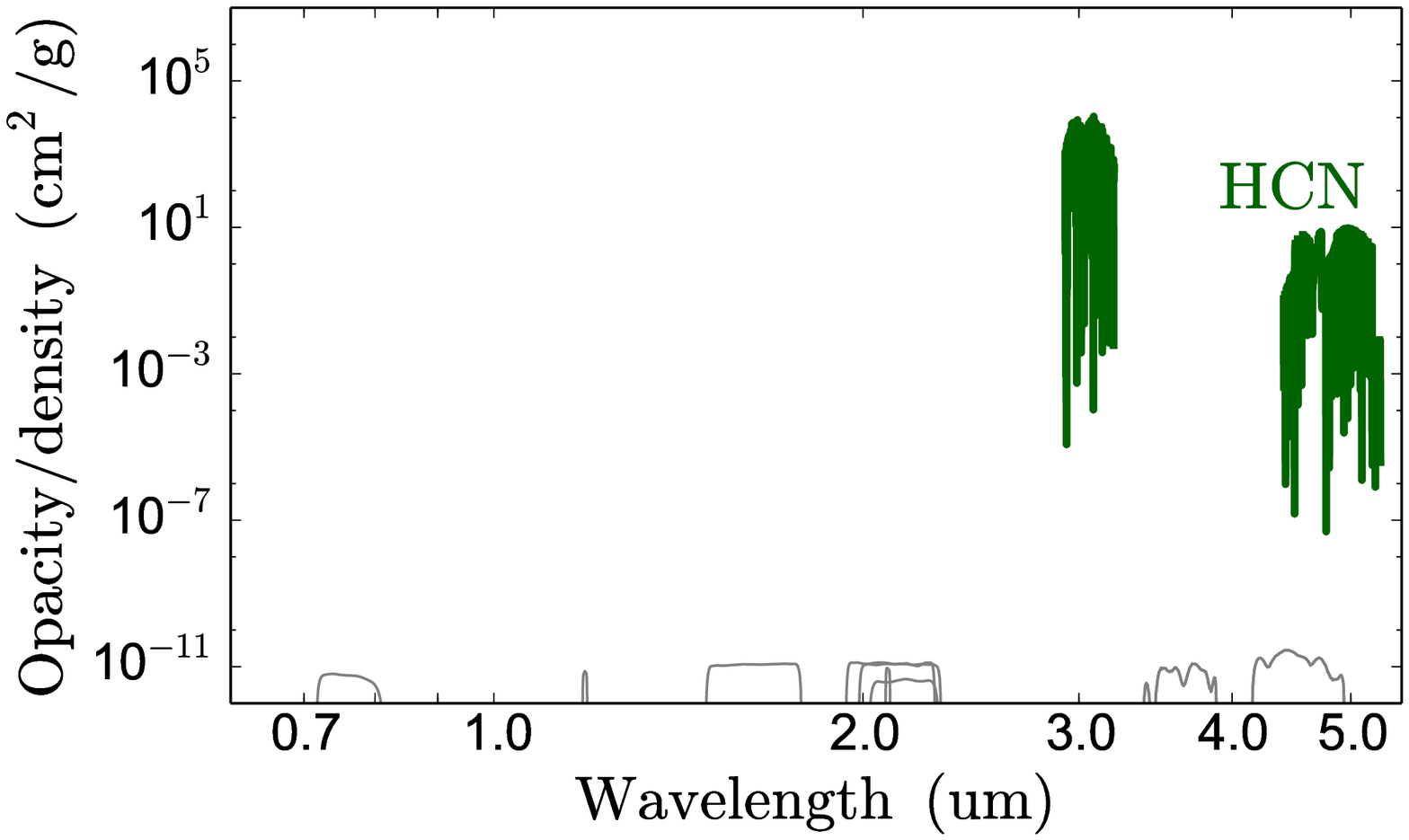}\hspace{-13pt}
\includegraphics[width=.25\textwidth, height=3.5cm]{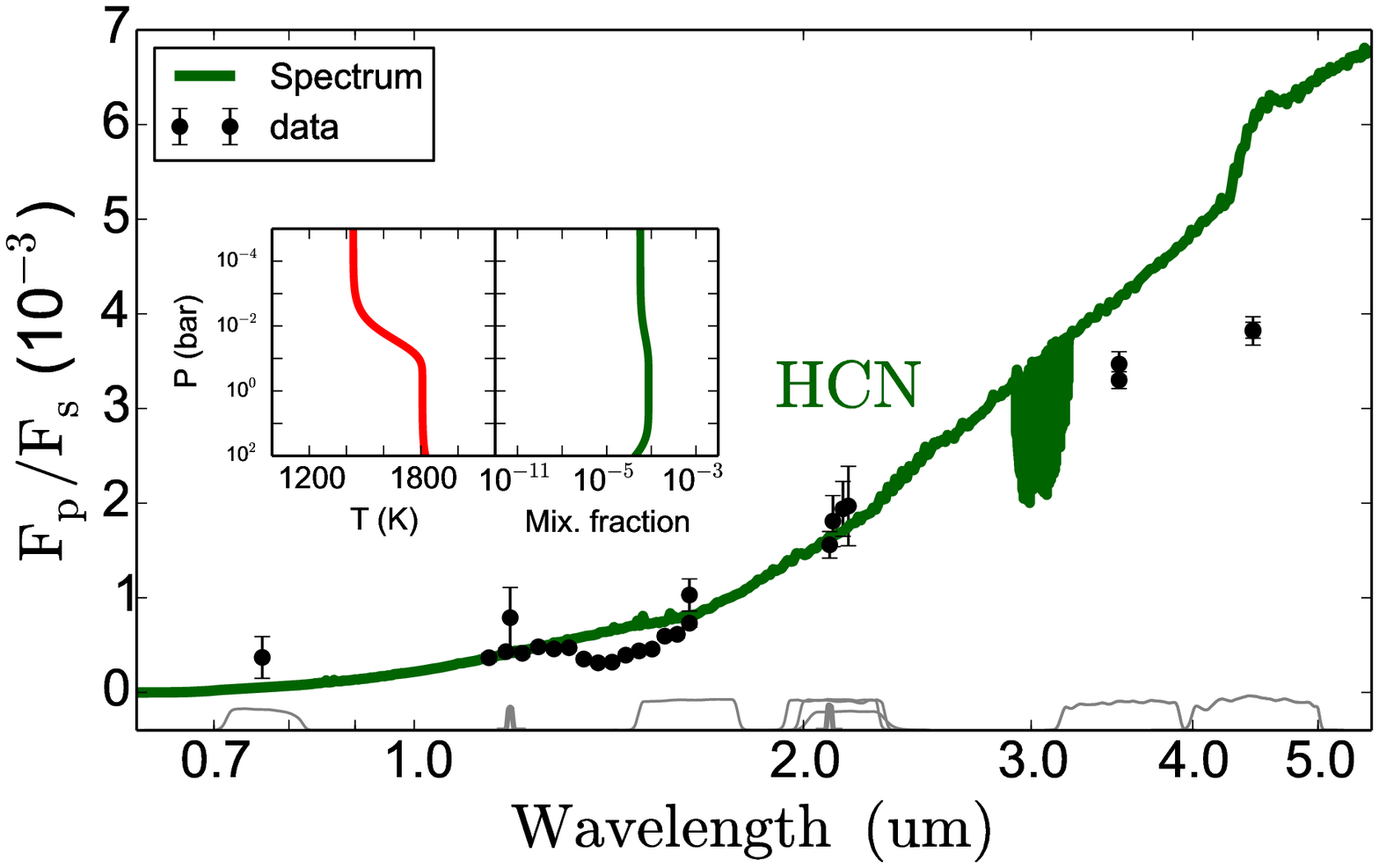}
\vspace{-2pt}

\includegraphics[width=.25\textwidth, height=3.5cm]{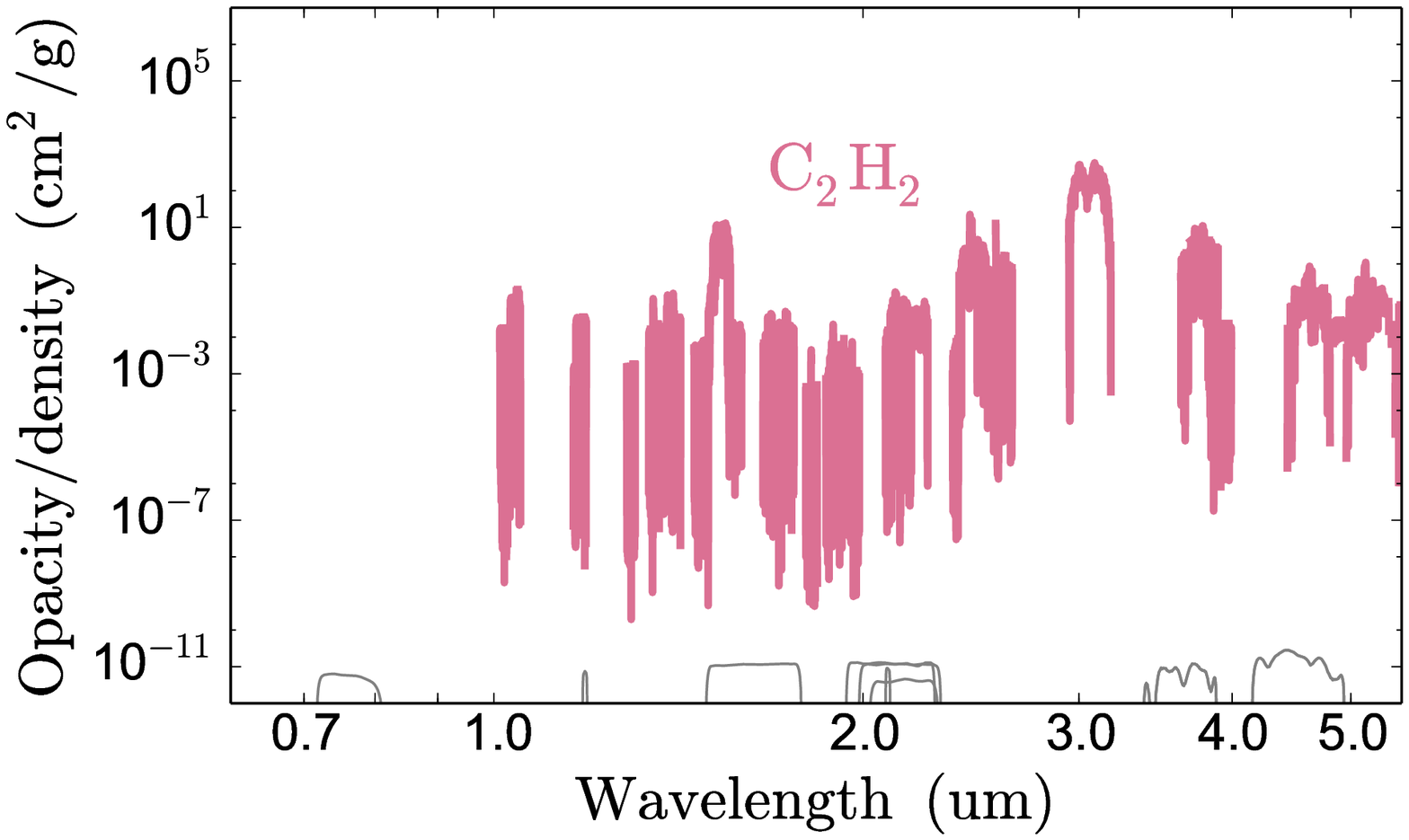}\hspace{-13pt}
\includegraphics[width=.25\textwidth, height=3.5cm]{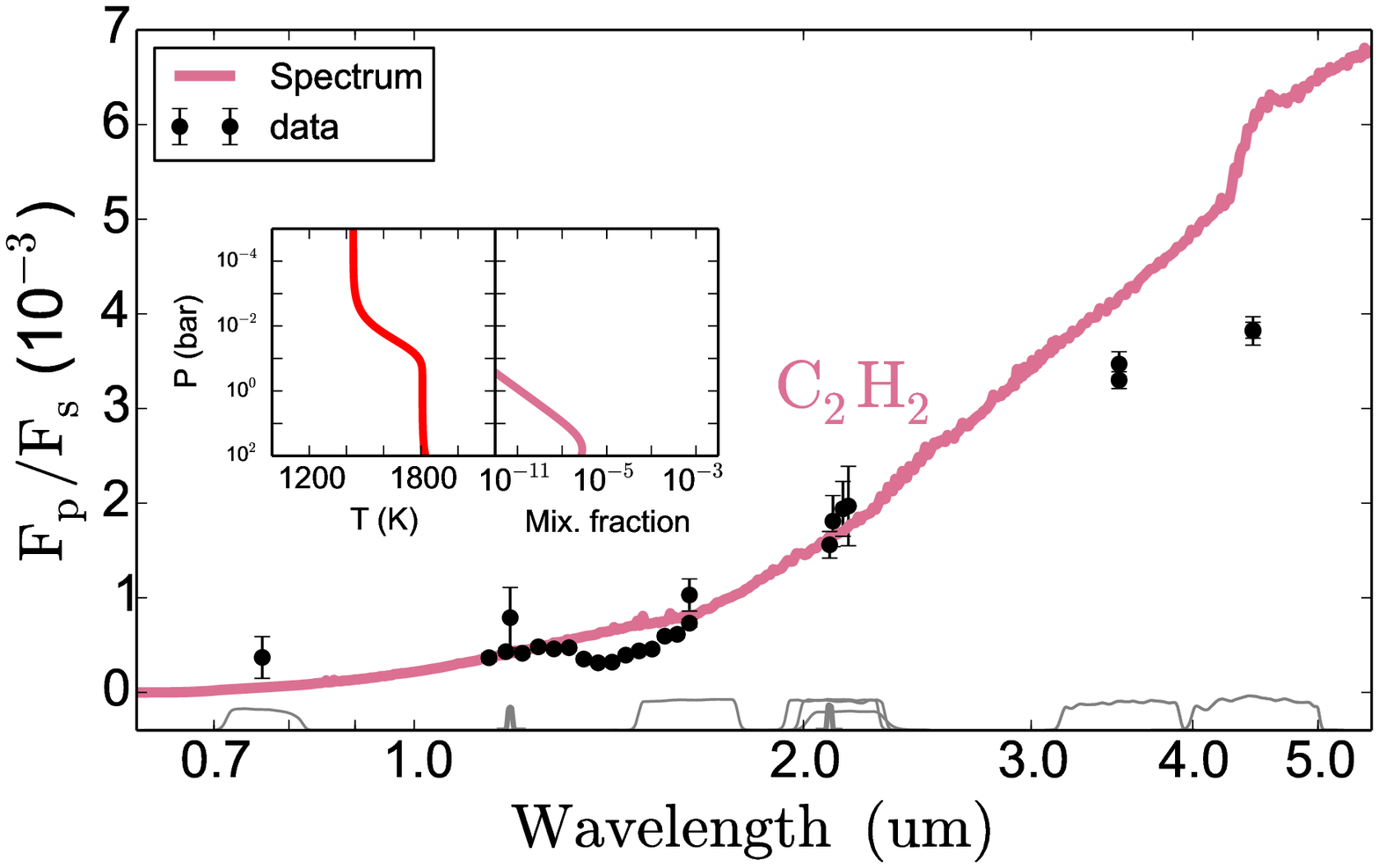}\hspace{-13pt}
\includegraphics[width=.25\textwidth, height=3.5cm]{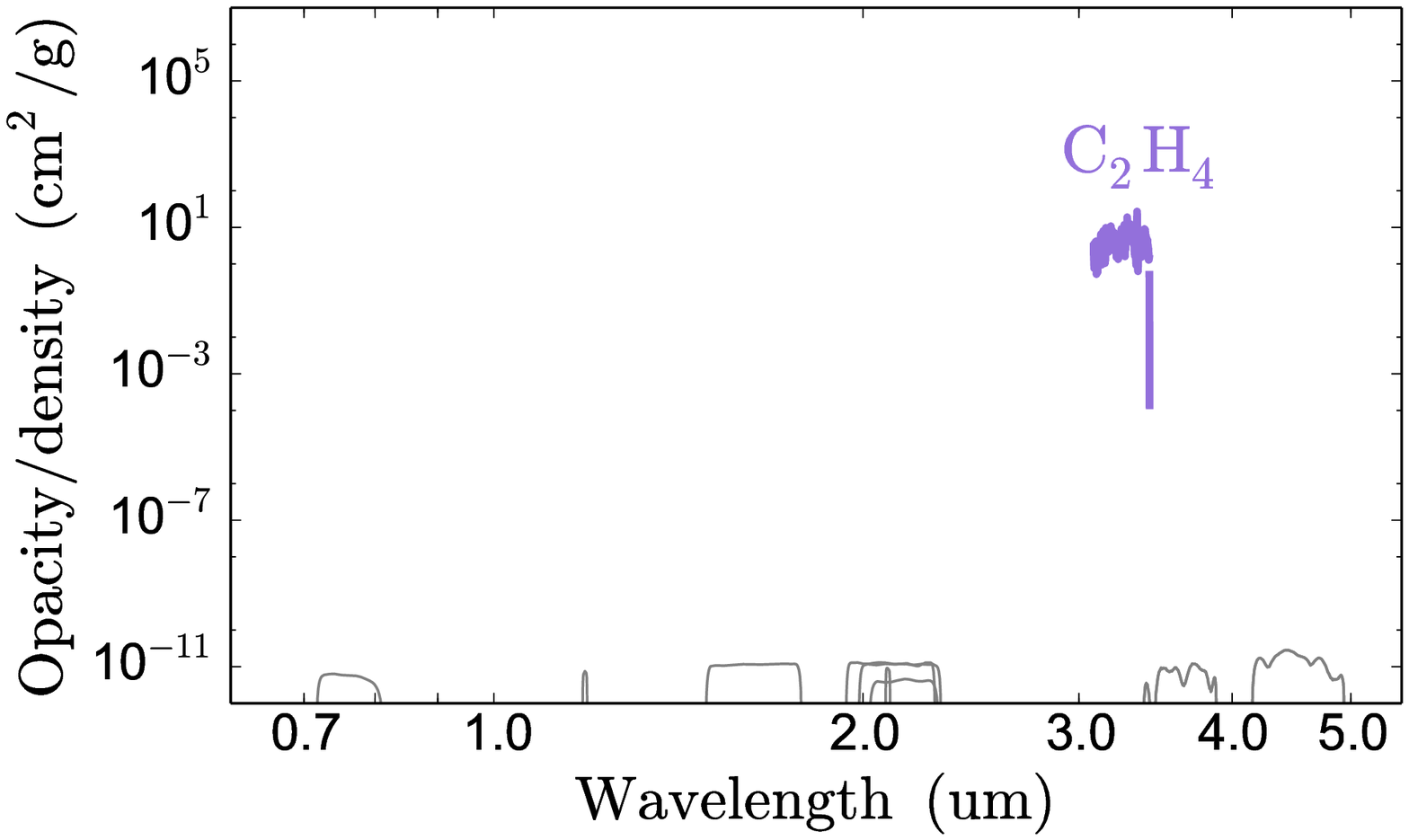}\hspace{-13pt}
\includegraphics[width=.25\textwidth, height=3.5cm]{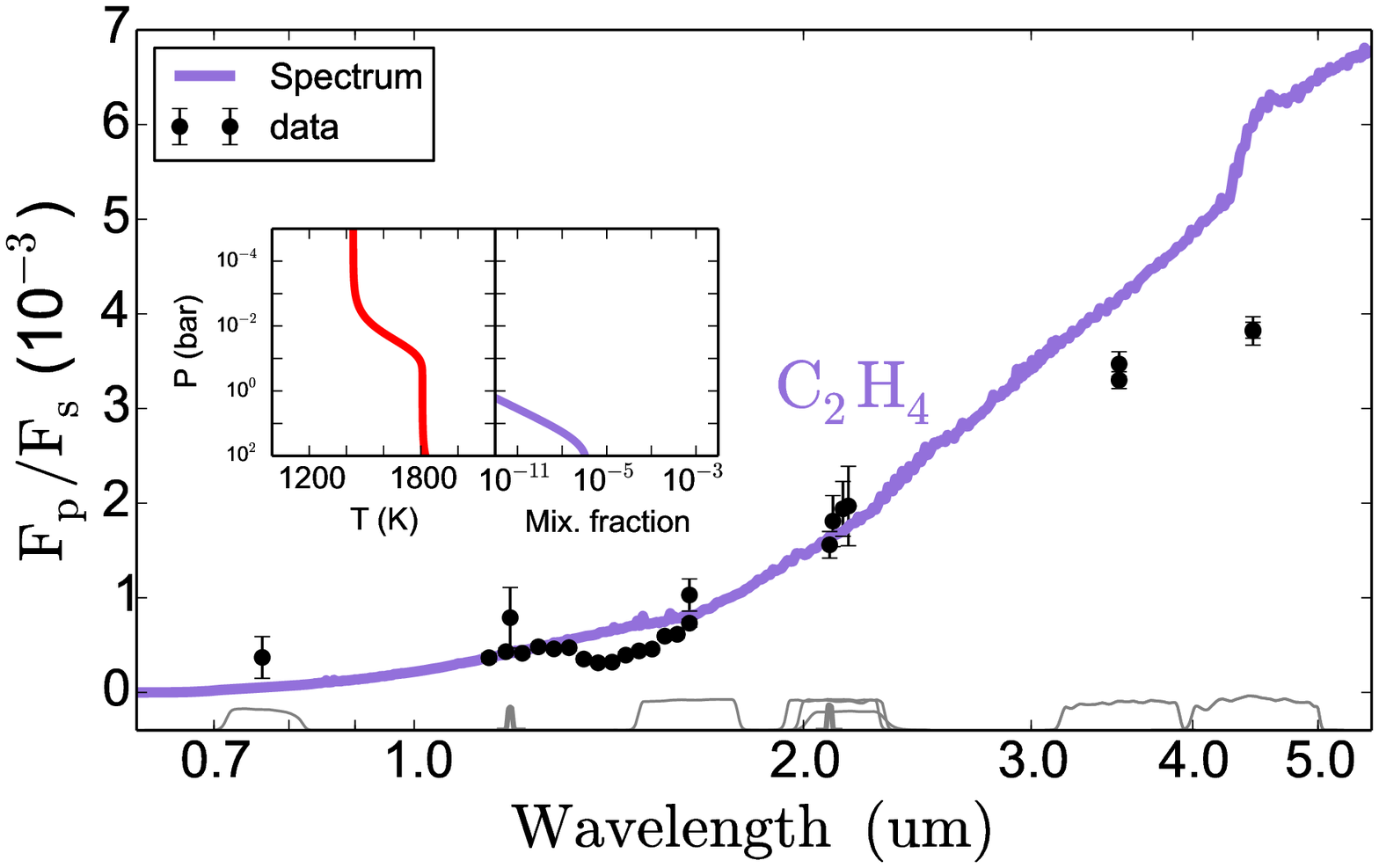}
\vspace{-2pt}

\includegraphics[width=.25\textwidth, height=3.5cm]{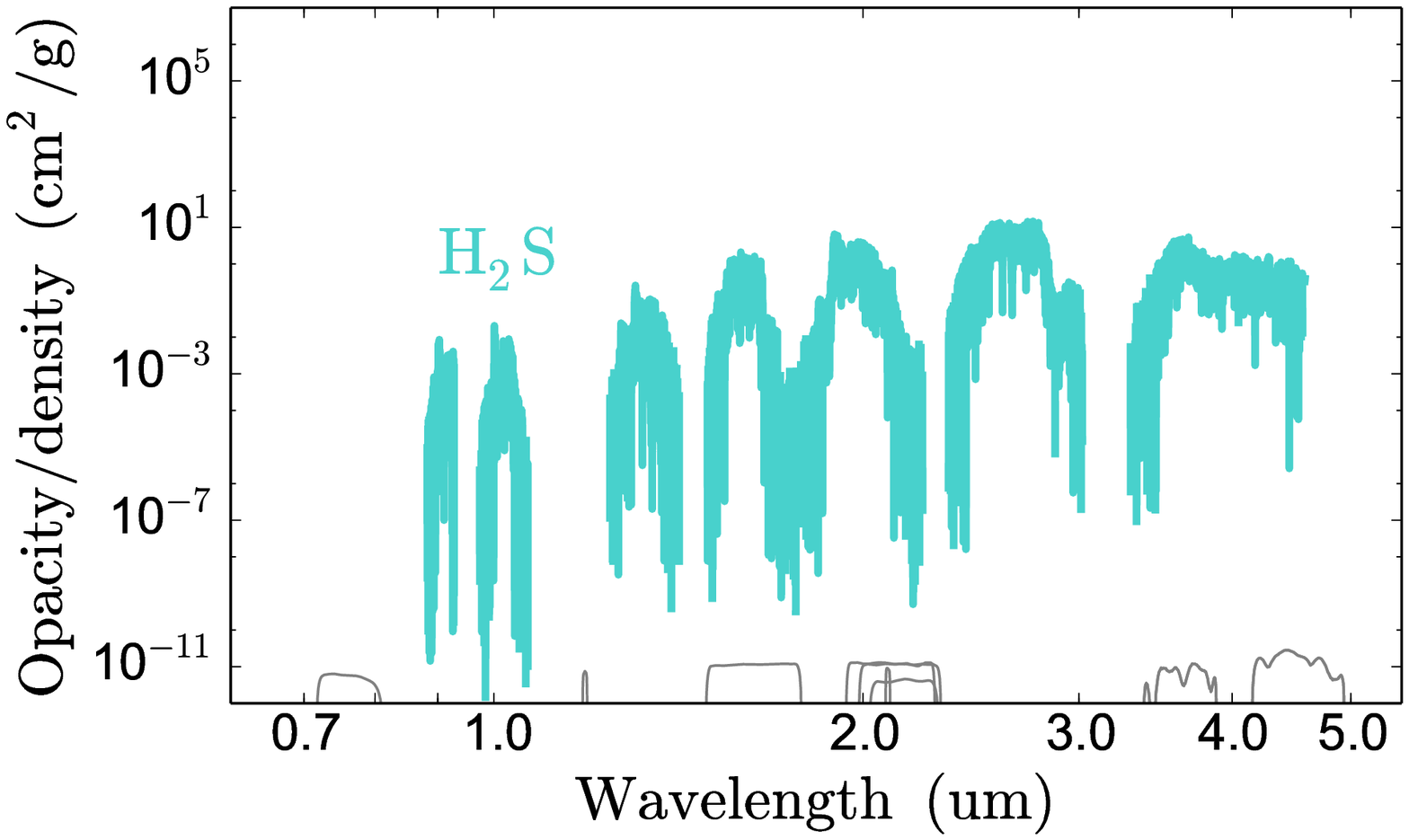}\hspace{-13pt}
\includegraphics[width=.25\textwidth, height=3.5cm]{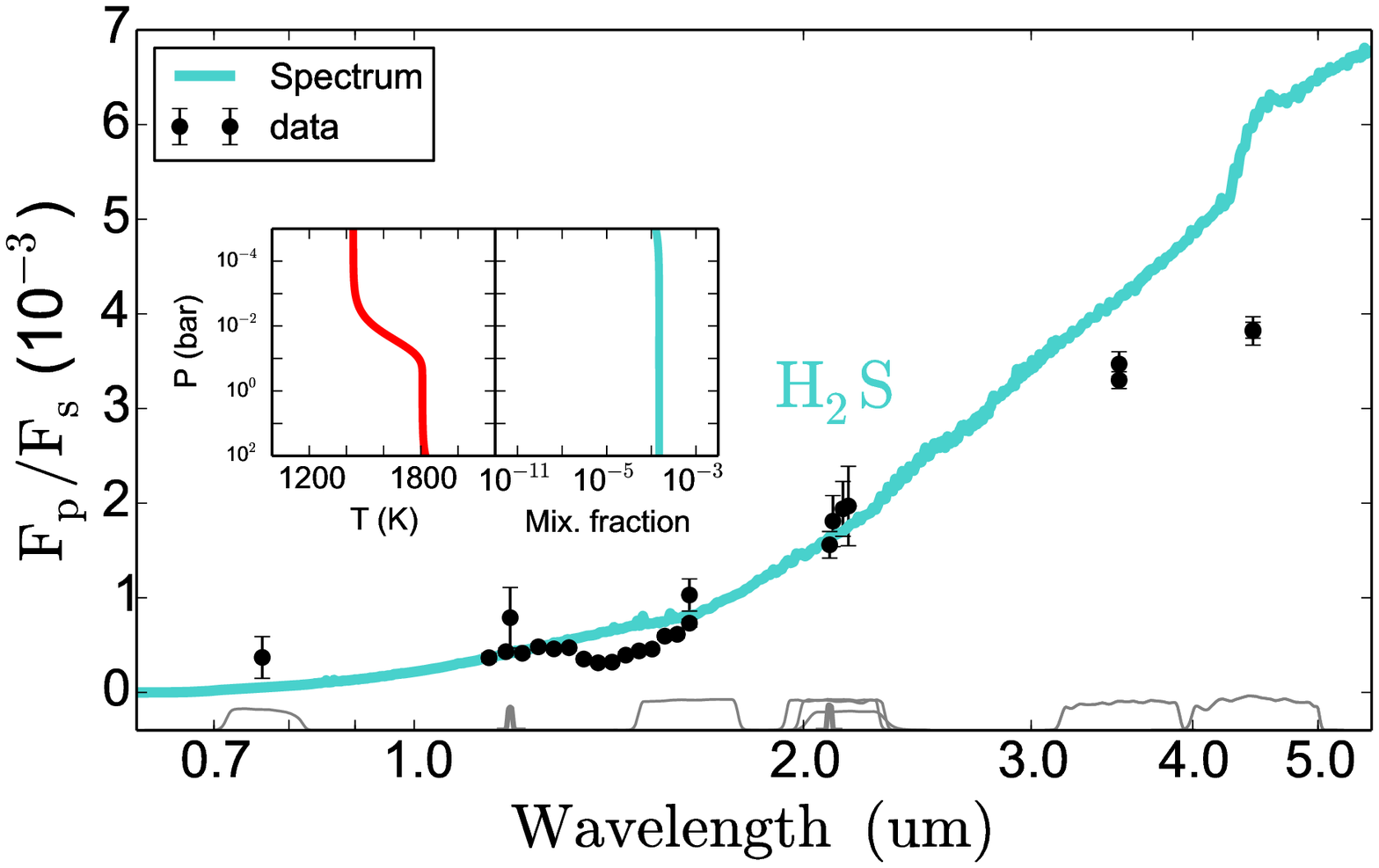}
\caption[Opacities of the HITRAN/HITEMP species and their influence on the equilibrium spectra of WASP-43b]{\scriptsize Opacities of the HITRAN/HITEMP species used in the retrieval at the temperature of 1500 K and pressure of 1 bar, and their influence on the equilibrium spectra of WASP-43b. The gaps in the opacity figures come from the limited \math{y}-axis range (the values not shown are well below 10\sp{-12} cm\sp{2}/g). The \math{T(p)} profile and species abundances used to generate WASP-43b spectra are given in the inset figures. Black dots represent the data points with uncertainties. At the bottom of the figures, we show the bandpasses for each of the observations used in the analysis. From the individual equilibrium WASP-43b models, we see that H\sb{2}O, CO, and NH\sb{3} are the only absorbers that significantly influence the WASP-43b emission spectra in the wavelength range of the available observations, with H\sb{2}O being the most apparent.}
\label{fig:opacs1}
\end{figure*}

\begin{figure*}[ht!]
\centering
\includegraphics[width=.26\textwidth, height=3.5cm]{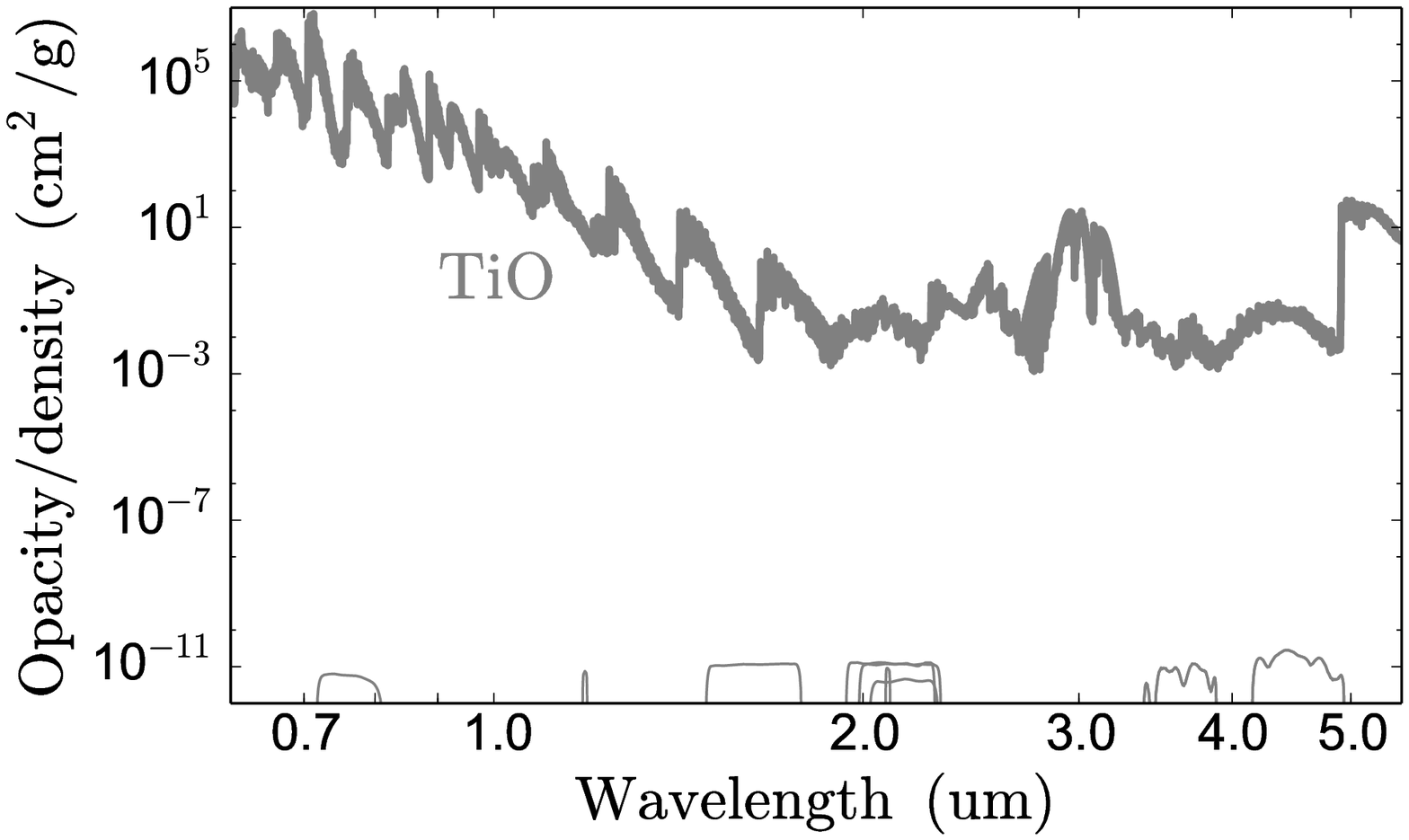}\hspace{-13pt}
\includegraphics[width=.26\textwidth, height=3.5cm]{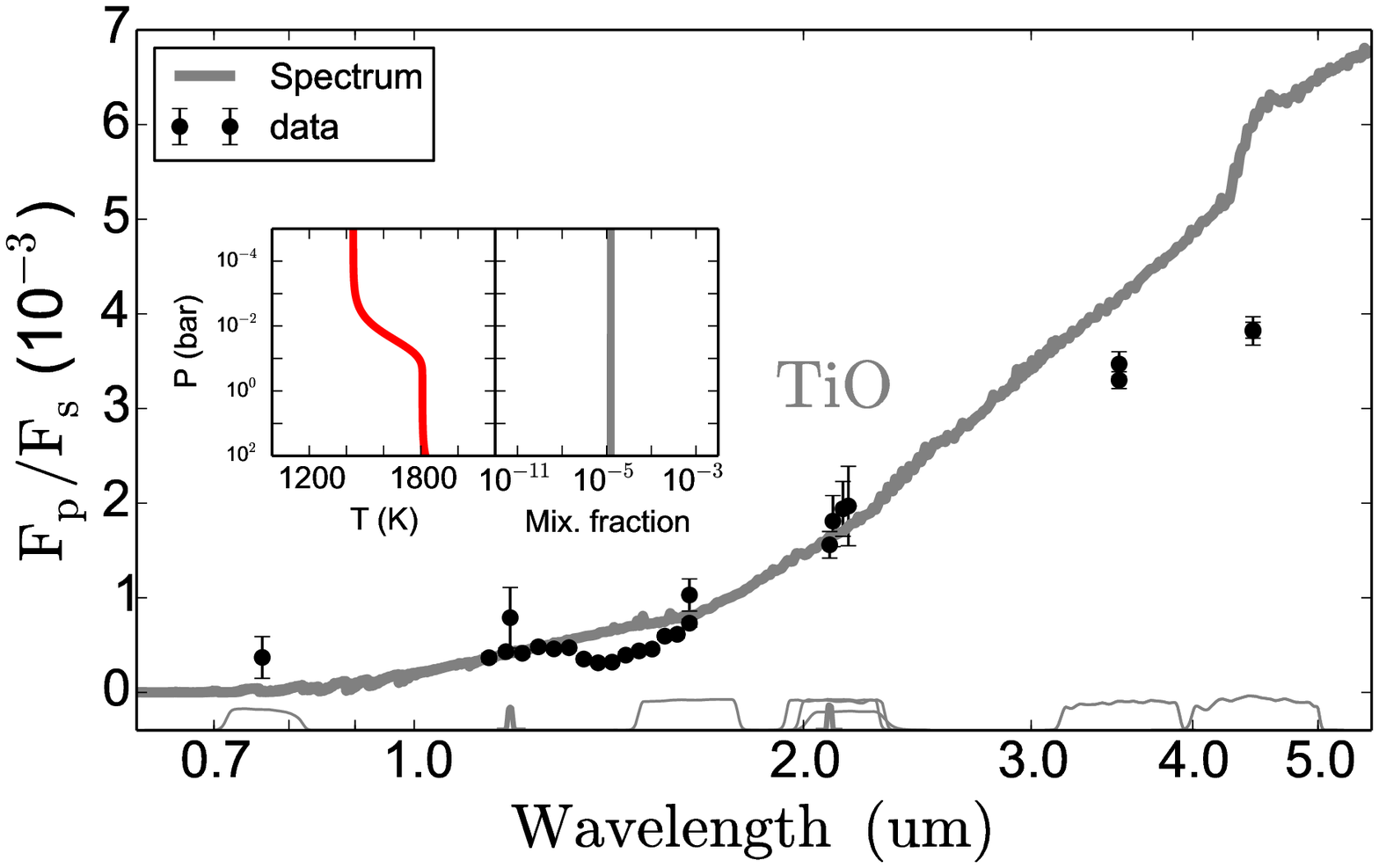}\hspace{-13pt}
\vspace{-2pt}
\includegraphics[width=.26\textwidth, height=3.5cm]{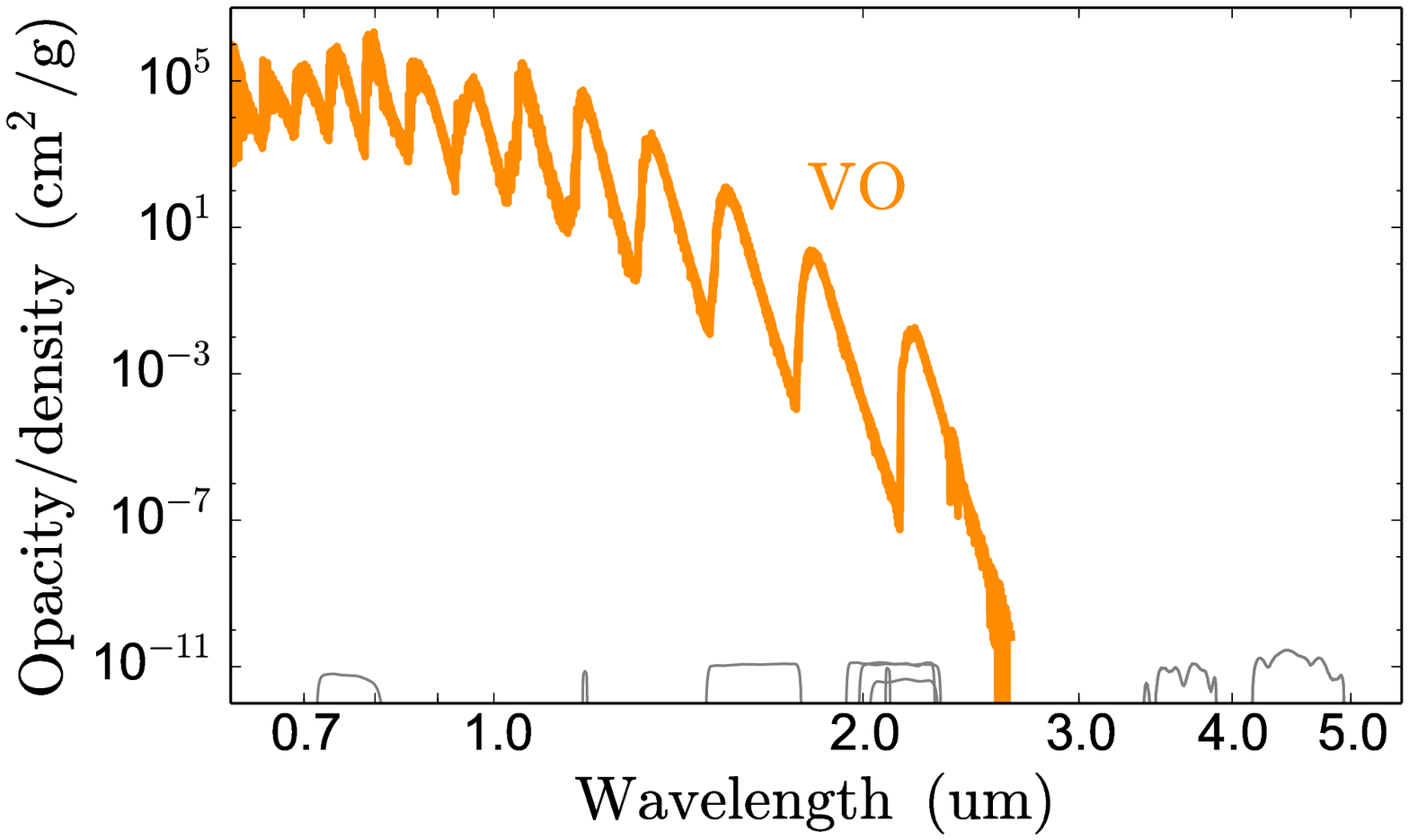}\hspace{-13pt}
\includegraphics[width=.26\textwidth, height=3.5cm]{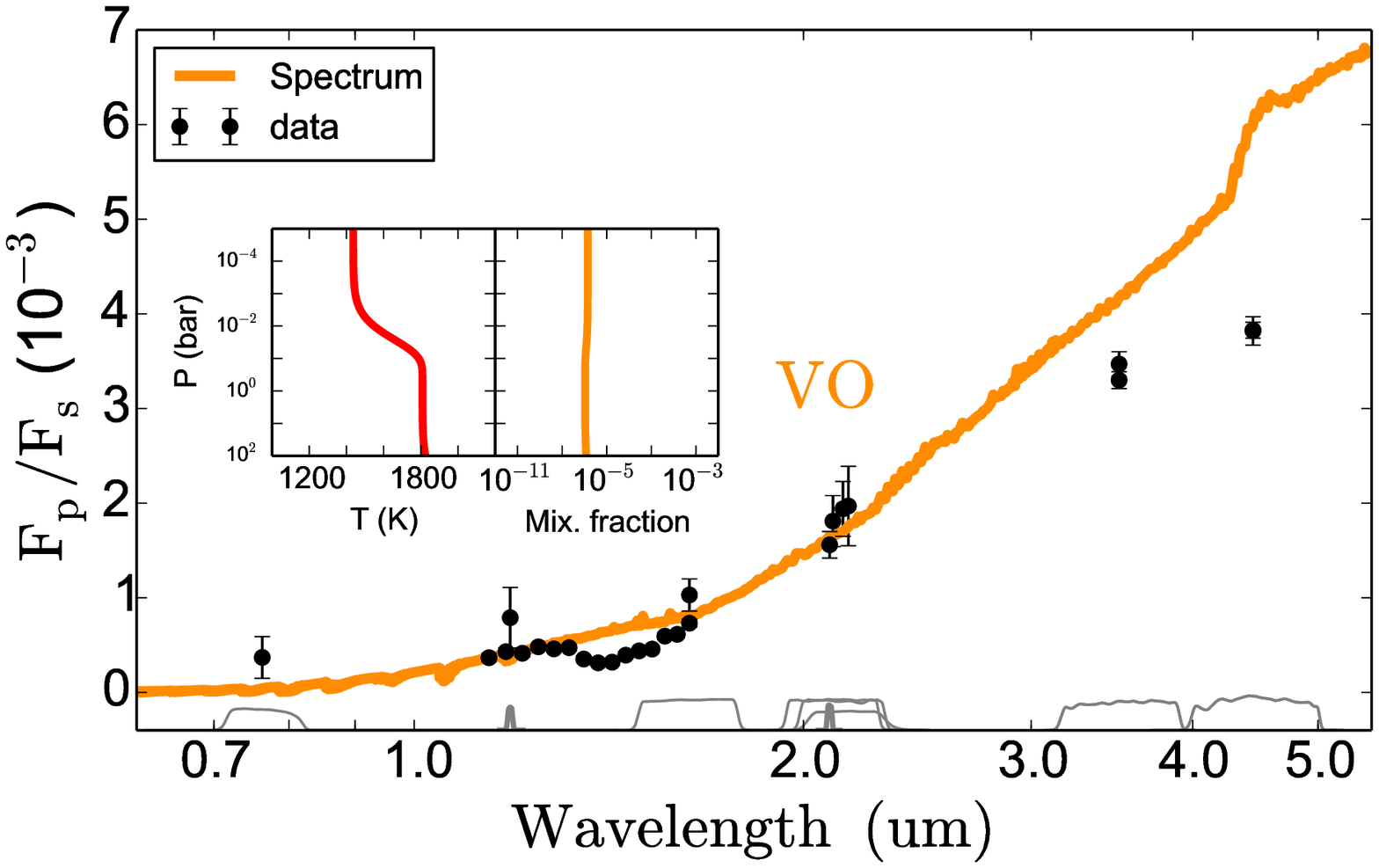}
\vspace{10pt}
\caption[Opacities of the TiO and VO species and their influence on the equilibrium spectra of WASP-43b]{\scriptsize Opacities of the TiO and VO species used in the retrieval at the temperature of 1500 K and pressure of 1 bar, and their influence on the equilibrium spectra of WASP-43b. The TiO line list comes from \citet{Schwenke1998TiO} and the VO line list from B. Plez (1999, private communication). }
\vspace{10pt}
\label{fig:opacs2}
\end{figure*}

Figures \ref{fig:opacs1} and \ref{fig:opacs2} show opacities for each of our species of interest for the wavelength range used in our analysis (0.6 - 5.5 {\micron}), temperature of 1500 K and pressure of 1 bar, as well as the effect of the opacity of an individual species on the equilibrium spectrum of WASP-43b. In addition, we plotted the filter bandpasses to show what spectral features are covered in the wavelength range of our observation. The atmospheric model used for this figure contains eleven molecular species. We used {\tt TEA} to calculate the species' thermochemical equilibrium mixing fractions for the \math{T(p)} profile given in the inset figure (the same for all the species).

We used flat priors on all parameters, with boundary limits set somewhat arbitrarily to allow {\tt MC\sp{3}} to explore the parameter phase space thoroughly. We imposed only one constraint: that the sum of fitted molecular species abundances must not exceed 15\%. This constraint forbids a random walk process, and guarantees physically plausible results. In addition, on each iteration, we rescaled the mixing ratios of H\sb{2} and He, preserving their original ratio such that the total sum of all species is unity.

We ran ten independent chains and enough iterations until the Gelman and Rubin convergence test for all free parameters dropped below 1\% \citep{GelmanRubin1992}. 

Our filter files, the transmission response functions, for VLT/HAWK I, GROND, and WIRCam observations were provided by \citet{ChenEtal2014-WASP43b}. The response functions of IRIS2/AAT observations were provided by \citet{ZhouEtal2014-WASP-43b}. The {\em Spitzer} response functions for the channel 1 and 2 subarray observations were found on the {\em Spitzer} website. For each of the HST observations, we used the top-hat response functions.

The system parameters (planetary mass and radius, star's metallicity, effective temperature, mass, radius, and gravity, and the semimajor axis) were taken from \citet{GillonEtal2012AA-WASP-43b}. These parameters were used in the parametrized temperature model and to generate the stellar spectrum by interpolating the stellar grid models from \citet{CastelliKurucz-2004new}.

Our initial \math{T(p)} profiles were chosen by running several short trial runs, or by taking the parameter values from the literature. Then, we produced the line-list data files for the species and the wavelength range of interest. From them, we generated the opacity tables for the case when we had four, seven, and eleven opacity sources separately. The opacity grid was generated between 300 to 3000 K in 100 K intervals. The wavelength range was generated between 0.6 - 5.5 {\microns} in 1 cm\sp{-1} intervals in the wavenumber space. The maximum optical depth was set to ten for all models ({\tt Transit} stops the extinction calculation at each wavelength when the optical depth reaches the user-defined value \math{\tau\sb{\rm max}}, \math{\tau \geq \tau\sb{\rm max}}, see collaborative paper by \citealp{CubillosEtal2015-BART}).

\subsection{Results - Four Fitted Species}
\label{sec:four}

In this section, we describe cases where we fit four major molecular species,  H\sb{2}O, CO\sb{2}, CO, and CH\sb{4}. We first wanted to compare our results with \citet{LineEtal2013-Retrieval-II} and \citet{KreidbergEtal2014-WASP43b}. Then, we investigated the goodness of fit between the models with a vertically-uniform abundances profile versus a profile produced by the thermochemical equilibrium calculations. We also explored how the inclusion of additional opacity sources affects the best-fit model. For these purposes, we generated four different cases:

\begin{enumerate}
\item We retrieved the \math{T(p)} profile and vertically uniform mixing ratios of the four major molecular species, using the best-fit \math{T(p)} profile and species abundances from \citet{LineEtal2013-Retrieval-II} as our initial guess. Following \citet{LineEtal2013-Retrieval-II}, we included only the four major molecular species and their opacity sources in the mean molecular mass and opacity calculation. 
\item We retrieved the \math{T(p)} profile and abundances using the equilibrium abundances profile calculated with {\tt TEA} as the initial guess. For the initial \math{T(p)} profile, we used the best parameters from several short trial runs. Again, we only included the four major molecular species and their opacities in the calculation.
\item Using the equilibrium abundances profile calculated with {\tt TEA} as the initial guess, and the same \math{T(p)} profile as in Case 2, we tested the statistical significance when additional molecules and their opacity sources were included in the calculation. In this case, in addition to H\sb{2}O, CO\sb{2}, CO, and CH\sb{4}, we included NH\sb{3}, HCN, and C\sb{2}H\sb{2}.
\item This case is the same as Case 3, with the inclusion of C\sb{2}H\sb{4}, H\sb{2}S, TiO, and VO.
\end{enumerate}

Figure \ref{fig:4mol-comp12} compares the vertically-uniform abundances profile and the {\tt TEA} abundances profile best-fit models. 

The initial species abundances for the uniform-model-atmosphere case were H=10\sp{-6}, He=0.15, C=10\sp{-20}, O=10\sp{-16}, H\sb{2}=0.85, H\sb{2}O=5x10\sp{-4}, CO\sb{2}=10\sp{-7}, CO=3x10\sp{-4}, and CH\sb{4}=10\sp{-6}. The initial free parameters were set to log\,\math{\kappa\sb{\rm IR}=-1.4} , log\math{\gamma\sb{1}=-0.74},  log\math{\gamma\sb{2}=0.0}, log\math{\alpha=0.0}, log\math{\beta=1.03}, log\,\math{{f}\sb{H\sb{2}O}=-1.07},  log\,\math{{f}\sb{CO\sb{2}}=-1.07}, log\,\math{{f}\sb{CO}=1.784}, and log\,\math{{f}\sb{CH\sb{4}}=1.784}. The first three parameters reproduce the best-fit temperature profile from \citet{LineEtal2013-Retrieval-II}, and the last four parameters are their reported best-fit scaling factors.

The initial species abundances for the equilibrium case were calculated using {\tt TEA}. The \math{T(p)} profile parameters were chosen from several short trial runs: log\,\math{\kappa\sb{\rm IR} = -0.6} , log\math{\gamma\sb{1} = -0.4},  log\math{\gamma\sb{2} = 0.0}, log\math{\alpha = 0.0}, and log\math{\beta = 1.09}.

\begin{figure*}[ht!]
\vspace{-8pt}
\centering
\includegraphics[width=0.75\textwidth, height=12cm]{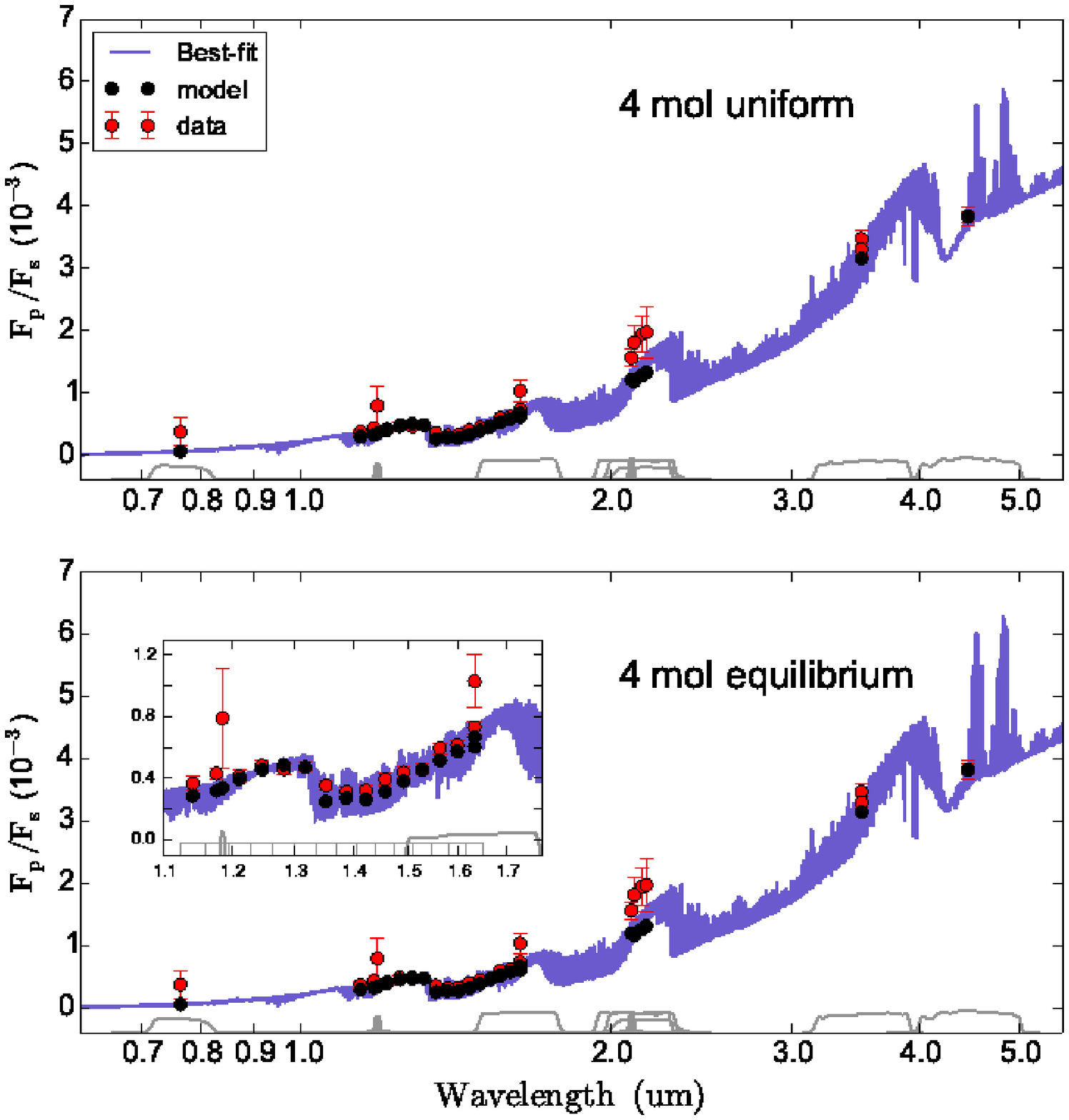}\hspace{-20pt}
\includegraphics[width=.28\textwidth, height=12cm]{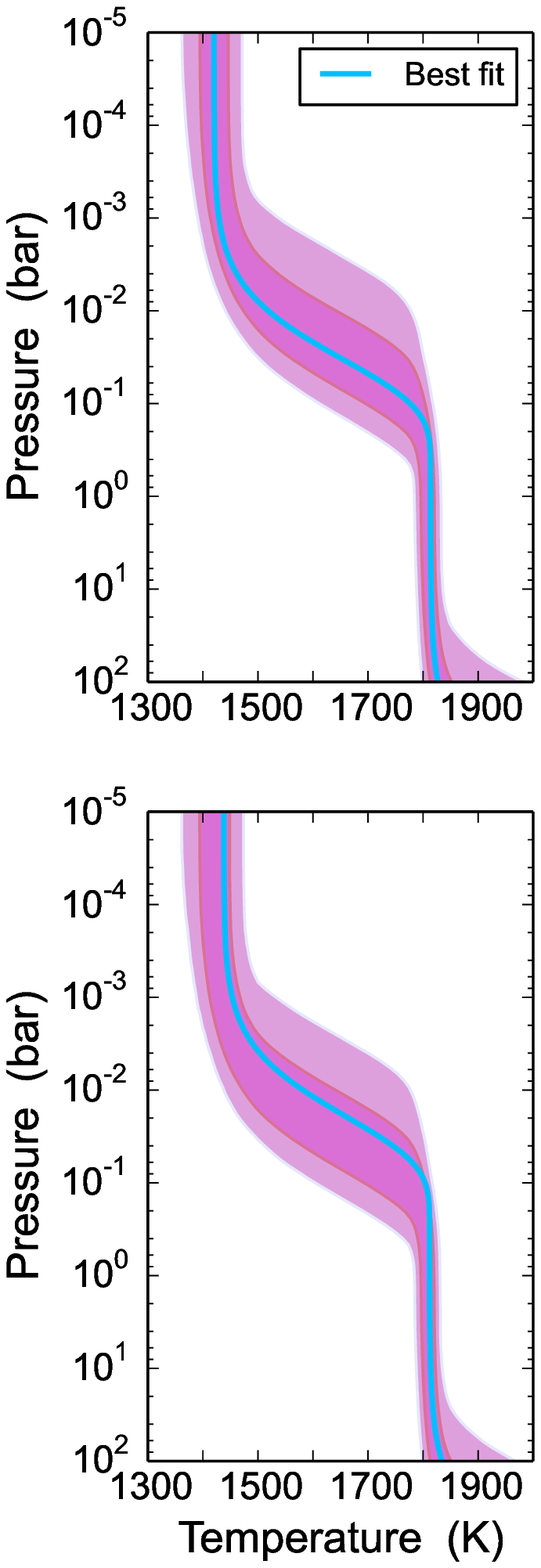}
\vspace{-8pt}
\caption[Best-fit models and \math{T(p)} profiles for Cases 1 and 2, four fitted species]{Best-fit models and \math{T(p)} profiles for Cases 1 and 2, four fitted species. The upper panel shows the model when vertically-uniform mixing ratios are used as the initial guess. The bottom panel uses the equilibrium mixing ratios calculated using {\tt TEA}. In red are the data points (eclipse depths) with error bars. In black are the integrated points of our model over the bandpasses shown in grey. The models are generated with the four major molecular species and their opacities.}
\label{fig:4mol-comp12}
\end{figure*}

Both cases were run with the initial 1.5x10\sp{5} iterations per chain. The equilibrium case took 1.2x10\sp{5} iterations until all parameters converged and the Gelman and Rubin statistics was satisfied. The uniform case needed additional iterations and converged after 1.8x10\sp{5} iterations.  \math{\chi\sp{2}\sb{\rm red}}, BIC, and SDR, for Cases 1 and 2 are reported in Table \ref{table:4mol-fit}.

According to \math{\chi\sp{2}\sb{\rm red}} and BIC, the equilibrium case provides a marginally better fit to the data. In the inset of the bottom panel of Figure \ref{fig:4mol-comp12}, we give a zoom-in detail of the wavelength range covered with the HST data.

\begin{table}[hb!]
\footnotesize{
\caption{\label{table:4mol-fit} 4 Species Goodness of Fit}
\atabon\strut\hfill\begin{tabular}{lcccc}
    \hline
    \hline
                          & \math{\chi\sp{2}\sb{\rm red}}  & BIC     & SDR          \\
    \hline
Case 1, Uniform           & 1.9856                           & 60.5324  & 0.000202731  \\
Case 2, Equilibrium       & 1.9709                           & 60.2547  & 0.000202287   \\
Case 3, 7 opacities       & 1.9849                           & 60.5198  & 0.000209638  \\
Case 4, 11 opacities      & 2.0542                           & 61.8365  & 0.000208284  \\
    \hline
\end{tabular}\hfill\strut\ataboff
}
\end{table}

\begin{figure*}[ht!]
\vspace{-5pt}
\centering
\includegraphics[width=.50\textwidth, height=2.5cm]{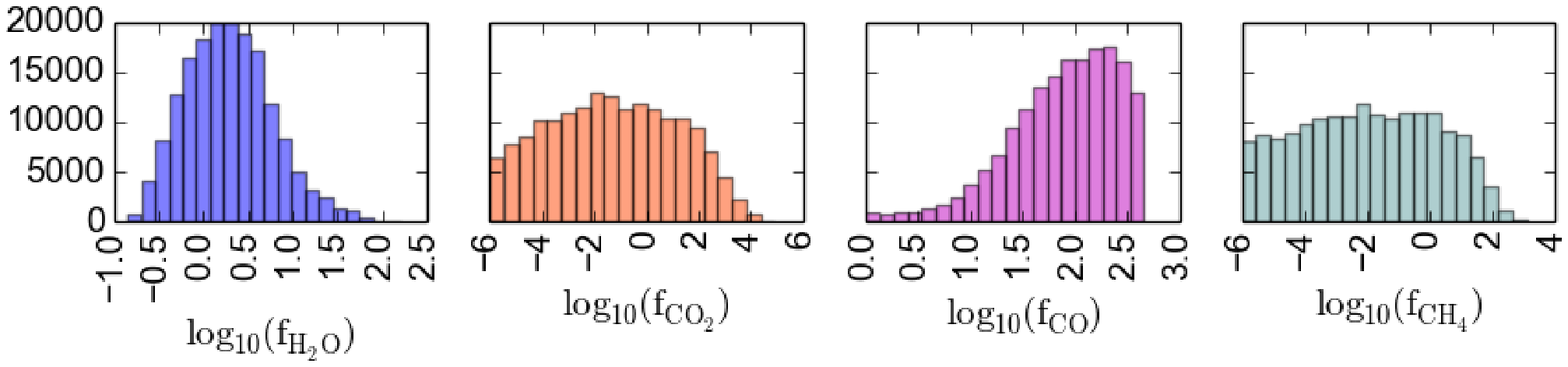}\hspace{-5pt}
\includegraphics[width=.50\textwidth, height=2.5cm]{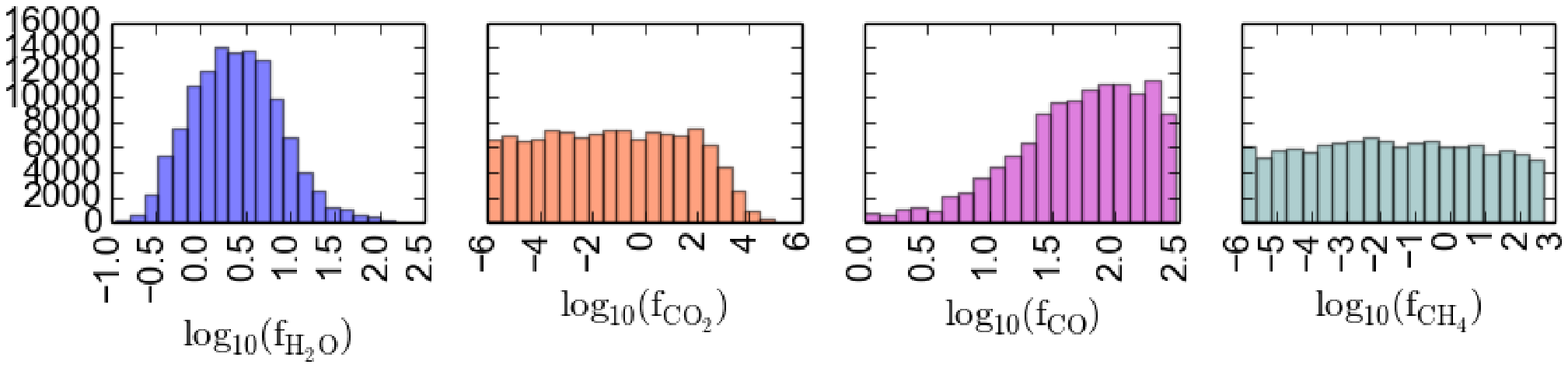}
\caption[Histograms for Cases 1 and 2, four fitted species]{Histograms for Cases 1 and 2, four fitted species. Figures show the species' scaling factors expressed as \math{log10(f\sb{X})}, where \math{X} is the species abundance.}
\label{fig:4mol-hist}
\vspace{-5pt}
\end{figure*}

\begin{figure*}[ht!]
\vspace{-5pt}
\centering
\includegraphics[width=.26\textwidth, height=3.5cm]{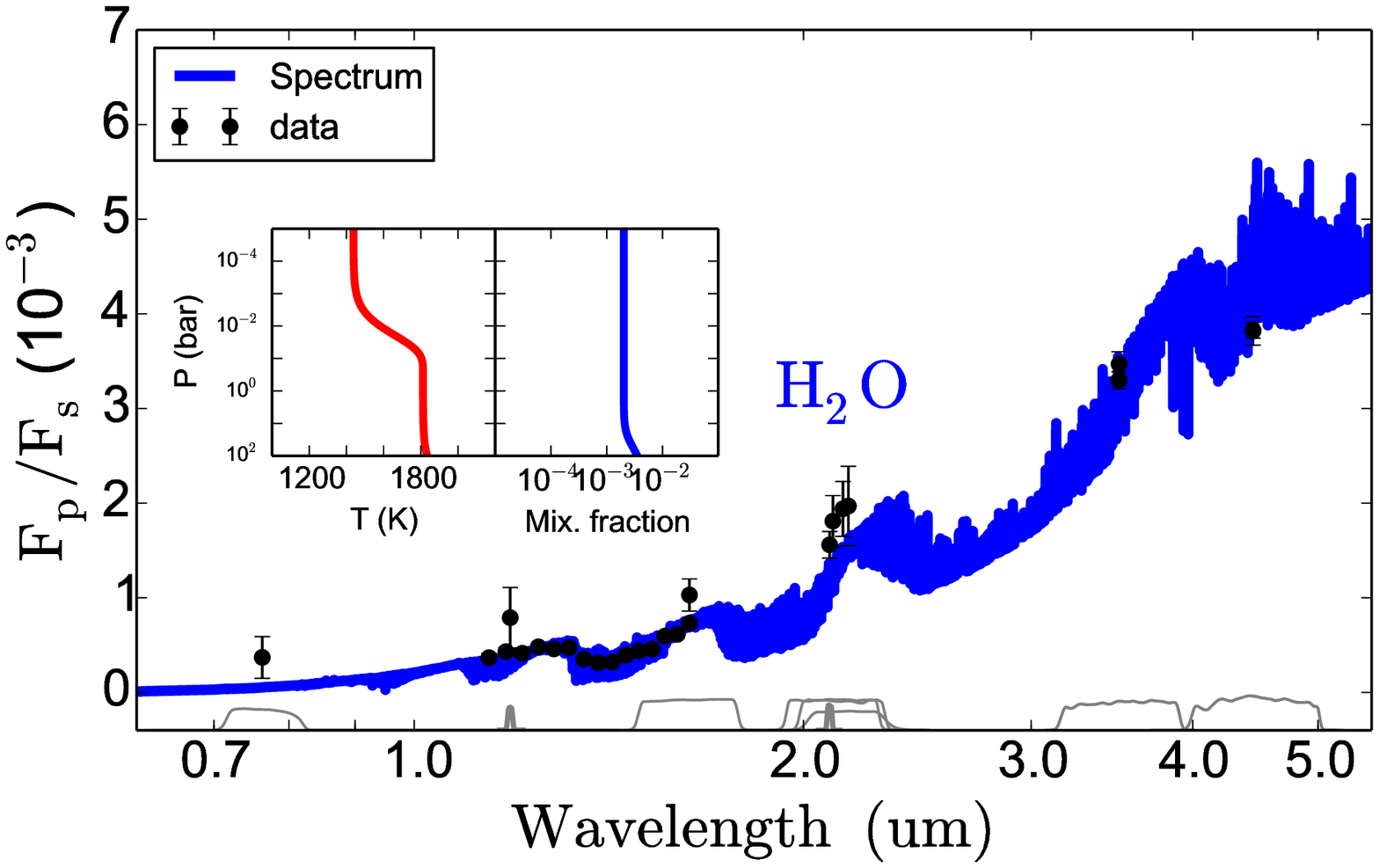}\hspace{-13pt}
\includegraphics[width=.26\textwidth, height=3.5cm]{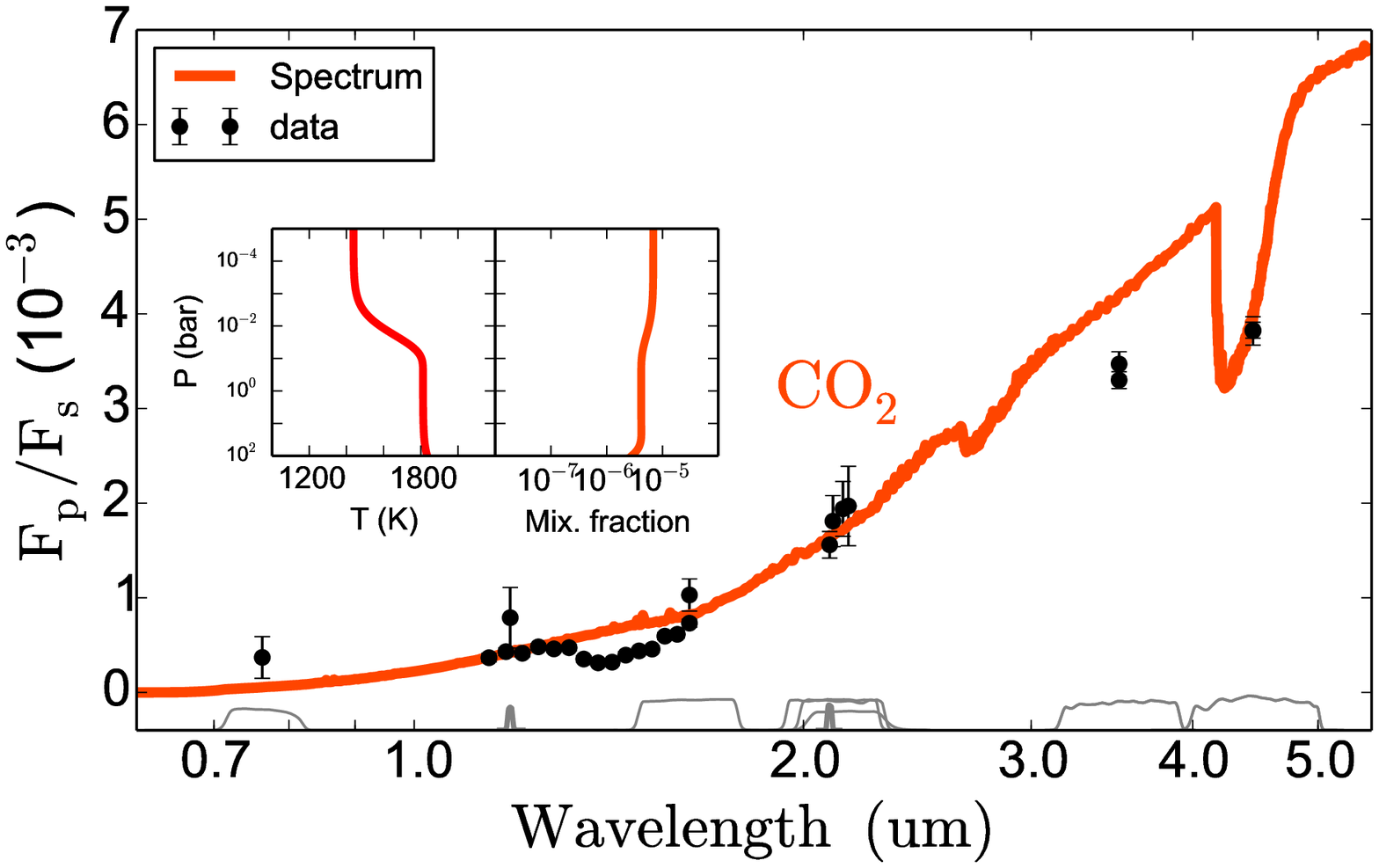}\hspace{-13pt}
\includegraphics[width=.26\textwidth, height=3.5cm]{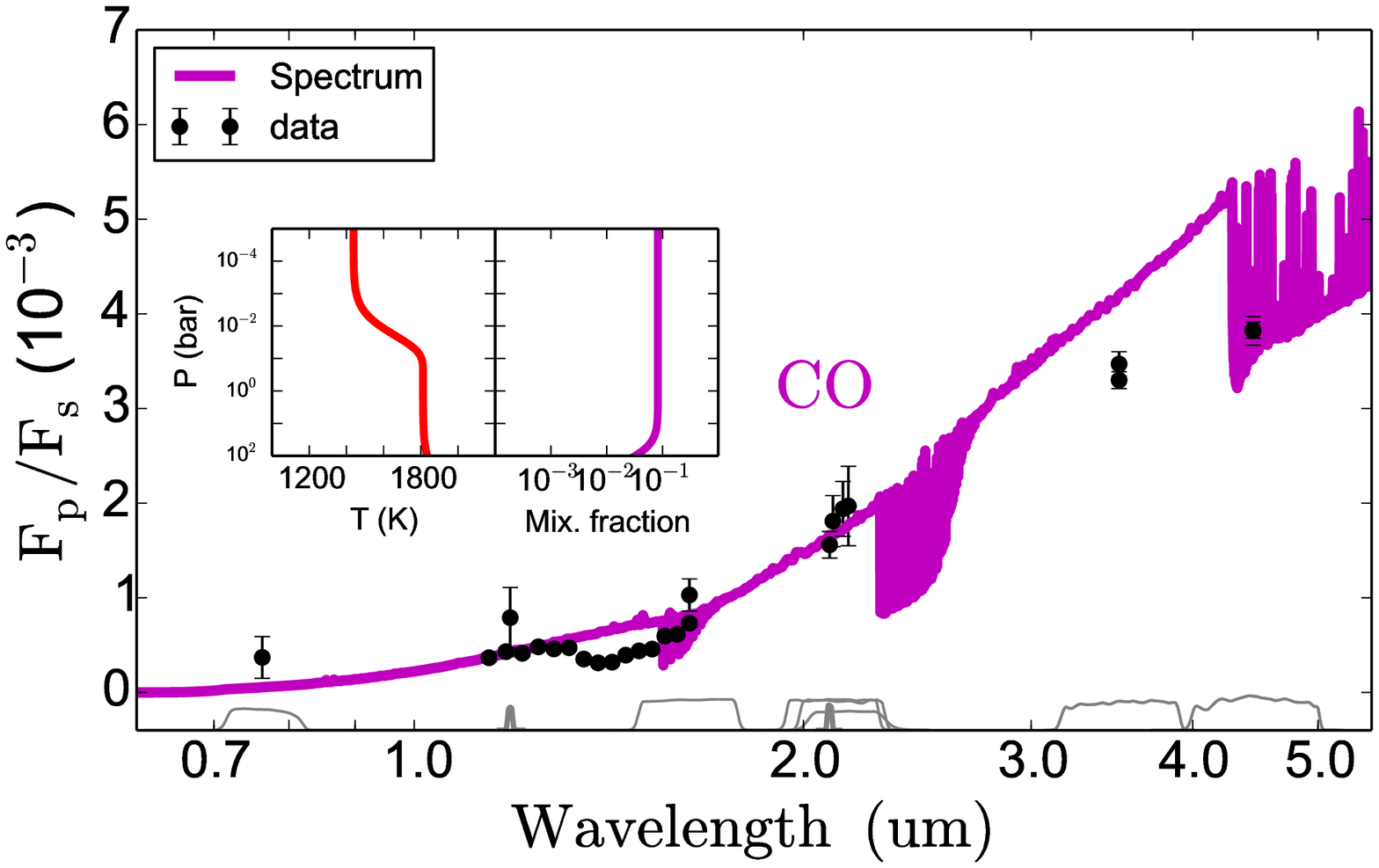}\hspace{-13pt}
\includegraphics[width=.26\textwidth, height=3.5cm]{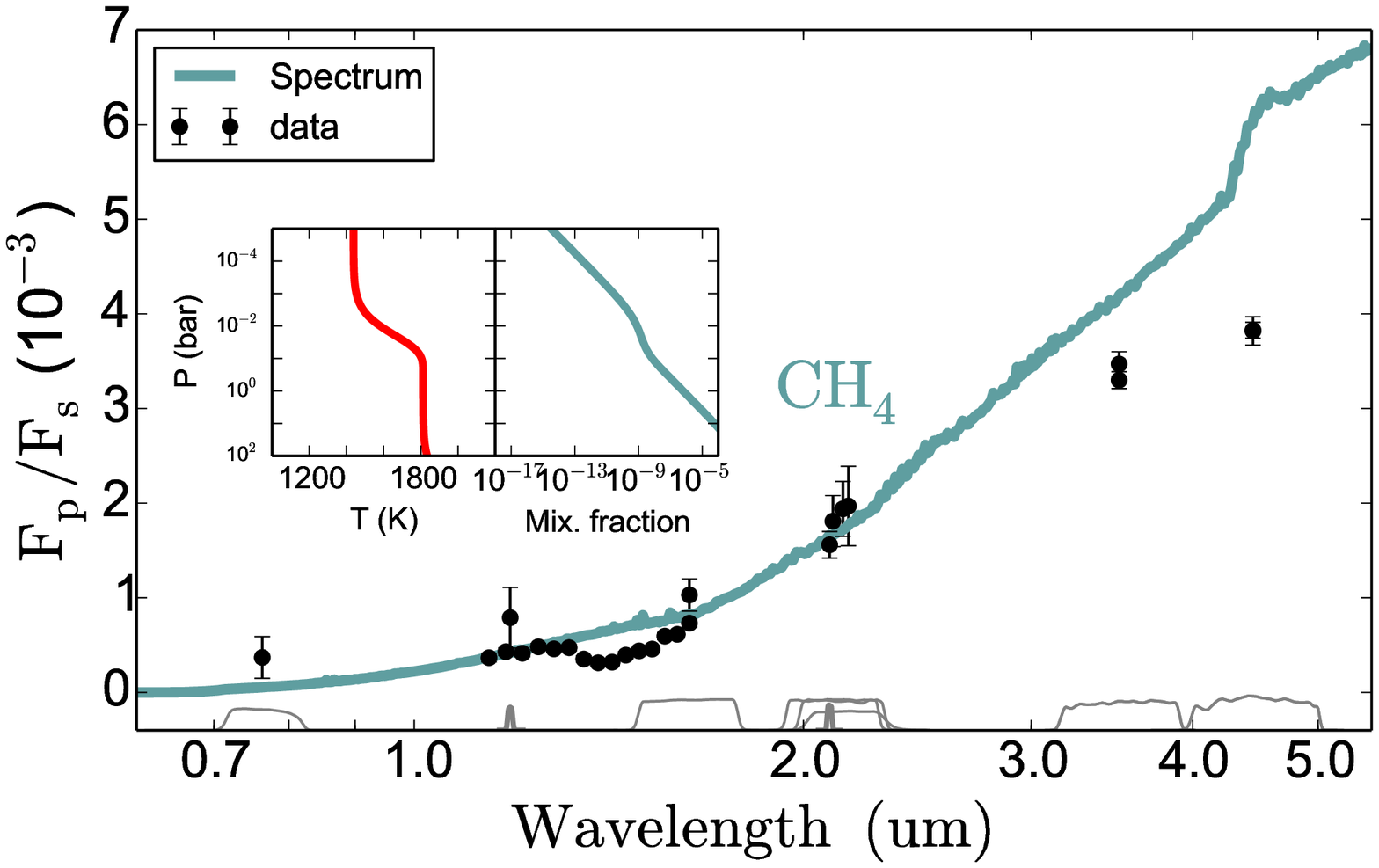}
\caption[Influence of each species to the Case 2 best-fit model, four fitted species]{Influence of each species to the best-fit model for Case 2, four fitted species. In the inset figures are the best-fit \math{T(p)} profile and the species's mixing ratio.}
\label{fig:4mol-indSpecs}
\vspace{-8pt}
\end{figure*}

Figure \ref{fig:4mol-hist} shows the histograms of the posterior distribution of the retrieved molecular species for both cases. Figure \ref{fig:4mol-indSpecs} shows the influence of the individual species on the best-fit model, Case 2. Figure \ref{fig:4mol-PT} shows the initial and the best-fit abundances for Case 2. As we can see, the water features are dominant in the spectrum; thus, water is the best constrained molecule in our analysis (Figure \ref{fig:4mol-hist}). CO\sb{2} and CH\sb{4} are fully unconstrained in our analysis. CO and CO\sb{2} show absorption features only above 4.2 {\microns}. Since CO\sb{2} and CH\sb{4} are unconstrained in our analysis, we tested a case where we fit only H\sb{2}O and CO; however, the results were unchanged.

\begin{figure}[ht!]
\vspace{-5pt}
\centering
\includegraphics[width=.5\textwidth, clip=True]{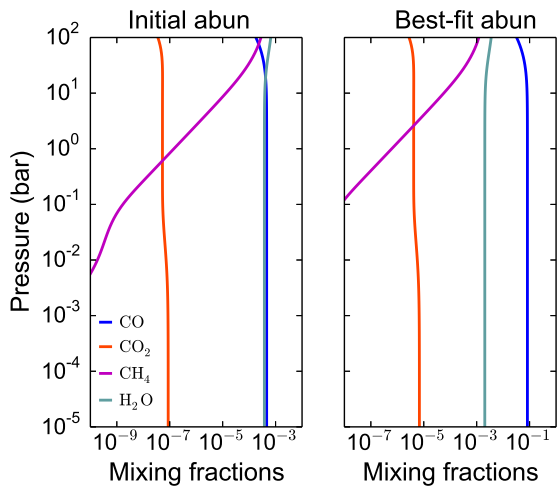}
\caption[Initial and best-fit abundances for Case 2, four fitted species]{Initial and best-fit abundances for Case 2, four fitted species.}
\label{fig:4mol-PT}
\vspace{-5pt}
\end{figure}

Considering that Case 2 provided a marginally better fit to the data, we continued to investigate the effect of the inclusion of additional species on this model atmosphere. We constructed the initial model atmosphere with seven molecular species, H\sb{2}O, CO\sb{2}, CO, CH\sb{4}, NH\sb{3}, HCN, and C\sb{2}H\sb{2}; used the same initial \math{T(p)} profile as in Case 2; and calculated the mixing ratios of the molecular species using {\tt TEA}. We ran BART including all seven opacity sources in the calculation. The procedure was repeated for eleven molecular species: H\sb{2}O, CO\sb{2}, CO, CH\sb{4}, NH\sb{3}, HCN, C\sb{2}H\sb{2}, C\sb{2}H\sb{4}, H\sb{2}S, TiO, and VO. These cases are referred to as Cases 3 and 4.

Table \ref{table:4mol-fit} compares the goodness of fit for Cases 3 and 4, and Table \ref{table:fourSpecs} lists the mixing fractions of the fitted molecular species for all the cases at the pressure level of 1 bar. Figure \ref{fig:4mol-comp34} shows the best-fit models for Cases 3 and 4. For these cases, the histograms and the influences of the individual fitted species to the best-fit spectrum are similar to Figures \ref{fig:4mol-hist} and \ref{fig:4mol-indSpecs}. As we see, according to BIC, the inclusion of additional opacity sources does not improve the fit.

\begin{figure*}[ht!]
\vspace{-8pt}
\centering
\includegraphics[width=.75\textwidth, height=12cm]{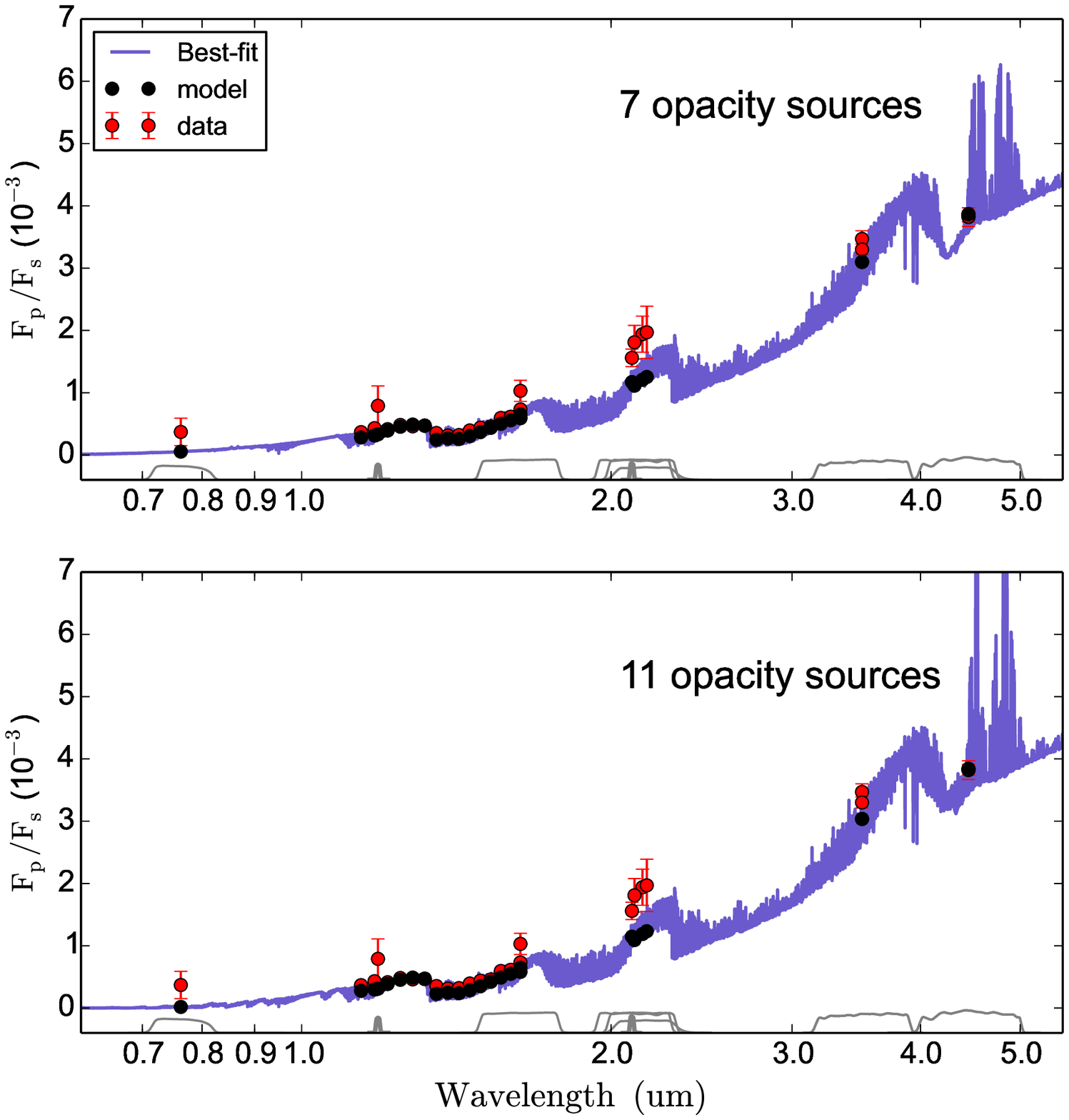}\hspace{-20pt}
\includegraphics[width=.28\textwidth, height=12cm]{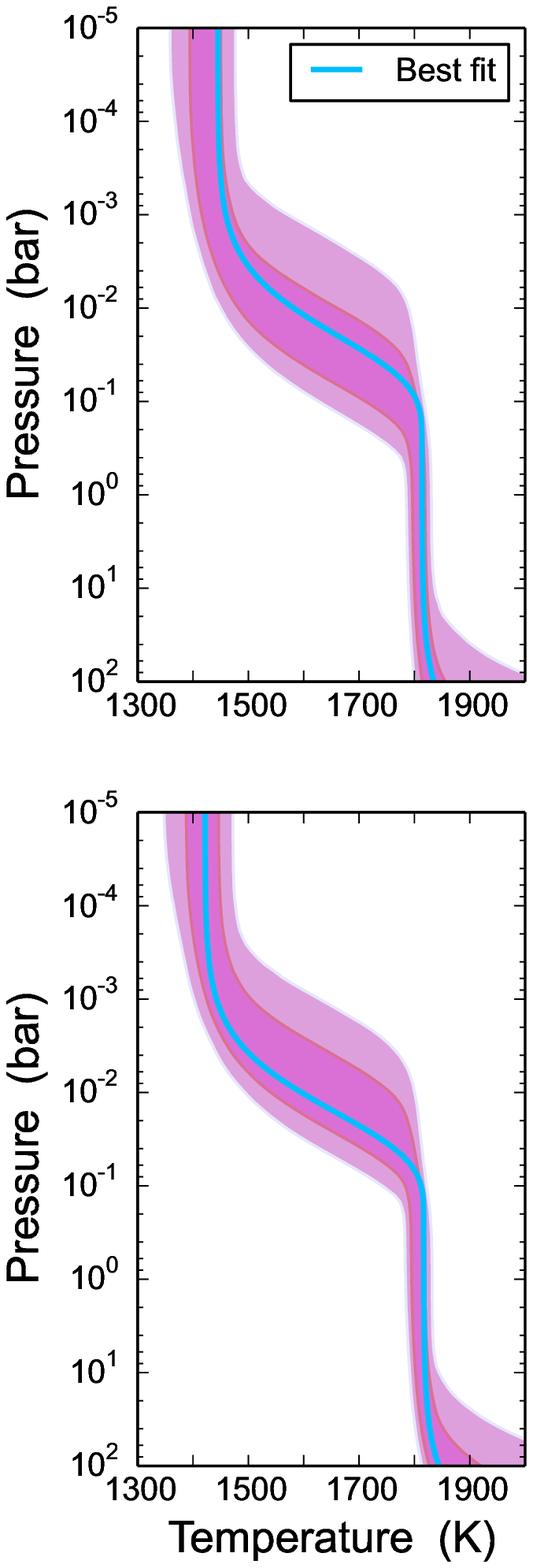}
\caption[Best-fit models and \math{T(p)} profiles for Cases 3 and 4, four fitted species]{Best-fit models and \math{T(p)} profiles for Cases 3 and 4, four fitted species. The upper panel shows the best-fit model when seven opacity sources are included in the calculation: H\sb{2}O, CO\sb{2}, CO, CH\sb{4}, NH\sb{3}, HCN, and C\sb{2}H\sb{2}. The bottom panel shows the best-fit model when eleven opacity sources are included in the calculation: H\sb{2}O, CO\sb{2}, CO, CH\sb{4}, NH\sb{3}, HCN, C\sb{2}H\sb{2}, C\sb{2}H\sb{4}, H\sb{2}S, TiO, and VO.}
\label{fig:4mol-comp34}
\end{figure*}

\begin{landscape}
\vspace*{\fill}
\begin{table*}[ht!]
\footnotesize{
\caption{\label{table:fourSpecs} 4 Species Mixing Ratios}
\atabon\strut\hfill\begin{tabular}{lccccc}
    \hline
    \hline
Case         &               &       H\sb{2}O      &      CO\sb{2}      &         CO          &  CH\sb{4}  \\
    \hline
Case 1       & Best Fit      &        8.22 x 10\sp{-4}      &       4.47 x 10\sp{-8}      &       3.90 x 10\sp{-2}      &       2.77 x 10\sp{-8}       \\
uniform      & 68\% interval & [2.70 x 10\sp{-4} -- 2.51 x 10\sp{-3}] & [1.52 x 10\sp{-10} -- 1.31 x 10\sp{-5}] & [1.17 x 10\sp{-2} -- 1.30 x 10\sp{-1}] & [1.82 x 10\sp{-10} -- 1.30 x 10\sp{-1}]  \\
    \hline
Case 2       & Best Fit      &       2.02 x 10\sp{-3}      &       4.15 x 10\sp{-6}      &       8.25 x 10\sp{-2}      &       6.77 x 10\sp{-7}       \\
equilibrium  & 68\% interval & [6.54 x 10\sp{-4} -- 6.27 x 10\sp{-3}] & [7.97 x 10\sp{-9} -- 2.16 x 10\sp{-3}] & [2.49 x 10\sp{-2} -- 2.73 x 10\sp{-1}] & [2.45 x 10\sp{-9} -- 2.73 x 10\sp{-1}]  \\
    \hline
Case 3       & Best Fit       &       2.14 x 10\sp{-3}         &       5.30 x 10\sp{-6}         &       1.01 x 10\sp{-1}         &       3.25 x 10\sp{-7}       \\
7 opacities  & 68\% interval  & [6.56 x 10\sp{-4} -- 7.00 x 10\sp{-3}] & [9.79 x 10\sp{-9} -- 2.86 x 10\sp{-3}] & [3.43 x 10\sp{-2} -- 2.97 x 10\sp{-1}] & [1.08 x 10\sp{-9} -- 9.76 x 10\sp{-5}]  \\
    \hline
Case 4      & Best Fit      &      2.96 x 10\sp{-3}      &       1.54 x 10\sp{-6}      &       1.23 x 10\sp{-1}      &       1.97 x 10\sp{-8}       \\
11 opacities & 68\% interval & [1.10 x 10\sp{-3} -- 7.96 x 10\sp{-3}] & [1.73 x 10\sp{-9} -- 1.36 x 10\sp{-3}] & [4.25 x 10\sp{-2} -- 3.57 x 10\sp{-1}] & [7.15 x 10\sp{-11} -- 3.57 x 10\sp{-1}]  \\
    \hline
\end{tabular}\hfill\strut\ataboff
}
\end{table*}
\vspace*{\fill}
\end{landscape}

\subsection{Results - Seven Fitted Species}
\label{sec:seven}

In this section, we describe cases where we fit seven molecular species,  H\sb{2}O, CO\sb{2}, CO, CH\sb{4}, NH\sb{3}, HCN, and C\sb{2}H\sb{2}. Again, we tested the uniform versus the equilibrium abundances profile case, and the effect of the inclusion of additional opacity sources. For these purposes, we generated three different cases:

\begin{enumerate}
\item Using the best-fit \math{T(p)} profile and vertically-uniform species abundances from \citet{LineEtal2013-Retrieval-II} as our initial guess, we modeled the atmosphere of WASP-43b including the opacity sources for all seven molecules. 
\item Using the \math{T(p)} profile from Case 5 and the equilibrium abundances profile calculated with {\tt TEA} as the initial guess (Section \ref{sec:four}), we retrieved the \math{T(p)} profile and mixing fractions including the opacity sources for all seven molecules.
\item Using the equilibrium abundances profile calculated with {\tt TEA} as the initial guess, and the same \math{T(p)} profile as in Case 6, we modeled WASP-43b atmosphere including all eleven opacity sources in the calculation (C\sb{2}H\sb{4}, H\sb{2}S, TiO, and VO in addition to the species from Cases 5 and 6).
\end{enumerate}

Figure \ref{fig:7mol-comp} shows the comparison between the vertically-uniform abundances profile and the {\tt TEA} abundances profile best-fit models for the case when seven molecular species are included in the mean molecular mass and the opacity calculation.

\begin{figure*}[ht!]
\centering
\includegraphics[width=.75\textwidth, height=12cm]{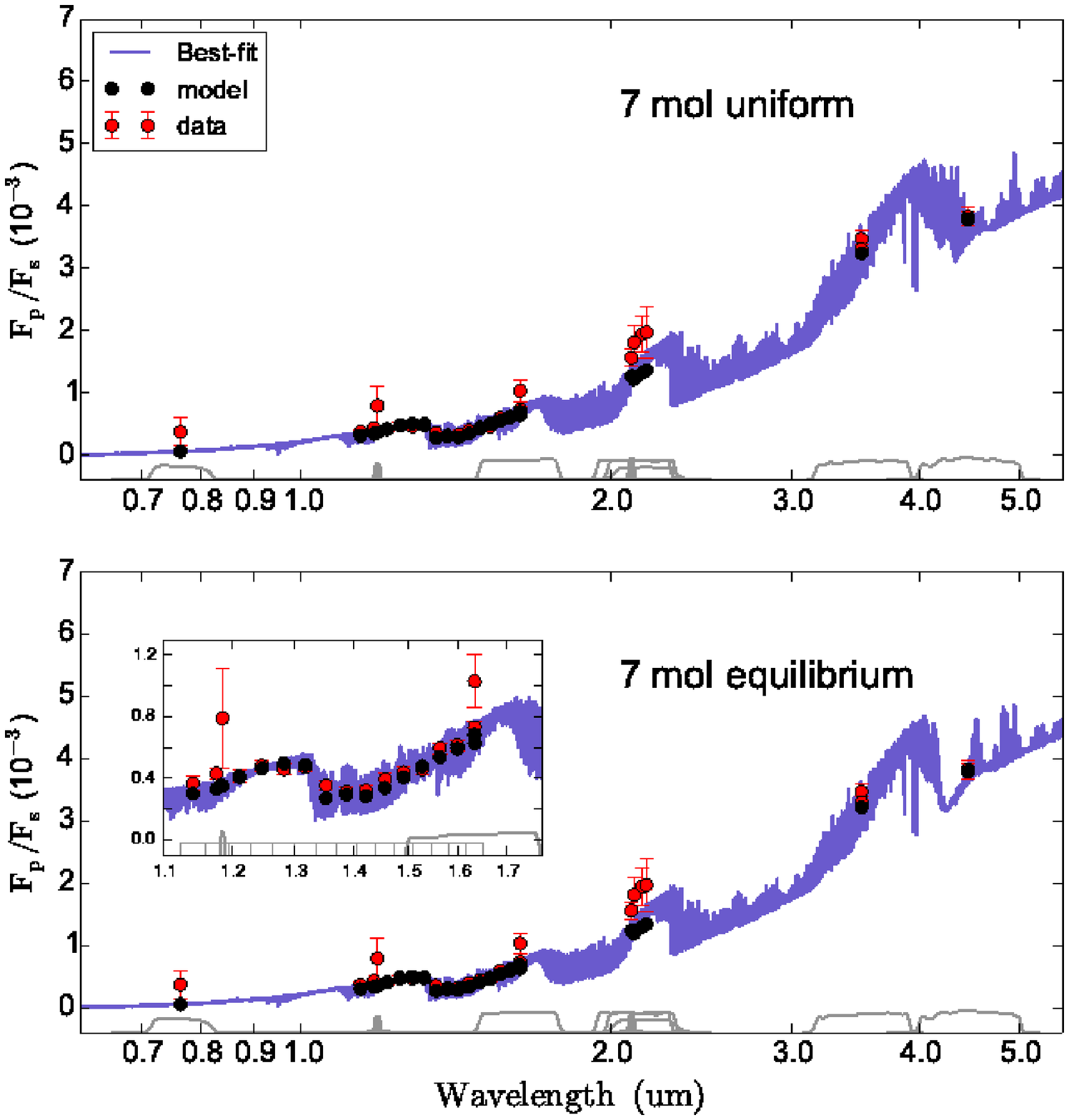}\hspace{-20pt}
\includegraphics[width=.28\textwidth, height=12cm]{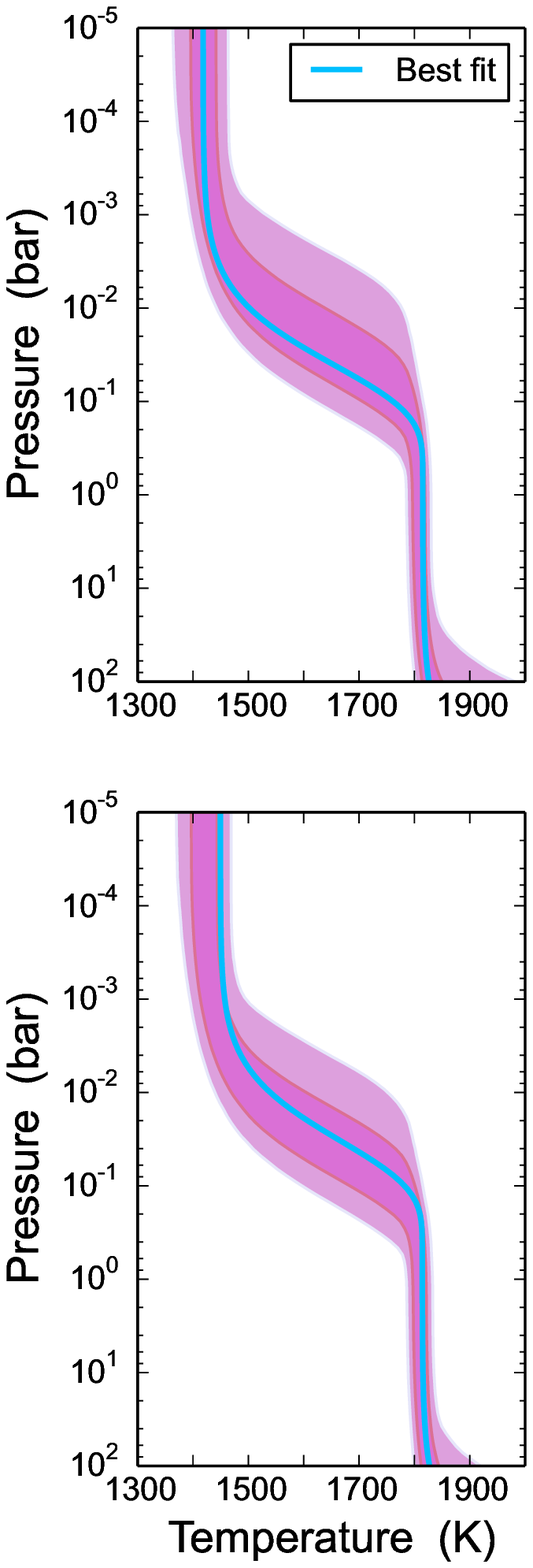}
\caption[Best-fit models and \math{T(p)} profiles for Cases 1 and 2, seven fitted species]{Best-fit models and \math{T(p)} profiles for Cases 5 and 6, seven fitted species. The upper panel shows the model when vertically-uniform mixing ratios are used as the initial guess. The bottom panel uses the equilibrium mixing ratios calculated using {\tt TEA}. In red are the data points (eclipse depths) with error bars. In black are the integrated points of our model over the bandpasses shown in grey. The models are generated with the seven major molecular species and their opacities.}
\label{fig:7mol-comp}
\end{figure*}

The initial species abundances for the uniform-model-atmosphere case were H = 10\sp{-4}, He = 0.145, C = 10\sp{-17}, N = 10\sp{-9}, O = 10\sp{-11}, H\sb{2} = 0.85, H\sb{2}O = 5x10\sp{-4}, CO\sb{2} = 10\sp{-8}, CO = 3x10\sp{-4}, CH\sb{4} = 10\sp{-7}, NH\sb{3} = 10\sp{-4}, HCN = 10\sp{-6}, and C\sb{2}H\sb{2} = 10\sp{-11}. The initial free parameters were set to log\,\math{\kappa\sb{\rm IR} = -1.4} , log\math{\gamma\sb{1} = -0.74},  log\math{\gamma\sb{2} = 0.0}, log\math{\alpha = 0.0}, log\math{\beta = 1.03}, log\,\math{{f}\sb{H\sb{2}O} = -1.07},  log\,\math{{f}\sb{CO\sb{2}} = -1.07}, log\,\math{{f}\sb{CO} = 1.784}, and log\,\math{{f}\sb{CH\sb{4}} = 1.784}. The first three parameters reproduce the best-fit temperature profile from \citet{LineEtal2013-Retrieval-II}, and the last four are their reported best-fit abundances values. The initial species abundances for the equilibrium case were calculated using {\tt TEA}. The \math{T(p)} profile was the same as in Section \ref{sec:four}.

\begin{figure*}[ht!]
\vspace{-5pt}
\centering
\includegraphics[width=.75\textwidth, height=3cm]{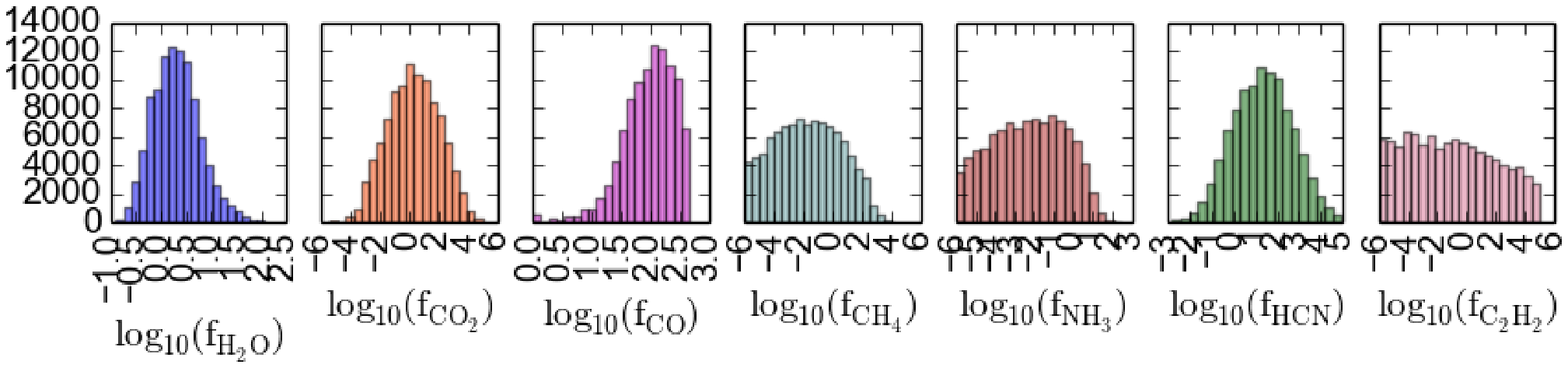}\hspace{-20pt}
\includegraphics[width=.75\textwidth, height=3cm]{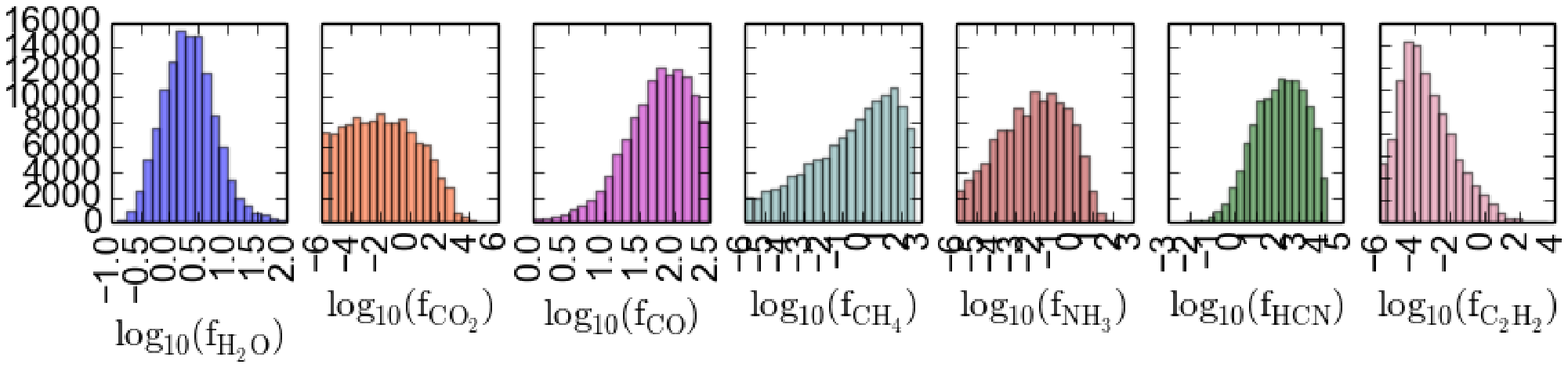}
\caption[Histograms for Cases 5 and 6, seven fitted species]{Histograms for Case 5 (top panel) and Case 6 (bottom panel), seven fitted species. Figures show the species' scaling factors expressed as \math{log10(f\sb{X})}, where \math{X} is the species abundance.}
\label{fig:7mol-hist}
\vspace{-5pt}
\end{figure*}

\begin{figure*}[ht!]
\vspace{-8pt}
\centering
\includegraphics[width=.26\textwidth, height=3.5cm]{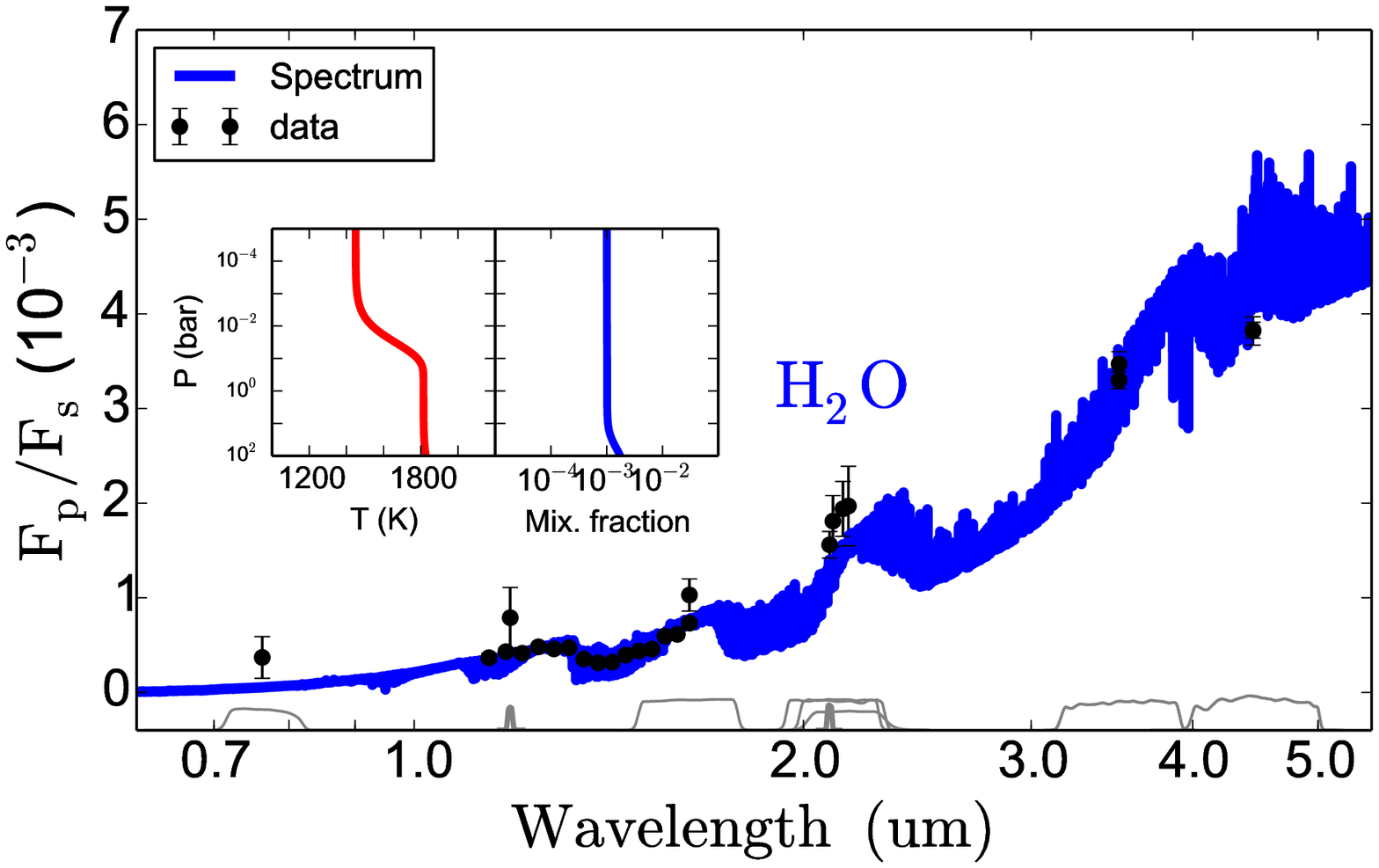}\hspace{-13pt}
\includegraphics[width=.26\textwidth, height=3.5cm]{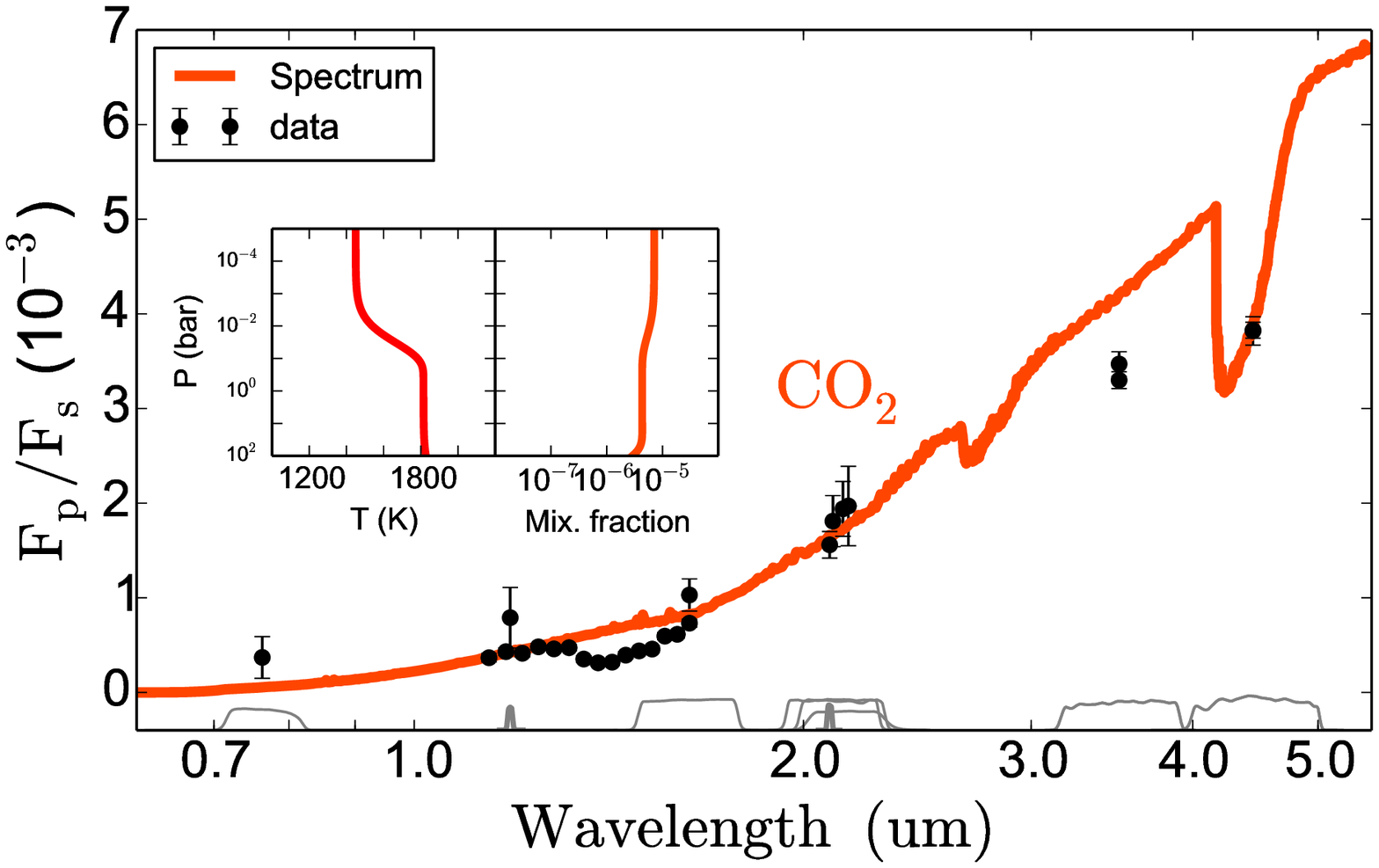}\hspace{-13pt}
\includegraphics[width=.26\textwidth, height=3.5cm]{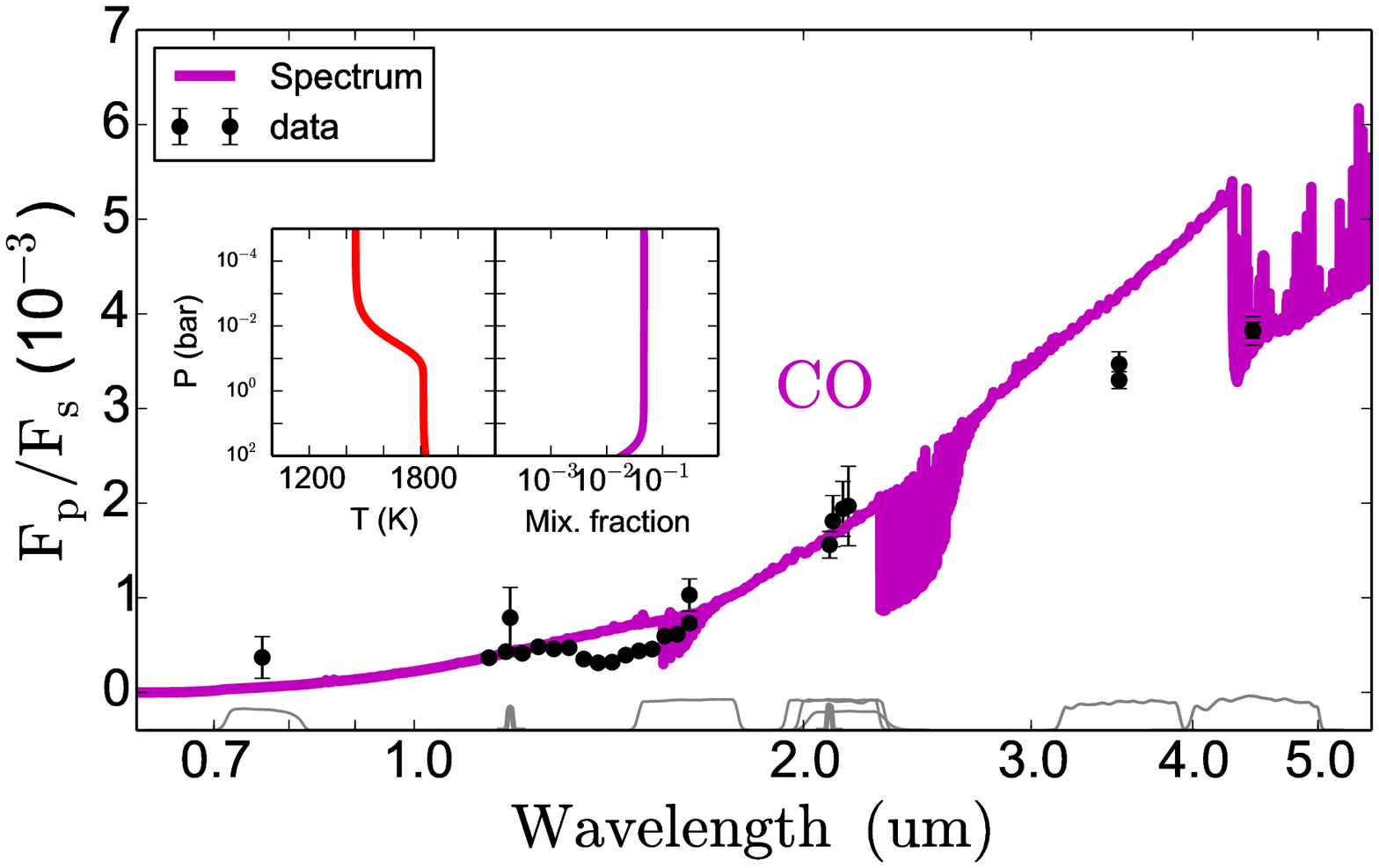}\hspace{-13pt}
\includegraphics[width=.26\textwidth, height=3.5cm]{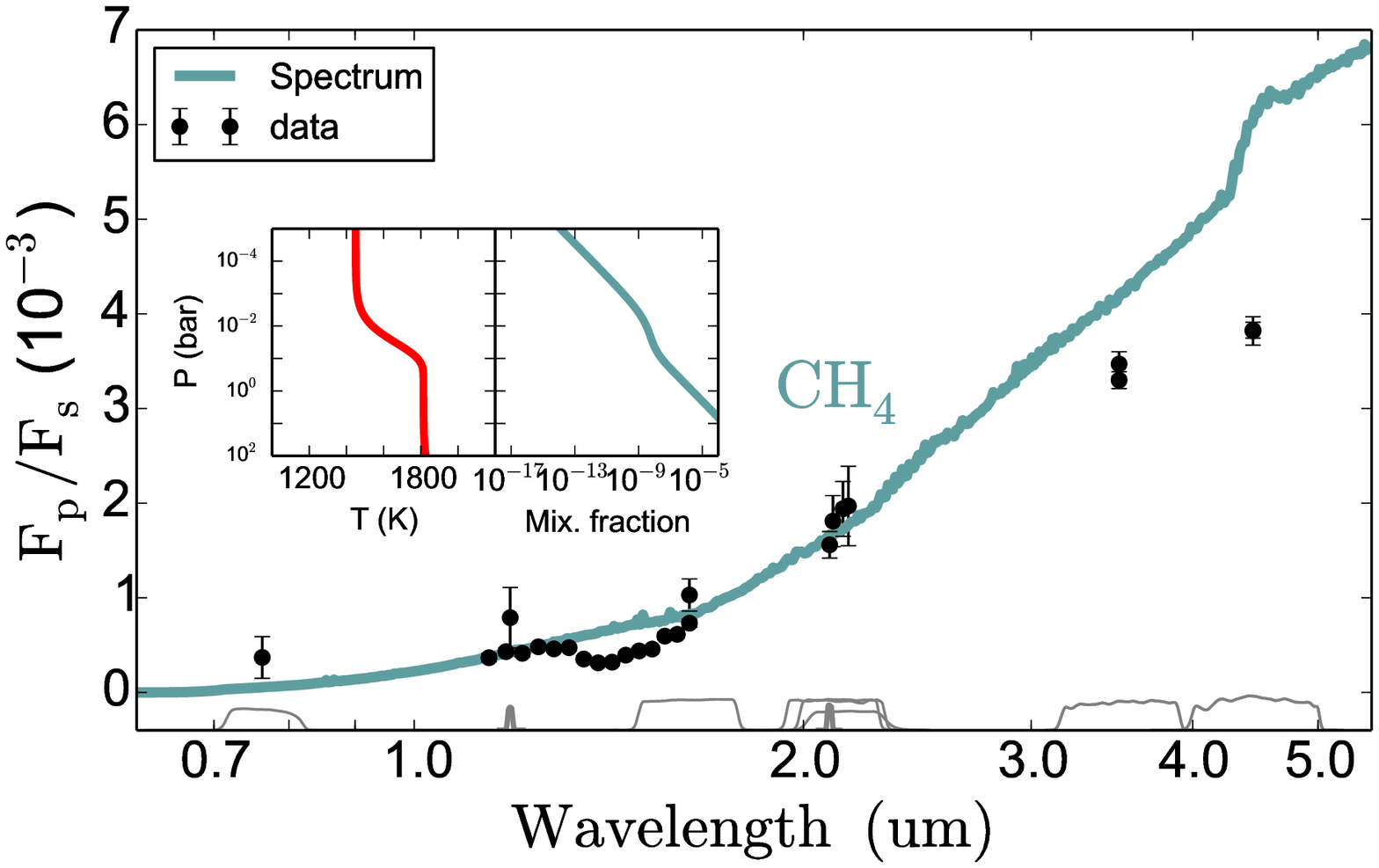}\hspace{-13pt}
\includegraphics[width=.26\textwidth, height=3.5cm]{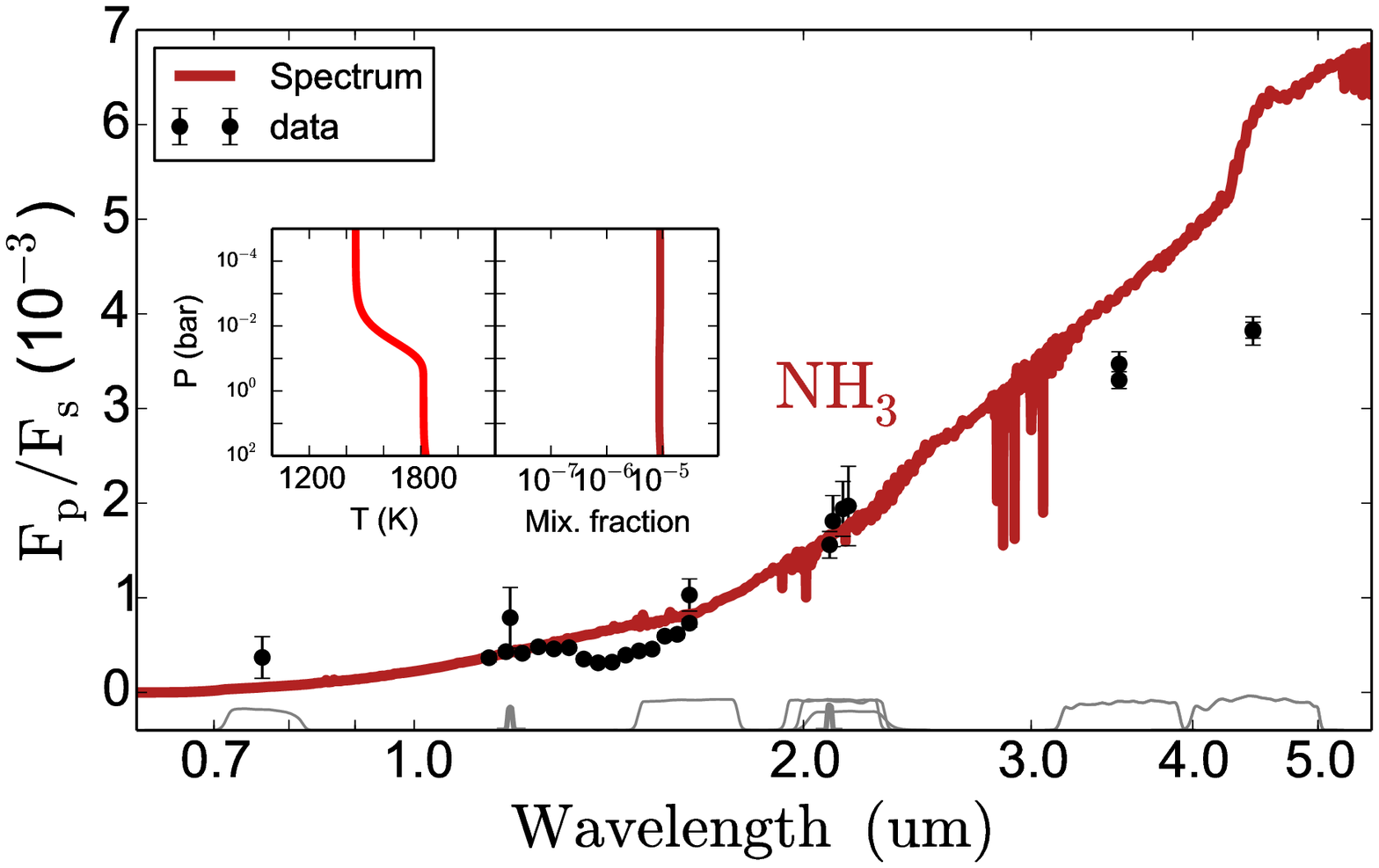}\hspace{-13pt}
\includegraphics[width=.26\textwidth, height=3.5cm]{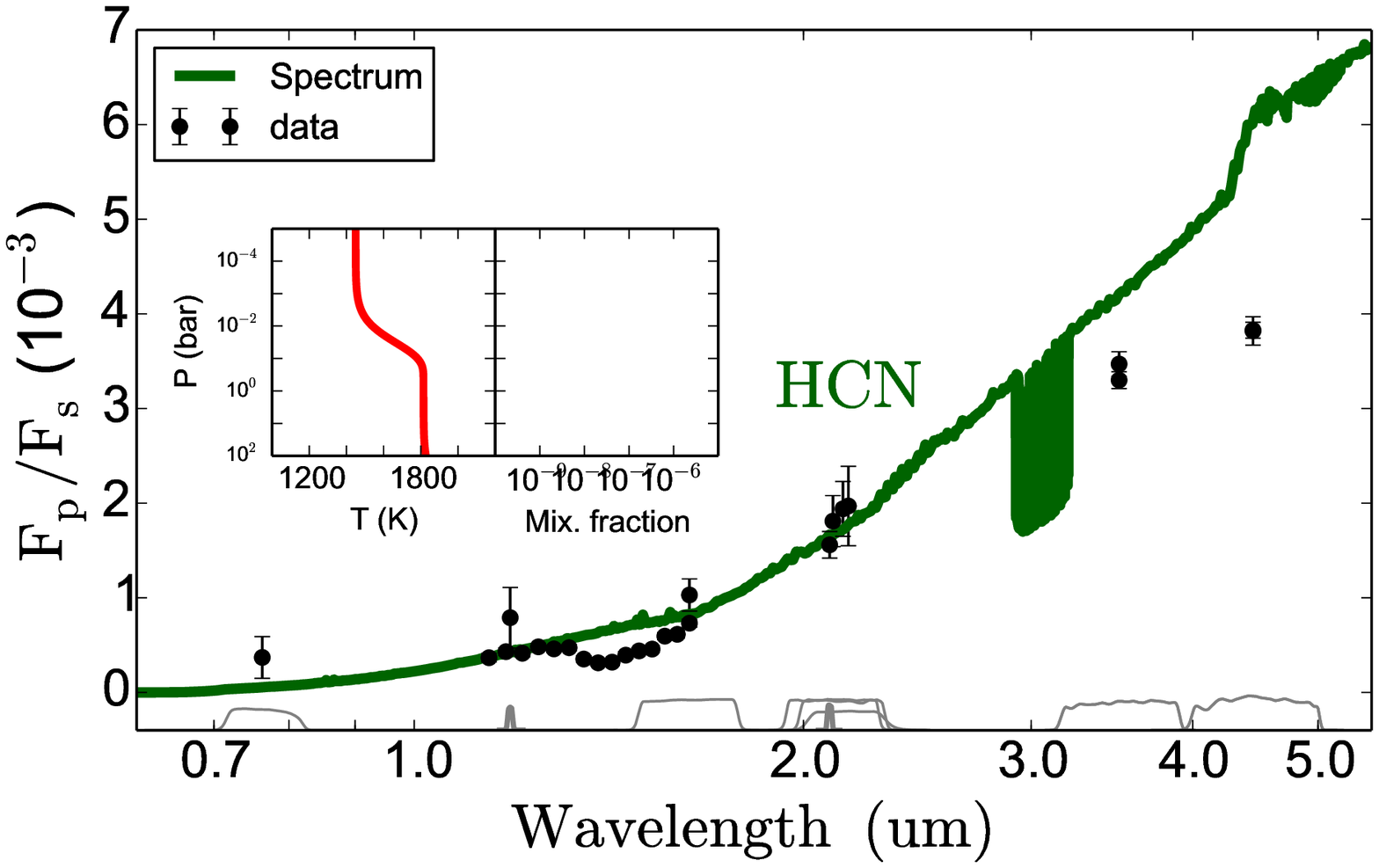}\hspace{-13pt}
\includegraphics[width=.26\textwidth, height=3.5cm]{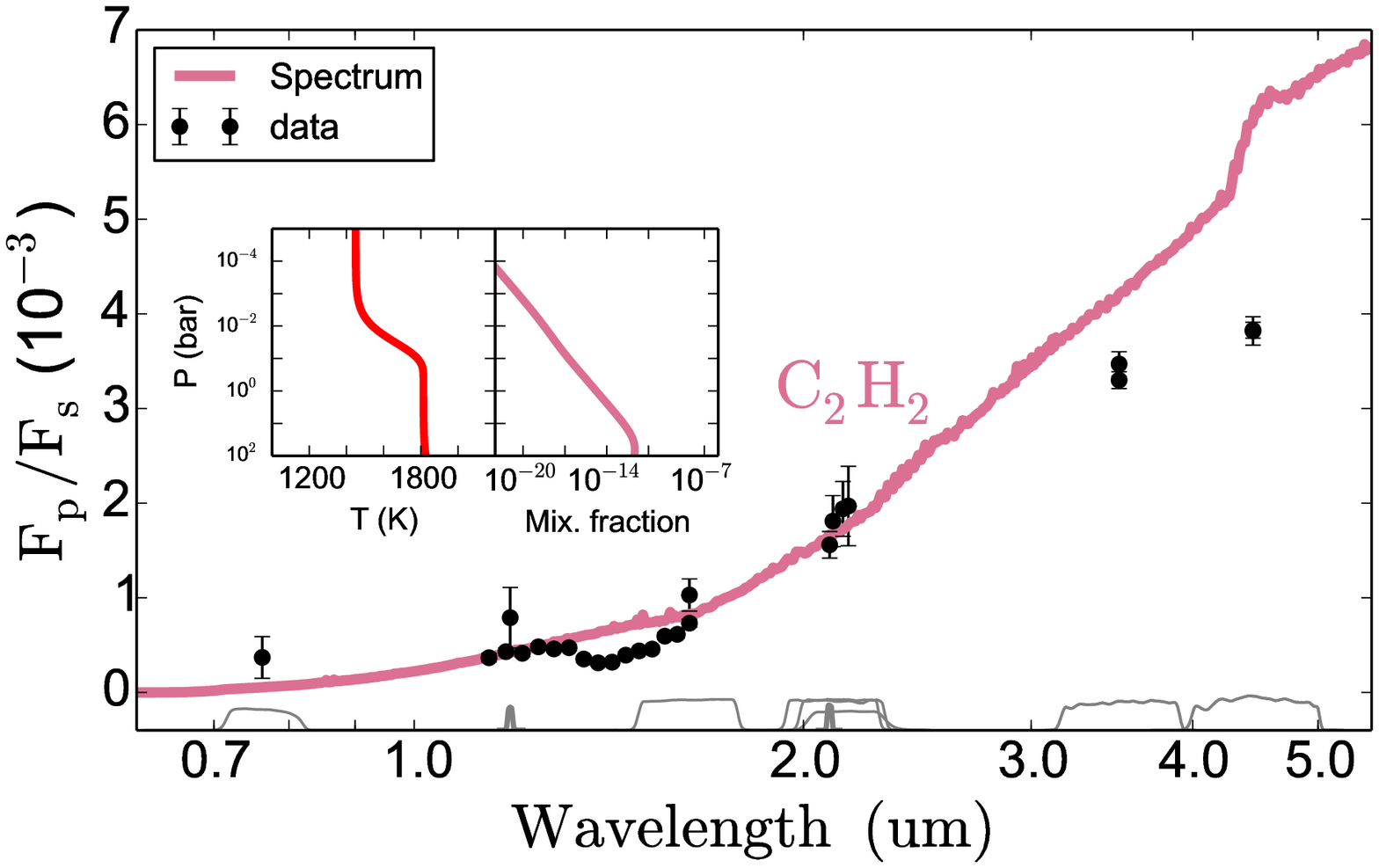}
\caption[Influence of each species to Case 6 best-fit model, seven fitted species]{Influence of each species to the best-fit model for Case 6, seven fitted species. In the inset figures are the best-fit \math{T(p)} profile and the species's mixing ratio.}
\label{fig:7mol-indSpecs}
\vspace{-5pt}
\end{figure*}

We ran both cases with an initial 1.5x10\sp{5} iterations per chain, and both converged for all parameters after 1.2x10\sp{5}.

\begin{table}[ht!]
\footnotesize{
\caption{\label{table:7mol-fit} 7 Species Goodness of Fit}
\atabon\strut\hfill\begin{tabular}{lcccc}
    \hline
    \hline
                          & \math{\chi\sp{2}\sb{\rm red}}  & BIC     & SDR      \\
    \hline
Case 5, Uniform           & 2.3588                           & 70.3210  & 0.000202230  \\
Case 6, Equilibrium       & 2.3385                           & 69.9968  & 0.000203473  \\
Case 7, 11 opacities      & 2.4337                           & 71.5197  & 0.000203565  \\
    \hline
\end{tabular}\hfill\strut\ataboff
}
\end{table}

\begin{figure}[hb!]
\centering
\includegraphics[width=.50\textwidth, clip=True]{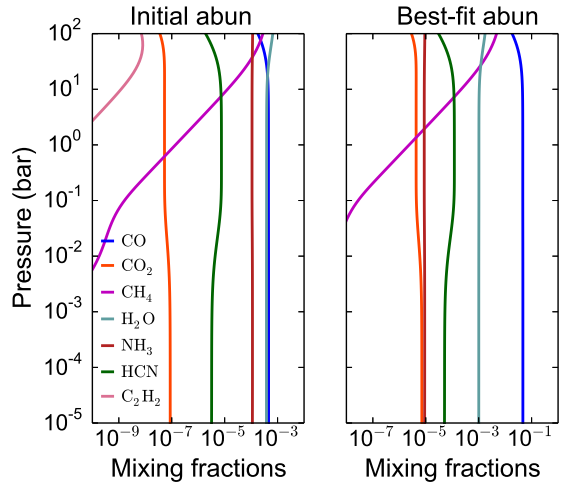}
\caption[Initial and best-fit abundances for Case 6, seven fitted species]{Initial and best-fit abundances for Case 6, seven fitted species.}
\label{fig:7mol-PT}
\end{figure}

Figure \ref{fig:7mol-hist} shows the histograms of the posterior distribution of the retrieved molecular species for Cases 5 and 6. Figure \ref{fig:7mol-indSpecs} shows the influence of the individual species on the best-fit model, Case 6. As in Section \ref{sec:four}, the most influence on the spectrum comes from the water spectral features; thus, water is constrained the most in our analysis (Figure \ref{fig:7mol-hist}). The additional retrieved species NH\sb{3}, HCN, and C\sb{2}H\sb{2} show no influence on the spectrum (Figure \ref{fig:7mol-indSpecs}), although they seem constrained to some extent in our analysis (Figure \ref{fig:7mol-hist}). Figure \ref{fig:7mol-PT} shows the initial and the best-fit abundances for Case 6.

According to the combination of the \math{\chi\sp{2}\sb{\rm red}} and BIC values, the equilibrium case provided a marginally better fit to the data. Using that model, we constructed Case 7 and included all eleven opacity sources into the retrieval, H\sb{2}O, CO\sb{2}, CO, CH\sb{4}, NH\sb{3}, HCN, C\sb{2}H\sb{2}, C\sb{2}H\sb{4}, H\sb{2}S, TiO, and VO. We give the best-fit species abundances in Table \ref{table:sevenSpecs} and the goodness of fit for all three cases in Table \ref{table:7mol-fit}. As we can see, according to BIC, the inclusion of additional opacity sources does not improve the fit.

\begin{landscape}
\vspace*{\fill}
\begin{table*}[ht!]
\footnotesize{
\caption{\label{table:sevenSpecs} 7 Species Mixing Ratios}
\atabon\strut\hfill\begin{tabular}{lccccc}
    \hline
    \hline
Case         &               &       H\sb{2}O                           &      CO\sb{2}                          &         CO                              &  CH\sb{4}                               \\  
    \hline
Case 5       & Best Fit       &       6.56 x 10\sp{-4}         &       1.16 x 10\sp{-9}         &       3.49 x 10\sp{-2}         &       5.40 x 10\sp{-9}         \\
uniform      & 68\% interval  & [2.25 x 10\sp{-4} -- 1.91 x 10\sp{-3}] & [2.01 x 10\sp{-11} -- 6.64 x 10\sp{-8}] & [1.27 x 10\sp{-2} -- 9.64 x 10\sp{-2}] & [2.79 x 10\sp{-11} -- 1.04 x 10\sp{-6}] \\
    \hline
Case 6       & Best Fit       &       1.02 x 10\sp{-3}                  &       4.31 x 10\sp{-6}                 &       4.64 x 10\sp{-2}                  &       2.58 x 10\sp{-6}    \\     
equilibrium  & 68\% interval  & [3.67 x 10\sp{-4} -- 2.82 x 10\sp{-3}] & [1.73 x 10\sp{-8} -- 1.08 x 10\sp{-3}] & [1.60 x 10\sp{-2} -- 1.34 x 10\sp{-1}] & [1.70 x 10\sp{-8} -- 3.92 x 10\sp{-4}]  \\
    \hline
Case 7       & Best Fit       &       2.52 x 10\sp{-3}                  &       4.81 x 10\sp{-10}                 &       1.17 x 10\sp{-1}                 &       1.72 x 10\sp{-11}     \\
11 opacities & 68\% interval  & [1.02 x 10\sp{-3} -- 6.22 x 10\sp{-3}] & [3.00 x 10\sp{-11} -- 7.69 x 10\sp{-9}] & [5.89 x 10\sp{-2} -- 2.31 x 10\sp{-1}] & [1.93 x 10\sp{-13} -- 1.53 x 10\sp{-9}] \\
    \hline
\end{tabular}\hfill\strut\ataboff
\centerline{
\atabon\strut\hfill\begin{tabular}{lcccc}
Case         &               &      NH\sb{3}                           &         HCN                             &  C\sb{2}H\sb{2}              \\
    \hline
Case 5       & Best Fit       &       1.67 x 10\sp{-6}         &       2.94 x 10\sp{-5}         &       9.57 x 10\sp{-15}       \\
uniform      & 68\% interval  & [2.28 x 10\sp{-8} -- 1.22 x 10\sp{-4}] & [1.32 x 10\sp{-6} -- 6.52 x 10\sp{-4}] & [9.33 x 10\sp{-18} -- 9.82 x 10\sp{-12}]  \\
    \hline
Case 6       & Best Fit      &       8.75 x 10\sp{-6}                  &       1.20 x 10\sp{-4}                  &       1.94 x 10\sp{-15}       \\
equilibirum  & 68\% interval & [1.57 x 10\sp{-7} -- 4.87 x 10\sp{-4}] & [8.53 x 10\sp{-6} -- 1.69 x 10\sp{-3}]  & [5.39 x 10\sp{-17} -- 6.98 x 10\sp{-14}]  \\
    \hline 
Case 7       & Best Fit      &       1.64 x 10\sp{-6}                  &       1.24 x 10\sp{-7}                  &       7.03 x 10\sp{-16}       \\ 
11 opacities & 68\% interval & [1.12 x 10\sp{-7} -- 2.38 x 10\sp{-5}] & [2.54 x 10\sp{-10} -- 6.04 x 10\sp{-5}] & [8.07 x 10\sp{-19} -- 6.12 x 10\sp{-13}]\\  
    \hline
\end{tabular}\hfill\strut\ataboff
}
}
\end{table*}
\vspace*{\fill}
\end{landscape}

\subsection{WASP-43\lowercase{b} Contribution Functions}
\label{sec:cfWASP43b}

Figure \ref{fig:cf} shows the \math{T(p)} profile and abundances, the weighting functions for all observations, and the normalized contribution functions for the HST, {\em Spitzer}, and ground-based observations, for the best-fit models from Sections \ref{sec:four} (upper panel) and \ref{sec:seven} (middle panel), as well as the model when all eleven opacity sources are included, Case 7, Section \ref{sec:seven} (bottom panel). The right panels show the maximum optical depth at each wavelength (upper panels), and the pressures where the maximum optical depth is reached (bottom panels).

From the right panels, we see that in the cases when we have four and seven opacity sources present in the retrieval, the wavelength range below 0.7 {\micron} is almost transparent to the outgoing radiation for the best-fit model. However, when all eleven species are included, we see a significant increase in opacity, but only in the region below 0.7 {\microns}. This contribution, based on Figures \ref{fig:opacs1} and \ref{fig:opacs2}, could come only from the TiO and VO opacities.

\begin{figure*}[t!]
\vspace{10pt}
\centering
\includegraphics[width=0.68\textwidth, height=4.8cm, trim=0 5 0 35]{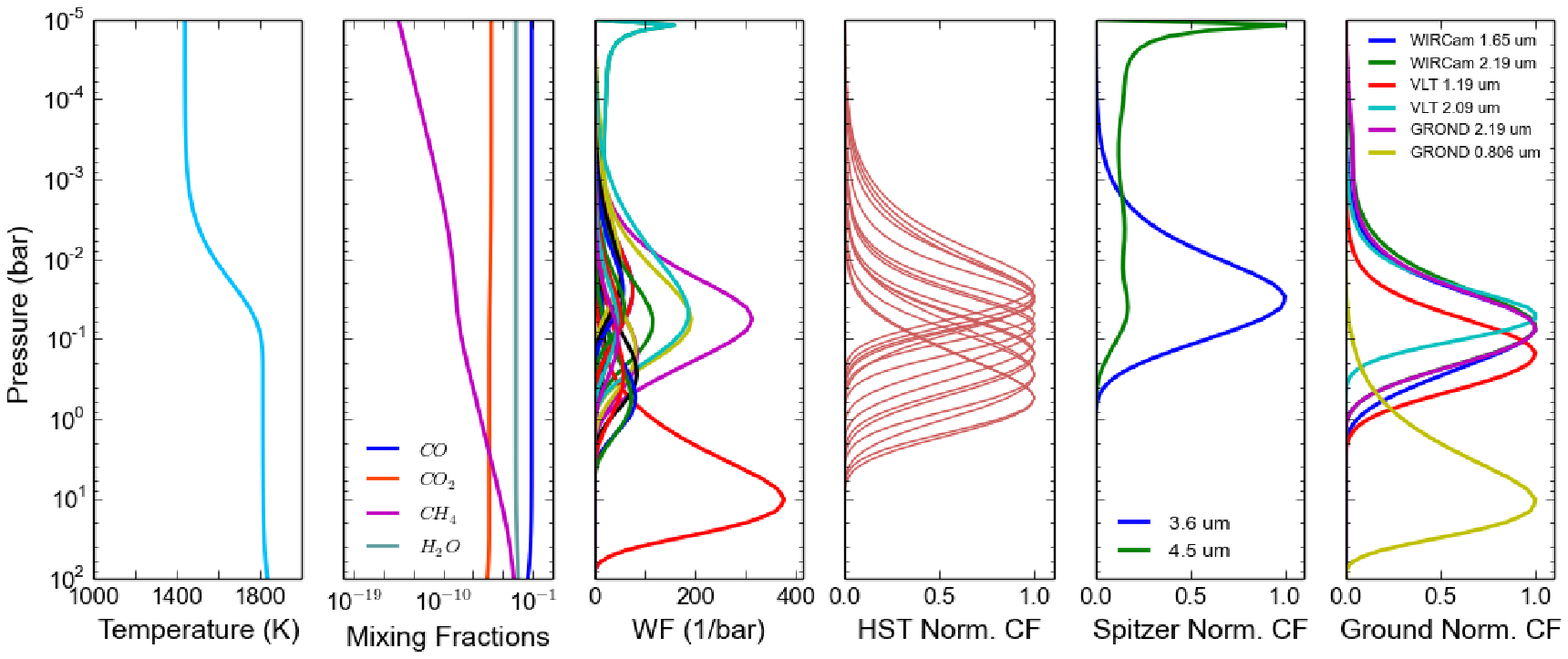}\hspace{-10pt}
\includegraphics[width=0.31\textwidth, clip=True]{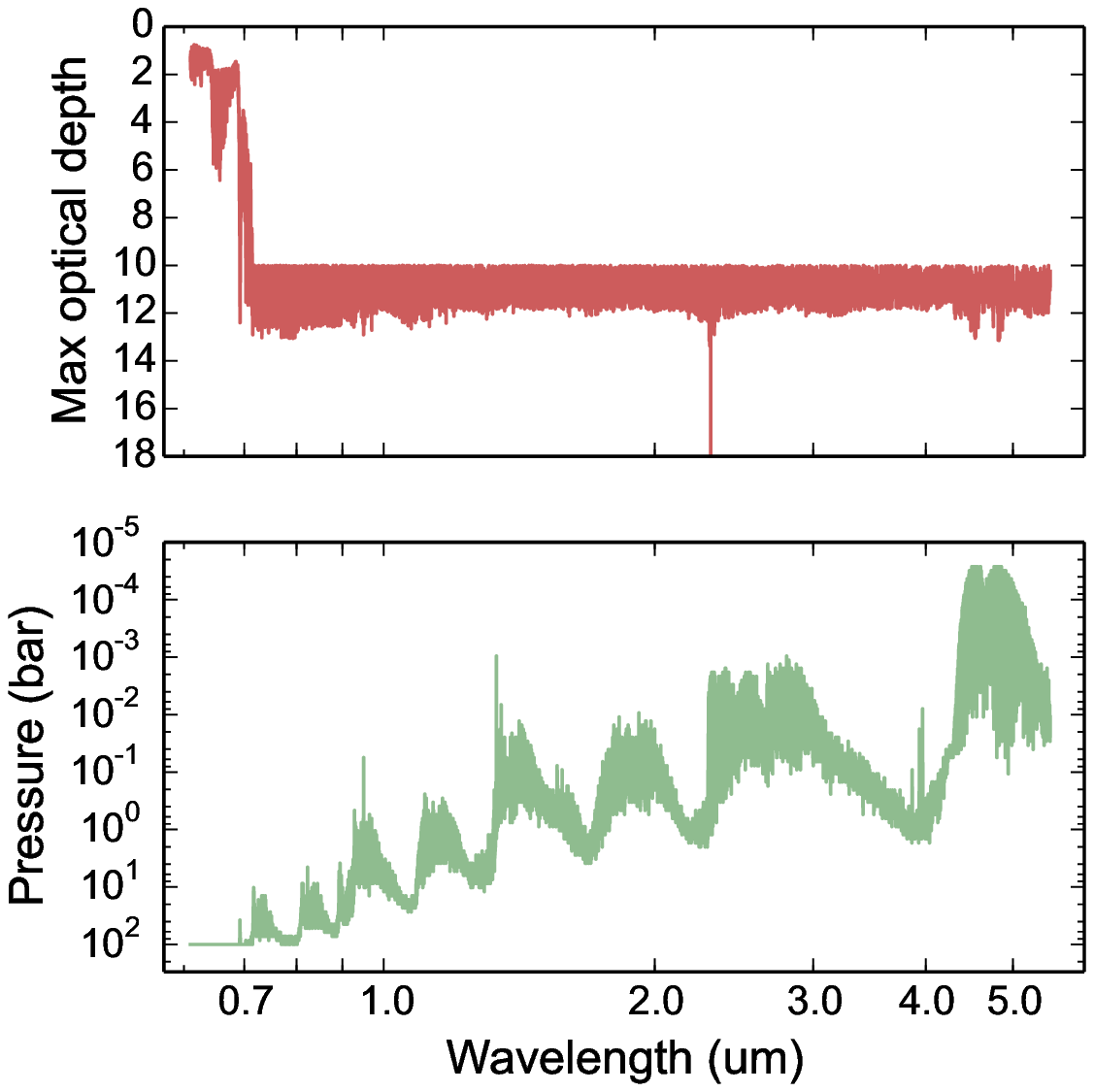}
\includegraphics[width=0.68\textwidth, height=4.8cm, trim=0 5 0 35]{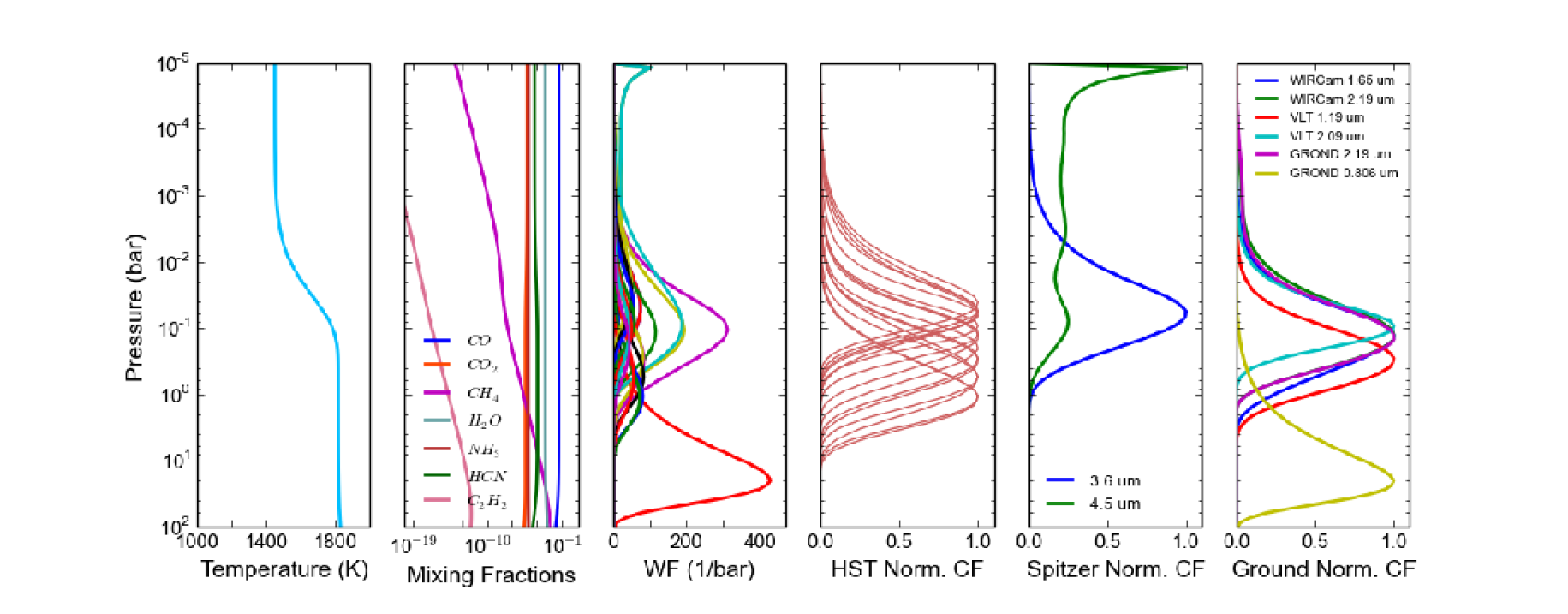}\hspace{-10pt}
\includegraphics[width=0.31\textwidth, clip=True]{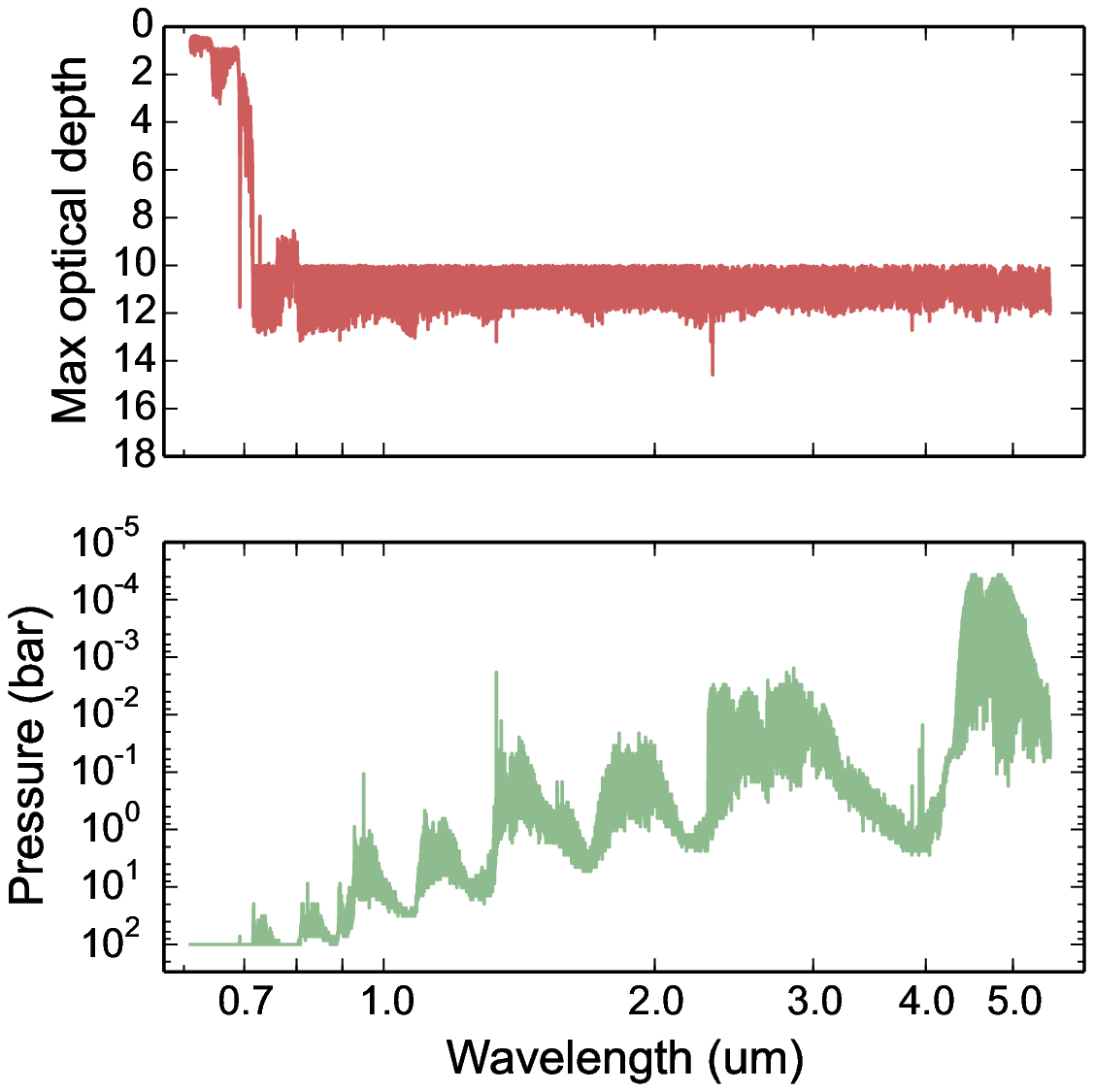}
\includegraphics[width=0.68\textwidth, height=4.8cm, trim=0 5 0 35]{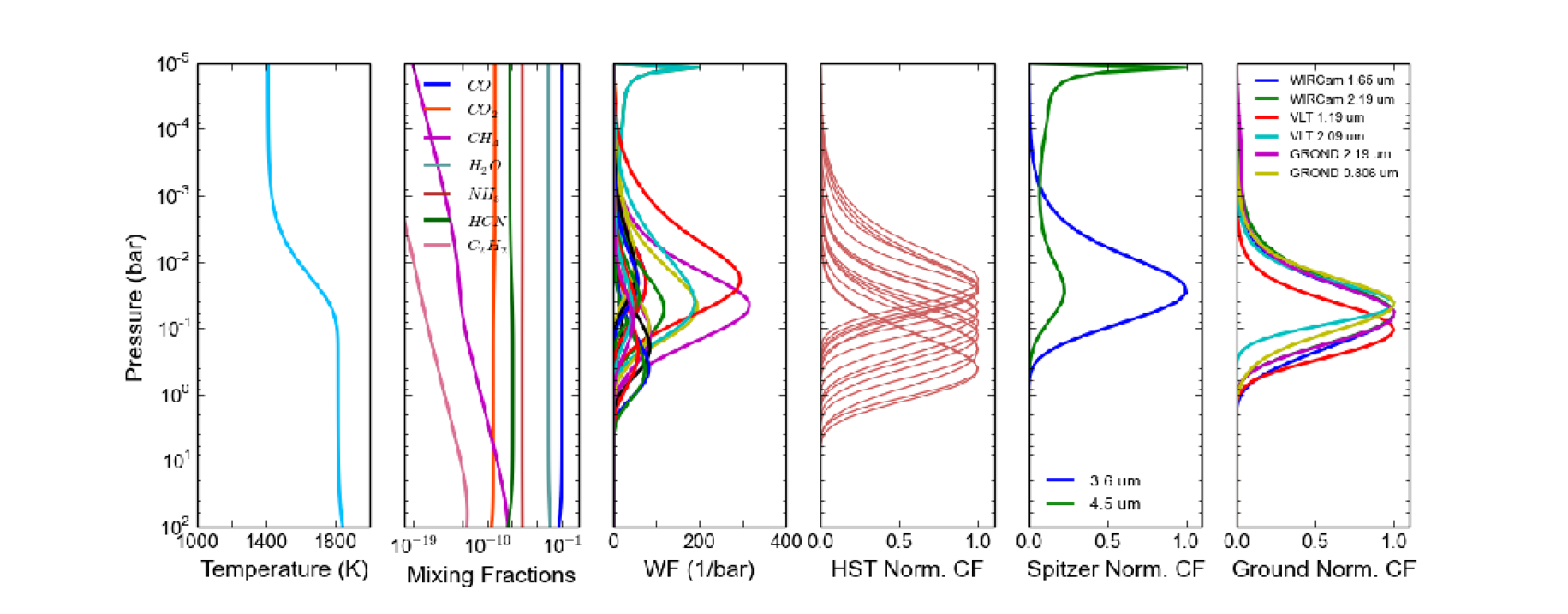}\hspace{-10pt}
\includegraphics[width=0.31\textwidth, clip=True]{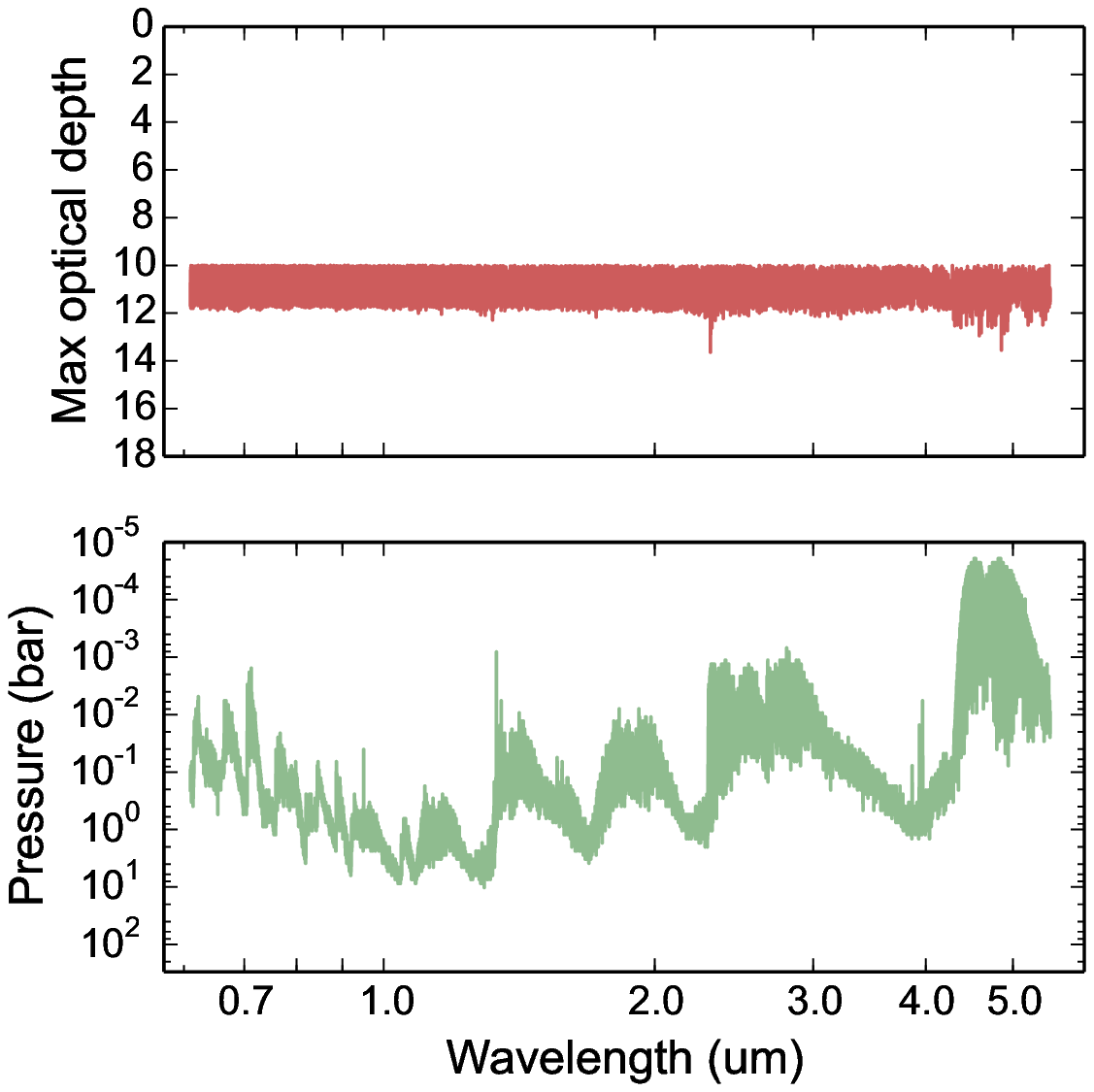}
\caption[WASP-43b contribution functions]{Contribution functions for the best-fit models from Sections \ref{sec:four} (upper panel), Section \ref{sec:seven} (middle panel), and Case 3 from Section \ref{sec:seven} (lower panel). Right panels show the maximum optical depth at each wavelength (upper panels), and the pressures where the maximum optical depth is reached for each model (bottom panels).}
\label{fig:cf}
\end{figure*}

From the left panels, we see that the atmospheric models with four and seven absorbers (when H\sb{2}S, TiO, and VO are not included) probe the thermal structure around 0.1 bars. This is not the case for the {\em GROND} observation at 0.809 {\microns} that probes deeper layers. In the bandpass of this observation, the atmosphere is mostly transparent to the outgoing radiation, and the peak of the contribution function sits at the deep pressure layers below 1 bar. 

The observations done by {\em Spitzer}'s channel 2 probe significantly lower pressures, explained by the presence of several opacity sources in this bandpass. The maximum optical depth for this bandpass is reached at very low pressures. Figures \ref{fig:4mol-indSpecs} and \ref{fig:7mol-indSpecs} show that most of the spectral features from H\sb{2}O, CO, and CO\sb{2} are concentrated in this region. However, this unusually high opacity on such low pressures must come from the high CO abundance (Tables \ref{table:fourSpecs} and \ref{table:sevenSpecs}).

\subsection{WASP-43\lowercase{b} C/O Ratio}
\label{sec:CO}

The C/O ratio in exoplanetary atmospheres is fundamental for understanding the underlying chemistry in our models \citep{Lodders02,MosesEtal2013-COratio, Madhusudhan2012-COratio}. We investigated the influence of a C/O ratio on the best-fit model by generating models with a C/O ratio of 0.5, 0.7, and 1.0 (Figure \ref{fig:CO}). To change the C/O ratio in our models, we swapped the elemental abundances for C and O, and decreased the C abundance until we reached the desired ratio (the O abundance is kept constant). For this analysis, we used all eleven opacity sources in our models and the species abundances calculated using {\tt TEA}, as our initial model atmosphere. The initial \math{T(p)} profile is the same as in Section \ref{sec:four}.

\begin{table}[ht!]
\footnotesize{
\caption{\label{tab:CO} C/O Ratio Goodness of Fit}
\atabon\strut\hfill\begin{tabular}{lcccc}
    \hline
    \hline
             & \math{\chi\sp{2}\sb{\rm red}}  & BIC     & SDR      \\
    \hline
C/O=0.5      & 2.4377                           & 71.5197  & 0.000203565  \\
C/O=0.7      & 2.4236                           & 71.3587  & 0.000202348  \\
C/O=1.0      & 2.4018                           & 71.0097  & 0.000202891  \\
    \hline
\end{tabular}\hfill\strut\ataboff
}
\end{table}

Table \ref{tab:CO} lists \math{\chi\sp{2}}, BIC, and SDR for all 3 models. We also tested the models with a C/O ratio \math{>} 1.0, which resulted in a bad fit to the data. We see that the \math{\chi\sp{2}} and BIC values differ only marginally between models, although some small decrease in the values is noticeable with the C/O ratio increase. However, the SDR value is the best for the model with a C/O ratio of 0.7. Although these differences are only marginal, our result is consistent with the previous work by \citet{ZhouEtal2014-WASP-43b, WangEtal2013-WASP43b}, and \citet{benneke2015strict}.

\begin{figure}[h!]
\centering
\includegraphics[width=.75\textwidth, trim=30 10 0 20]{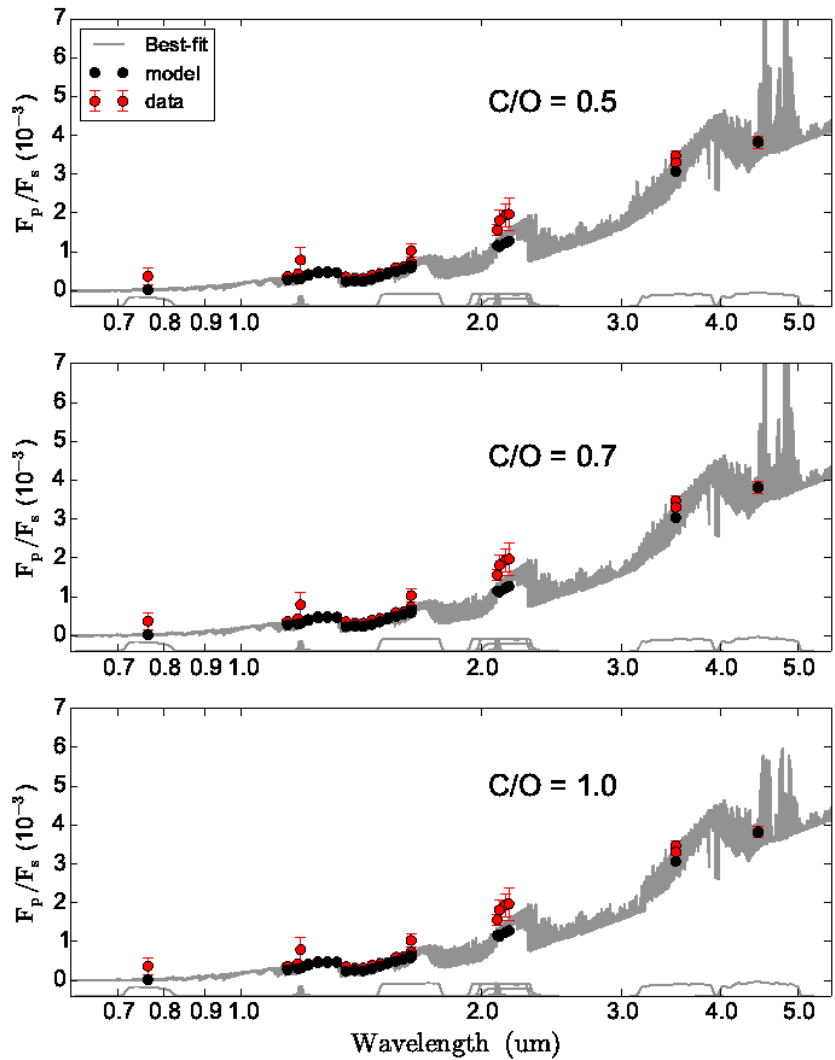}
\caption[Best-fit models for C/O ratio 0.5, 0.7, and 1.0]{Best-fit models for C/O ratio 0.5, 0.7, and 1.0.}
\label{fig:CO}
\end{figure}

\subsection{Results - Discussion}
\label{sec:diss}

Additional opacity sources besides the four major molecules H\sb{2}O, CO\sb{2}, CO, and CH\sb{4}, have proven to be fundamental in retrievals \citep{MadhusudhanEtal2011natWASP12batm, CrossfieldEatl2012ApJWASP12b-reavaluation}. H\sb{2}O is one of the most abundant species in oxygen-rich atmospheres, showing strong absorption features in the near infrared (see Figures \ref{fig:opacs1} and \ref{fig:opacs2}). However, hydrocarbons, CH\sb{4}, HCN, C\sb{2}H\sb{2}, and C\sb{2}H\sb{4} become important in carbon-rich atmospheres as they are more abundant than H\sb{2}O \citep[e.g.,][]{MadhusudhanEtal2011-Cplanets, MosesEtal2013-COratio, BlecicEtal2015apjsTEA}.

The inclusion of TiO and VO is justified since both species are strong absorbers in the visible that could lead to thermal inversion around \sim1 mbar level \citep{Hubeny2003, FortneyEtal2008-TiO, Spiegel2009apjInversion}. However, questions have been raised whether high stellar activity may cause them to be photochemically destroyed \citep{KnutsonHowardIsaacson2010ApJ-CorrStarPlanet} or absent in the carbon-rich atmospheres \citep[e.g.,][]{Madhusudhan2012-COratio}. There are also other species \citep[like sulfur, ][]{Zahnle09-SulfurPhotoch} that absorb highly in visible and could cause thermal inversion (the H\sb{2}S species included in our models is the only sulfur species available in the HITRAN database).

We see no evidence of thermal inversion in any of our model cases. WASP-43b shows a distinctive decrease in temperature with altitude in all of our models, consistent with previous work. In addition, our best \math{T(p)} profiles for all of our cases are consistent with the \math{T(p)} profiles from \citet{StevensonEtal2014-WASP43b, LineEtal2013-Retrieval-II}, and \citet{BlecicEtal2014-WASP43b}.

The inclusion of additional opacity sources only marginally influences the shape of the best-fit spectrum (Figure \ref{fig:4mol-comp34}). The difference is only noticeable in the region above 4.2 {\microns}. In addition, Table \ref{table:4mol-fit} does not show a major difference in the \math{\chi\sp{2}\sb{\rm red}} and BIC values. This is similar for the cases when we fit seven molecular species (Section \ref{sec:seven}). These results suggest that the inclusion of additional chemical species does not have a major impact on the best-fit model. The results are almost unchanged. In our analysis, the spectrum is dominated by the H\sb{2}O, and to some extent CO, CO\sb{2}, NH\sb{3}, and HCN absorption features. However, only H\sb{2}O, CO, and CO\sb{2} show spectral features in the bandpasses of our observations, with water being the most dominant (CO and CO\sb{2} show absorption features only above 4.2 {\microns}). According to the posterior histograms (Figures \ref{fig:4mol-hist} and \ref{fig:7mol-hist}), H\sb{2}O, CO, and surprisingly HCN show some level of constraint, with again water being the most constrained. CO\sb{2} and CH\sb{4} are fully unconstrained in our analysis.

Based on these conclusions, we calculated the water abundance on the dayside of WASP-43b using our best-fit model (Section \ref{sec:four}, Case 2). Assuming the same solar water abundance of 6.1x10\sp{-4} (for a solar composition gas in thermochemical equilibrium at planetary temperatures) as \citet{KreidbergEtal2014-WASP43b}, we constrained the water abundance to 1-10\math{\times}solar at 1 bar pressure level, similar to the conclusions made by \citet{BlecicEtal2014-WASP43b}. 

CO shows unusually high abundances in all of our models (Tables \ref{table:fourSpecs} and \ref{table:sevenSpecs}). Although CO is the most abundant carbon-bearing species in the oxygen-rich and carbon-rich hot-Jupiter atmospheres (especially in the high-metallicity atmospheres), this level of enhancement is surprising. \citet{kataria2014atmospheric} find, using a self-consistent 3D circulation approach, that 5\math{\times}solar metallicity model is the best match to the data. \citet{ZhouEtal2014-WASP-43b} find the best fit with the 2\math{\times}solar metallicity model. However, neither metallicity level supports the amount of CO that we are seeing in our models.

We also compared a vertically-uniform abundances profile versus a thermochemical equilibrium abundances profile goodness of fit. Although questions have been raised whether the data provide enough constraints on the vertical abundances profile \citep{LeeEtal2012-CF, LineEtal2012-Retrieval-Intro}, and vertical mixing tends to quench minor chemical species \citep{MosesEtal2011-diseq}, the major molecular species already have close to vertically-uniform abundances profiles in the thermochemical equilibrium (CO, CO\sb{2}, H\sb{2}O, NH3, HCN, see Figures \ref{fig:4mol-PT} and \ref{fig:7mol-PT}). As the thermochemical equilibrium calculations provide more realistic initial atmospheric models, we assessed them against the uniform cases with the goodness-of-fit statistics. In all cases, the thermochemical equilibrium abundances models provide a marginally better fit to the data according to the combination of the \math{\chi\sp{2}\sb{\rm red}} and BIC values (see Sections \ref{sec:four} and \ref{sec:seven}, Tables \ref{table:4mol-fit} and \ref{table:7mol-fit}).

\newpage
\section{CONCLUSIONS}
\label{sec:conc}

This paper is a part of three contributed papers that present a novel retrieval framework, the Bayesian Atmospheric Radiative Transfer (BART) code (see also \citealp{HarringtonEtal2015-BART, CubillosEtal2015-BART}). BART is an open-source open-development Bayesian, thermochemical, radiative-transfer code under a reproducible-research license available at \href{https://github.com/exosports/BART}{https://github.com/exosports/BART}. It consists of three self-sufficient modules: {\tt TEA}, the Thermochemical Equilibrium Abundances module that calculates the mixing fractions of gaseous species; {\tt Transit}, the radiative-transfer code; and {\tt MC\sp{3}}, the Multi-core Markov-chain Monte Carlo package.

In this paper, we presented the implementation and the underlying theory of the initialization routines, TEA module, atmospheric profile generator, eclipse module, best-fit routines, and the contribution functions module. Other modules and packages are described in \citet{CubillosEtal2015-BART}. We also presented an atmospheric analysis of WASP-43b using BART.

We performed a comprehensive analysis of all available space and ground secondary eclipse data to constrain the dayside atmosphere of WASP-43b. Data confirmed a decreasing temperature with pressure, i.e., an absence of thermal inversion, consistent with previous analyses. The inclusion of additional opacity sources does not improve the fit. According to BIC, the atmospheric model with only four opacity sources is the best match to the data. However, this result is only marginal. We find no significant difference in the best-fit models produced using the uniform-abundances profiles versus the models with thermochemical equilibrium profiles as the initial atmospheric model. We also put a marginal constraint on the WASP-43b C/O ratio of 0.7.

We see an unusually high abundance of CO in our models with the peak of the Spitzer channel 2 contribution function is located high in the planetary atmosphere. This high amount of CO suggests that the data drove {\tt MC\sp{3}} into a possibly physically unplausable phase space (our retrieval technique is fully tested on a synthetic spectra, see collaborative paper by \citealp{CubillosEtal2015-BART}). Compared to the retrieval results from \citet{LineEtal2013-Retrieval-II}, the difference could come from the line-list databases used in the analysis, number of data points included, and/or missing important opacity sources. 

What we are seeing in our analysis could support conclusions from recent studies by \citet{hansen2014features, LaughlinLissauer2015, Burrows2014-review}, and \citet{Swain2013-WASP12b} that complex atmospheric models might not be supported by the data. \citet{hansen2014features} state that the features in the broadband emission spectra are due to astrophysical and instrumental noise rather than the molecular bands, and any claims about chemistry or C/O are premature. These statements are supported by \citet{LaughlinLissauer2015}, their Section 5.3, Figure 8, and their analysis of the {\em Spitzer} channel 2 observations of HD 80606b (\href{http://oklo.org/2013/08/21/central-limit-theorem/}{http://oklo.org/2013/08/21/central-limit-theorem/}). Similar conclusions came from \citet{Burrows2014-review} that states that theorists and observers have a tendency to overinterpret the exoplanetary measurements.

In light of these studies and our conclusions, we are unsure what caused the high CO abundance in our models. Our other results are consistent with the results of previous analyses. Efforts have been made to provide accurate databases of gaseous species at high temperatures that occur in hot-Jupiter atmospheres \citep[e.g., for NH\sb{3}, CH\sb{4}, H\sb{2}S, PH\sb{3}, ][]{HargreavesEtal2011-hotAmonia, HargreavesEtal2013-hotMethane, Yurchenko2014PNAS-LineListCH4, AzzamEtal2013-H2S, Sousa-SilvaEtal2015hotPH3}, see also \citet{TennysonYurchenko2014-ExoMol} for the list of the ExoMol hot molecules. We are hoping that with the inclusion of various species opacities in the retrieval (supporting more realistic atmospheric solution) and the spectroscopic phase curve observations of WASP-43b using the {\em James Webb Space Telescope}, we will be able to address any inconsistencies between the results of various groups.

\newpage
\section{ACKNOWLEDGMENTS}

This project was completed with the support of the
NASA Earth and Space Science Fellowship Program, grant NNX12AL83H,
held by Jasmina Blecic, PI Joseph Harrington, and through the Science
Mission Directorate's Planetary Atmospheres Program, grant
NNX12AI69G. Part of this work is based on observations made with the {\em Spitzer Space Telescope}, which is operated by the Jet Propulsion Laboratory, California Institute of Technology under a contract with NASA. We would like to thank Kevin B. Stevenson for providing the 3.6 and 4.5 {\micron} {\em Spitzer}\/ data, Guo Chen and George Zhou for the transmission response functions, and Jonathan Fortney for a useful discussion. We also thank contributors to SciPy, NumPy, Matplotlib, and the Python Programming Language; the open-source development website GitHub.com; and other contributors to the free and open-source community. \\

\newpage
\bibliographystyle{apj}
\bibliography{chap-BART}

\fi



\appendix

\end {document}